\documentclass[12pt,notitlepage,fleqn]{article}
\usepackage{amssymb}
\usepackage{graphicx}
\usepackage{amsmath}
\usepackage{indentfirst}

\input{tcilatex}

\begin{document}

\begin{center}
{\Huge Phenomenological Formulae for }

{\Huge \ Quarks, Baryons and Mesons }

\bigskip

{\normalsize Jiao Lin Xu }

{\small The Center for Simulational Physics, The Department of Physics and
Astronomy}

{\small University of Georgia, Athens, GA 30602, U. S. A.}

E- mail: {\small \ jxu@hal.physast.uga.edu}

\ \ \ \ \ \ \ \ \ \ \ \ \ \ \ \ \ \ \ \ \ \ \ \ \ \ \ \ \ \ \ \ \ \ \ \ \ \
\ \ \ \ \ \ \ \ \ \ 

\textbf{Abstract}
\end{center}

{\small We assume that there is only} {\small one unflavored elementary
quark family }$\epsilon $ {\small with three colors and two isospin states (}%
$\epsilon _{u}${\small \ with I}$_{z}$=$\frac{1}{2}${\small \ and Q=}$\frac{2%
}{3},${\small \ }$\epsilon _{d}${\small \ with I}$_{z}$=$\frac{-1}{2}$%
{\small \ and Q=}$\frac{-1}{3}${\small ) in the vacuum. Using
phenomenological formulae, we can deduce the rest masses and intrinsic
quantum numbers (I, S, C, b and Q) of the excited quarks, from the
symmetries of the regular rhombic dodecahedron. The five deduced ground
quarks\ correspond to the five current quarks [u(313)}$\leftrightarrow $%
{\small u, d(313)}$\leftrightarrow ${\small d, d}$_{S}${\small (493)}$%
\leftrightarrow $s{\small , u}$_{C}${\small (1753)}$\leftrightarrow $c%
{\small \ and d}$_{b}${\small (4913)}$\leftrightarrow ${\small b]. We can
then deduce the baryon spectrum and the meson spectrum, from the excited
quarks, using sum laws and the phenomenological binding energy formulae of
baryons (qqq) and mesons (q}$\overline{\text{{\small q}}}$){\small . The
deduced intrinsic quantum numbers (I, S, C, b and Q) of the baryons and the
mesons are the same as those of the experimental results. The deduced rest
masses of the baryons and the mesons agree with about 98\% of the
experimental results. Experiments have already discovered almost all of the
deduced quarks in Table 11. This paper infers that there are large constant
binding energies (\TEXTsymbol{>}\TEXTsymbol{>}M}$_{\Pr oton}$){\small \ in
baryons and mesons. These large binding energies provide a possible
foundation for the confinement of the quarks.} {\small This paper predicts
many new hadrons (}$\Lambda _{c}^{+}${\small (6599), }$\Lambda _{\text{b}%
}^{0}$({\small 9959}), {\small D(6231), B(9503) and }$\Upsilon ${\small %
(17868)) and a ``fine structure''\ phenomenon of baryons and mesons with
large widths. The experimental investigation of the ``fine structure''\ (for
example f}$_{0}${\small (600) with }$\Gamma ${\small = 600-1000) provides a
crucial test.\ PACS Numbers: 12.39.-x, 14.65.-q, 14.40.-n, 14.20.-c}

\section{Introduction}

The Quark Model \cite{Quark Model} has been in use for 40 years. Already it
has greatly advanced and pushed forward the particle physics. During the
past 40 years, experimental physicists have discovered many new baryons and
new mesons with the Quark Model guidance. Today we know of more than three
times the number of new particles \cite{Particle(04)} than we knew of 40
years ago \cite{Particle(1964)}. We think that many new particles will be
found, as the development of the experimental equipment and techniques
advances, just as astronomers have discovered many new stars after
developing more powerful telescopes.\ In order to account for new
experimental discoveries and to advance new experiments, original
theoretical models usually need modification. In this paper, we will use
phenomenological formulae to modify the current Quark Model. The following
problems of the current Quark Model will be addressed in this paper.

\subsection{What Problems\ Need to Be Solved?\ \ \ \ \ \ \ \ \ \ \ \ \ \ \ \
\ \ \ \ \ \ \ \ \ \ \ \ \ \ \ \ \ \ \ \ \ \ \ \ \ \ \ \ \ \ \ \ \ \ \ \ \ \
\ \ \ \ \ \ \ \ \ \ \ \ \ \ \ \ \ \ \ \ \ \ \ \ \ \ \ \ \ \ \ \ \ \ \ \ \ \
\ \ \ \ \ \ \ \ \ \ \ \ \ \ \ \ \ \ \ \ \ \ \ \ \ \ \ \ \ \ \ \ \ \ \ \ \ \
\ \ \ \ \ \ \ \ \ \ \ \ \ \ \ \ \ \ \ \ \ \ \ \ \ \ \ \ \ \ \ \ \ \ \ \ \ \
\ \ \ \ \ \ \ \ \ \ \ \ \ \ \ \ \ \ \ \ \ \ \ \ \ \ \ \ \ \ \ \ \ \ \ \ \ \
\ \ \ \ \ \ \ \ \ \ \ \ \ \ \ \ \ \ \ \ \ \ \ \ \ \ \ \ \ \ \ \ \ \ \ \ \ \
\ \ \ \ \ \ \ \ \ \ \ \ \ \ \ \ \ \ \ \ \ \ \ \ \ \ \ \ \ \ \ \ \ \ \ \ \ \
\ \ \ \ \ \ \ \ \ \qquad\ \ \ \ \ \ \ \ \ \ \ \ \ \ \ \ \ \ \ \ \ \ \ \ \ \
\ \ \ \ \ \ \ \ \ \ \ \ \ \ \ \ \ \ \ \ \ \ \ \ \ \ \ \ \ \ \ \ \ \ \ \ \ \
\ \ \qquad}

A1. A formula that can deduce the rest masses of the quarks is necessary; a
formula that can deduce the rest masses of all baryons from the rest masses
of the quarks and a formula that can deduce the rest masses of all mesons
from the rest masses of the quarks are also necessary. Unfortunately, there
is no united mass formula that can deduce the rest masses of the quarks, the
baryons or the mesons in the current Quark Model. There is not even a
phenomenological formula in the elated literature that can do so.

A2. In order to explain the masses of the light unflavored mesons ($\eta $, $%
\omega $, $\phi $, h and f) with I = 0, the Quark Model has to depart from
the principle that a meson is made of a quark and an antiquark and allow a
meson to be a mixture of three quark-antiquark pairs (u$\overline{u}$, d$%
\overline{d}$ and s$\overline{s})$. (Note: the meson is not a superposition
of three quark-antiquark pairs (u$\overline{u}$, d$\overline{d}$ and s$%
\overline{s})$ because the three pairs are independent elementary particle
pairs in the current Quark Model). For example \cite{3Pair-Mixture}:

\begin{equation}
\begin{tabular}{|l|}
\hline
$\eta $(548) = $\eta _{8}\cos \theta _{p}$-$\eta _{1}\sin \theta _{p}$, \\ 
\hline
$\eta ^{\prime }$(958) = $\eta _{8}\sin \theta _{p}$+$\eta _{1}\cos \theta
_{p}$, \\ \hline
$\eta _{1}$ = (u$\overline{u}$+d$\overline{d}$+s$\overline{s}$)/$\sqrt{3}$,
\\ \hline
$\eta _{8}$ = (u$\overline{u}$+d$\overline{d}$-2s$\overline{s}$)/$\sqrt{6}$.
\\ \hline
\end{tabular}
\ \ \ \ \ \ \ \   \label{Mixture Moson}
\end{equation}
This \textquotedblleft mixture\textquotedblright\ violates the principle
that a meson is made of a quark and an antiquark. It is difficult to
understand what physical force could unite the three different independent
quark pairs to form one meson. This also causes the model to need a
parameter $\theta _{p}$. There are more than 20 such mixture mesons {\small %
\ }[$\omega ${\small (782), }$\phi ${\small (1020)], }[{\small h}$_{1}$%
{\small (1170), h}$_{1}${\small (1380)], ... , }[$f_{2}${\small (1810),f}$%
_{2}${\small (2010)]}\ \cite{3Pair-Mixture}. Therefore, according to the
principle that a meson is made of a quark and an antiquark, more quarks are
needed to compose these \textquotedblleft mixture\textquotedblright\ mesons.
This paper will show that we actually have the quarks to compose these
mesons without the three independent-quark-pair mixture.

A3. According to the current Quark Model, there are 12 experimental heavy
mesons that are composed of b$\overline{\text{b}}$ \cite{Heavy-Meson}: \ 
\begin{align*}
& \text{ \ \ \ \ \ \ \ \ \ \ \ \ \ \ \ \ \ \ \ \ \ \ \ \ \ \ \ \ \ Table 1.\
\ The Heavy Mesons of b}\overline{\text{b}}\text{\ \ } \\
& 
\begin{tabular}{|l|l|l|l|l|l|l|l|}
\hline
q$_{i}\overline{\text{q}_{j}}$ & Meson(M) & $\Gamma $ & I$^{G}$( J$^{PC}$) & 
* & q$_{i}\overline{\text{q}_{j}}$ & Meson(M) & I$^{G}$( J$^{PC}$) \\ \hline
b$\overline{\text{b}}$ & $\Upsilon $(1S)(9460) & 53kev & 0$^{\text{-}}$(1$^{%
\text{- -}}$) & * & b$\overline{\text{b}}$ & $\chi _{b0}$(1P) (9860) & 0$^{%
\text{+}}$(0$^{\text{+ +}}$) \\ \hline
b$\overline{\text{b}}$ & $\Upsilon $(2S)(10023) ) & 44kev & 0$^{\text{-}}$(1$%
^{\text{- -}}$) & * & b$\overline{\text{b}}$ & $\chi _{b1}$(1P) (9893) & 0$^{%
\text{+}}$(1$^{\text{+ +}}$) \\ \hline
b$\overline{\text{b}}$ & $\Upsilon $(3S)(10355) & 26kev & 0$^{\text{-}}$(1$^{%
\text{- -}}$) & * & b$\overline{\text{b}}$ & $\chi _{b2}$(1P)(9913) & 0$^{%
\text{+}}$(2$^{\text{+ +}}$) \\ \hline
b$\overline{\text{b}}$ & $\Upsilon $(4S)(10580) & 14Mev & 0$^{\text{-}}$(1$^{%
\text{- -}}$) & * & b$\overline{\text{b}}$ & $\chi _{b0}$(2P)(10232) & 0$^{%
\text{+}}$(0$^{\text{+ +}}$) \\ \hline
b$\overline{\text{b}}$ & $\Upsilon $(10860) & 110Mev & 0$^{\text{-}}$(1$^{%
\text{- -}}$) & * & b$\overline{\text{b}}$ & $\chi _{b1}$(2P)(10255) & 0$^{%
\text{+}}$(1$^{\text{+ +}}$) \\ \hline
b$\overline{\text{b}}$ & $\Upsilon $(11020) & 79Mev & 0$^{\text{-}}$(1$^{%
\text{- -}}$) & * & b$\overline{\text{b}}$ & $\chi _{b2}$(2P)(10269) & 0$^{%
\text{+}}$(2$^{\text{+ +}}$) \\ \hline
\end{tabular}%
\end{align*}
The mesons $\Upsilon $(1S)(9460), $\Upsilon $(2S)(10023), $\Upsilon $%
(3S)(10355), $\Upsilon $(4S)(10580), $\Upsilon $(10860) and $\Upsilon $%
(11020), all have the same I$^{G}$( J$^{PC}$) = 0$^{\text{-}}$(1$^{\text{- -}%
}$) showing that these mesons are not all b$\overline{\text{b}}$ with
different I$^{G}$( J$^{PC}$). They need more quarks to be explained. This
paper will show that there are other quarks that can explain these mesons.

A4. The intrinsic quantum numbers (I, S, C, b and Q) of the quarks need to
be deduced \cite{Handin}. This paper will do so using phenomenological
formulae [(\ref{S-Number}), (\ref{IsoSpin}), (\ref{S+DS}) and (\ref{DaltaS}%
)] from the symmetries of the regular rhombic dodecahedron.

A5. The s-quark, the c-quark and the b-quark can only compose unstable
baryons and mesons, giving reason to doubt that these quark are independent
elementary particles. Experimental results show that higher mass quarks can
decay into lower mass ones, such as b$\rightarrow $c , c$\rightarrow $s and s%
$\rightarrow $u (or s$\rightarrow $d). This might indicate that the five
quarks (u, d, s, c and b) are not all independent elementary particles. This
paper will show that there may be only two elementary quarks [$\epsilon _{u}$
and $\epsilon _{d}$] in the vacuum and that all quarks inside hadrons are
excited states of these two elementary quarks.

A6. Why do SU(3) and SU(4) work well, and how many quarks will there be
inside hadrons? This paper attempts to explain.

A7. A reduction of the number of parameters in the Quark Model is necessary 
\cite{Standard}.

A8. There are very large full widths in some baryons and mesons, such as $%
\Gamma $ = 600-1000 of f$_{0}$(600), $\Gamma $ = 360 of h$_{1}$(1170), $%
\Gamma $ = 200-600 of $\pi $(1300) and $\Gamma $ = 450 of N(2190). The very
large full widths need an explanation.

With all of the above problems, a modification of the Quark Model is very
difficult to imagine. On one hand, it needs to increase many high mass
quarks to explain the high mass baryons and mesons; on the other hand, it
needs to reduce the quark numbers to decrease the number of parameters. At
the same time, it needs to deduce the intrinsic quantum numbers and the rest
masses of the quarks, baryons and mesons. At first, it looks almost
impossible.

These high mass baryons and new mesons, however, not only require a further
development of the Quark Model, but also create the conditions for this
development. On the basis of the experimental baryon and meson spectra, we
have found some phenomenological formulae that can reduce the number of
elementary quarks and increase the number of excited\ (from the vacuum)
quarks\ that compose baryons and mesons, as well as deduce the rest masses
and intrinsic quantum numbers (I, S, C, b and Q) of the quarks, baryons and
mesons.

\subsection{How to Solve the Above Problems \ \ \ \ \ \ \ \ \ \ \ \ \ \ \ \
\ \ \ \ \ \ \ \ \ \ \ \ \ \ \ \ \ \ \ \ \ \ \ \ \ \ \ \ \ \ \ \ \ \ \ \ \ \
\ \ \ \ \ \ \ \ \ \ \ \ \ \ \ \ \ \ \ \ \ \ \ \ \ \ \ \ \ \ \ \ \ \ \ \ \ \
\ \ \ \ \ \ \ \ \ \ \ \ \ \ \ \ \ \ \ \ \ \ \ \ \ \ \ \ \ \ \ \ \ \ \ \ \ \
\ \ \ \ \ \ \ \ \ \ \ \ \ \ \ \ \ \ \ \ \ \ \ \ \ \ \ \ \ \ \ \ \ \ \ \ \ \
\ \ \ \ \ \ \ \ \ \ \ \ \ \ \ \ \ \ \ \ \ \ \ \ \ \ \ \ \ \ \ \ \ \ \ \ \ \
\ \ \ \ \ \ \ \ \ \ \ \ \ \ \ \ \ \ \ \ \ \ \ \ \ \ \ \ \ \ \ \ \ \ \ \ \ \
\ \ \ \ \ \ \ \ \ \ \ \ \ \ \ \ \ \ \ \ \ \ \ \ \ \ \ \ \ \ \ \ \ \ \ \ \ \
\ \ \ \ \ \ \ \qquad\ \ \ \ \ \ \ \ \ \ \ \ \ \ \ \ \ \ \ \ \ \ \ \ \ \ \ \
\ \ \ \ \ \ \ \ \ \ \ \ \ \ \ \ \ \ \ \ \ \ \ \ \ \ \ \ \ \ \ \ \ \ \ \ \ \
\qquad\ \ \ \ \ \ }

B1. Using only one unflavored (S = C = b = 0) elementary quark family ($%
\epsilon $) with three colors and two isospin states ($\epsilon _{u}$ wit I$%
_{z}$=$\frac{1}{2}$ and Q = $\frac{2}{3}$, $\epsilon _{d}$ with I$_{z}$=$%
\frac{-1}{2}$ and Q = $\frac{-1}{3}$) in the vacuum and a phenomenological
mass formula (\ref{Rest Mass}) as well as the symmetries of a regular
rhombic dodecahedron (see Fig. 1), we can deduce an excited quark spectrum
(m, I, S, C, b and Q), shown in Table 10 and Table 11. The rest masses of
the excited quarks are deduced with the mass formula (\ref{Rest Mass}). The
quantum numbers are deduced with the symmetries of the regular rhombic
dodecahedron [(\ref{S-Number}), (\ref{IsoSpin}), (\ref{S+DS}), (\ref{DaltaS}%
), (\ref{Charmed}) and (\ref{Battom})]. The electric charge Q of the excited
quarks is determined by the elementary quark [$\epsilon _{u}$ or $\epsilon
_{d}$]. Q = $\frac{2}{3}$ for the excited states of the $\epsilon _{u}$; Q = 
$\frac{-1}{3}$ for the excited states of the $\epsilon _{d}$.

B2. Using the sum laws (\ref{Sum(SCbQ)}), we can deduced the intrinsic
quantum numbers\ (I, S, C, b and Q) of all known baryons (qqq), from the
quark spectrum. The deduced quantum numbers of the baryons match the
experimental results. With the sum law (\ref{Baryon M}), we can also deduce
the rest masses of the baryons, except\ for charmed baryons. We can deduce
the masses of the charmed baryons using a phenomenological binding energy
formula (\ref{CB-Eb}). The deduced rest masses of all known baryons (see
Table 16 - Table 21) are about 98\% consistent with experimental results 
{\small .}

B3. Using the sum laws and the intrinsic quantum numbers\ (I, S, C, b and Q)
of the quarks, we deduce the intrinsic quantum numbers of all discovered
mesons (q$_{i}\overline{\text{q}_{j}}$). The deduced quantum numbers of the
mesons are exactly the same as the experimental results. Using a
phenomenological binding energy formula (\ref{M-Ebin}), we deduce the rest
masses of all discovered mesons (see Table 25-Table 31). The deduced masses
of the mesons agree within a 2\% error of the experimental results.

\subsection{Experimental Evidence, Predictions\ and\ Crucial Test\ \ \ \ \ \
\ \ \ \ \ \ \ \ \ \ \ \ \ \ \ \ \ \ \ \ \ \ \ \ \ \ \ \ \ \ \ \ \ \ \ \ \ \
\ \ \ \ \ \ \ \ \ \ \ \ \ \ \ \ \ \ \ \ \ \ \ \ \ \ \ \ \ \ \ \ \ \ \ \ \ \
\ \ \ \ \ \ \ \ \ \ \ \ \ \ \ \ \ \ \ \ \ \ \ \ \ \ \ \ \ \ \ \ \ \ \ \ \ \
\ \ \ \ \ \ \ \ \ \ \ \ \ \ \ \ \ \ \ \ \ \ \ \ \ \ \ \ \ \ \ \ \ \ \ \ \ \
\ \ \ \ \ \ \ \ \ \ \ \ \ \ \ \ \ \ \ \ \ \ \ \ \ \ \ \ \ \ \ \ \ \ \ \ \ \
\ \ \ \ \ \ \ \ \ \ \ \ \ \ \ \ \ \ \ \ \ \ \ \ \ \ \ \ \ \ \ \ \ \ \ \ \ \
\ \ \ \ \ \ \ \ \ \ \ \ \ \ \ \ \ \ \ \ \ \ \ \ \ \ \ \ \ \ \ \ \ \ \ \ \ \
\ \ \ \ \ \ \ \ \ \ \ \ \ \ \ \ \ \ \qquad\ \ \ \ \ \ \ \ \ \ \ \ \ \ \ \ \
\ \ \ \ \ \ \ \ \ \ \ \ \ \ \ \ \ \ \ \ \ \ \ \ \ \ \ \ \ \ \ \ \ \ \ \ \ \
\ \ \ \ \ \ \ \ \ \ \ \qquad}

This paper deduces many new excited quarks shown in Table 11. Already
experiments have discovered almost all of the new quarks inside the baryons
or the mesons (see Table 16--Table 21 and Table 25--Table 31). At the same
time, this paper predicts many new baryons ($\Lambda _{c}^{+}$(6599) and $%
\Lambda _{\text{b}}^{0}$(9959), ...) and new mesons (D(6231), B(9503), $%
\Upsilon $(17868), ...){\small . }From the large full widths of some mesons
and baryons, it also predicts a \textquotedblleft fine
structure\textquotedblright\ phenomenon in particle physics. Several mesons
or baryons (different constitutions of quarks) that have the same intrinsic
quantum numbers (I, S, C, b and Q), the same angular momentums and parities,
but different rest masses form a meson (baryon) with large width. The
experimental investigations of the \textquotedblleft fine
structure\textquotedblright\ (for example f$_{0}$(600) with $\Gamma $=
600-1000 \cite{Particle(04)}), provide a crucial test for our
phenomenological formulae.

Now let us present the applicable phenomenological formulae.

\section{Phenomenological Formulae}

1. We assume that there is only one unflavored elementary quark family ($%
\epsilon $) with three colors that have two isospin states ($\epsilon _{u}$
with I$_{Z}$ = $\frac{1}{2}$ and Q = $\frac{2}{3}$, $\epsilon _{d}$ with I$%
_{Z}$ = $\frac{-1}{2}$ and Q = -$\frac{1}{3}$) for each color in the vacuum.
Thus there are six Fermi (s = $\frac{1}{2}$) elementary quarks in the vacuum
(S = C = b = 0).

2. As a colored elementary quark $\epsilon $ ($\epsilon _{u}$ or $\epsilon
_{d}$) is excited from the vacuum, its color, electric charge and spin
remain unchanged, but it receives energy and intrinsic quantum numbers. In
order to explain the experimental results related to quarks-baryons-mesons,
we propose a phenomenological formula to determine\ the strong interaction
excited energy of a colored (red, yellow or blue) elementary quark \cite%
{0505184}, as follows: 
\begin{equation}
\text{E(}\vec{k}\text{,}\vec{n}\text{) = V}_{0}\text{ + }\alpha \text{[(n}%
_{1}\text{-}\xi \text{)}^{2}\text{+(n}_{2}\text{-}\eta \text{)}^{2}\text{+(n}%
_{3}\text{-}\zeta \text{)}^{2}]  \label{E(n,k)}
\end{equation}
where $\vec{k}$ = ($\xi $, $\eta $, $\zeta $) is a\ vector in a regular
rhombic dodecahedron (see Fig. 1) (including its surfaces) in $\vec{k}$
-space. $V_{0}$ is the minimum energy that a elementary quark $\epsilon $ is
excited from the vacuum. $\alpha $ is a constant. n$_{1}$, n$_{2}$ and n$%
_{3} $ are integers. This formula is the same for any colored elementary
quark, so we do not mark the quark's color.

\begin{figure}[h]
\vspace{5.8in} \includegraphics{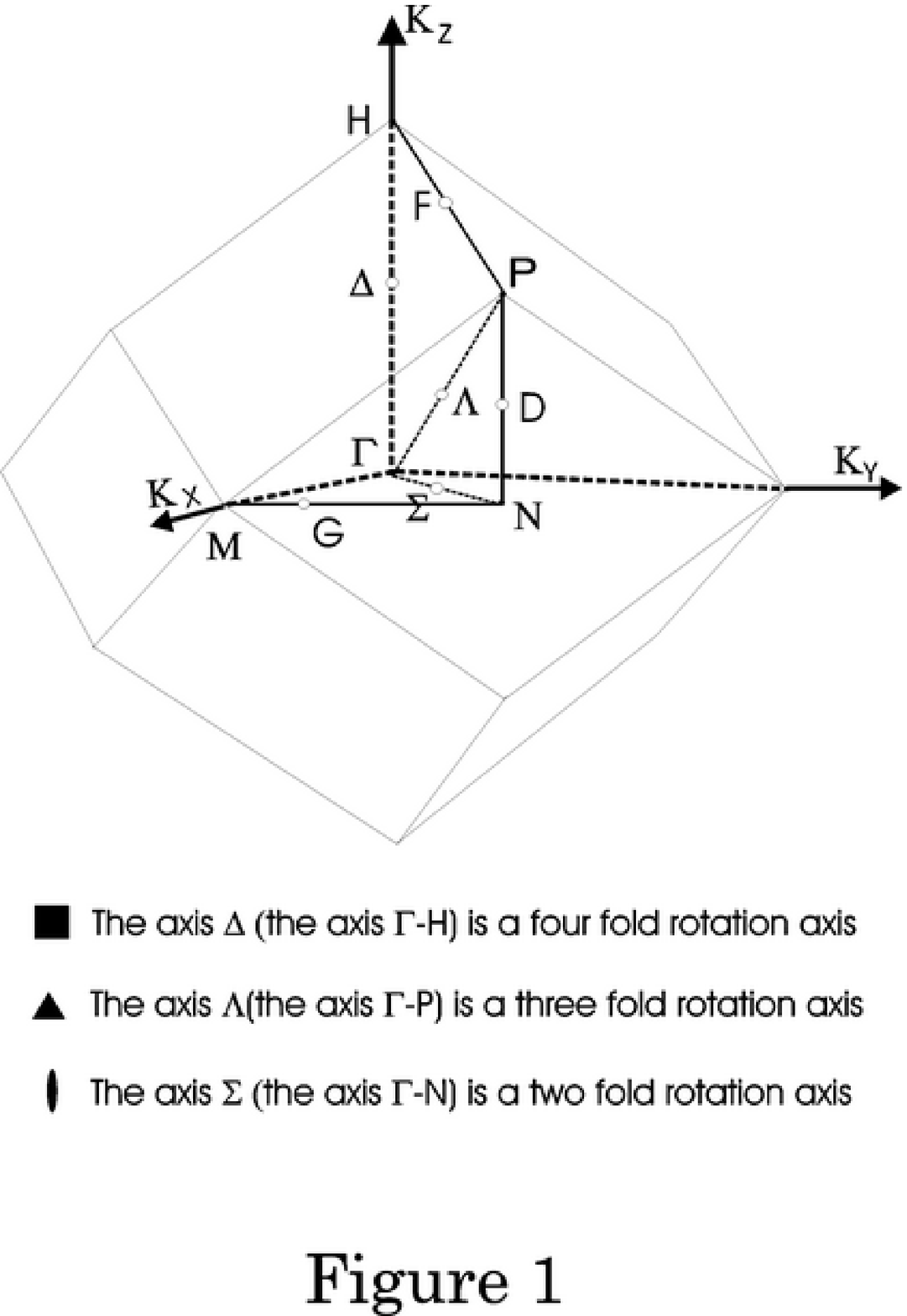} \label{Fig1}
\caption{{\protect\small The regular rhombic dodecahedron. The symmetry
points and axes are indicated. The }$\Delta ${\protect\small -axis is a
fourfold rotation axis, the strange number S = 0 and the fourfold baryon
family }$\Delta ${\protect\small \ (}$\Delta ^{++},${\protect\small \ }$%
\Delta ^{+},${\protect\small \ }$\Delta ^{0},${\protect\small \ }$\Delta
^{-} ${\protect\small ) will appear on the axis. The axes }$\Lambda $%
{\protect\small \ and F are threefold rotation axes, the strange number S =\
-1 and the threefold baryon family }$\Sigma ${\protect\small \ (}$\Sigma
^{+},${\protect\small \ }$\Sigma ^{0},${\protect\small \ }$\Sigma ^{-})$%
{\protect\small \ will appear on the axes. The axes }$\Sigma $%
{\protect\small \ and G are twofold rotation axes, the strange number S = -2
and the twofold baryon family }$\Xi ${\protect\small \ (}$\Xi ^{0},$%
{\protect\small \ }$\Xi ^{-}${\protect\small ) will appear on the axes. The
axis D is parallel to the axis }$\Delta ${\protect\small , S = 0. And it is
a twofold rotation axis, the twofold baryon family N (N}$^{+}$%
{\protect\small , N}$^{0}${\protect\small ) will be on the axis.}}
\end{figure}

In order to satisfy the symmetries of the regular rhombic dodecahedron, we
give a condition for the $\overrightarrow{n}$ = (n$_{1}$, n$_{2}$, n$_{3})$
values. If assuming n$_{1}$ = \textit{l}$_{2}$ \textit{+ l}$_{3}$, n$_{2}$ =%
\textit{\ l}$_{3}$ \textit{+ l}$_{1}$ and n$_{3}$ =\textit{\ l}$_{1}$ 
\textit{+ l}$_{2},$ we have 
\begin{equation}
\begin{tabular}{l}
\textit{l}$_{1}$ = $\frac{1}{2}$(-n$_{1}$ + n$_{2}$ + n$_{3}$), \\ 
\textit{l}$_{2}$ = $\frac{1}{2}$(+n$_{1}$ - n$_{2}$ + n$_{3}$), \\ 
\textit{l}$_{3}$ = $\frac{1}{2}$(+n$_{1}$ + n$_{2}$ - n$_{3}$).%
\end{tabular}
\label{l-n}
\end{equation}
The condition is that only those values of $\overrightarrow{n}$ = (n$_{1}$, n%
$_{2}$, n$_{3}$) are allowed that make $\overrightarrow{l}$ = \textit{(l}$%
_{1}$\textit{, l}$_{2}$\textit{, l}$_{3}$\ ) an integer vector. For example, 
$\vec{n}$ cannot take the values $(0,0,1)$ or $\left( 1,1,-1\right) $, but
can take $(0,0,2)$ and $(1,-1,2)$. This is a result of the symmetries of the
regular rhombic dodecahedron. The low level allowed n = (n$_{1}$, n$_{2}$, n$%
_{3})$ values are shown in (\ref{Equ-N}) of Appendix A.

3. The energy (\ref{E(n,k)}) of excited quarks satisfies the symmetries of
the regular rhombic dodecahedron. From Fig. 1, we can see that there are
four kinds of symmetry points ($\Gamma $, H, P and N) and six kinds of
symmetry axes ($\Delta $, $\Lambda $, $\Sigma $, D, F and G) in the regular
rhombic dodecahedron. The coordinates ($\xi $, $\eta $, $\varsigma $) of the
symmetry points are as follows: 
\begin{equation}
\Gamma \text{ }\text{= (0, 0, 0), H = (0, 0, 1), P = (}\frac{\text{1}}{\text{%
2}}\text{, }\frac{\text{1}}{\text{2}}\text{, }\frac{\text{1}}{\text{2}}\text{%
) and N }\text{= (}\frac{\text{1}}{\text{2}}\text{, }\frac{\text{1}}{\text{2}%
}\text{, 0). }  \label{Sym-Point}
\end{equation}%
The coordinates ($\xi $, $\eta $, $\varsigma $) of the symmetry axes are: 
\begin{equation}
\begin{tabular}{ll}
$\Delta \text{ }\text{= (0, 0, }\zeta \text{),\ 0 \TEXTsymbol{<}}\zeta \text{
\TEXTsymbol{<}1; }$ & $\Lambda \text{ = (}\xi \text{, }\xi \text{, }\xi 
\text{), 0 \TEXTsymbol{<}}\xi \text{ \TEXTsymbol{<}}\frac{\text{1}}{\text{2}}%
\text{;}$ \\ 
$\Sigma \text{{}}\text{= (}\xi \text{, }\xi \text{, 0), 0 \TEXTsymbol{<}}\xi 
\text{ \TEXTsymbol{<}}\frac{\text{1}}{\text{2}}\text{;}$ & $\text{D }\text{=
(}\frac{\text{1}}{\text{2}}\text{, }\frac{\text{1}}{\text{2}}\text{, }\xi 
\text{), 0 \TEXTsymbol{<}}\xi \text{ \TEXTsymbol{<}}\frac{\text{1}}{\text{2}}%
\text{;}$ \\ 
$\text{G}\text{= (}\xi \text{, 1-}\xi \text{, 0), }\frac{\text{1}}{\text{2}}%
\text{ \TEXTsymbol{<}}\xi \text{ \TEXTsymbol{<}1;}$ & $\text{F = (}\xi \text{%
, }\xi \text{, 1-}\xi \text{), 0 \TEXTsymbol{<}}\xi \text{ \TEXTsymbol{<}}%
\frac{\text{1}}{\text{2}}\text{.}$%
\end{tabular}
\label{Axes}
\end{equation}%
Thus, the energy of the excited quarks has six kinds of symmetry axes (see
Table A2). The energy with a $\overrightarrow{n}$\ = (n$_{1}$, n$_{2}$, n$%
_{3}$) along a symmetry axis (from the lowest energy to the highest energy)
forms an energy band. The energy bands on different symmetry axes have
different symmetries with different energies and intrinsic quantum numbers
(S, C, b and I) that\ mean different (excited) quarks. The minimum energy of
the energy band is the rest mass of the excited quark. Each energy band with
a rest mass (m) and the intrinsic quantum numbers (I, S, C, b and Q)
corresponds to a quark with the same rest mass and the same intrinsic
quantum numbers. All quarks inside baryons and mesons are the excited states
of the elementary $\epsilon $-quark.

4. The strange number S of an excited quark that lies on an axis inside the
regular rhombic dodecahedron is 
\begin{equation}
\text{S = R - 4.}  \label{S-Number}
\end{equation}
For the three axes (the P-N axis, the P-H axis and the M-N axis) on the
surface of the regular rhombic dodecahedron, the strange numbers are as
follows: 
\begin{equation}
\begin{tabular}{l}
the P-N axis parallels with the $\Gamma $-H axis, \ S$_{\text{P-N}}$ = S$%
_{\Gamma \text{-H}}$ = 0; \\ 
the P-H axis parallels with the $\Gamma $-P$^{^{\prime }}\text{(}\frac{\text{%
1}}{\text{2}}\text{, }\frac{\text{1}}{\text{2}}\text{, }\frac{-\text{1}}{%
\text{2}}\text{)}$ axis, S$_{\text{P-H}}$ = S$_{\Gamma \text{-P}^{\prime }}$
= - 1; \\ 
the M-N axis parallels with the $\Gamma $-N$^{^{\prime }}$($\frac{\text{1}}{%
\text{2}}\text{, }\frac{-\text{1}}{\text{2}}\text{, }$0$\text{)}$ axis, S$_{%
\text{M-N}}$ = S$_{\Gamma \text{-N}^{\prime }}$ = - 2.%
\end{tabular}
\label{S-S-Numb}
\end{equation}
Since the $\Gamma $-P$^{^{\prime }}$($\frac{\text{1}}{\text{2}}$, $\frac{%
\text{1}}{\text{2}}$, $\frac{-\text{1}}{\text{2}}$) axis inside the regular
rhombic dodecahedron, R = 3, is an equivalent axis\ (Table A2) of the $%
\Gamma $-P axis, S = -1 . The $\Gamma $-N$^{^{\prime }}$ axis inside the
regular rhombic dodecahedron, R = 2, is an equivalent axis (Table A2) of the 
$\Gamma $-N axis, thus S = -2.

5. For any valid value of the integers $\overrightarrow{\text{n}}$ (\ref%
{Equ-N}) in Appendix A, substituting the ($\xi $, $\eta $, $\zeta $)
coordinates (\ref{Axes}) of an axis into formula (\ref{E(n,k)}), we can get
the energy bands that are shown in Table B1--Table B7 (see Appendix B).

From these tables, we can see that there are eightfold, sixfold, fourfold,
threefold and twofold energy bands. A group of energy bands (number=deg)
with the same energy and equivalent $\overrightarrow{\text{n}}$ values (\ref%
{Equ-N}) are called the \textbf{degenerate energy bands}. For a group of
degenerate energy bands (number = deg), its isospin I is determined by 
\begin{equation}
\text{deg = 2I + 1.}  \label{IsoSpin}
\end{equation}
The formal z-components of the isospin are as follows:

\begin{equation}
\begin{tabular}{|l|l|}
\hline
for I = $\frac{3}{2},$ & I$_{z}^{^{\prime }}$ = $\frac{3}{2}$, $\frac{1}{2}$%
, $\frac{-1}{2}$, $\frac{-3}{2}$; \\ \hline
for I = 1, & I$_{z}^{^{\prime }}$ = 1, 0, -1; \\ \hline
for I = $\frac{1}{2}$, & I$_{z}^{^{\prime }}$ = $\frac{1}{2}$, $\frac{-1}{2}$%
; \\ \hline
for I = 0, & I$_{z}^{^{\prime }}$ = 0. \\ \hline
\end{tabular}
\label{Formal-Iz}
\end{equation}

If a group of g-fold energy bands with the same energy and g \TEXTsymbol{>}
R - the rotary fold\ of the symmetry axis (see Table A3 in Appendix A) 
\begin{equation}
\text{g \ \TEXTsymbol{>} R,}  \label{deg > R}
\end{equation}
the g-fold bands will be divided into $\gamma $ sub-fold energy bands (each
has R-bands) first (we call it the first division, K = 0; the energy and the
strange numbers are not changed in the first division), 
\begin{equation}
\gamma \text{ = g/R.}  \label{Subdeg}
\end{equation}
If the sub-fold energy bands are degenerate, using (\ref{IsoSpin}), we can
find the isospin value for the sub-fold degenerate energy bands. Since each
sub-degeneracy group has R-fold degenerate bands 
\begin{equation}
\text{I = (R-1)/2.}  \label{(R-1)/2}
\end{equation}
For sub-fold energy bands with non-equivalent n values, the sub-fold energy
bands will be divided into sub-subgroups with equivalent (or single) $%
\overrightarrow{n}$ values (the second kind of division, K = 1) again (see
Table A3). Then, using (\ref{IsoSpin}), we can find the isospin values of
the sub-subgroups.

6. From (\ref{S-Number}) and (\ref{IsoSpin}), the single energy bands on the 
$\Gamma $-H axis (see Table B1) will have I = S = 0 and the single energy
bands on the $\Gamma $-N axis (see Table B2) will have I = 0 and S = -2.
Since\ there is not any quark that has I = 0 and S = 0 or S = -2 in the
current Quark Model, we need a new formula to deduce the correct S values
for the single bands.

For the single energy bands on the $\Gamma $-H and the $\Gamma $-N axes, the
strange number 
\begin{equation}
\text{S = S}_{axis}\text{ + }\Delta \text{S}  \label{S+DS}
\end{equation}
\begin{equation}
\Delta \text{S = }\delta \text{(}\widetilde{n}\text{) + [1-2}\delta \text{(S}%
_{axis}\text{)]Sign(}\widetilde{n}\text{),\ \ }  \label{DaltaS}
\end{equation}
where $\delta $($\widetilde{n}$)\ and $\delta $(S$_{axis}$) are Dirac
functions, and S$_{axis}$ is the strange number of the axis (see Table A3).
For an energy band with $\overrightarrow{n}$ = (n$_{1}$, n$_{2}$, n$_{3})$, $%
\widetilde{n}$ \ is defined as 
\begin{equation}
\widetilde{n}\text{ }\equiv \frac{\text{n}_{1}\text{+n}_{2}\text{+n}_{3}}{%
\left\vert \text{n}_{1}\right\vert \text{+}\left\vert \text{n}%
_{2}\right\vert \text{+}\left\vert \text{n}_{3}\right\vert }\text{.}
\label{n/n}
\end{equation}
\begin{equation}
\text{Sign(}\widetilde{n}\text{) = }\left[ \text{%
\begin{tabular}{l}
+1 for $\widetilde{n}$ \TEXTsymbol{>} 0 \\ 
0 \ \ for $\widetilde{n}$ = 0 \\ 
-1 \ for $\widetilde{n}$ \TEXTsymbol{<} 0%
\end{tabular}
}\right] \text{.}  \label{Sin(N)}
\end{equation}

\begin{equation}
\text{If }\widetilde{n}\text{ = 0 \ \ \ \ }\Delta \text{S = }\delta \text{%
(0) = +1.}  \label{n=0-DaltaS=+1}
\end{equation}
If $\qquad \widetilde{n}$ $=$ $\frac{0}{0},$

\begin{equation}
\Delta \text{S = - S}_{Axis}\text{.}  \label{DaltaS=-Sax}
\end{equation}
Thus, for $\overrightarrow{n}$ = (0, 0, 0), from (\ref{n/n}) and (\ref%
{DaltaS=-Sax}), we have \ 
\begin{equation*}
\text{Sign(}\widetilde{n}\text{)}=\frac{\text{n}_{1}\text{+n}_{2}\text{+n}%
_{3}}{\left\vert \text{n}_{1}\right\vert \text{+}\left\vert \text{n}%
_{2}\right\vert \text{+}\left\vert \text{n}_{3}\right\vert }\text{ = }\frac{0%
}{0}\text{.}
\end{equation*}
\begin{equation}
\text{S =S}_{Axis}\text{+}\Delta \text{S = S}_{Axis}\text{- S}_{Axis}\text{
= 0.}  \label{S=0 of n=0}
\end{equation}
\qquad \qquad \qquad\ \ \ \ 

\qquad 
\begin{tabular}{l}
\ \ \ Table 2. The S-Number of the Bands with $\overrightarrow{n}$ = (0, 0,
0) \\ 
\begin{tabular}{|l|l|l|l|l|l|l|}
\hline
$\overrightarrow{n}$ & Axis & Energy Band & $\text{S}_{Axis}$ & $\Delta 
\text{S}$ & S & $\Delta $E$^{\#}$ \\ \hline
(0, 0, 0) & $\ \Delta $ & E$_{\Gamma }$=0$\rightarrow $E$_{H}$=1 & 0 & 0 & 0
& 0 \\ \hline
(0, 0, 0) & $\ \Lambda $ & E$_{\Gamma }$=0$\rightarrow $E$_{P}$=$\frac{3}{4}$
& -1 & +1 & 0 & 0 \\ \hline
(0, 0, 0) & $\ \Sigma $ & E$_{\Gamma }$=0$\rightarrow $E$_{N}$=$\frac{1}{2}$
& -2 & +2 & 0 & 0 \\ \hline
(0, 0, 0) & \ D & E$_{N}$=$\frac{1}{2}\rightarrow $E$_{P}$=$\frac{3}{4}$ & 0
& 0 & 0 & 0 \\ \hline
(0, 0, 0) & \ F & E$_{P}$=$\frac{3}{4}\rightarrow $E$_{H}$=1 & -1 & +1 & 0 & 
0 \\ \hline
(0, 0, 0) & \ G & E$_{N}$=$\frac{1}{2}\rightarrow $E$_{M}$=1 & -2 & +2 & 0 & 
0 \\ \hline
\end{tabular}
\\ 
\ \ \ \ \ \ {\small \# Since S = C = b =0, from (\ref{Dalta-E}), }$\Delta $%
{\small E = 0}%
\end{tabular}

\ \ \ \ \ \ \ \ \ \ \ \ \ \ \ \ \ \ \ \ \ \ \ \ \ \ \ \ \ \ \ \ \ \ \ \ \ \ 

7. We believe that the $\Delta $S is the result of the fluctuation of
strange numbers of excited energy bands. Some energy bands have the
fluctuations of the strange numbers ($\Delta $S $\neq $ 0), while some
energy bands do not have the fluctuation of strange numbers ($\Delta $S =
0). \ It is necessary to find the conditions that determine the fluctuation
of the strange numbers.

For the R-degenerate bands of an axis with rotary fold R, $\Delta S$ = 0.
Such as: the fourfold bands of the $\Delta $-axis, the threefold bands of
the $\Lambda $-axis and the twofold degenerate bands of the D-axis, $\Sigma $%
-axis and G-axis. For a group of energy bands with fold = $\gamma $R ($%
\gamma $ is an integer number 2 or 3 or 4), after first division, there may
be $\gamma $ (R-degenerate) energy bands that have $\Delta $S = 0. For a
group of degenerate energy bands with the deg \TEXTsymbol{<} R, using (6),
we can get S; with (8), we can get I. If it has the same S and I as a quark
of the current Quark Model, $\Delta $S = 0. For example, for the single
energy bands of the $\Lambda $-axis (or the F-axis), it has S = -1 and I = 0
that is the same as the s-quark has, $\Delta $S = 0. The results are shown
in Table A4 of Appendix A.

Otherwise, $\Delta $S $\neq $ 0, for a group with the deg \TEXTsymbol{<} R
that has S and I that no quark of the current Quark Model has. For the
single bands of the $\Delta $-axis, $\Sigma $-axis, D-axis and G-axis, and
the double degenerate bands of the F-axis, $\Delta $S\ $\neq $ 0. Detailed
results are shown in Table A5.\ \ \ \ \ \ \ \ \ \ \ 

8. The \textquotedblleft strange number\textquotedblright\ from (\ref{S+DS})
is not exactly the same as the strange number from (\ref{S-Number}). In
order to compare them with the experimental results, we would like to give
new names under certain circumstances. The new names will be the charmed
number and the bottom number. If S = +1, we call it the charmed number C: 
\begin{equation}
\text{if }\Delta \text{S = +1}\rightarrow \text{S =S}_{Ax}\text{+}\Delta 
\text{S = +1, }C\text{ }\equiv \text{ +1.}  \label{Charmed}
\end{equation}
If S = -1$,$ which originates from $\Delta S=+1$ on a single energy band,
and there is an energy fluctuation,$\ $we call it the bottom number\ b:$\ \
\ \ \ \ \ \ \ \ \ \ \ $%
\begin{equation}
\text{for single bands, if\ }\Delta \text{S = +1}\rightarrow \text{S = -1
and }\Delta \text{E}\neq \text{0, b }\equiv \text{ -1.}  \label{Battom}
\end{equation}
Similarly, we can obtain charmed strange quarks q$_{\Xi _{C}}$\ and q$%
_{\Omega _{C}}$.

9. The sixfold energy bands of the F-axis (R = 3) need two divisions. In the
first division, the sixfold band divides into two ($\frac{6}{R}$=2)
threefold bands; the energy and the strange number do not change (S = -1
steel). In the second division (K = 1), the threefold bands with
non-equivalent n values divide into a twofold band and a $\sin $gle band.
For the twofold energy band, if 
\begin{equation}
\text{\lbrack (for sixfold bands) }\Delta \text{S = +1 and \ E \TEXTsymbol{>}
m}_{u_{c}}\text{(1753 Mev)]}\rightarrow \text{q}_{_{\Xi _{C}}}\text{,}
\label{Kersa-C}
\end{equation}
the twofold energy band represents a twofold family q$_{\Xi _{C}}$-quark
with S = -1 and C = +1.

The sixfold energy bands of the G-axis (R = 2) need two divisions also. In
the first division, the sixfold band divides into three ($\frac{6}{R}$=3)
twofold bands. The energy and the strange number do not change (S = -2). In
the second division (K = 1), the twofold band with non-equivalent $%
\overrightarrow{n}$ values divides into two single bands. For a single
energy band, if 
\begin{equation}
\text{\lbrack (for sixfold bands) }\Delta \text{S = +1 and \ E \TEXTsymbol{>}
m}_{u_{c}}\text{(1753 Mev)]}\rightarrow \text{d}_{\Omega _{C}}\text{,}
\label{Omiga-C}
\end{equation}
the single energy band represents a d$_{\Omega _{C}}$-quark with S = -2 and
C = +1.

10. The elementary quark $\epsilon _{u}$ (or $\epsilon _{d}$) determines the
electric charge Q of an excited quark $\epsilon _{u}$ (or $\epsilon _{d}$).
For an excited quark of $\epsilon _{u}$ (or $\epsilon _{d}$), Q = +$\frac{2}{%
3}$ (or -$\frac{1}{3}$). For a quark with isospin I , there are 2I +1
members . Since the $\epsilon _{u}$-quark has I$_{z}$ = $\frac{1}{2}$ 
\TEXTsymbol{>} 0 and the $\epsilon _{d}$-quark has I$_{z}$ = -$\frac{1}{2}$ 
\TEXTsymbol{<} 0, for an excited quark with \ I$_{z}^{^{\prime }}$ 
\TEXTsymbol{>} 0 from (\ref{Formal-Iz}), it is an excited quark of $\epsilon
_{u}$, its electric charge Q$_{q}$ is 
\begin{equation}
\text{Q}_{q}\text{ = Q}_{\epsilon _{u}}\text{ = }\frac{2}{3}.  \label{2/3}
\end{equation}
For an excited quark with \ I$_{z}^{^{\prime }}$ \TEXTsymbol{<} 0 from (\ref%
{Formal-Iz}), it is an excited quark of $\epsilon _{d}$, its electric charge
Q$_{q}$ is

\begin{equation}
\text{Q}_{q}\text{ = \ Q}_{\epsilon _{d}}\text{ = \ - }\frac{1}{3}.
\label{-1/3}
\end{equation}
For I$_{z}^{^{\prime }}$ = 0, if S + C + b \TEXTsymbol{>} 0, 
\begin{equation}
\text{Q}_{q}\text{ = Q}_{\epsilon _{u}}\text{ = }\frac{2}{3}\text{;}
\label{2/3ofIz=0}
\end{equation}
if S + C + b \TEXTsymbol{<} 0, 
\begin{equation}
\text{Q}_{q}\text{ = Q}_{\epsilon _{d}}\text{ = -}\frac{1}{3}\text{.}
\label{-1/3of Iz=0}
\end{equation}
There is not any quark with I$_{z}^{^{\prime }}$ = 0 and S + C + b = 0.

\ 11. After getting S, C, b and Q$_{\text{q}_{z^{\prime }}}$ (B = $\frac{1}{3%
}$) of a quark, we can deduce a physical I$_{z}$ of the quark using the
generalized Gell-Mann-Nishijima relationship \cite{GMN}:

\begin{equation}
\text{Q}_{\text{q}_{z^{\prime }}}\text{ = I}_{z}\text{ + }\frac{1}{2}\text{%
(B + S + C + b),}  \label{GMN}
\end{equation}
where B is the baryon number (B = $\frac{1}{3}$) and Q$_{\text{q}_{z^{\prime
}}}$ is the electric charge of the quark. For the d$_{S}$-quark, B = $\frac{1%
}{3}$, S = -1, C = b = 0 and Q = -$\frac{1}{3},$ from (\ref{GMN}), I$_{Z}$ =
0; for the u$_{C}$-quark, B = $\frac{1}{3}$, C = 1, S = b = 0 and Q = +$%
\frac{2}{3},$ from (\ref{GMN}), I$_{Z}$ = 0. The physical I$_{Z}$-values of
all low energy quarks will be shown in Table 10.\ \ \ \ \ \ \ \ \ \ \ \ \ \
\ \ \ \ \ \ \ \ \ \ \ \ \ \ \ \ \ \ \ \ \ \ \ \ \ \ \ \ \ \ \ \ \ \ \ \ \ \
\ \ \ \ \ \ \ \ \ \ \ \ \ \ \ \ \ \ \ \ \ \ \ \ \ \ \ \ \ \ \ \ \ \ \ \ \ \
\ \ \ \ \ \ \ \ \ \ \ \ \ \ \ \ \ \ \ \ \ \ \ \ \ \ \ \ \ \ \ \ \ \ \ \ \ \
\ \ \ \ \ \ \ \ \ \ \ \ \ \ \ \ \ \ \ \ \ \ \ \ \ \ \ \ \ \ \ \ \ \ \ \ \ \
\ \ \ \ \ \ \ \ \ \ \ \ \ \ \ \ \ \ \ \ \ \ \ \ \ \ \ \ \ \ \ \ \ \ \ \ \ \
\ \ \ \ \ \ \ \ \ \ \ \ \ \ \ \ \ \ \ \ \ \ \ \ \ \ \ \ \ \ \ \ \ \ \ \ \ \
\ \ \ \ \ \ \ \ \ \ \ \ \ \ \ \ \ \ \ \ \ \ \ \ \ \ \ \ \ \ \ \ \ \ \ \ \ \
\ \ \ \ \ \ \ \ \ \ \ \ \ \ \ \ \ \ \ \ \ \ \ \ \ \ \ \ \ \ \ \ \ \ \ \ \ \
\ \ \ \ \ \ \ \ \ \ \ \ \ \ \ \ \ \ \ \ \ \ \ \ \ \ \ \ \ \ \ \ \ \ \ \ \ \
\ \ \ \ \ \ \ \ \ \ \ \ \ \ \ \ \ \ \ \ \ \ \ \ \ \ \ \ \ \ \ \ \ \ \ \ \ \
\ \ \ \ \ \ \ \ \ \ \ \ \ \ \ \ \ \ \ \ \ \ \ \ \ \ \ \ \ \ \ \ \ \ \ \ \ \
\ \ \ \ \ \ \ \ \ \ \ \ \ \ \ \ \ \ \ \ \ \ \ \ \ \ \ \ \ \ \ \ \ \ \ \ \ \
\ \ \ \ \ \ \ \ \ \ \ \ \ \ \ \ \ \ \ \ \ \ 

12. The rest masses m$^{\ast }$ of the excited quarks are the minimum energy
of the energy band (\ref{E(n,k)}), 
\begin{equation}
\text{m}^{\ast }\text{ = \{V}_{0}\text{+}\alpha \times \text{minimum[(n}_{1}%
\text{-}\xi \text{)}^{2}\text{+(n}_{2}\text{-}\eta \text{)}^{2}\text{+(n}_{3}%
\text{-}\zeta \text{)}^{2}\text{]+}\Delta \text{E\},}  \label{Q-Mass}
\end{equation}
where $\Delta $E is the fluctuate energy of the excited quark. The
fluctuations will increase as the energies increase. In order to take this
effect into account, we define a fluctuation order number J. J depends on
the symmetry axis, the symmetry point, energy bands and $\Delta $S. If $%
\Delta $S = 0, J = 0.\ For the energy bands on the $\Lambda $-axis, $\Delta $%
S = 0, J = 0. For a symmetry axis, the same fold energy bands with the same $%
\Delta $S and the same intrinsic quantum numbers (S, C and b), the
fluctuation numbers are positive order integers 1, 2, 3, ... from the lowest
energy band to higher bands. We give the J values for all fluctuation energy
bands sufficient to cover the experimental data in Table B1-Table B7. The
fluctuation energy $\Delta $E can be found with a phenomenological formula
as follows:

\begin{equation}
\begin{tabular}{l}
$\Delta $E = 100\{$\Theta $C[2(J$_{C}$- 3.5I) -S] + KS $\text{+ }$(S +b)[(1
+S$_{Ax}$)(J$_{S,b}$ +S$_{Ax}$)]$\Delta S$\} \\ 
\ \ \ \ \ J$_{C}$ = 1, 2, 3, ... \ J$_{S,b}$= $\Theta $(1- S$_{Ax})$+ (1, 2,
3,...) ($\Delta $E = 0, for J $\leq \Theta $(1-S$_{Ax}$)).%
\end{tabular}
\label{Dalta-E}
\end{equation}
From Table A3, for the axes $\Delta ,$ $\Sigma $ and $\Lambda $, $\Theta $
=0; for the axes D, F and G, $\Theta $ = 1. S$_{Ax}$ is the strange number
of the symmetry axis. J$_{C}$ is the order number of the charmed quarks from
low energy to high energy in a symmetry axis (see Table B5, Table B6 and
Table B7). J$_{S.b}$ is the fluctuation order number of the fluctuation
energy bands with S $\neq $ 0\ or b $\neq $\ 0. As an example, we can deduce
the $\Delta $E of the energy bands with $\overrightarrow{n}$ = (0, 0, 0)
using (\ref{Dalta-E}). From Table 2, for these energy bands, C = S = b = 0,
we have $\Delta $E = 0.

We find the parameters V$_{0}$ and $\alpha $ now. According to the Quark
Model \cite{Quark Model}, a baryon is composed of three different colored
quarks. In particular the proton and the neutron are given by 
\begin{equation*}
\text{p = uud and \ n = udd,}
\end{equation*}
so that, if E$_{bind}$ represents the binding energy of a proton or a
neutron, then 
\begin{equation}
\begin{tabular}{l}
M$_{P}$ = m$_{u}+$ m$_{u}+$ m$_{d}$ - $\left\vert \text{E}_{bind}\right\vert
,$ \\ 
M$_{n}$ = m$_{u}+$ m$_{d}+$ m$_{d}$ - $\left\vert \text{E}_{bind}\right\vert 
$.%
\end{tabular}
\label{Mp, Mn}
\end{equation}
If omitting electric masses, we have an approximation, 
\begin{equation}
\begin{tabular}{l}
M$_{P}\sim $M$_{n}$= 939 Mev, \\ 
m$_{u}=$ m$_{d}.$%
\end{tabular}
\label{939Mev}
\end{equation}
Thus, from (\ref{Mp, Mn}) and (\ref{939Mev}), we have 
\begin{equation}
\text{m}_{u}\text{ = m}_{d}\text{ = }\frac{1}{3}\text{(939 +}\left\vert 
\text{E}_{bind}\right\vert \text{) = 313 +}\frac{1}{3}\left\vert \text{E}%
_{bind}\right\vert \text{,}  \label{313Mev}
\end{equation}
where E$_{bind}$ is the total binding energy of the three quarks (colors) in
a baryon. We use $\Delta $ = $\frac{1}{3}\left\vert \text{E}%
_{bind}\right\vert $ that is the phenomenological approximations of the
color's strong interaction energies. Since the three colors are the same for
all baryons, $\Delta $ is an unknown positive constant for all baryons. It
originates from the three colors of the three quark inside the baryons.
Since high energy scattering experiments (an uncounted number) have not
separated the quarks inside the baryons, it means that $\left\vert
E_{bind}\right\vert $ (3$\Delta $) is much larger than M$_{p}$, 
\begin{equation}
\Delta \text{ = }\frac{1}{3}\left\vert \text{E}_{bind}\right\vert \text{ 
\TEXTsymbol{>}\TEXTsymbol{>} M}_{p}\text{.}  \label{Dalta}
\end{equation}
The lowest mass of the quarks is the rest mass of the u-quark (or the
d-quark) (\ref{313Mev}). From (\ref{Q-Mass}), (\ref{313Mev}) and (\ref{Dalta}%
), the lowest quark mass is 
\begin{equation}
\text{m}_{_{\text{Lowest}}}^{\ast }\text{ = V}_{0}\text{ = (313+}\Delta 
\text{) Mev.}  \label{V=313}
\end{equation}
Fitting experimental results, we can get 
\begin{equation}
\alpha \text{ = 360 Mev.}  \label{Alpha=360}
\end{equation}
The rest mass (m$^{\ast }$) of a quark, from (\ref{Q-Mass}), is 
\begin{equation}
\begin{tabular}{l}
$\text{m}^{\ast }\text{ = \{313+ 360 minimum[(n}_{1}\text{-}\xi \text{)}^{2}%
\text{+(n}_{2}\text{-}\eta \text{)}^{2}\text{+(n}_{3}\text{-}\zeta \text{)}%
^{2}\text{]+}\Delta \text{E+}\Delta \text{\} (Mev)}$ \\ 
\ \ \ \ = m + $\Delta $ \ (Mev),%
\end{tabular}
\label{Rest Mass}
\end{equation}
where $\Delta $E is in (\ref{Dalta-E}). $\Delta $ is an unknown large
constant ($\Delta $ \TEXTsymbol{>}\TEXTsymbol{>} M$_{P}$). Since (-3$\Delta $%
) is the three colors interaction energy, from (\ref{313Mev}) and (\ref%
{Dalta}), the $\Delta $ part in (\ref{Rest Mass}) of quark mass is from the
color. This formula (\ref{Rest Mass}) is the united rest mass formula for
quarks.

In the next section, we will use the above phenomenological formulae to
deduce the rest masses and the intrinsic quantum numbers (S, C, b, I and Q)
of the quarks (a quark spectrum).

\section{The Quark Spectrum\ \ \ \ \ \ \ \ \ \ \ \ \ \ \ \ \ \ \ \ \ \ \ \ \
\ \ \ \ \ \ \ \ \ \ \ \ \ \ \ \ \ \ \ \ \ \ \ \ \ \ \ \ \ \ \ \ \ \ \ \ \ \
\qquad \qquad \qquad \qquad\ \ \ \ \ \ \ \ \ \ \ \ \ \ \ \ \ \ \ \ \ \ \ \ \
\ \ \ \ \ \ \ \ \ \ \ \ \ \ \ \ \ \ \ \ \ \ \ \ \ \ \ \ \ \ \ \ \ \ \ \ \ \
\ \ \ \ \ \ \ \ \ \ \ \ \ \ \ \ \ \ \ \ \ \ \ \ \ \ \ \ \ \ \ \ \ \ \ \ \ \
\ \ \ \ \ \ \ \ \ \ \ \ \ \ \ \ \ \ \ \ \ \ \ \ \ \ \ \ \ \ \ \ \ \ \ \ \ \
\ \ \ \ \ \ \ \ \ \ \ \ \ \ \ \ \ \ \ \ \ \ \ \ \ \ \ \ \ \ \ \ \ \ \ \ \ \
\ \ \ \ \ \ \ \ \ \ \ \ \ \ \ \ \ \ \ \ \ \ \ \ \ \ \ \ \ \ \ \ \ \ \ \ \ \
\ \ \ \ \ \ \ \ \ \ \ \ \ \ \ \ \ \ \ \ \ \ \ \ \ \ \ \ \ \ \ \ \ \ \ \
\qquad\ \ \ \ \ \ \ \ \ \ \ \ \ \ \ \ \ \ \ \ \ \ \ \ \ \ \ \ \ \ \ \ \ \ \
\ \ \ \ \ \ \ \ \ \ \ \ \ \ \ \ \ \ \ \ \ \ \ \ \ \ \ \ \ \ \ \ \ \ \ \ \ \
\ \ \ \ \ \ \ \ \ \ \ \ \ \ \ \ \ \ \ \ \ \ \ \ \ \ \ \ \ \ \ \ \ \ \ \ \ \
\ \ \ \ \ \ \ \ \ \ \ \ \ \ \ \ \ \ \ \ \ \ \ \ \ \ \ \ \ \ \ \ \ \ \ \ \ \
\ \ \ \ \ \ \ \ \ \ \ \ \ \ \ \ \ \ \ \ \ \ \ \ \ \ \ \ \ \ \ \ \ \ \ \ \ \
\ \ \ \ \ \ \ \ \ \ \ \ \ \ \ \ \ \ \ \ \ \ \ \ \ \ \ \ \ \ \ \ \ \ \ \ \ \
\ \ \ \ \ \ \ \ \ \ \ \ \ \ \ \ \ \ \ \ \ \ \ \ \ \ \ \ \ \ \ \ \ \ \ \ \ \
\ \ \ \ \ \ \ \ \ \ \ \ \ \ \ \ \ \ \ \ \ \ \ \ \ \ \ \ \ \ \ \ \ \ \ \ \ \
\ \ \ \ \ \ \qquad\ \ \ \ \ \ \ \ \ \ \ \ \ \ \ \ \ \ \ \ \ \ \ \ \ \ \ \ \
\ \ \ \ \ \ \ \ \ \ \ \ \ \ \ \ \ \ \ \ \ \ \ \ \ \ \ \ \ \ \ \ \ \ \ \ \ }

Using the above formulae, we can deduce an excited (from the vacuum) quark
spectrum. We will find the energy band excited states of the elementary $%
\epsilon $-quarks first.

\subsection{The Energy Bands\ \ \ \ \ \ \ \ \ \ \ \ \ \ \ \ \ \ \ \ \ \ \ \
\ \ \ \ \ \ \ \ \ \ \ \ \ \ \ \ \ \ \ \ \ \ \ \ \ \ \ \ \ \ \ \ \ \ \ \ \ \
\ \ \ \ \ \ \ \ \ \ \ \ \ \ \ \ \ \ \ \ \ \ \ \ \ \ \ \ \ \ \ \ \ \ \ \ \ \
\ \ \ \ \ \ \ \ \ \ \ \ \ \ \ \ \ \ \ \ \ \ \ \ \ \ \ \ \ \ \ \ \ \ \ \ \ \
\ \ \ \ \ \ \ \ \ \ \ \ \ \ \ \ \ \ \ \ \ \ \ \ \ \ \ \ \ \ \ \ \ \ \ \ \ \
\ \ \ \ \ \ \ \ \ \ \ \ \ \ \ \ \ \ \ \ \ \ \ \ \ \ \ \ \ \ \ \ \ \ \ \ \ \
\ \ \ \ \ \ \ \ \ \ \ \ \ \ \ \ \ \ \ \ \ \ \ \ \ \ \ \ \ \ \ \ \ \ \ \ \ \
\ \ \ \ \ \ \ \ \ \ \ \ \ \ \ \ \ \ \ \ \ \ \ \ \ \ \ \ \ \ \ \ \ \ \ \ \ \
\ \ \qquad\ \ \ \ \ \ \ \ \ \ \ \ \ \ \ \ \ \ \ \ \ \ \ \ \ \ \ \ \ \ \ \ \
\ \ \ \ \ \ \ \ \ \ \ \ \ \ \ \ \ \ \ \ \ \ \ \ \ \ \ \ \ \ \ \ \ \qquad}

In order to show how to calculate the energy bands, we give the calculation
of some low energy bands in the $\Delta $-axis as an example.

First we find the formulae for the E($\vec{k}$, $\vec{n}$) at the points $%
\Gamma $, H and $\Delta $ of the $\Delta $-axis (see Fig. 1). From (\ref%
{E(n,k)}), (\ref{Sym-Point}), (\ref{V=313}) and (\ref{Alpha=360}), we get: 
\begin{equation}
\text{E(}\vec{k}\text{,}\vec{n}\text{) = 313+}\Delta \text{+ }360\text{ E}_{%
\vec{k}}(\text{n}_{1},\text{n}_{2},\text{n}_{3})\text{;}  \label{Epoint}
\end{equation}
\begin{equation*}
\text{E}_{\vec{k}}(\text{n}_{1},\text{n}_{2},\text{n}_{3})\text{ = [(n}_{1}%
\text{-}\xi \text{)}^{2}\text{+(n}_{2}\text{-}\eta \text{)}^{2}\text{+(n}_{3}%
\text{-}\zeta \text{)}^{2}\text{].}
\end{equation*}
\begin{equation}
\text{For }\Gamma \text{ = (0, 0, 0), E(}\vec{k}\text{,}\vec{n}\text{) = 313+%
}\Delta \text{+ 360E}_{\Gamma }\text{, \ \ E}_{\Gamma }\text{ = (n}_{1}^{2}%
\text{ + n}_{2}^{2}\text{ + n}_{3}^{2}\text{);}  \label{Gama}
\end{equation}
\begin{equation}
\text{for H = (0, 0, 1), E(}\vec{k}\text{,}\vec{n}\text{) = 313+}\Delta 
\text{+ 360E}_{\text{H}}\text{, \ \ E}_{\text{H}}\text{ = [n}_{1}^{2}\text{
+ n}_{2}^{2}\text{ + (n}_{3}\text{-1)}^{2}\text{];}  \label{H-E}
\end{equation}
\begin{equation}
\text{for }\Delta \text{ = (0, 0, }\zeta \text{), E(}\vec{k}\text{,}\vec{n}%
\text{) = 313+}\Delta \text{+360E}_{\Delta },\text{ \ \ E}_{\Delta }\text{=
[n}_{1}^{2}\text{ + n}_{2}^{2}\text{ + (n}_{3}\text{-}\zeta \text{)}^{2}%
\text{].}  \label{E-Dalta}
\end{equation}

Then, using (\ref{Gama})--(\ref{E-Dalta}) and beginning from the lowest
energy, we get:

A1. The lowest E($\vec{k}$, $\vec{n}$) is at ($\xi $, $\eta $, $\zeta $) = 0
(the $\Gamma $-point) and $\vec{n}$ = (0, 0, 0).\ From (\ref{Gama}) and (\ref%
{Epoint}) , we have 
\begin{equation}
\vec{n}\text{ = (0, 0, 0), E}_{\Gamma }\text{(0, 0, 0) = 0, E(}%
\overrightarrow{0}\text{,}\overrightarrow{0}\text{) = 313 + }\Delta \text{.}
\end{equation}

A2. Starting from E$_{\Gamma }$= 0 [E($\overrightarrow{0}$,$\overrightarrow{0%
}$) = 313 +$\Delta $ (Mev)], from (\ref{V=313}) and (\ref{Alpha=360}), we
find that there is one energy band (the lowest energy band) E$_{\Delta }$ = $%
\zeta ^{2}$ ($\zeta $ = 0$\rightarrow $1) along the $\Delta $-axis, with n$%
_{1}$ = n$_{2}$ = n$_{3}$ = 0 (see(\ref{E-Dalta})) ended at the point E$_{H}$
= 673 +$\Delta $: 
\begin{gather}
\vec{n}\text{ = (0, 0, 0) (single band), \ } \\
\text{E}_{\Gamma }\text{\ = 0\ }\rightarrow \text{\ E}_{\Delta }\text{ = }%
\zeta ^{2}\text{ (}\zeta \text{ = 0}\rightarrow \text{1)}\rightarrow \text{E}%
_{\text{H}}\text{ =1,} \\
\text{E(}\Gamma \text{,}0\text{) = 313+}\Delta \rightarrow \text{E(}\Delta ,0%
\text{) = 313+}\Delta \text{+}\zeta ^{2}\rightarrow \text{E(H,}0\text{)= 673+%
}\Delta \text{.\ \ \ \ \ \ \ }
\end{gather}

A3. At the end point H, the energy E(H,0) = 673 +$\Delta $. When $n=(\pm
1,0,1)$, $(0,\pm 1,1)$, and (0,0,2), E(H,$\overrightarrow{n}$) = 673 +$%
\Delta $ also (see (\ref{H-E})). Starting from E$_{H}$ = 673 +$\Delta $,
along the $\Delta $-axis, we find there are three energy bands ending at the
points E$_{\Gamma }=$313 +$\Delta $, $E_{\Gamma }=$ 1033 +$\Delta $ and $%
E_{\Gamma }=$ 1753 +$\Delta $, respectively: 
\begin{gather*}
\vec{n}\text{ = (0, 0, 0) (single band), } \\
\text{E}_{H}\text{ =1}\rightarrow \text{E}_{\Delta }\text{= }\zeta ^{2}\text{%
\ (\ }\zeta \text{=1}\rightarrow \text{0)}\rightarrow \text{E}_{\Gamma }%
\text{\ = 0,} \\
\text{E{\small (H, 0) }=673+}\Delta \rightarrow \text{E}(\Delta ,0)\text{
=313+}\Delta \text{+}\alpha \text{ }\zeta ^{2}\rightarrow \text{E{\small (}}%
\Gamma \text{, {\small 0) }= 313+}\Delta \text{\ ; }
\end{gather*}
\begin{gather*}
\vec{n}\text{ = (0, 0, 2) (single band), \ \ } \\
\text{E}_{H}\text{ =1}\rightarrow \text{E}_{\Delta }\text{= (2-}\zeta \text{)%
}^{2}\text{\ (\ }\zeta \text{=1}\rightarrow \text{0)}\rightarrow \text{E}%
_{\Gamma }\text{\ = 4,} \\
\text{E{\small (H,002)}= 673+}\Delta \rightarrow \text{E}_{\Delta }\text{=
313+}\Delta \text{+}\alpha \text{(2-}\zeta \text{)}^{2}\rightarrow \text{E}%
{\small (\Gamma },{\small 002)}\text{ =1753+}\Delta \text{; }
\end{gather*}
\begin{gather*}
\vec{n}\text{ = (}{\small \pm }\text{1, 0, 1) and (0,}\pm \text{1, 1)
(fourfold degeneracy), \ \ } \\
\text{E}_{H}\text{ =1}\rightarrow \text{E}_{\Delta }\text{= 313+}\Delta
+\alpha \text{[1+(1- }\zeta \text{)}^{2}\text{]\ (\ }\zeta \text{=1}%
\rightarrow \text{0)}\rightarrow \text{E}_{\Gamma }\text{\ = 2,} \\
\text{E(H;}{\small \pm }\text{{\small 1,0,1)} =673+}\Delta \rightarrow \text{%
E}_{\Delta }\text{=673+}\Delta \text{+}\alpha \text{(1-}\zeta \text{)}%
^{2}\rightarrow \text{E}{\small (\Gamma }\text{{\small ;}}{\small \pm 1}%
\text{{\small ,0,1)}=1033+}\Delta \text{,} \\
\text{E(H;{\small 0}}{\small ,\pm }\text{{\small 1,1}}{\small )}\text{ =673+}%
\Delta \rightarrow \text{E}_{\Delta }\text{=673+}\Delta \text{+}\alpha \text{%
(1-}\zeta \text{)}^{2}\rightarrow \text{E}{\small (\Gamma }\text{{\small ;0,}%
}{\small \pm }\text{{\small 1,1}}{\small )}\text{=1033+}\Delta \text{.}
\end{gather*}
Continuing this process, we can find all low energy bands of the $\Delta $%
-axis. We show the energy bands in Table B1 of the Appendix B.\ \ 

Similarly, we deduce all low energy bands on the $\Lambda $-axis, the\ $%
\Sigma $-axis, the D-axis, the F-axis and the G-axis. These low energy bands
are sufficient to cover experimental data. We show these energy bands in
Table B2--Table B7 of the Appendix B.\ \ \ \ \ \ \ \ \ \ \ \ \ \ \ \ \ \ \ \
\ \ \ \ \ \ \ \ \ \ \ \ \ \ \ \ \ \ \ \ \ \ \ \ \ \ \ \ \ \ \ \ \ \ \ \ \ \
\ \ \ \ \ \ \ \ \ \ \ \ \ \ \ \ \ \ \ \ \ \ \ \ \ \ \ 

\subsection{The Excited Quarks of the Elementary $\protect\epsilon $-Quarks
\ \ \ \ \ \ \ \ \ \ \ \ \ \ \ \ \ \ \ \ \ \ \ \ \ \ \ \ \ \ \ \ \ \ \ \ \ \
\ \ \ \ \ \ \ \ \ \ \ \ \ \ \ \ \ \ \ \ \ \ \ \ \ \ \ \ \ \ \ \ \ \ \ \ \ \
\ \ \ \ \ \ \ \ \ \ \ \ \ \ \ \ \ \ \ \ \ \ \ \ \ \ \ \ \ \ \ \ \ \ \ \ \ \
\ \ \ \ \ \ \ \ \ \ \ \ \ \ \ \ \ \ \ \ \ \ \ \ \ \ \ \ \ \ \ \ \ \ \ \ \ \
\ \ \ \ \ \ \ \ \ \ \ \ \ \ \ \ \ \ \ \ \ \ \ \ \ \ \ \ \ \ \ \ \ \ \ \ \ \
\ \ \ \ \ \ \ \ \ \ \ \ \ \ \ \ \ \ \ \ \ \ \ \ \ \ \ \ \ \ \ \ \ \ \ \ \ \
\ \ \ \ \ \ \ \ \ \ \ \ \ \ \ \ \ \ \ \ \ \ \ \ \ \ \ \ \ \ \ \ \ \ \ \ \ \
\ \ \ \ \ \ \ \ \ \ \ \ \ \ \ \ \ \ \ \ \ \ \ \ \ \ \qquad\ \ \ \ \ \ \ \ \
\ \ \ \ \ \ \ \ \ \ \ \ \ \ \ \ \ \ \ \ \ \ \ \ \ \ \ \ \ \ \ \ \ \ \ \ \ \
\ \ \ \ \ \ \ \ \ \ \ \ \ \ \ \ \ \ \ \qquad}

\subsubsection{The Quarks on the $\Delta$-Axis (the $\Gamma$-H axis)\textbf{%
\qquad}}

Since the $\Delta $-axis is a fourfold rotatory axis (see Fig. 1), R = 4.
From (\ref{S-Number}), we get S = 0. Because the axis has $R=4$, we can use (%
\ref{deg > R}) and (\ref{Subdeg}) to determine that the energy bands with
8-fold degeneracy will be divided into two fourfold degenerate bands (K = 0
from Table A3). \ \ \ \ \ \ \ \ \ \ \ \ \ \ \ \ \ \ \ \ \ \ 

1. The Quarks of the Fourfold Degenerate Bands on the $\Delta $-Axis

For fourfold degenerate bands, using (\ref{IsoSpin}), we get I = $\frac{3}{2}
$ and I$_{Z}^{^{\prime }}=\frac{\text{3}}{2}$, $\frac{\text{1}}{2}$, -$\frac{%
\text{1}}{2}$, -$\frac{\text{3}}{2}$, from (\ref{Formal-Iz}). Thus each
fourfold degenerate band represents a fourfold quark family, q$_{\Delta }$(=q%
$_{\Delta }^{\frac{3}{2}}$, q$_{\Delta }^{\frac{1}{2}}$, q$_{\Delta }^{\frac{%
-1}{2}}$, q$_{\Delta }^{\frac{-3}{2}}$), with 
\begin{equation}
\text{B = }\frac{\text{1}}{3}\text{, S = 0, I = }\frac{\text{3}}{2}\text{, I}%
_{z}^{^{\prime }}\text{ = }\frac{\text{3}}{2}\text{, }\frac{\text{1}}{2}%
\text{, -}\frac{\text{1}}{2}\text{, -}\frac{\text{3}}{2}\text{.}
\label{3/2-Quark}
\end{equation}
Using Table B1, we can get E$_{\Gamma }$, E$_{\text{H}}$ and $\vec{n}$
values. When we put the values of E$_{\Gamma }$ and E$_{\text{H}}$ into the
rest mass formula (\ref{Rest Mass}), we can find the rest mass m$_{q_{\Delta
}}^{\ast }$. $\func{Si}$nce $\Theta $ = K = $\Delta $S = 0 \{see Table A3
and A4\}, from (\ref{Dalta-E}), $\Delta $E =0. Thus we have [q$_{\Delta }$ =
(q$_{\Delta }^{\frac{3}{2}}$, q$_{\Delta }^{\frac{1}{2}}$, q$_{\Delta }^{%
\frac{-1}{2}}$, q$_{\Delta }^{\frac{-3}{2}}$)]:

\ \ \ \ \ \ \ \ \ \ \ \ \ \ \ \ \ \ \ \ \ \ \ \ \ \ \ \ \ \ \ \ 

\begin{tabular}{l}
\ \ \ \ \ \ \ \ \ \ \ \ \ \ \ \ \ \ \ \ \ \ \ \ \ \ \ \ \ Table 3. The q$%
_{\Delta }$(m$^{\ast }$)-quarks \\ 
\begin{tabular}{|l|l|l|l|l|l|l|l|}
\hline
E$_{Point}$ & $\text{(n}_{1}\text{n}_{2}\text{n}_{3}\text{, ... )}$ & E($%
\overrightarrow{k,}\overrightarrow{n}$) & I & $\Delta $S & J & $\Delta E$ & $%
q_{\Delta }${\small (m}$^{\ast }${\small (Mev))} \\ \hline
$\text{E}_{H}\text{=1}$ & $\text{(101,-101,011,0-11)}$ & $\text{673}$ & $%
\frac{3}{2}$ & 0 & 0 & 0 & $\text{q}_{\Delta }\text{(673+}\Delta \text{)}$
\\ \hline
$\text{E}_{\Gamma }\text{=2}$ & $\text{(110,1-10,-110,-1-10,}$ & $\text{1033}
$ & $\frac{3}{2}$ & 0 & 0 & 0 & $\text{q}_{\Delta }\text{(1033+}\Delta \text{%
{\small )}}$ \\ \hline
& $\text{10-1,-10-1,01-1,0-1-1)}$ & $\text{1033}$ & $\frac{3}{2}$ & 0 & 0 & 0
& $\text{q}_{\Delta }\text{(1033+}\Delta \text{{\small )}}$ \\ \hline
$\text{E}_{H}\text{=3}$ & $\text{(112,1-12,-112,-1-12)}$ & $\text{1393}$ & $%
\frac{3}{2}$ & 0 & 0 & 0 & $\text{q}_{\Delta }\text{(1393+}\Delta \text{)}$
\\ \hline
$\text{E}_{\Gamma }\text{=4}$ & $\text{(200,-200,020,0-20)}$ & $\text{1753}$
& $\frac{3}{2}$ & 0 & 0 & 0 & $\text{q}_{\Delta }\text{(1753+}\Delta \text{)}
$ \\ \hline
$\text{E}_{H}\text{=5}$ & $\text{(121,1-21,-121,--1-21,}$ & $\text{2113}$ & $%
\frac{3}{2}$ & 0 & 0 & 0 & $\text{q}_{\Delta }\text{(2113+}\Delta \text{)}$
\\ \hline
& $\text{{\small \ }211,2-11,-211,-2-11)}$ & $\text{2113}$ & $\frac{3}{2}$ & 
0 & 0 & 0 & $\text{q}_{\Delta }\text{(2113+}\Delta \text{)}$ \\ \hline
$\text{E}_{H}\text{=5}$ & $\text{(202,-202,022,0-22)}$ & $\text{2113}$ & $%
\frac{3}{2}$ & 0 & 0 & 0 & $\text{q}_{\Delta }\text{(2113+}\Delta \text{)}$
\\ \hline
$\text{E}_{H}\text{=5}$ & $\text{(013,0-13,103,-103)}$ & $\text{2113}$ & $%
\frac{3}{2}$ & 0 & 0 & 0 & $\text{q}_{\Delta }\text{(2113+}\Delta \text{)}$
\\ \hline
$\text{E}_{\Gamma }\text{=6}$ & $\text{(12}\overline{\text{1}}\text{,1}%
\overline{\text{2}}\overline{\text{1}}\text{,}\overline{\text{1}}\text{21,}%
\overline{\text{1}}\overline{\text{2}}\overline{\text{1}},$ & $\text{2473}$
& $\frac{3}{2}$ & 0 & 0 & 0 & $\text{q}_{\Delta }\text{({\small 2473}+}%
\Delta \text{)}$ \\ \hline
& {\small \ 21}$\overline{\text{1}}${\small ,2}$\overline{\text{1}}\overline{%
\text{1}}${\small ,}$\overline{\text{2}}${\small 1}$\overline{\text{1}}$%
{\small ,}$\overline{\text{2}}\overline{\text{1}}\overline{\text{1}}${\small %
)} & $\text{2473}$ & $\frac{3}{2}$ & 0 & 0 & 0 & $\text{q}_{\Delta }\text{(%
{\small 2473}+}\Delta \text{)}$ \\ \hline
$\text{E}_{\Gamma }\text{=6}$ & $\text{(11}\overline{\text{2}}\text{,1}%
\overline{\text{1}}\overline{\text{2}}\text{,}\overline{\text{1}}\text{1}%
\overline{\text{2}}\text{,}\overline{\text{1}}\overline{\text{1}}\overline{%
\text{2}}\text{)}$ & $\text{2473}$ & $\frac{3}{2}$ & 0 & 0 & 0 & $\text{q}%
_{\Delta }\text{({\small 2473}+}\Delta \text{)}$ \\ \hline
\end{tabular}%
\end{tabular}

\ \ \ \ \ \ \ \ \ \ \ \ \ \ \ \ \ \ \ \ \ \ \ \ \ \ \ \ \ \ \ \qquad \qquad

2. The Quarks of the Single Energy Bands on the $\Delta $-Axis

For the single bands on the $\Delta $-axis (see Table B1), R = 4, S$_{\Delta
}$ = 0 from (\ref{S-Number}); deg = 1, I = 0 from (\ref{IsoSpin}). For the
single bands, using (\ref{S+DS}) instead of (\ref{S-Number}), we find that $%
\Delta $S = +1 at E$_{\Gamma }$\ = 4, 16, 36, ... from (\ref{DaltaS}) and $%
\Delta $S = -1 at E$_{\text{H}}$= 1, 9, 25, 49, ... from (\ref{DaltaS}). For 
$\overrightarrow{n}$ = (0, 0, 0), from Table 2, S = C = b = 0; from (\ref%
{Dalta-E}), $\Delta $E = 0. For other energy bands, $\Theta $\ = K = b = 0
(see Table A3), from (\ref{Dalta-E}), $\Delta $E = 100$\text{ }$S J$%
_{S}\Delta S$,\ J$_{S}$ = 1, 2, 3,.... Using (\ref{Charmed}), we have:

\ \ \ \ \ \ \ \ \ \ \ \ \ \ \ \ \ \ \ \ \ \ 

\begin{tabular}{l}
\ \ \ \ \ \ \ \ \ \ \ \ \ \ Table 4. The u$_{C}$(m$^{\ast }$)-quarks and the
d$_{S}$(m$^{\ast }$)-quarks \\ 
\begin{tabular}{|l|l|l|l|l|l|l|l|l|l|}
\hline
$\text{E}_{Point}$ & E($\overrightarrow{k}$,$\overrightarrow{n}$) & $\text{n}%
_{1,}\text{n}_{2,}\text{n}_{3}$ & $\Delta \text{S}$ & \ \ J & I & S & C & $%
\Delta \text{E}$ & $q_{\text{{\small Name}}}${\small (m}$^{\ast }${\small %
(Mev))} \\ \hline
$\text{E}_{\Gamma }\text{=0}$ & 313 & $\text{{\small 0,\ \ 0, \ 0}}$ & 0 & J$%
\text{ = 0}$ & $\frac{1}{2}$ & 0 & 0 & 0 & $\text{u(313+}\Delta \text{)}$ \\ 
\hline
$\text{E}_{H}\text{=1}$ & 673 & $\text{{\small 0, \ 0, \ 2}}$ & -1 & J$_{%
\text{S,H}}\text{=1}$ & 0 & -1 & 0 & 100 & $\text{d}_{S}\text{(773+}\Delta 
\text{)}$ \\ \hline
$\text{E}_{\Gamma }\text{=4}$ & 1753 & $\text{{\small 0, \ 0, -2}}$ & +1 & J$%
_{\text{C,}\Gamma }\text{=1}$ & 0 & 0 & 1 & 0 & $\text{u}_{C}\text{(1753+}%
\Delta \text{)}$ \\ \hline
$\text{E}_{H}\text{=9}$ & 3553 & $\text{{\small 0, \ 0, \ 4}}$ & -1 & J$_{%
\text{S,H}}\text{=2}$ & 0 & -1 & 0 & 200 & $\text{d}_{S}\text{(3753+}\Delta 
\text{)}$ \\ \hline
$\text{E}_{\Gamma }\text{=16}$ & 6073 & $\text{{\small 0, \ 0, -4}}$ & +1 & J%
$_{\text{C,}\Gamma }\text{=2}$ & 0 & 0 & 1 & 0 & $\text{u}_{C}\text{(6073+}%
\Delta \text{)}$ \\ \hline
$\text{E}_{H}\text{=25}$ & $\text{9313}$ & $\text{{\small 0, \ 0, \ 6}}$ & -1
& J$_{\text{S,H}}\text{=3}$ & 0 & -1 & 0 & 300 & $\text{d}_{S}\text{(9613+}%
\Delta \text{)}$ \\ \hline
$\text{E}_{\Gamma }\text{=36}$ & $\text{13273}$ & $\text{{\small 0, \ 0, -6}}
$ & +1 & J$_{\text{C,}\Gamma }\text{=3}$ & 0 & 0 & 1 & 0 & $\text{u}_{C}%
\text{(13273+}\Delta \text{)}$ \\ \hline
$\text{E}_{H}\text{=49}$ & $\text{17953}$ & $\text{{\small 0, \ 0, \ 8}}$ & 
-1 & J$_{\text{S,H}}\text{=4}$ & 0 & -1 & 0 & 400 & $\text{d}_{S}\text{%
(18353+}\Delta \text{)}$ \\ \hline
\end{tabular}%
\end{tabular}

\subsubsection{The Quarks on the $\Sigma $-Axis ($\Gamma $-N)}

The $\Sigma $-axis is a twofold rotatory axis, R = 2, from (\ref{S-Number})
S = - 2 .

1. The Quarks of the Twofold Degenerate Energy Bands on the $\Sigma $-axis ($%
\Gamma $-N)

For the twofold degenerate energy bands, each represents a quark family q$%
_{\Xi }$ (q$_{\Xi }^{\frac{1}{2}}$, q$_{\Xi }^{-\frac{1}{2}}$) with B = 1/3,
S = -2, I = 1/2 from (\ref{IsoSpin}), I$_{z}^{^{\prime }}$ = $\frac{1}{2}$, -%
$\frac{1}{2}$ from (\ref{Formal-Iz}) and Q =$\frac{2}{3}$, $\frac{-1}{3}$
from (\ref{2/3}) and (\ref{-1/3})$.$\ $\func{Si}$nce $\Theta $ = K = $\Delta 
$S = 0 (see Table A3 and Table A4), from (\ref{Dalta-E}), $\Delta $E =0.
Similar to Table 3, we have [$q_{\Xi }$ (q$_{\Xi }^{\frac{1}{2}}$, q$_{\Xi
}^{-\frac{1}{2}}$)]: $\ $\ \ \ \ \ \ \ \ \ \ \ \ \ \ \ \ \ \ \ 

\bigskip 
\begin{tabular}{l}
\ \ \ \ \ \ \ \ \ \ \ \ Table 5. The $\text{q}_{\Xi }\text{(m}^{\ast }\text{%
)-}$quarks on Twofold Energy Bands \\ 
$%
\begin{tabular}{|l|l|l|l|l|l|l|l|l|}
\hline
E$_{Point}$ & $\text{\ \ n}_{1}\text{n}_{2}\text{n}_{3}$ & $S_{\Xi }$ & I & 
E($\overrightarrow{k}$,$\overrightarrow{n}$) & $\Delta $S & J & $\Delta $E & 
$\text{q}_{\Xi }\text{(m}^{\ast }\text{(Mev))}$ \\ \hline
$E_{\Gamma }=2$ & $(\text{1-10,-110)}$ & -2 & $\frac{1}{2}$ & $\text{1033}$
& 0 & 0 & 0 & $\text{q}_{\Xi }\text{(1033+}\Delta \text{)}$ \\ \hline
$E_{N}=\frac{5}{2}$ & $\text{(200,020)}$ & -2 & $\frac{1}{2}$ & $\text{1213}$
& 0 & 0 & 0 & $\text{q}_{\Xi }\text{(1213+}\Delta \text{)}$ \\ \hline
$E_{\Gamma }=4$ & $\text{(002,00-2}$ & -2 & $\frac{1}{2}$ & $\text{1753}$ & 0
& 0 & 0 & $\text{q}_{\Xi }\text{(1753+}\Delta \text{)}$ \\ \hline
& $\text{-200,0-20)}$ & -2 & $\frac{1}{2}$ & $\text{1753}$ & 0 & 0 & 0 & $%
\text{q}_{\Xi }\text{(1753+}\Delta \text{)}$ \\ \hline
$E_{N}=\frac{9}{2}$ & $\text{(112,11-2)}$ & -2 & $\frac{1}{2}$ & $\text{1933}
$ & 0 & 0 & 0 & $\text{q}_{\Xi }\text{(1933+}\Delta \text{)}$ \\ \hline
... & ... & ... & ... & ... & ... & ... & ... & .... \\ \hline
\end{tabular}
\ \ $%
\end{tabular}
\ \ \ \ \ \ \ \ \ \ \ \ \ \ \ 

2. The\ Quarks of the Fourfold Degenerate Energy Bands on the $\Sigma -$Axis

According to (\ref{deg > R}) and (\ref{Subdeg}), each fourfold degenerate
energy band on the $\Sigma $-axis with R = 2 divides into two twofold
degenerate bands. From Table 5, each of them represents a quark family q$%
_{\Xi }$ (q$_{\Xi }^{\frac{1}{2}}$, q$_{\Xi }^{-\frac{1}{2}}$) with B = 1/3,
S = -2, I = 1/2, I$_{z}^{^{\prime }}$ = $\frac{1}{2}$, -$\frac{1}{2}$ and Q
= $\frac{2}{3}$, $\frac{-1}{3}$.\ Thus we have [since $\Theta $ = K = $%
\Delta $S = 0, from (\ref{Dalta-E}), $\Delta $E =0]:

\ \ \ \ \ \ \ \ \ \ \ \ \ \ \ \ 

\begin{tabular}{l}
\ \ \ \ \ \ \ \ \ \ \ \ \ \ \ \ \ Table 6. The $\text{q}_{\Xi }\text{(m}%
^{\ast }\text{)-}$quarks on Fourfold Energy Bands \\ 
$%
\begin{tabular}{|l|l|l|l|l|l|l|l|}
\hline
E$_{Point}$ & $\text{(n}_{1}\text{n}_{2}\text{n}_{3}\text{, ... )}$ & I & S
& $\Delta $S & J & $\Delta $E & $\text{q}_{\Xi }${\small (m}$^{\ast }$%
{\small (Mev))} \\ \hline
$E_{N}=\frac{3}{2}$ & $(\text{101,10-1,011,01-1})$ & $\frac{1}{2}$ & -2 & 0
& 0 & 0 & $\text{2}\times \text{q}_{\Xi }\text{(853+}\Delta \text{)}$ \\ 
\hline
$E_{\Gamma }=2$ & $(\text{-101,-10-1,0-11,0-1-1})$ & $\frac{1}{2}$ & -2 & 0
& 0 & 0 & $\text{2}\times \text{q}_{\Xi }\text{(1033+}\Delta \text{)}$ \\ 
\hline
$E_{N}=\frac{7}{2}$ & $(\text{121,12-1,211,21-1})$ & $\frac{1}{2}$ & -2 & 0
& 0 & 0 & $\text{2}\times \text{q}_{\Xi }\text{(1573+}\Delta $ \\ \hline
... & ... & ... & ... & ... & ... & ... & ... . \\ \hline
\end{tabular}
\ \ \ \ $%
\end{tabular}

\ \ \ \ \ \ \ \ \ \ \ \ \ \ \ \ \ \ \ \ \ \ \ \ 

3. The Quarks of the Single Energy Bands on the $\Sigma $-axis ($\Gamma $-N)

For the single bands on the $\Sigma $-axis (see Table B2), R = 2, S$_{\Sigma
}$ = -2 from (\ref{S-Number}); deg = 1, I = 0 from (\ref{IsoSpin}). Using (%
\ref{S+DS}) instead of (\ref{S-Number}), we get $\Delta $S = +1 at E$_{N}$ = 
$\frac{1}{2}$, $\frac{9}{2}$, $\frac{25}{2}$, $\frac{49}{2}$, $\frac{81}{2}$%
, \ ... from (\ref{DaltaS}); at E$_{\Gamma }$\ = 2, 8, 18, 32, ... $\Delta $%
S = -1 from (\ref{DaltaS}) . For $\overrightarrow{n}$ = (0, 0, 0), from
Table 2, S = b = $\Theta $ = K = 0, $\Delta $E = 0 (\ref{Dalta-E}). For
other bands, since $\Theta $ = K = 0, from (\ref{Dalta-E}), $\Delta $E = -
100(S+b)(J$_{S,b}$-2)$\Delta $S, J$_{S}$ = 3, 4, 5, ...; $\Delta $E = 0 J$%
_{S}$ \TEXTsymbol{<} 3. Using (\ref{Battom}), similar to Table 4, we have:\ 

\ \ \ \ \ \ \ \ \ \ \ \ \ \ \ \ \ \ \ \ \ \ \ \ \ \ \ \ \ \ \ \ \ \ \ \ \ \ 

\begin{tabular}{l}
\ \ \ \ \ \ \ \ \ \ \ \ \ \ \ \ \ \ Table 7. The d$_{\Omega }$(m$^{\ast }$%
)-quarks and d$_{S}$(m$^{\ast }$)-quarks \\ 
$%
\begin{tabular}{|l|l|l|l|l|l|l|l|l|l|}
\hline
E$_{Point}$ & $\text{\ n}_{1}\text{n}_{2}\text{n}_{3}$ & $\Delta \text{S}$ & 
S & b & $\ \ \ \text{J}$ & I & E($\overrightarrow{k}$,$\overrightarrow{n}$)
& $\Delta \text{E}$ & $d${\small (m}$^{\ast }${\small (Mev))} \\ \hline
$\text{E}_{\Gamma }\text{=0}$ & (0, 0, 0) & +2 & 0 & 0 & J$_{\text{S,}\Gamma
}\text{ =0}$ & $\frac{1}{2}$ & 313 & \ \ 0 & d(313$\text{+}\Delta $) \\ 
\hline
$\text{E}_{N}\text{=}\frac{\text{1}}{2}$ & $(\text{1,1,0})$ & +1 & -1 & 0 & J%
$_{\text{S,N}}\text{ =1}$ & 0 & $\text{493}$ & \ \ 0 & $\text{d}_{S}\text{%
(493+}\Delta \text{)}$ \\ \hline
$\text{E}_{\Gamma }\text{=2}$ & $(\text{-1,-1,0})$ & -1 & -3 & 0 & J$_{\text{%
S,}\Gamma }\text{ =1}$ & 0 & $\text{1033}$ & \ \ 0 & $\text{d}_{\Omega }%
\text{(1033+}\Delta \text{)}$ \\ \hline
$\text{E}_{N}\text{=}\frac{\text{9}}{2}$ & $(\text{2,2,0})$ & +1 & -1 & 0 & J%
$_{\text{S,N}}\text{ =2}$ & 0 & $\text{1933}$ & \ \ 0 & $\text{d}_{S}\text{%
(1933+}\Delta \text{)}$ \\ \hline
$\text{E}_{\Gamma }\text{=8}$ & $(\text{-2,-2,0})$ & -1 & -3 & 0 & J$_{\text{%
S,}\Gamma }\text{ =2}$ & 0 & $\text{3193}$ & \ \ 0 & $\text{d}_{\Omega }%
\text{(3193+}\Delta \text{)}$ \\ \hline
$\text{E}_{N}\text{=}\frac{\text{25}}{2}$ & $(\text{3,3,0})$ & +1 & 0 & -1 & 
J$_{\text{S,N}}\text{ =3}$ & 0 & $\text{4813}$ & 100 & $\text{d}_{b}\text{%
(4913+}\Delta \text{)}$ \\ \hline
$\text{E}_{\Gamma }\text{=18}$ & $(\text{-3,-3,0})$ & -1 & -3 & 0 & J$_{%
\text{S,}\Gamma }\text{ =3}$ & 0 & $\text{6793}$ & -300 & $\text{d}_{\Omega }%
\text{(6493+}\Delta \text{)}$ \\ \hline
$\text{E}_{N}\text{=}\frac{\text{49}}{2}$ & $(\text{4,4,0})$ & +1 & 0 & -1 & 
J$_{\text{S,N}}\text{ =4}$ & 0 & $\text{9133}$ & 200 & $\text{d}_{b}\text{%
(9333+}\Delta \text{)}$ \\ \hline
$\text{E}_{\Gamma }\text{=32}$ & $\text{(-4,-4,0)}$ & -1 & -3 & 0 & J$_{%
\text{S,}\Gamma }\text{ =4}$ & 0 & $\text{11833}$ & -600 & $\text{d}_{\Omega
}\text{(11233+}\Delta \text{)}$ \\ \hline
$\text{E}_{N}\text{=}\frac{8\text{1}}{2}$ & $(\text{5,5,0})$ & +1 & 0 & -1 & 
J$_{\text{S,N}}\text{ =5}$ & 0 & $\text{14893}$ & 300 & $\text{d}_{b}\text{%
(15193+}\Delta \text{)}$ \\ \hline
\end{tabular}
\ \ \ \ \ $%
\end{tabular}

\subsubsection{The Quarks on the $\Lambda $-Axis ($\Gamma $-P)}

Since the $\Lambda $-axis is a threefold rotatory axis (see Fig. 1), R = 3,
from (\ref{S-Number}) we have S = -1. From Table B3, we see that there are
two single energy bands with $\vec{n}$ = (0, 0, 0) and\ $\vec{n}$ = (2, 2,
2), and all other bands are either threefold degenerate energy bands (deg =
3) or sixfold degenerate bands (deg = 6). From (\ref{deg > R}) and (\ref%
{Subdeg}), the sixfold degenerate energy bands will divide into two
threefold energy bands.

1. The Quarks of the Threefold Degenerate Energy Bands on the $\Lambda $%
-Axis ($\Gamma $-P)

For the threefold degenerate energy bands, using (\ref{IsoSpin}), (\ref%
{Formal-Iz}), (\ref{2/3}) and (\ref{-1/3}), we have I = 1 and I$%
_{z}^{^{\prime }}$ = 1, 0, -1. Thus we get a three-member quark family q$%
_{\Sigma }$(q$_{\Sigma }^{1}$, q$_{\Sigma }^{0}$, q$_{\Sigma }^{-1}$) with B
= 1/3, S = -1 and\ I = 1. $\func{Si}$nce $\Theta $ = K = $\Delta $S = 0,
from (\ref{Dalta-E}), $\Delta $E =0.$\ $Using Table B3, we have:

\ \ \ \ \ \ \ \ \ \ \ \ \ \ \ \ \ \ \ \ 

\begin{tabular}{l}
\ \ \ \ \ \ \ \ \ \ \ \ \ \ \ Table 8. The q$_{\Sigma }$(m$^{\ast }$)-quarks
(S = S$_{Ax}$ + $\Delta $S = -1) \\ 
$%
\begin{tabular}{|l|l|l|l|l|l|l|l|}
\hline
E$_{Point}$ & ($\text{n}_{1}\text{n}_{2}\text{n}_{3}$, ...) & E($%
\overrightarrow{k}$,$\overrightarrow{n}$) & I & $\Delta $S & J & $\Delta $E
& $\text{q}_{\Sigma }\text{(m}^{\ast }\text{)}$ \\ \hline
$\text{E}_{\text{P}}\text{=}\frac{\text{3}}{4}\text{ }$ & $\text{{\small %
(101,011,110)}}$ & {\small 583} & 1 & 0 & 0 & 0 & $\text{q}_{\Sigma }\text{%
(583+}\Delta \text{)}$ \\ \hline
$\text{E}_{\Gamma }\text{=2 }$ & $\text{{\small (1-10,-110,01-1,}}$ & $\text{%
{\small 1033}}$ & 1 & 0 & 0 & 0 & $\text{q}_{\Sigma }\text{({\small 1033}+}%
\Delta \text{)}$ \\ \hline
& $\text{{\small \ \ 0-11,10-1,-101})}$ & $\text{{\small 1033}}$ & 1 & 0 & 0
& 0 & $\text{q}_{\Sigma }\text{({\small 1033}+}\Delta \text{)}$ \\ \hline
$\text{E}_{\Gamma }\text{=2 }$ & $\text{{\small (-10-1,0-1-1,-1-10)}}$ & $%
\text{{\small 1033}}$ & 1 & 0 & 0 & 0 & $\text{q}_{\Sigma }\text{({\small %
1033}+}\Delta \text{)}$ \\ \hline
$\text{E}_{\text{P}}\text{=}\frac{\text{11}}{4}\text{ }$ & $\text{{\small %
(020,002,200) \ }}$ & $\text{{\small 1303}}$ & 1 & 0 & 0 & 0 & $\text{q}%
_{\Sigma }\text{(1303+}\Delta \text{)}$ \\ \hline
$\text{E}_{\text{P}}\text{=}\frac{\text{11}}{\text{4}}$ & $\text{{\small %
(121,211,112) }}$ & $\text{{\small 1303}}$ & 1 & 0 & 0 & 0 & $\text{q}%
_{\Sigma }\text{(1303+}\Delta \text{)}$ \\ \hline
$\text{E}_{\Gamma }\text{=4}$ & $\text{{\small (0-20,-200,00-2)}}$ & $\text{%
{\small 1753}}$ & 1 & 0 & 0 & 0 & $\text{q}_{\Sigma }\text{(1753+}\Delta 
\text{)}$ \\ \hline
$\text{E}_{\text{P}}\text{=}\frac{\text{19}}{\text{4}}$ & $\text{{\small %
(1-12,-112,21-1,}}$ & $\text{{\small 2023}}$ & 1 & 0 & 0 & 0 & $\text{q}%
_{\Sigma }\text{({\small 2023}+}\Delta \text{)}$ \\ \hline
& $\text{{\small \ \ 2-11,12-1,-121)}}$ & $\text{{\small 2023}}$ & 1 & 0 & 0
& 0 & $\text{q}_{\Sigma }\text{({\small 2023}+}\Delta \text{)}$ \\ \hline
$\text{E}_{\text{P}}\text{=}\frac{\text{19}}{\text{4}}$ & $\text{{\small %
(202,022,220)}}$ & $\text{{\small 2023}}$ & 1 & 0 & 0 & 0 & $\text{q}%
_{\Sigma }\text{({\small 2023}+}\Delta \text{)}$ \\ \hline
$\text{E}_{\Gamma }\text{=6 }$ & $\text{{\small (-211,2-1-1,-1-12,}}$ & $%
\text{{\small 2473}}$ & 1 & 0 & 0 & 0 & $\text{q}_{\Sigma }\text{({\small %
2473}+}\Delta \text{)}$ \\ \hline
$\text{ }$ & $\text{{\small \ \ \ 11-2,-12-1,1-21)}}$ & $\text{{\small 2473}}
$ & 1 & 0 & 0 & 0 & $\text{q}_{\Sigma }\text{({\small 2473}+}\Delta \text{)}$
\\ \hline
$\text{E}_{\Gamma }\text{=6 }$ & $\text{{\small (-1-21,1-2-1,-11-2,}}$ & $%
\text{{\small 2473}}$ & 1 & 0 & 0 & 0 & $\text{q}_{\Sigma }\text{({\small %
2473}+}\Delta \text{)}$ \\ \hline
& $\text{{\small \ \ 1-1-2,-21-1,-2-11)}}$ & $\text{{\small 2473}}$ & 1 & 0
& 0 & 0 & $\text{q}_{\Sigma }\text{({\small 2473}+}\Delta \text{)}$ \\ \hline
$\text{E}_{\Gamma }\text{=6 }$ & $\text{{\small (-1-2-1,-1-1-2,-2-1-1)}}$ & $%
\text{{\small 2473}}$ & 1 & 0 & 0 & 0 & $\text{q}_{\Sigma }\text{({\small %
2473}+}\Delta \text{)}$ \\ \hline
... & ... & ... & ... & ... & ... & ... & .... \\ \hline
\end{tabular}
\ $%
\end{tabular}

\ \ \ \ \ \ \ \ \ \ \ \ \ \ \ \ \ \ \ \ \ \ \ \ 

2. The Single Energy Bands

For the single bands, S$_{Ax}$ = -1, I = 0 and $\Delta $E = 0. From Table 2,
for $\overrightarrow{n}$= (0, 0, 0), $\Delta $S = -S$_{axis}\rightarrow $ S
= 0. At E$_{\text{P}}$ = $\frac{27}{4},\overrightarrow{n}$= (2, 2, 2), S$%
_{Ax}$ = -1, I = 0 and $\Delta $S = $\Delta $E = 0:

\ \ \ \ \ \ \ \ \ \ \ \ \ \ \ \ \ \ \ \ \ \ \ \ 

\begin{tabular}{l}
\ \ \ \ \ \ \ \ \ \ \ Table 9. The d$_{S}(m^{\ast })$-quarks on the Single
Bands of the $\Lambda $-axis \\ 
\begin{tabular}{|l|l|l|l|l|l|l|l|l|l|l|}
\hline
E$_{\text{Point}}$ & $\text{n}_{1},\text{n}_{2},\text{n}_{3}$ & E($%
\overrightarrow{k}$,$\overrightarrow{n}$) & I & S$_{\text{Axis}}$ & $\Delta $%
S & J & S & C & b & d{\small (m}$^{\ast }${\small (Mev))} \\ \hline
$\text{E}_{\Gamma }\text{= 0 }$ & $\text{{\small (0, 0, 0)}}$ & 313 & $\frac{%
1}{2}$ & -1 & 1 & 0 & 0 & 0 & 0 & $\text{d(313+}\Delta \text{)}$ \\ \hline
$E_{\text{P}}=\frac{27}{4}$ & $\text{{\small (2, 2, 2)}}$ & $\text{{\small %
2743}}$ & 0 & -1 & 0 & 0 & -1 & 0 & 0 & $\text{d}_{S}\text{(2743+}\Delta 
\text{)}$ \\ \hline
... & ... & ... & ... & ... &  & ... & ... & ... & ... & ... . \\ \hline
\end{tabular}%
\end{tabular}

\ \ \ \ \ \ \ \ \ \ \ \ \ \ \ \ \ \ \ \ \ \ \ \ \ \ \ \ \ \ \ \ \ \ \ \ \ 

Continuing the above procedure, we\ can deduce low energy excited quarks on
the D-axis, the F-axis and the G-axis. Using Table B1-Table B7, we show the
excited quarks of low energies that are sufficient to cover experimental
data (see Appendix B).\ 

\subsection{The Quark Spectrum\ \ \ \ \ \ \ \ \ \ \ \ \ \ \ }

In this section, from the energy bands in Table B1-Table B7, using
phenomenological formulae S = R - 4, deg = 2I + 1, S = S$_{Ax}$+$\Delta $S \{%
$\Delta $S = $\delta $($\widetilde{n}$) + [1-2$\delta $(S$_{axis}$)]Sign($%
\widetilde{n}$)\}, (\ref{2/3}), (\ref{-1/3}), (\ref{Charmed}), (\ref{Battom}%
), (\ref{Kersa-C}) and (\ref{Omiga-C}), we have deduced the strange numbers
(S), the isospins (I), the electric charges (Q), the charmed numbers (C) and
the bottom numbers (b) of the quarks. For each deduced quark there are
always three different colored members. These three different colored quarks
have exactly the same rest masses and intrinsic quantum numbers. We can omit
the colors as we show the deduced intrinsic quantum numbers in Table 10

\begin{tabular}{l}
\ \ \ \ \ Table 10. The Quantum Numbers of the Quarks \\ 
$%
\begin{tabular}{|l|l|l|l|l|l|l|l|l|l|}
\hline
q$_{\text{Name}}^{\text{I}_{Z}^{^{\prime }}}$ & q$_{N}^{\frac{1}{2}}$ & q$%
_{N}^{\frac{-1}{2}}$ & q$_{\Delta }^{\frac{3}{2}}$ & q$_{\Delta }^{\frac{1}{2%
}}$ & q$_{\Delta }^{\frac{-1}{2}}$ & q$_{\Delta }^{\frac{-3}{2}}$ & q$%
_{\Sigma }^{1}$ & q$_{\Sigma }^{0}$ & q$_{\Sigma }^{-1}$ \\ \hline
S & 0 & 0 & 0 & 0 & 0 & 0 & -1 & -1 & -1 \\ \hline
C & 0 & 0 & 0 & 0 & 0 & 0 & 0 & 0 & 0 \\ \hline
b & 0 & 0 & 0 & 0 & 0 & 0 & 0 & 0 & 0 \\ \hline
I & $\frac{1}{2}$ & $\frac{1}{2}$ & $\frac{3}{2}$ & $\frac{3}{2}$ & $\frac{3%
}{2}$ & $\frac{3}{2}$ & 1 & 1 & 1 \\ \hline
I$_{Z}^{^{\prime }}$ & $\frac{1}{2}$ & -$\frac{1}{2}$ & $\frac{3}{2}$ & $%
\frac{1}{2}$ & -$\frac{1}{2}$ & -$\frac{3}{2}$ & 1 & 0 & -1 \\ \hline
$\text{Q}_{q}$ & $\frac{2}{3}$ & -$\frac{1}{3}$ & $\frac{2}{3}$ & $\frac{2}{3%
}$ & $\frac{-1}{3}$ & $\frac{-1}{3}$ & $\frac{2}{3}$ & $\frac{-1}{3}$ & $%
\frac{-1}{3}$ \\ \hline
$\epsilon _{\text{Name}}^{\text{I}_{\text{Z}}}$ & u$_{N}^{\frac{1}{2}}$ & d$%
_{N}^{\frac{-1}{2}}$ & u$_{\Delta }^{\frac{1}{2}}$ & u$_{\Delta }^{\frac{1}{2%
}}$ & d$_{\Delta }^{\frac{-1}{2}}$ & d$_{\Delta }^{\frac{-1}{2}}$ & u$%
_{\Sigma }^{1}$ & \ d$_{\Sigma }^{0}$ & d$_{\Sigma }^{0}$ \\ \hline
I$_{Z}$ & $\frac{1}{2}$ & -$\frac{1}{2}$ & $\frac{1}{2}$ & $\frac{1}{2}$ & -$%
\frac{1}{2}$ & -$\frac{1}{2}$ & \ 1 & \ 0 & 0 \\ \hline
{*}**** & *** & *** & *** & *** & *** & *** & *** & *** & *** \\ \hline
q$_{\text{Name}}^{\text{I}_{Z}^{^{\prime }}}$ & q$_{\Xi }^{\frac{1}{2}}$ & q$%
_{\Xi }^{\frac{-1}{2}}$ & q$_{S}^{0}$ & q$_{\Omega }^{0}$ & q$_{C}^{0}$ & q$%
_{b}^{0}$ & q$_{\Omega _{C}}^{0}$ & q$_{\Xi _{C}}^{\frac{1}{2}}$ & q$_{\Xi
_{C}}^{\frac{-1}{2}}$ \\ \hline
S & -2 & -2 & -1 & -3 & 0 & 0 & -2 & -1 & -1 \\ \hline
C & 0 & 0 & 0 & 0 & 1 & 0 & 1 & 1 & 1 \\ \hline
b & 0 & 0 & 0 & 0 & 0 & -1 & 0 & 0 & 0 \\ \hline
I & $\frac{1}{2}$ & $\frac{1}{2}$ & 0 & 0 & 0 & 0 & 0 & $\frac{1}{2}$ & $%
\frac{1}{2}$ \\ \hline
I$_{Z}^{\text{'}}$ & $\frac{1}{2}$ & -$\frac{1}{2}$ & 0 & 0 & 0 & 0 & 0 & $%
\frac{1}{2}$ & -$\frac{1}{2}$ \\ \hline
$\text{Q}_{q}$ & $\frac{2}{3}$ & -$\frac{1}{3}$ & -$\frac{1}{3}$ & -$\frac{1%
}{3}$ & $\frac{2}{3}$ & -$\frac{1}{3}$ & -$\frac{1}{3}$ & $\frac{2}{3}$ & -$%
\frac{1}{3}$ \\ \hline
$\epsilon _{\text{Name}}^{\text{I}_{Z}}$ & u$_{\Xi }^{\frac{3}{2}}$ & d$%
_{\Xi }^{\frac{1}{2}}$ & d$_{S}^{0}$ & d$_{\Omega }^{1}$ & u$_{C}^{0}$ & d$%
_{b}^{0}$ & d$_{\Omega _{C}}^{0}$ & u$_{\Xi _{C}}^{\frac{1}{2}}$ & d$_{\Xi
_{C}}^{\frac{-1}{2}}$ \\ \hline
I$_{z}$ & $\frac{3}{2}$ & $\frac{1}{2}$ & 0 & 1 & $0$ & 0 & 0 & $\frac{1}{2}$
& -$\frac{1}{2}$ \\ \hline
\end{tabular}
$%
\end{tabular}

{\small The\ name of q}$_{\text{Name}}^{\text{I}_{z}^{^{\prime }}}${\small \
is the name of the excited quark; I}$_{Z}^{^{\prime }}${\small \ is the
formal z-component of the isospin of the q}$_{\text{Name}}^{\text{I}%
_{z}^{^{\prime }}}${\small \ (\ref{Formal-Iz}). The I}$_{z}${\small \ of }$%
\epsilon _{\text{Name}}^{\text{I}_{Z}}${\small \ is the physical z-component
of the isospin of the excited quark }$\epsilon _{\text{Name}}^{\text{I}_{Z}}$%
{\small \ from I}$_{z}${\small \ = Q}$_{q}${\small \ -}$\frac{1}{2}${\small %
(B+S+C+b). The Q}$_{q}${\small \ is the electric charge of the excited quark
q}$_{\text{Name}}^{\text{I}_{Z}^{^{\prime }}}${\small .\ }

Using the phenomenological united rest mass formula (\ref{Rest Mass}), we
have deduced the rest masses of the quarks. We show the low rest masses
quarks that are sufficient to cover experimental data in Table 11. Since the
three different colored quarks with the same flavor have completely the same
rest mass and quantum numbers, we can omit the colors in Table 11.

\begin{tabular}{|l|}
\hline
\ \ \ \ \ \ \ \ \ \ \ \ \ \ \ \ \ \ \ \ \ \ \ \ \ \ \ \ \ \ \ \ \ Table 11.
The Quark Spectrum \\ \hline
\ \ 
\begin{tabular}{|l|}
\hline
\ \ \ \ \ \ \ \ \ \ \ \ \ \ \ \ \ \ \ The Elementary Quark $\epsilon $ [$%
\epsilon _{u}$ and $\epsilon _{d}$] \\ \hline
$\epsilon _{u}$: I =$\frac{1}{2}$, s =$\frac{1}{2}$ , S = C = b =0, I$_{z}$=$%
\frac{1}{2}$, m$_{\epsilon _{u}(0)}$ = 0 and one of 3 colors \\ \hline
$\epsilon _{d}$: I =$\frac{1}{2}$, s =$\frac{1}{2}$ , S = C = b =0, I$_{z}$=$%
\frac{-1}{2}$, m$_{\epsilon _{d}(0)}$ = 0 and one of 3 colors \\ \hline
\end{tabular}
\\ \hline
\ \ \ \ \ \ \ \ \ \ \ \ \ \ \ \ \ \ \ \ The Excited Quarks q$_{\text{Name}%
}^{\Delta \text{S}}$((m+$\Delta $)(Mev))\ of $\epsilon $ \\ \hline
\begin{tabular}{|l|}
\hline
1.\ Unflavored Quarks{\small \ \ [The Ground Quarks are}\textbf{\ u}$^{0}$%
\textbf{(313}$\text{+}\Delta $\textbf{)}{\small \ and }\textbf{d}$^{0}$%
\textbf{(313}$\text{+}\Delta $\textbf{)}{\small ]} \\ \hline
D-Axis: \textbf{u}$^{0}$\textbf{(313}$\text{+}\Delta $\textbf{), }{\small q}$%
_{\text{N}}^{0}${\small (583}$\text{+}\Delta ${\small ), q}$_{\text{N}}^{0}$%
{\small (853}$\text{+}\Delta ${\small )}$,${\small 2q}$_{\text{N}}^{0}$%
{\small (1213}$\text{+}\Delta ${\small )}$,${\small 2q}$_{\text{N}}^{0}$%
{\small (1303}$\text{+}\Delta ${\small ), } \\ \hline
\ \ \ \ \ \ \ \ \ \ \ \ \ \ {\small 2q}$_{\text{N}}^{0}${\small (1573}$\text{%
+}\Delta ${\small )}$,${\small 3q}$_{\text{N}}^{0}${\small (2023}$\text{+}%
\Delta ${\small ), q}$_{\text{N}}^{0}${\small (2293}$\text{+}\Delta ${\small %
), 2q}$_{\text{N}}^{0}${\small (2653}$\text{+}\Delta $). \\ \hline
F-Axis: \textbf{d}$^{0}$\textbf{(313}$\text{+}\Delta $\textbf{), }{\small q}$%
_{\text{N}}^{1}${\small (583}$\text{+}\Delta ${\small ), q}$_{\text{N}}^{1}$%
{\small (673}$\text{+}\Delta ${\small )}$,${\small 3q}$_{N}^{1}${\small (1303%
}$\text{+}\Delta ${\small )}$,${\small q}$_{\text{N}}^{1}${\small (1393}$%
\text{+}\Delta ${\small )}$,${\small \ } \\ \hline
\ \ \ \ \ \ \ \ \ \ \ \ {\small q}$_{\text{N}}^{1}${\small (2023}$\text{+}%
\Delta ${\small ).} \\ \hline
$\Delta $-Axis: \textbf{u}$^{0}$\textbf{(313}$\text{+}\Delta $\textbf{), }%
{\small q}$_{\Delta }^{0}${\small (673}$\text{+}\Delta ${\small )}, {\small %
2q}$_{\Delta }^{0}${\small (1033+}$\Delta ${\small ), q}$_{\Delta }^{0}\text{%
(}${\small 1393}$\text{+}\Delta \text{)}$, {\small q}$_{\Delta }^{0}\text{%
(1753+}\Delta \text{)},$ \\ \hline
$\ \ \ \ $\ \ \ \ \ \ \ \ \ 4{\small q}$_{\Delta }^{0}\text{(2113+}\Delta 
\text{)}$, $\text{{\small 3}}${\small q}$_{\Delta }^{0}\text{(2473+}\Delta 
\text{).}$ \\ \hline
$\Lambda $ and F-Axes : \textbf{d}$^{0}$\textbf{(313}$\text{+}\Delta $%
\textbf{) (}$\text{d}_{S}^{1}\text{({\small 313}+}\Delta \text{)); }\Sigma $
and G-Axes \textbf{d}$^{0}$\textbf{(313}$\text{+}\Delta $\textbf{) } \\ 
\hline
\end{tabular}
\\ \hline
\begin{tabular}{|l|}
\hline
{\small \ \ \ \ \ \ \ \ \ \ \ \ \ \ \ \ }2. Strange Quarks{\small \ [ The
Ground Quark is }\textbf{d}$_{S}^{1}$\textbf{(493}$\text{+}\Delta $\textbf{)}%
$_{\Sigma }$] \\ \hline
{\small G-Axis: }$2${\small d}$_{S}^{1}${\small (493}$\text{+}\Delta $%
{\small ), d}$_{S}^{1}${\small (773}$\text{+}\Delta ${\small ), d$_{S}^{1}$%
(1413$\text{+}\Delta $), d}$_{S}^{1}${\small (1513}$\text{+}\Delta $), 
{\small d}$_{S}^{1}${\small (2513}$\text{+}\Delta $). \\ \hline
$\Delta $-Axis: {\small d}$_{S}^{\text{-1}}${\small (773}$\text{+}\Delta $%
{\small ), d}$_{S}^{\text{-1}}${\small (3753}$\text{+}\Delta ${\small ), d}$%
_{S}^{\text{-1}}${\small (9613}$\text{+}\Delta ${\small ), d}$_{S}^{\text{-1}%
}${\small (18353}$\text{+}\Delta $). \\ \hline
F-Axis: {\small d}$_{S}^{0}${\small (493}$\text{+}\Delta ${\small ), d}$%
_{S}^{0}${\small (773}$\text{+}\Delta $), {\small 2d}$_{S}^{0}${\small (1203}%
$\text{+}\Delta ${\small ), d}$_{S}^{0}${\small (1303}$\text{+}\Delta $), 
{\small 2d}$_{S}^{0}${\small (1393}$\text{+}\Delta ${\small ),}\  \\ \hline
\ \ \ \ {\small \ \ \ \ \ \ \ \ \ \ }2{\small d}$_{\text{S}}^{0}${\small %
(1923}$\text{+}\Delta $), {\small 4d}$_{S}^{0}${\small (2013}$\text{+}\Delta 
${\small ), d}$_{S}^{0}${\small (2023}$\text{+}\Delta $), {\small d}$%
_{S}^{0} ${\small (2643}$\text{+}\Delta ${\small ); } \\ \hline
D-Axis: {\small d}$_{\text{S}}^{-1}${\small (493}$\text{+}\Delta ${\small ),
d}$_{\text{S}}^{-1}${\small (1503}$\text{+}\Delta ${\small ), d}$_{\text{S}%
}^{-1}${\small (1603}$\text{+}\Delta ${\small ), d}$_{S}^{-1}$(2333+$\Delta $%
). \\ \hline
$\Lambda $-Axis: $\text{d}_{S}^{0}\text{({\small 2743}+}\Delta \text{), }$%
{\small d}$_{\text{S}}^{0}${\small (4633}+$\Delta ${\small )}; $\text{q}%
_{\Sigma }\text{({\small 583}+}\Delta \text{)}${\small , 3}$\text{q}_{\Sigma
}\text{({\small 1033}+}\Delta \text{)}${\small , }$\text{2q}_{\Sigma }\text{(%
{\small 1303}+}\Delta \text{),}$ \\ \hline
$\ \ \ \ \ \ \ \ \ \ \ \ \text{ }${\small ,}$\text{q}_{\Sigma }\text{(%
{\small 1753}+}\Delta \text{)}${\small , }$3\text{q}_{\Sigma }\text{({\small %
2023}+}\Delta \text{), }${\small 5}$\text{q}_{\Sigma }\text{({\small 2473}+}%
\Delta \text{), 2q}_{\Sigma }\text{({\small 2743}+}\Delta \text{).}$ \\ 
\hline
$\Sigma $-Axis: {\small \ }\textbf{d}$_{S}^{1}$\textbf{(493}$\text{+}\Delta $%
\textbf{)}$_{\Sigma }$, {\small d}$_{S}^{1}$(1933); 2{\small q}$_{\Xi }^{0}%
\text{({\small 853}+}\Delta \text{), }${\small 3q}$_{\Xi }^{0}${\small (1033}%
$\text{+}\Delta ${\small ), q}$_{\Xi }^{0}${\small (1213}$\text{+}\Delta $),%
{\small \ \ }$\text{.}$ \\ \hline
\ \ \ \ \ \ \ \ \ \ \ \ \ \ {\small 2q}$_{\Xi }^{0}${\small (1573}$\text{+}%
\Delta ${\small ), }$2${\small q}$_{\Xi }^{0}${\small (1753}$\text{+}\Delta $%
{\small ), q}$_{\Xi }^{0}${\small (1933}$\text{+}\Delta ${\small ), 2q}$%
_{\Xi }^{0}${\small (2293}$\text{+}\Delta ${\small ).} \\ \hline
$G$-Axis: {\small q}$_{\Xi }^{0}${\small (673}$\text{+}\Delta ${\small ), q}$%
_{\Xi }^{0}${\small (853}$\text{+}\Delta ${\small )}, {\small 3q}$_{\Xi
}^{0} ${\small (1393}+$\Delta ${\small )}, {\small 2q}$_{\Xi }^{0}${\small %
(1573}+$\Delta ${\small ), }2{\small q}$_{\Xi }^{0}${\small (1733}$\text{+}%
\Delta ${\small )}, \\ \hline
F-Axis: {\small q}$_{\Xi }^{-1}${\small (1823}$\text{+}\Delta ${\small ), 2q}%
$_{\Xi }^{-1}${\small (1913}$\text{+}\Delta ${\small ).}$\ $G:\ 2{\small q}$%
_{\Xi }^{-1}${\small (1913}$\text{+}\Delta ${\small ), }$2${\small q}$_{\Xi
}^{-1}${\small (2113}$\text{+}\Delta ${\small ), } \\ \hline
$\Sigma $-Axis: {\small d}$_{\Omega }^{-1}${\small (1033}$\text{+}\Delta $%
{\small )}, $\text{d}_{\Omega }^{-1}\text{({\small 3193}+}\Delta \text{); G:
d}_{\Omega }^{-1}${\small (1633}$\text{+}\Delta ${\small )}, $\text{d}%
_{\Omega }^{-1}${\small (1813}$\text{+}\Delta ${\small )}, \\ \hline
\end{tabular}
\\ \hline
\ \ \ \ \ \ 
\begin{tabular}{|l|}
\hline
{\small \ \ \ \ \ \ \ \ \ \ }3. Charmed Quarks{\small \ \ [The Ground Quark
is}\textbf{\ u}$_{C}^{1}$\textbf{(1753+}$\Delta $\textbf{)}$\text{]}$ \\ 
\hline
$\Delta $-Axis{\small : }\textbf{u}$_{C}^{1}$\textbf{(1753+}$\Delta $)$\text{%
, u}_{\text{C}}^{1}\text{(6073+}\Delta \text{), u}_{C}^{1}\text{(13273+}%
\Delta \text{).}$ \\ \hline
{\small D-Axis: }$\text{u}_{\text{C}}^{1}\text{(2133+}\Delta \text{)},\text{u%
}_{\text{C}}^{1}\text{(2333+}\Delta \text{)},\text{u}_{\text{C}}^{1}\text{%
(2533+}\Delta \text{)},\text{u}_{\text{C}}^{1}\text{(}${\small 3543+}$\Delta 
\text{).}$ \\ \hline
{\small F-Axis: }q$_{\Xi _{C}}^{1}\text{(1873+}\Delta \text{)},\text{ q}%
_{\Xi _{C}}^{1}\text{(2163+}\Delta \text{)},\text{ q}_{\Xi _{C}}^{1}\text{%
(2363+}\Delta \text{), }$q$_{\Xi _{C}}^{1}${\small (3193+}$\Delta ${\small )}%
$.$ \\ \hline
{\small G-Axis: }$\text{d}_{\Omega _{C}}^{1}\text{(2133+}\Delta \text{), d}%
_{\Omega _{C}}^{1}\text{(2513+}\Delta \text{), d}_{\Omega _{C}}^{1}\text{(3%
{\small 253}+}\Delta \text{)}.$ \\ \hline
\end{tabular}
\\ \hline
\ \ \ \ \ \ \ \ 
\begin{tabular}{|l|}
\hline
\ \ \ \ \ \ 4. Bottom Quarks$\ $[{\small The Ground Quark is }\textbf{d}$%
_{b}^{1}$\textbf{(4913+}$\Delta $\textbf{)}$\text{]}$ \\ \hline
$\Sigma $-Axis: \textbf{d}$_{b}^{1}$\textbf{(4913+}$\Delta $\textbf{), }%
{\small d}$_{b}^{1}${\small (9333}$\text{+}\Delta ${\small ), d}$_{b}^{1}$%
{\small (15193}$\text{+}\Delta ${\small ), d}$_{b}^{1}${\small (22493}$\text{%
+}\Delta ${\small ), ....} \\ \hline
\end{tabular}
\\ \hline
\end{tabular}

In fact, there are three completely identical tables (as Table 11) with the
red, the yellow and the blue colors respectively. Omitting the color, we use
one of the three tables to represent all three tables.

For the four flavored quarks, there are five ground quarks [u(313+$\Delta $%
), d(313+$\Delta $), d$_{S}$(493+$\Delta $), u$_{C}$(1753+$\Delta $) and d$%
_{b}$(4913+$\Delta $)] in the quark spectrum. They correspond with the five
quarks \cite{Quarks} of the current Quark Model: u $\leftrightarrow $ u(313+$%
\Delta $), d $\leftrightarrow $ d(313+$\Delta $), s $\leftrightarrow $ d$%
_{S} $(493+$\Delta $), c $\leftrightarrow $ u$_{C}$(1753+$\Delta $) and b $%
\leftrightarrow $ d$_{b}$(4913+$\Delta $). They are all the energy band
excited states of the elementary quarks $\epsilon $. The quarks u(313+$%
\Delta $) and u$_{C}$(1753+$\Delta $) (with Q = +$\frac{2}{3}$) are the
excited states of the elementary $\epsilon _{u}$-quark. The quarks d(313+$%
\Delta $), d$_{S}$(493+$\Delta $) and d$_{b}$(4913+$\Delta $) (with Q = -$%
\frac{1}{3}$) are the excited states of the elementary $\epsilon _{d}$-quark.

There are the values of the current quarks that can compare with the deduced
values of the five ground quarks. The deduced intrinsic quantum numbers (I,
S, C, b and Q ) of the five ground quarks are exactly the same as the five
current quarks (see Table 10 and \cite{Quarks}). Table 11$^{\ast }$ shows
that 
\begin{eqnarray*}
&&\text{ \ \ \ \ \ \ \ \ {\small Table 11* The Comparison of Deduce Ground
Quarks and Current Quarks}} \\
&& 
\begin{tabular}{|l|l|l|l|l|l|}
\hline
& {\small u(m (Mev))} & {\small d(m (Mev))} & {\small u(m (Mev))} & {\small %
u(m (Mev))} & {\small u(m (Mev))} \\ \hline
{\small Current} & {\small u(1.5 to 4)} & {\small d(4 to 8)} & s{\small (80
to 130)} & {\small c(1250 to 1350)} & {\small b(4100 to 4900)} \\ \hline
{\small Deduced} & {\small u(313)} & {\small d(313)} & d$_{s}${\small (493)}
& {\small u}$_{c}${\small (1753)} & {\small d}$_{b}${\small (4913)} \\ \hline
$\Delta m$ & 310 & 307 & 388 & 493 & 487 \\ \hline
\end{tabular}%
\end{eqnarray*}
the deduced rest masses of the five ground quarks are roughly a constant
(about 400 Mev) larger than the masses of the current quarks. These can be
originated from different energy reference systems. If we use the same
energy reference system, the deduced masses of ground quarks are roughly
consistent with the masses of the corresponding current quarks. Of course
the ultimate test is whether or not the baryons and mesons that are composed
by the quarks are consistent with the experimental results.

Now we have deduced a quark (excited from the vacuum) spectrum. We have
found the intrinsic quantum numbers (S, C, b and Q) and rest masses of the
quarks. From these rest masses and the quantum numbers, we can deduce a
baryon spectrum and a meson spectrum using sum laws and phenomenological
binding energy formulae,

\section{The Baryon Spectrum}

According to the Quark Model \cite{Quark Model}, a baryon is composed of
three quarks with different colors. For each flavor, the three different
colored quarks have the same I, S, C, b, Q and rest mass. Thus, we can omit
the color when we deduce the rest masses and intrinsic quantum numbers of
the baryons. We must remember, however, that three colored quarks (red,
yellow and blue) compose a colorless baryon. Since we have already found the
quarks, we can use sum laws to find the intrinsic\ quantum numbers (I, S, C,
b and Q) of baryons (qqq) and the rest masses of all baryons, except those
of the charmed baryons. We can deduce the rest masses of the charmed baryons
with a phenomenological formula (\ref{CB-Eb}). There are more than 80 quarks
in Table 11. They can make many possible baryons, much more than the
experimental baryons \cite{Baryon04}. Since the probabilities that three
quarks make a baryon are different, the experimentally discovered baryons
are the possible baryons with larger observable probabilities. The
observable probability depends on the following factors: 1) the ground
quarks have higher probabilities of occurrence than other quarks have, 2)
the lower rest mass quarks have higher probabilities of occurrence, 3) the
quarks with lower isospin have more possibilities of occurrence than the
higher isospin quarks, 4) the quarks born on the symmetry axes with more
symmetry operations have higher probabilities of occurrence. If we use P(q$%
_{name}$(m)) to represent the probability of a q$_{name}$(m) forming a
baryon, we assume: \ \ \ 
\begin{equation}
\begin{tabular}{l}
P(u(313)) $\sim $ P(d(313))\TEXTsymbol{>}\TEXTsymbol{>} \\ 
P(d$_{S}$(493))\TEXTsymbol{>}P(u$_{C}$(1753))\TEXTsymbol{>}P(d$_{b}$(4913))%
\TEXTsymbol{>}\TEXTsymbol{>} \\ 
P({\small u}$_{C}${\small (m)})\TEXTsymbol{>}P(d$_{b}$(m))\TEXTsymbol{>} P(d$%
_{S}$(m))\TEXTsymbol{>} P(q$_{N}$(m))\TEXTsymbol{>} P(d$_{\Sigma }$(m))%
\TEXTsymbol{>} \\ 
P(q$_{\Delta }$(m))\TEXTsymbol{>} P(d$_{\Omega }$(m)) \TEXTsymbol{>} P(d$%
_{\Xi _{_{C}}}${\small (m)})\TEXTsymbol{>} P(d$_{\Omega _{_{C}}}${\small (m)}%
)\TEXTsymbol{>} P(d$_{\Xi }$(m))\TEXTsymbol{>}.%
\end{tabular}
\label{P(Qk)}
\end{equation}%
From (\ref{P(Qk)}) we find that the possible baryons (with none or only one
ground q(313)) have much lower observable probabilities than baryons with
two ground quarks (or three ground q(313) quarks). Thus, we omit the baryons
that have only one ground quark q(313) or do not have any ground quark
q(313). We will show that the possible baryons with two ground quarks [q$%
_{1} $(m)q(313)(q(313)] can explain the experimental baryon spectrum.

\subsection{The Intrinsic Quantum Numbers of the Baryons\ \ \ \ \ \ \ \ \ \
\ \ }

A1. The intrinsic quantum numbers of a baryon are the sums of the three
constituent quarks (q$_{1}$(m) + q$_{2}$(313) + q$_{3}$(313)). We can find
the strange number S$_{B}$, the charmed number C$_{\text{B}}$, the bottom
number b$_{\text{B}}$ and the electric charge Q$_{\text{B}}$ of the baryons
by 
\begin{equation}
\begin{tabular}{l}
$\text{S}_{\text{B}}\text{ }\text{= S}_{q_{1}}\text{+ S}_{q_{_{N}(313)}}%
\text{+ S}_{q_{_{N}(313)}}=\text{ S}_{q_{1(m)}}\text{,}$ \\ 
$\text{C}_{\text{B\ }}\text{ }\text{= C}_{q_{1}}\text{ + C}_{q_{_{N}(313)}}%
\text{ + C}_{q_{_{N}(313)}}\text{= C}_{q_{1}(m)}\text{,}$ \\ 
$\text{b}_{\text{B}}\text{=}\text{ b}_{q_{1}}\text{+ b}_{q_{_{N}(313)}}\text{%
+ b}_{q_{_{N}(313)}}\text{= b}_{q_{1(m)}}\text{,}$ \\ 
$\text{Q}_{\text{B}}\text{ = Q}_{q_{1}}\text{+ Q}_{q_{_{N}(313)}}\text{+ Q}%
_{_{q_{_{N}(313)}}}.$%
\end{tabular}
\label{Sum(SCbQ)}
\end{equation}
Since the ground quark q(313) has S = C = b = 0 , the baryon [q$_{1}$(m)+q$%
_{2}$(313)+q$_{3}$(313)] has the same S, C and b as the q$_{1}$(m)-quark
has.\qquad \qquad

A2. The Isospin I$_{\text{B}}$ of the baryon (q$_{1}$q$_{2}$q$_{3}$) is
found by 
\begin{equation}
\overrightarrow{I_{B}}\text{ = }\overrightarrow{I_{q_{1}(m)}}\text{ + }%
\overrightarrow{I_{q_{2}}}\text{ + }\overrightarrow{I_{q_{3}}}\text{.}
\label{I of qqq}
\end{equation}
Since I$_{q_{2}}$= I$_{q_{3}}$= I$_{q(313)}$ = $\frac{1}{2}$ and the top
limit of the experimental isospin values of baryons is $\frac{3}{2}$, the
isospins of the baryons (q$_{1}$q$_{2}$ q$_{3}$) are:

\begin{equation}
\ 
\begin{tabular}{l}
for I$_{q_{1}}$ = $\frac{3}{2}\rightarrow $I$_{B}$ = $\frac{1}{2}$, $\frac{3%
}{2}$; \\ 
for I$_{q_{1}}$ = 1 $\rightarrow $I$_{B}$ = 0, 1; \\ 
for I$_{q_{1}}$ = $\frac{1}{2}\rightarrow $I$_{B}$ = $\frac{1}{2}$, $\frac{3%
}{2}$; \\ 
for I$_{q_{1}}$ = 0 $\rightarrow $I$_{B}$ = 0, 1.%
\end{tabular}
\ \ \ \ \ \   \label{I of B}
\end{equation}
After having the baryon isospin I$_{\text{B}}$, we can get the I$_{\text{B, z%
}}$ using the follow formula 
\begin{equation}
\begin{tabular}{ll}
for I$_{B}$ = $\frac{3}{2}$, & I$_{\text{B, z}}$ = $\ \frac{3}{2}$, $\frac{1%
}{2},$-$\frac{1}{2}$, -$\frac{3}{2};$ \\ 
for I$_{B}$ =\ 1$,$ & I$_{\text{B, z}}$ = \ 1, 0, -1; \\ 
for I$_{B}$ = $\frac{1}{2},$ & I$_{\text{B, z}}$ = $\frac{1}{2},$ -$\frac{1}{%
2}$; \\ 
for I$_{B}$ = 0$,$ & I$_{\text{B, z}}$ = 0.%
\end{tabular}
\ \ \text{. }  \label{Iz,b}
\end{equation}

Using R values (see Table A3) and equivalent $\overrightarrow{n}$ values (%
\ref{Equ-N}), we can find the maximum isospin value of quarks for each
symmetry point. The probability that a quark (q$_{1}$) with lower isospin
forms a baryon [q$_{1}$(m)q$_{2}$(313)q$_{3}$(313)] with higher isospin is
low. Thus, there is very low possibility that a baryon has a higher isospin
than the maximum I of quarks have on the point. Since all symmetry axes at
the point N are two fold, the maximum I of the baryon is $\frac{1}{2}$ at
the N-point. We show the maximum I values of the quarks and baryons in Table
A6.

A3. For the baryon [q$_{1}$(m)q$_{2}$(313)q$_{3}$(313)], we can deduce the
isospin I$_{\text{B}}$ and I$_{\text{B,Z}}$using (\ref{I of B}) and (\ref%
{Iz,b}).\ For\ the q$_{1}$(m)-quark, I$_{\text{Z}}$ can be found in Table
10. The quark q$_{1}$(m) selects two ground quarks q$_{\text{2}}$ and q$_{%
\text{3}}$ from u(313)u(313), u(313)d(313) and d(313)d(313), using the
follow formulae:

\begin{equation}
\text{I}_{\text{Z,B}}\text{ = I}_{\text{Z,q}_{1}\text{(m)}}\text{+ I}_{\text{%
Z},}\text{{\small q}}_{\text{2}}\text{ + I}_{\text{Z},}\text{{\small q}}_{%
\text{3}}\text{,}  \label{Izb}
\end{equation}
\begin{equation}
\text{Q}_{\text{B}}\text{ = Q}_{\text{q}_{1}}\text{ + Q}_{\text{q}_{2}}\text{
+ Q}_{\text{q}_{3}}\text{,}  \label{QQQ}
\end{equation}
to get the correct I$_{\text{Z,B}}$ and Q$_{\text{B}}$. We show selected
results in Table 12:

: 
\begin{tabular}{l}
\ \ \ \ \ \ \ \ \ \ \ \ \ \ Table 12. \ The Quark Constitutions of the
Baryons \\ 
$%
\begin{tabular}{|l|l|l|l|l|l|l|l|l|l|}
\hline
B$_{aryon}^{I_{\text{z, }B}}$ & N$_{{}}^{\frac{1}{2}}$ & N$_{N}^{\frac{-1}{2}%
}$ & $\Delta _{\Delta }^{\frac{3}{2}}$ & $\Delta _{\Delta }^{\frac{1}{2}}$ & 
$\Delta _{\Delta }^{\frac{-1}{2}}$ & $\Delta _{\Delta }^{\frac{-3}{2}}$ & $%
\Sigma _{\Sigma }^{1}$ & $\Sigma _{\Sigma }^{0}$ & $\Sigma _{\Sigma }^{-1}$
\\ \hline
I$_{B}$ & 1/2 & 1/2 & 3/2 & 3/2 & 3/2 & 3/2 & 1 & 1 & 1 \\ \hline
q$_{\text{1}}^{_{\text{I}_{Z}}}$(u$^{_{\text{I}_{Z}}}$or d$^{_{\text{I}%
_{z}}} $) & u$_{N}^{\frac{1}{2}}$ & d$_{N}^{\frac{-1}{2}}$ & u$_{\Delta }^{%
\frac{1}{2}}$ & u$_{\Delta }^{\frac{1}{2}}$ & d$_{\Delta }^{\frac{-1}{2}}$ & 
\ d$_{\Delta }^{\frac{-1}{2}}$ & u$_{\Sigma }^{1}$ & d$_{\Sigma }^{0}$ & d$%
_{\Sigma }^{0}$ \\ \hline
q$_{2}^{I_{z}}$ & u$^{\frac{1}{2}}$ & u$^{\frac{1}{2}}$ & u$^{\frac{1}{2}}$
& u$^{\frac{1}{2}}$ & u$^{\frac{1}{2}}$ & d$^{\frac{-1}{2}}$ & u$^{\frac{1}{2%
}}$ & u$^{\frac{1}{2}}$ & d$^{\frac{-1}{2}}$ \\ \hline
q$_{3}^{Iz}$ & d$^{\frac{-1}{2}}$ & d$^{\frac{-1}{2}}$ & u$^{\frac{1}{2}}$ & 
d$^{\frac{-1}{2}}$ & d$^{\frac{-1}{2}}$ & \ d$^{\frac{-1}{2}}$ & d$^{\frac{-1%
}{2}}$ & d$^{\frac{-1}{2}}$ & d$^{\frac{-1}{2}}$ \\ \hline
I$_{Z,B}$=$\Sigma $I$_{z,q_{i}}$ & $\frac{1}{2}$ & $\frac{-1}{2}$ & $\frac{3%
}{2}$ & $\frac{1}{2}$ & $\frac{-1}{2}$ & $\frac{-3}{2}$ & \ 1 & 0 & -1 \\ 
\hline
Q$_{B}$=$\Sigma $Q$_{q_{i}}$ & 1 & 0 & 2 & 1 & 0 & -1 & \ 1 & 0 & -1 \\ 
\hline
S$_{\text{B}}$= S$_{q_{1}}$ & 0 & 0 & 0 & 0 & 0 & \ 0 & -1 & -1 & -1 \\ 
\hline
C$_{\text{B}}$= C$_{q_{1}}$ & 0 & 0 & 0 & 0 & 0 & \ 0 & \ 0 & 0 & 0 \\ \hline
b$_{\text{B}}$= b$_{q_{1}}$ & 0 & 0 & 0 & 0 & 0 & \ 0 & \ 0 & 0 & 0 \\ \hline
B$_{aryon}^{Q_{B}}$ & N$^{+}$ & N$^{0}$ & $\Delta ^{++}$ & $\Delta ^{+}$ & $%
\Delta ^{0}$ & $\ \Delta ^{-}$ & $\ \Sigma ^{+}$ & $\ \Sigma ^{0}$ & $\Sigma
^{-}$ \\ \hline
{*}******** & *** & *** & *** & *** & *** & *** & *** & *** & *** \\ \hline
B$_{aryon}^{I_{z},B}$ & $\Xi _{\Xi }^{\frac{1}{2}}$ & $\Xi _{\Xi }^{\frac{-1%
}{2}}$ & $\Lambda _{S}^{0}$ & $\Lambda _{b}^{0}$ & $\Omega ^{0}$ & $\Xi
_{C}^{\frac{1}{2}}$ & $\Xi _{C}^{\frac{-1}{2}}$ & $\Lambda _{C}^{0}$ & $%
\Omega _{C}^{0}$ \\ \hline
I$_{B}$ & $\frac{1}{2}$ & $\frac{1}{2}$ & 0 & 0 & 0 & $\frac{1}{2}$ & $\frac{%
1}{2}$ & 0 & 0 \\ \hline
q$_{q_{1}}^{_{\text{I}_{Z}}}$(u$^{_{\text{I}_{Z}}}$,d$^{_{\text{I}_{z}}}$) & 
u$_{\Xi }^{\frac{3}{2}}$ & d$_{\Xi }^{\frac{1}{2}}$ & d$_{S}^{0}$ & d$%
_{b}^{0}$ & d$_{\Omega }^{1}$ & u$_{\Xi _{C}}^{\frac{1}{2}}$ & d$_{\Xi
_{C}}^{\frac{-1}{2}}$ & u$_{C}^{0}$ & d$_{\Omega _{C}}^{0}$ \\ \hline
q$_{2}^{I_{z}}$ & d$^{\frac{-1}{2}}$ & d$^{\frac{-1}{2}}$ & u$^{\frac{1}{2}}$
& u$^{\frac{1}{2}}$ & d$^{\frac{-1}{2}}$ & u$^{\frac{1}{2}}$ & u$^{\frac{1}{2%
}}$ & u$^{\frac{1}{2}}$ & u$^{\frac{1}{2}}$ \\ \hline
q$_{3}^{Iz}$ & d$^{\frac{-1}{2}}$ & d$^{\frac{-1}{2}}$ & d$^{\frac{-1}{2}}$
& d$^{\frac{-1}{2}}$ & d$^{\frac{-1}{2}}$ & d$^{\frac{-1}{2}}$ & d$^{\frac{-1%
}{2}}$ & d$^{\frac{-1}{2}}$ & d$^{\frac{-1}{2}}$ \\ \hline
I$_{Z,B}$=$\Sigma $I$_{z,q_{i}}$ & $\frac{1}{2}$ & $\frac{-1}{2}$ & 0 & 0 & 0
& $\frac{1}{2}$ & $\frac{-1}{2}$ & 0 & 0 \\ \hline
Q$_{B}$=$\Sigma $Q$_{q_{i}}$ & 0 & -1 & 0 & 0 & -1 & 1 & 0 & 1 & 0 \\ \hline
S$_{\text{B}}$= S$_{q_{1}}$ & -2 & -2 & -1 & 0 & -3 & -1 & -1 & 0 & -2 \\ 
\hline
C$_{\text{B}}$= C$_{q_{1}}$ & 0 & 0 & 0 & 0 & 0 & 1 & 1 & 1 & 1 \\ \hline
b$_{\text{B}}$= b$_{q_{1}}$ & 0 & 0 & 0 & -1 & 0 & 0 & 0 & 0 & 0 \\ \hline
B$_{aryon}^{Q_{B}}$ & $\Xi ^{0}$ & $\Xi ^{-}$ & $\Lambda _{S}^{0}$ & $%
\Lambda _{b}^{0}$ & $\ \Omega ^{^{-1}}$ & $\Xi _{c}^{+}$ & $\Xi _{c}^{0}$ & $%
\Lambda _{C}^{+}$ & $\Omega _{C}^{0}$ \\ \hline
\end{tabular}
\ $%
\end{tabular}

\ \ \ \ \ \ \ \ \ \ \ \ \ \ \ \ \ \ \ \ \ \ \ \ \ \ \ \ \ \ \ \ \ \ \ \ \ \
\ \ \ \ \ \ \ \ 

Table 12 also shows the intrinsic quantum numbers (I, S, C, b and Q) of the
baryons. The baryons that we deduced from the phenomenological formulae have
exactly the same intrinsic quantum numbers as the experimental results.

\subsection{The Binding Energy of the Three Quarks Inside a Charmed Baryon\
\ \ \ \ \ \ \ \ \ \ \ \ \ \ \ \ \ \ \ \ \ \ \ \ \ \ \ \ \ \ \ \ \ \ \ \ \ \
\ \ \ \ \ \ \ \ \ \ \ \ \ \ \ \ \ \ \ \ \ \ \ \ \ \ \ \ \ \ \ \ \ \ \ \ \ \
\ \ \ \ \ \ \ \ \ \ \ \ \ \ \ \ \ \ \ \ \ \ \ \ \ \ \ \ \ \ \ \ \ \ \ \ \ \
\ \ \ \ \ \ \ \ \ \ \ \ \ \ \ \ \ \ \ \ \ \ \ \ \ \ \ \ \ \ \ \ \ \ \ \ \ \
\ \ \ \ \ \ \ \ \ \ \ \ \ \ \ \ \ \ \ \ \ \ \ \ \ \ \ \ \ \ \ \ \ \ \ \ \ \
\ \ \ \ \ \ \ \ \ \ \ \ \ \ \ \ \ \ \ \ \ \ \ \ \ \ \ \ \ \ \ \ \ \ \ \ \ \
\ \ \ \ \ \ \ \ \ \ \ \ \ \ \ \ \ \ \ \ \ \ \ \ \ \ \ \ \ \ \ \ \ \ \ \ \ \
\ \ \ \ \ \ \ \ \ \ \ \ \ \ \ \ \ \ \ \ \ \ \ \ \qquad\ \ \ \ \ \ \ \ \ \ \
\ \ \ \ \ \ \ \ \ \ \ \ \ \ \ \ \ \ \ \ \ \ \ \ \ \ \ \ \ \ \ \ \ \ \ \ \ \
\ \ \ \ \ \ \ \ \ \ \ \ \ \ \ \ \ \qquad}

Since all quarks inside baryons are the excited states of the same
elementary quarks $\epsilon $, the binding energies of the quarks inside the
baryons are the strong interaction energy of the three colors. Because the
three colors are exaetly the same for all baryons, the baryon binding
energies are the same for all baryons. The baryons binding energy of all
baryons is a roughly unknown constant (-3$\Delta $, $\Delta $ =$\frac{1}{3}%
\left\vert E_{B-B}\right\vert $ (\ref{Dalta})). Thus, the rest mass (M) of a
baryon will be 
\begin{equation*}
\text{M = m}_{q_{1}}^{\ast }\text{ + m}_{q_{2}}^{\ast }\text{ + m}%
_{q_{3}}^{\ast }\text{- 3}\Delta \text{.}
\end{equation*}%
Because a quark has mass m$_{q}^{\ast }$ (from Table 11), 
\begin{equation*}
\text{m}_{q}^{\ast }\text{=m}_{q}\text{+}\Delta \text{,}
\end{equation*}%
we have a baryon mass, 
\begin{equation}
\begin{tabular}{l}
$\text{M}_{B}\text{ }\text{=}\text{ m}_{q_{1}}^{\ast }\text{ + m}%
_{q_{2}}^{\ast }\text{ + m}_{q_{3}}^{\ast }\text{- 3}\Delta $ \\ 
$\ \ $\ = $\text{m}_{q_{1}}+\text{m}_{q_{2}}+\text{m}_{q_{3}},$%
\end{tabular}
\label{Baryon M}
\end{equation}%
where m$_{q}$ = m$_{q}^{\ast }$ - $\Delta $. The above formula (\ref{Baryon
M}) means that the rest mass of a baryon equals the sum of the three quark
rest masses (m). Since the binding energy (-3$\Delta $) is always cancelled
by the three ($\Delta $) of the three quark masses, we can omit the $\Delta $
in the rest mass of the three quarks inside the baryon and omit the (-3$%
\Delta $) in the binding energy of the baryon when we count the rest masses
of the baryons. In fact, the interaction energies of the colors are very
complex; (-3$\Delta $) is only a phenomenological approximation of the
binding energy.

The mass sum law (\ref{Baryon M}) is valid for most baryons except the
charmed baryons. For charmed baryons, however, we shall add a small amount
of adjustment in energy, $\Delta $e, to the large unknown constant (3$\Delta 
$): 
\begin{equation}
\Delta \text{e = 100C [2I - 1 - }\frac{\text{1}}{\text{2}}{\small \Theta }%
\text{(1+S}_{Ax}\text{)]\ \ (Mev),}  \label{CB-Eb}
\end{equation}
where C is the charmed number of \ the baryon, I is the isospin of the
baryon and S$_{Ax}$ is the strange number of the axis of the charmed quark (q%
$_{1}$). The in-out number $\Theta $ is defined in (\ref{f(se)}). From Table
A3, $\Theta $ = 0 for the $\Delta $-axis, the $\Lambda $-axis and the $%
\Sigma $-axis, while $\Theta $ = 1 for the D-axis, the F-axis and the
G-axes. From (\ref{Baryon M}) and (\ref{CB-Eb}), for charmed baryons, 
\begin{equation}
\text{M}_{B}\text{ }\ \text{= m}_{q_{1}}\text{+ m}_{q_{2}}\text{+ m}_{q_{3}}%
\text{+ }\Delta \text{e}  \label{C-B-Mass}
\end{equation}

From the quark spectrum (Table 11), using the sum laws (\ref{Sum(SCbQ)}), (%
\ref{Baryon M}) and (\ref{I of qqq}) [and the binding energy formula\ (\ref%
{C-B-Mass}) for the charmed baryons only], we can deduce the baryons.

\subsection{Deduction of the Baryons}

We have already found the intrinsic quantum numbers of the baryons that are
shown in Table 12. If we can deduce the rest masses of baryons, we will
deduce the baryon spectrum. We give three types of examples to show how to
deduce the baryons.\ \ \ \ \ \ \ \ \ \ \ \ \ \ \ \ \ \ \ \ \ \ \ \ \ \ \ \ \
\ \ \ \ \ \ \ \ \ \ \ \ \ \ \ \ \ \ \ \ \ \ \ \ \ \ \ \ \ \ \ \ \ \ \ \ \ \
\ \ \ \ \ \ \ \ \ \ \ \ \ \ \ \ \ \ \ \ \ \ \ \ \ \ \ \ \ \ \ \ \ \ \ \ \ \
\ \ \ \ \ \ \ \ \ \ \ \ \ \ \ \ \ \ \ \ \ \ \ \ \ \ \ \ \ \ \ \ \ \ \ \ \ \
\ \ \ \ \ \ \ \ \ \ \ \ \ \ \ \ \ \ \ \ \ \ \qquad\ \ \ \ \ \ \ \ \ \ \ \ \
\ \ \ \ \ \ \ \ \ \ \ \ \ \ \ \ \ \ \ \ \ \ \ \ \ \ \ \ \ \ \ \ \ \ \ \ \ \
\ \ \ \ \ \ \ \ \ \ \ \ \ \ \ \qquad\ \ \ \ \ \ \ \ \ \ \ \ \ \ \ \ \ \ \ 

C1. The Deduction of Charmed Baryons on the $\Delta $-axis ($\Gamma $-H)

In this example, we will deduce the charmed baryons on the $\Delta $-axis,
the $\Sigma $-axis and the $\Lambda $-axis. In fact, all charmed baryons are
only on the $\Delta $-axis. There is not any charmed baryons on the other
two axes.

For the $\Delta $-axis, S$_{\Delta }$ = 0 and $\Theta $ = 0 from Table 3;
from (\ref{CB-Eb}),\ \ 
\begin{equation}
\Delta \text{e =\ 100}\times \text{C(2I-1) \ (Mev)}  \label{S-C-B}
\end{equation}
\ From Table 4, we have d$_{S}$(m) and u$_{C}$(m). From (\ref{C-B-Mass}) and
(\ref{S-C-B}), we have:

for the quark u$_{C}$(m)$\rightarrow $ $\left\{ 
\begin{tabular}{l}
u$_{C}$(m)+u({\small 313})+d({\small 313})] = $\Lambda _{C}$(m+626+$\Delta $%
e) and \\ 
u$_{C}$(m)+q$_{N}$({\small 313})+q$_{N}$({\small 313})] = $\Sigma _{c}$%
(m+626+$\Delta $e)%
\end{tabular}
\right\} $;

for the quark d$_{S}$(m)$\rightarrow $ \ $\left\{ 
\begin{tabular}{l}
d$_{S}$(m)+u({\small 313})+ d({\small 313}) = $\Lambda $(m+626) and \\ 
d$_{S}$(m)+q$_{N}$({\small 313})+ q$_{N}$({\small 313}) = $\Sigma $(m+626)%
\end{tabular}%
\ \ \right\} $.

Table 13 shows the $\Lambda _{c}$-baryons, the\ $\Sigma _{c}$-baryons, the $%
\Lambda $-baryons and the\ $\Sigma $-baryons:\qquad \qquad \qquad \qquad
\qquad \qquad\ \ \ 

\begin{tabular}{l}
\ \ \ \ \ \ \ \ \ \ \ \ \ \ \ \ Table 13. The $\Lambda _{c}$-Baryons and
the\ $\Sigma _{c}$-Baryons \\ 
\begin{tabular}{|l|l|l|l|l|l|l|}
\hline
$\text{E}_{Point}$ & $\text{q}_{\text{name}}\text{(m})$ & {\small m}$_{q_{2}%
\text{+}q_{_{3}}}$ & {\small \ }$\Delta ${\small e}$_{\Lambda }$ & $\Lambda $%
(M{\small (Mev)}) & {\small \ }$\Delta ${\small e} & $\Sigma $(M{\small (Mev)%
}) \\ \hline
$\text{E}_{H}\text{=1}$ & $\text{d}_{S}\text{(773)}$ & {\small 626} & \ \ 
{\small 0} & $\Lambda \text{(1399)}$ & \ {\small 0} & $\Sigma \text{(1399)}$
\\ \hline
$\text{E}_{\Gamma }\text{=4}$ & $\text{u}_{C}\text{(1753)}$ & {\small 626} & 
{\small -100} & $\Lambda _{c}\text{(2279)}$ & $\text{100}$ & $\Sigma _{c}%
\text{({\small 2479})}$ \\ \hline
$\text{E}_{H}\text{=9}$ & $\text{d}_{S}\text{(3753)}$ & {\small 626} & \ \ 
{\small 0} & $\Lambda \text{(4379)}$ & \ {\small 0} & $\Sigma \text{(4379)}$
\\ \hline
$\text{E}_{\Gamma }\text{=16}$ & $\text{u}_{C}\text{(6073)}$ & {\small 626}
& {\small -100} & $\Lambda _{c}\text{(6599)}$ & {\small 1}$\text{00}$ & $%
\Sigma _{c}\text{({\small 6799})}$ \\ \hline
$\text{E}_{H}\text{=25}$ & $\text{d}_{S}\text{(9613)}$ & {\small 626} & \ \ 
{\small 0} & $\Lambda \text{(10239)}$ & \ {\small 0} & $\Sigma \text{(%
{\small 10239})}$ \\ \hline
$\text{E}_{\Gamma }\text{=36}$ & $\text{u}_{C}\text{({\small 13273})}$ & 
{\small 626} & {\small -100} & $\Lambda _{c}\text{({\small 13799})}$ & 
{\small 1}$\text{00}$ & $\Sigma _{c}\text{({\small 13999})}$ \\ \hline
\end{tabular}%
\end{tabular}

\ \ \ \ \ \ \ \ \ \ \ \ \ \ \ \ \ 

C2. \ Deduction of the Charmed Baryons on the D-axis, the F-axis and the
G-axis

In this example, we will deduce low-rest-mass charmed baryons on the D-axis,
F-axis and G-axis. From Table B5 (the D-axis), we obtain the charmed quarks u%
$_{C}$(2133), u$_{C}$(2333), u$_{C}$(2533) and u$_{C}$(3543). From Table B6
(the F-axis), we obtain\ the charmed strange quarks q$_{\Xi _{C}}$(1873), q$%
_{\Xi _{C}}$(2163), q$_{\Xi _{C}}$(2363) and q$_{\Xi _{C}}$(3163). From
Table B7 (the G-axis), we obtain the charmed strange quarks \ q$_{\Omega
_{C}}$(2133) and q$_{\Omega _{C}}$(2513). Using the sum laws (\ref{Sum(SCbQ)}%
) and the binding energies of the charmed baryons\ (\ref{C-B-Mass}), we
have:\ 

\ \ \ \ \ \ \ \ \ \ \ \ \ \ \ \ \ \ \ \ \ \ \ \ \ \ \ \ \ \ \ \ 

\begin{tabular}{l}
\ \ \ \ \ \ \ \ Table 14.\ The Charmed Baryons on the D, F and G Axis \\ 
\begin{tabular}{|l|l|l|l|l|l|l|l|l|}
\hline
Axes & S$_{AX}$ & $\Theta $ & C & I & \ Quark$_{1}$ & m$_{q_{2}}$+m$_{q_{3}}$
& $\Delta $e & Baryon \\ \hline
D & 0 & 1 & 1 & 0 & 
\begin{tabular}{l}
{\small u}$_{C}${\small (2133)} \\ 
{\small u}$_{C}${\small (2333)} \\ 
{\small u}$_{C}${\small (2533)} \\ 
{\small u}$_{C}${\small (3543)}%
\end{tabular}
& 626 & -150 & 
\begin{tabular}{l}
$\Lambda _{C}${\small (2609)} \\ 
$\Lambda _{C}${\small (2809)} \\ 
$\Lambda _{C}${\small (3009)} \\ 
$\Lambda _{C}${\small (4019)}%
\end{tabular}
\\ \hline
F & -1 & 1 & 1 & $\frac{1}{2}$ & 
\begin{tabular}{l}
{\small q}$_{\Xi _{C}}${\small (1873)} \\ 
{\small q}$_{\Xi _{C}}${\small (2163)} \\ 
{\small q}$_{\Xi _{C}}${\small (2363)} \\ 
{\small q}$_{\Xi _{C}}${\small (3193)}%
\end{tabular}
& 626 & \ \ 0 & 
\begin{tabular}{l}
$\Xi _{C}${\small (2499)} \\ 
$\Xi _{C}${\small (2789)} \\ 
$\Xi _{C}${\small (2989)} \\ 
$\Xi _{C}${\small (3819)}%
\end{tabular}
\\ \hline
G & -2 & 1 & 1 & 0 & 
\begin{tabular}{l}
{\small q}$_{\Omega _{C}}${\small (2133)} \\ 
{\small q}$_{\Omega _{C}}${\small (2513)}%
\end{tabular}
& 626 & -50 & 
\begin{tabular}{l}
$\Omega _{C}${\small (2709)} \\ 
$\Omega _{C}${\small (3089)}%
\end{tabular}
\\ \hline
\end{tabular}%
\end{tabular}

\ \ \ \ \ \ \ \ \ \ \ \ \ \ \ \ \ \ \ \ \ \ \ \ \ \ \ \ \ \ \ \ \ \ \ \ \ \
\ 

C3. Deduction of the Uncharmed Baryons (C = 0)

For uncharmed baryons, Table 12 has already given the intrinsic quantum
numbers. Since C = 0$\rightarrow \Delta $e =0, from (\ref{Baryon M}), the
rest masses of the baryons M$_{\text{B}}$= m$_{q_{1}}$+ m$_{q_{2}}$+ m$%
_{q_{3}}$. \ \ \ \ \ \ \ \ \ \ \ \ \ \ \ \ \ \ 

Example 1. We deduce the baryons on the fourfold energy bands of the $\Delta 
$-axis. For the fourfold energy bands, from Table B1, C = 0; thus, $\Delta $%
e = 0 from (\ref{CB-Eb}). With (\ref{I of B}), we can get two kinds of
baryons (the N-baryons with I = $\frac{1}{2}$ and the $\Delta $-baryons with
I = $\frac{3}{2}$). From Table 3, we have q$_{\Delta }$(m) (q$_{1}$ of q$%
_{1} $q$_{2}$q$_{3}$). Using Table 12, we get q$_{2}$ and q$_{3}$. From (\ref%
{Baryon M}), we get the baryon mass M = m$_{\text{q}_{\Delta }}$+ {\small m}$%
_{q_{2}}$+$\ $m$_{q_{_{3}}}$ = m$_{\text{q}_{\Delta }}$+626(Mev) $%
\rightarrow \Delta $(M=m$_{\text{q}_{\Delta }}$+ 626) and N(M=m$_{\text{q}%
_{\Delta }}$+ 626):

\ \ \ \ \ \ \ \ \ \ \ \ \ \ \ \ \ \ \ \ \ \ \ \ 

\begin{tabular}{l}
\ \ \ \ \ \ \ \ Table 15 A. The $\Delta $-Baryons and the N-Baryons on the $%
\Delta $-axis \\ 
\begin{tabular}{|l|l|l|l|l|l|l|l|l|}
\hline
E$_{Point}$ & E($\overrightarrow{k,}\overrightarrow{n}$) & I$_{q_{1}}$ & $q_{%
\text{Name}}$(m) & {\small m}$_{q_{2}+q_{_{3}}}$ & I$_{\Delta }$ & $\Delta 
\text{(M) }$ & I$_{\text{N}}$ & $\text{N(M) }$ \\ \hline
$\text{E}_{H}\text{=1}$ & $\text{673}$ & $\frac{3}{2}$ & $\text{q}_{\Delta }%
\text{(673)}$ & 626 & $\frac{3}{2}$ & $\Delta \text{(1299)}$ & $\frac{1}{2}$
& $\text{N(1299)}$ \\ \hline
$\text{E}_{\Gamma }\text{=2}$ & $\text{1033}$ & $\frac{3}{2}$ & $\text{q}%
_{\Delta }\text{(1033{\small )}}$ & 626 & $\frac{3}{2}$ & $\Delta \text{%
(1659)}$ & $\frac{1}{2}$ & $\text{N(1659)}$ \\ \hline
& $\text{1033}$ & $\frac{3}{2}$ & $\text{q}_{\Delta }\text{(1033{\small )}}$
& 626 & $\frac{3}{2}$ & $\Delta \text{(1659)}$ & $\frac{1}{2}$ & $\text{%
N(1659)}$ \\ \hline
$\text{E}_{H}\text{=3}$ & $\text{1393}$ & $\frac{3}{2}$ & $\text{q}_{\Delta }%
\text{(1393)}$ & 626 & $\frac{3}{2}$ & $\Delta \text{(2019)}$ & $\frac{1}{2}$
& $\text{N(2019)}$ \\ \hline
$\text{E}_{\Gamma }\text{=4}$ & $\text{1753}$ & $\frac{3}{2}$ & $\text{q}%
_{\Delta }\text{(1753)}$ & 626 & $\frac{3}{2}$ & $\Delta \text{(2379)}$ & $%
\frac{1}{2}$ & $\text{N(2379)}$ \\ \hline
$\text{E}_{H}\text{=5}$ & $\text{2113}$ & $\frac{3}{2}$ & $\text{q}_{\Delta }%
\text{(2113)}$ & 626 & $\frac{3}{2}$ & $\Delta \text{(2739)}$ & $\frac{1}{2}$
& $\text{N(2739)}$ \\ \hline
& $\text{2113}$ & $\frac{3}{2}$ & $\text{q}_{\Delta }\text{(2113)}$ & 626 & $%
\frac{3}{2}$ & $\Delta \text{(2739)}$ & $\frac{1}{2}$ & $\text{N(2739)}$ \\ 
\hline
$\text{E}_{H}\text{=5}$ & $\text{2113}$ & $\frac{3}{2}$ & $\text{q}_{\Delta }%
\text{(2113)}$ & 626 & $\frac{3}{2}$ & $\Delta \text{(2739)}$ & $\frac{1}{2}$
& $\text{N(2739)}$ \\ \hline
$\text{E}_{H}\text{=5}$ & $\text{2113}$ & $\frac{3}{2}$ & $\text{q}_{\Delta }%
\text{(2113)}$ & 626 & $\frac{3}{2}$ & $\Delta \text{(2739)}$ & $\frac{1}{2}$
& $\text{N(2739)}$ \\ \hline
$\text{E}_{\Gamma }\text{=6}$ & $\text{2473}$ & $\frac{3}{2}$ & $\text{q}%
_{\Delta }\text{({\small 2473})}$ & 626 & $\frac{3}{2}$ & $\Delta \text{%
(3099)}$ & $\frac{1}{2}$ & $\text{N(3099)}$ \\ \hline
& $\text{2473}$ & $\frac{3}{2}$ & $\text{q}_{\Delta }\text{({\small 2473}+}%
\Delta \text{)}$ & 626 & $\frac{3}{2}$ & $\Delta \text{(3099)}$ & $\frac{1}{2%
}$ & $\text{N(3099)}$ \\ \hline
$\text{E}_{\Gamma }\text{=6}$ & $\text{2473}$ & $\frac{3}{2}$ & $\text{q}%
_{\Delta }\text{({\small 2473}+}\Delta \text{)}$ & 626 & $\frac{3}{2}$ & $%
\Delta \text{(3099)}$ & $\frac{1}{2}$ & $\text{N(3099)}$ \\ \hline
\end{tabular}%
\end{tabular}

\ \ \ \ \ \ \ \ \ \ \ \ \ \ \ \ \ \ \ \ \ \ \ \ \ \ \ \ \ \ \ \ \ 

Example 2. We deduce the N(M)-baryons and the $\Delta $(M)-baryons on the
D-axis. From Table B5, we get the q$_{N}$(m)-quarks{\small .} As with
example 1, q$_{N}$(m)$\rightarrow $ N(M=m$_{\text{q}_{\Delta }}$+626) and $%
\Delta $(M=m$_{\text{q}_{\Delta }}$+ 626) at point p; from Table A6, q$_{N}$%
(m)$\rightarrow $N(M=m$_{\text{q}_{\Delta }}$+626) at point N:

\begin{tabular}{l}
Table 15B. The N(m)-baryons and the $\Delta $(M)-baryons on the D-axis \\ 
$%
\begin{tabular}{|l|l|l|l|l|l|l|l|}
\hline
$\text{E}_{Start}$ & {\small E} & \ {\small q}$_{N}${\small (m)} & {\small m}%
$_{q_{2}+q_{_{3}}}$ & I$_{\text{B}}$ & $\text{N(M)}$ & I$_{\text{B}}$ & $%
\Delta \text{(M)}$ \\ \hline
{\small E}$_{\text{P}}${\small \ = }$\frac{3}{4}$ & {\small 583} & {\small \
\ q}$_{N}^{0}${\small (583)} & 626 & $\frac{1}{2}$ & N(1209) & $\frac{3}{2}$
& $\ \Delta \text{(1209)}$ \\ \hline
{\small E}$_{\text{N}}${\small \ = }$\frac{3}{2}$ & {\small 853} & {\small \
\ q}$_{N}^{0}${\small (853)} & 626 & $\frac{1}{2}$ & N(1479) &  &  \\ \hline
{\small E}$_{\text{N}}${\small \ = }$\frac{5}{2}$ & {\small 1213} & 2{\small %
q}$_{N}^{0}${\small (1213)} & 626 & $\frac{1}{2}$ & N(1839) &  &  \\ \hline
{\small E}$_{\text{P}}${\small \ = }$\frac{11}{4}$ & {\small 1303} & 2%
{\small q}$_{N}^{0}${\small (1303)} & 626 & $\frac{1}{2}$ & N(1929) & $\frac{%
3}{2}$ & 2$\Delta \text{(1929)}$ \\ \hline
{\small E}$_{\text{N}}${\small \ = }$\frac{7}{2}$ & {\small 1573} & 2{\small %
q}$_{N}^{0}${\small (1573)} & 626 & $\frac{1}{2}$ & N(2199) &  &  \\ \hline
{\small E}$_{\text{p}}${\small \ = }$\frac{19}{4}$ & {\small 2023} & {\small %
\ \ q}$_{N}^{0}${\small (2023)} & 626 & $\frac{1}{2}$ & N(2649) & $\frac{3}{2%
}$ & $\ \Delta \text{(2649)}$ \\ \hline
{\small E}$_{\text{P}}${\small \ = }$\frac{19}{4}$ & {\small 2023} & 2%
{\small q}$_{N}^{0}${\small (2023)} & 626 & $\frac{1}{2}$ & N(2649) & $\frac{%
3}{2}$ & 2$\Delta \text{(2649)}$ \\ \hline
{\small E}$_{\text{N}}${\small \ = }$\frac{11}{2}$ & {\small 2293} & {\small %
\ \ q}$_{N}^{0}${\small (2293)} & 626 & $\frac{1}{2}$ & N(2919) &  &  \\ 
\hline
{\small E}$_{\text{N}}${\small \ = }$\frac{13}{2}$ & {\small 2653} & 2%
{\small q}$_{N}^{0}${\small (2653)} & 626 & $\frac{1}{2}$ & N(3279) &  &  \\ 
\hline
\end{tabular}
\ $%
\end{tabular}

\ \ \ \ \ \ \ \ \ \ \ \ \ \ \ \ \ \ \ \ \ \ \ \ \ \ \ \ \ \ \ \ \ \ \ \ \ \
\ 

We have already deduced low mass charmed baryons in Table 13 and Table 14.
Using sum laws, we deduced some uncharmed baryons in Table 15A and Table
15b. Continuing the above procedure, we can deduce a baryon spectrum that is
shown in Table 16-Table 21.\ \ \ \ \ \ \ \ \ \ \ \ \ \ \ \ \ \ \ \ \ \ \ \ \
\ \ \ \ \ \ \ \ \ \ \ \ \ \ \ \ \ \ \ \ \ \ \ \ 

\subsection{Comparing with the Experimental Results\ of Baryons\ \ \ \ \ \ \
\ \ \ \ \ \ \ \ \ \ \ \ \ \ }

Using Table 16 -- Table 21, we can compare the deduced baryon spectrum with
the experimental results \cite{Baryon04}. In this comparison, we do not take
into account the angular momenta of the experimental results. We assume that
the small differences of the masses in the same group of baryons with the
same quantum numbers (I, S, C, b and Q) are from their different angular
momenta. If we ignore this effect, their masses would be essentially the
same. Since all baryons in the group have the same intrinsic quantum numbers
with the same name, we use the baryon name to represent the intrinsic
quantum numbers, as shown in the second column of Table 16. If the name is
the same between the deduced baryon and the experimental baryon, this means
that the intrinsic quantum numbers (I, S, C, b and Q) are exactly the same.
We use the baryons with the average rest mass of the group of baryons (see
Table C1) to represent the group of the baryons. The mass units of quarks
and baryons as well as the widths are ``Mev''\ :

\begin{tabular}{l}
\ \ \ \ \ \ \ \ \ \ \ \ \ \ \ \ \ \ \ \ \ Table 16. \ The Ground Baryons. \\ 
\begin{tabular}{|l|l|l|l|l|}
\hline
Deduced & Quantum No. & Experiment & $\frac{\Delta \text{M}}{\text{M}}\%$ & 
Lifetime \\ \hline
Name({\small M}) & \ S, \ C,\ b, \ \ \ I, \ \ Q & Name({\small M}) &  &  \\ 
\hline
p(939) & \ 0,\ \ 0, \ 0, \ $\frac{1}{2}${\small , \ \ }1 & p(938) & 0.11 & 
\TEXTsymbol{>}10$^{29}years$ \\ \hline
n(939) & 0, \ 0, \ 0, \ $\frac{1}{2}${\small , \ \ }0 & n(940) & 0.11 & 
885.7 s \\ \hline
$\Lambda ^{0}$(1119) & -1, \ 0, \ 0, \ \ 0, \ \ 0 & $\Lambda ^{0}(1116)$ & 
0.27 & 2.6$\times $ \ 10$^{-10}$s \\ \hline
$\Sigma ^{+}$(1209) & -1, \ 0, \ 0, \ \ 1, \ \ 1 & $\Sigma ^{+}(1189)$ & 1.7
& .80$\times $ \ 10$^{-10}$s \\ \hline
$\Sigma ^{0}$(1209) & -1, \ 0, \ 0, \ \ 1, \ \ 0 & $\Sigma ^{0}(1193)$ & 1.4
& 7.4$\times $ \ 10$^{-20}$s \\ \hline
$\Sigma ^{-}$(1209) & -1, \ 0, \ 0, \ \ 1, \ -1 & $\Sigma ^{-}(1197)$ & 1.0
& 1.5$\times $10$^{-10}$s \\ \hline
$\Xi ^{0}$(1299) & -2, \ 0, \ 0, \ $\frac{1}{2}$, \ \ 0 & $\Xi ^{0}(1315)$ & 
1.2 & 2.9$\times $10$^{-10}$s \\ \hline
$\Xi ^{-}$(1299) & -2, \ 0, \ 0, \ $\frac{1}{2}$, -1 & $\Xi ^{-}(1321)$ & 1.7
& 1.6$\times $10$^{-10}$s \\ \hline
$\Omega ^{-}$(1659) & -3, \ 0, \ 0. \ \ 0,\ \ -1 & $\Omega ^{-}(1672)$ & 0.78
& .82$\times $10$^{-10}$s \\ \hline
$\Lambda _{c}^{+}$(2279) & 0, \ 1, \ 0, \ \ 0, \ \ 1 & $\Lambda
_{c}^{+}(2285)$ & 0.26 & 200$\times $10$^{-15}$s \\ \hline
$\Xi _{c}^{+}$(2499) & -1, \ 1, \ 0, \ $\frac{1}{2}$, \ 1 & $\Xi
_{c}^{+}(2466)$ & 1.4 & 442$\times $10$^{-15}$ \\ \hline
$\Xi _{c}^{0}$(2499) & -1, \ 1, \ 0, \ $\frac{1}{2}$, \ 0 & $\Xi
_{c}^{0}(2472)$ & 1.2. & 112$\times $10$^{-15}$s \\ \hline
$\Omega _{C}^{0}$(2709) & 0, \ 0, -1, \ \ 0, \ \ 0 & $\Omega _{c}^{0}(2698)$
& 0.41 & 69$\times $10$^{-15}s$ \\ \hline
$\Lambda _{b}^{0}\text{(5539)}$ & 0, \ 0, -1, \ \ 0, \ \ 0 & $\Lambda
_{b}^{0}(5624)$ & 1.5 & 1.2310$^{-12}s$ \\ \hline
$\Sigma _{C}^{++}$(2479) & -1, \ 1, \ 0, \ \ 1, \ \ 2 & $\Sigma
_{C}^{++}(2453)$ & 1.1 & $\Gamma $=2.2 Mev \\ \hline
$\Sigma _{C}^{+}$(2479) & -1, \ 1, 0, \ \ 1, \ \ 1 & $\Sigma _{C}^{++}(2451)$
& 1.2 & $\Gamma $\TEXTsymbol{<}4.6 Mev \\ \hline
$\Sigma _{C}^{0}$(2479) & -1, \ 1, \ 0, \ \ 1, \ \ 0 & $\Sigma
_{C}^{++}(2452)$ & 1.2 & $\Gamma $=2.1 Mev \\ \hline
\end{tabular}%
\end{tabular}

\ \ \ \ \ \ \ \ \ \ \ \ \ \ \ \ \ \ \ \ \ \ \ \ \ \ \ \ \ \ \ \ \ \ \ \ \ \
\ \ 

The most important baryons are shown in Table 16. \ These baryons have
relatively long lifetimes. They are the most important experimental results
of the baryons. Their deduced intrinsic quantum numbers are the same as the
experimental results. The deduced mass values are over 98\% consistent with
the experimental values.\ \ 

\ \ \ \ \ \ \ \ \ \ \ \ \ \ \ \ \ \ \ \ \ \ \ \ \ \ \ \ \ \ \ \ \ \ \ \ \ \
\ \ \ \ \ \ \ \ \ \ \ \ \ \ \ \ \ \ \ \ 

Two kinds of the strange baryons $\Lambda $ and $\Sigma $ are compared in
Table 17. Their deduced intrinsic quantum numbers are the same as the
experimental results.\ The deduced masses of the baryons $\Lambda $ and $%
\Sigma $ are about 98\% consistent with the experimental results.

\ \ \ \ \ \ \ \ \ \ \ \ \ \ \ \ \ \ \ \ \ \ \ \ \ \ \ \ \ \ \ \ \ \ 

\bigskip 
\begin{tabular}{l}
\ \ \ \ \ \ \ \ \ Table 17. Two Kinds of Strange Baryons $\Lambda $ and $%
\Sigma $ ($S=-1$) \\ 
\begin{tabular}{|l|l|l||l|l|l|}
\hline
Deduced & Experiment, $\Gamma $ & $\frac{\Delta \text{M}}{\text{M}}\%$ & 
Deduced & Experiment, $\Gamma $ & $\frac{\Delta \text{M}}{\text{M}}\%$ \\ 
\hline
$\Lambda $(1119) & $\Lambda $(1116) & 0.36 & $\Sigma $(1209) & $\Sigma $%
(1193) & 1.4 \\ \hline
$\Lambda $(1399) & $\Lambda $(1405),50\ \ \  & 0.43 & $\Sigma $(1399) & $%
\Sigma $(1385),37 & 1.0 \\ \hline
$\Lambda $($\overline{\text{1659}}$) & $\Lambda $($\overline{\text{1620}}$),$%
\overline{\text{65}}$ & 2.4 & $\Sigma $($\overline{\text{1727}}$) & $\Sigma $%
($\overline{\text{1714}}$),$\overline{\text{93}}$ & 0.76 \\ \hline
$\Lambda $($\overline{\text{1889}}$) & $\Lambda $($\overline{\text{1830}}$),$%
\overline{\text{145}}$ & 3.2 & $\Sigma $($\overline{\text{1929}}$) & $\Sigma 
$($\overline{\text{1928}}$),$\overline{\text{170}}$ & 0.05 \\ \hline
$\Lambda $($\overline{\text{2079}}$) & $\Lambda $($\overline{\text{2105}}$),$%
\overline{\text{200}}$ & 1.24 & $\Sigma $($\overline{\text{2019}}$) & $%
\Sigma $($\overline{\text{2045}}$),$\overline{\text{220}}$ & 1.3 \\ \hline
$\Lambda $($\overline{\text{2339}}$) & $\mathbf{\Lambda (}\overline{\text{%
2338}}$\textbf{),} $\overline{\text{159}}$ & 0.04 & $\Sigma $($\overline{%
\text{2249}}$) & $\Sigma $(2250),100 & 0.05 \\ \hline
$\Lambda $($\overline{\text{2615}}$) & $\Lambda $(2585)$^{\ast }$,300 & 1.2
& $\Sigma $($\overline{\text{2492}}$) & $\Sigma $(2455)$^{\ast }$,140 & 1.1
\\ \hline
$\Lambda $(3099) & Prediction &  & $\Sigma $($\overline{\text{2644}}$) & $%
\Sigma $(2620)$^{\ast }$,200 & 0.84 \\ \hline
$\Lambda $(3369) & Prediction &  & $\Sigma $(3099) & $\Sigma $(3085)$^{\ast
} $, \ ? & 0.45 \\ \hline
&  &  & $\Sigma $(3369) & Prediction &  \\ \hline
\end{tabular}
\\ 
{\small \ *Evidences of existence for these baryons are only fair; they are
not listed} \\ 
{\small in the Baryon Summary Table \cite{Baryon04}.}%
\end{tabular}

\ \ \ \ \ \ \ \ \ \ \ \ \ \ \ \ \ \ \ \ \ \ \ \ \ \ \ \ \ \ \ \ \ 

Table 18 compares the deduced results with the experimental results of the
unflavored baryons $N$ and $\Delta $:

\ \ \ \ \ \ \ \ \ \ \ \ \ \ \ \ \ \ \ \ \ \ \ \ \ \ \ \ \ \ 

\ 
\begin{tabular}{l}
\ \ \ \ \ \ \ \ \ \ \ \ Table 18. The Unflavored Baryons $N$ and $\Delta $ ($%
S$= $C$=$b$ = 0) \\ 
\begin{tabular}{|l|l|l||l|l|l|}
\hline
Deduced & Experiment, $\Gamma $ & $\frac{\Delta \text{M}}{\text{M}}$\% & 
Deduced & Experiment, $\Gamma $ & $\frac{\Delta \text{M}}{\text{M}}$\% \\ 
\hline
N(939) & N(939) & 0.0 &  &  &  \\ \hline
N($\overline{\text{1254}}$) & Covered by $\Delta $(1232) &  & $\Delta $(1254)
& $\Delta $(1232),120 & 1.8 \\ \hline
N(1479) & N($\overline{\text{1498}}$),$\overline{\text{207}}$ & 1.3 &  &  & 
\\ \hline
N($\overline{\text{1650}}$) & N($\overline{\text{1689}}$),$\overline{\text{%
130}}$ & 2.3 & $\Delta $($\overline{\text{1659}}$) & $\Delta $($\overline{%
\text{1640}}$),$\overline{\text{267}}$ & 1.2 \\ \hline
N($\overline{\text{1929}}$) & N($\overline{\text{1912}}$),440 & 0.31 & $%
\Delta $($\overline{\text{1955}}$) & $\Delta $($\overline{\text{1923}}$),$%
\overline{\text{264}}$ & 1.7 \\ \hline
N($\overline{\text{2199}}$) & N($\overline{\text{2220}}$),$\overline{\text{%
\textit{417}}}$ & 0.95 &  &  &  \\ \hline
N(2379) & Covered by $\Delta $(2420) &  & $\Delta $(2379) & $\Delta $%
(2420),400 & 1.7 \\ \hline
N($\overline{\text{2649}}$) & N(2600),650 & 1.9 &  &  &  \\ \hline
N(2739) & N(2700)$^{\ast }$,600 & 1.5 & $\Delta $($\overline{\text{2694}}$)
& $\Delta $(2750)$^{\ast }$,400 & 2.0 \\ \hline
N(2919) & Prediction &  & $\Delta $(3099) & $\Delta $($\overline{\text{2975}}
$)$^{\ast }$,$\overline{\text{750}}$ &  \\ \hline
N(3099) & Prediction &  & $\Delta $(3369) & Prediction &  \\ \hline
\end{tabular}
\\ 
{*}{\small Evidences are fair, they are not listed in the Baryon Summary
Table }\cite{Baryon04}.%
\end{tabular}

\ \ \ \ \ \ \ \ \ \ \ \ \ \ \ \ \ \ \ \ \ \ \ \ \ \ \ \ \ \ \ \ \ \ \ \ \ \
\ \ \ \qquad\ 

\ The deduced masses of the baryons $N$ and $\Delta $\ are 98\% consistent
with the experimental results. We do not find the deduced N(1209) and
N(1299) in the experiment results. We believe that they are covered up by
the experimental baryon $\Delta (1232)$ because of the following reasons:
(1) they are unflavored baryons with the same S, C and b; (2) the width (120
Mev) of $\Delta (1232)$ is very large, and the baryons $N(1209)$ and $%
N(1299) $ both fall within the width region of $\Delta (1232)$; (3) the
average mass (1255 Mev) of $N(1209)$ and $N(1299)$ is essentially the same
as the mass (1232 Mev) of $\Delta (1232)$ $(\Gamma $ = 120 Mev).

\ \ 

The deduced intrinsic quantum numbers of the baryons $\Xi $ and $\Omega $
are the same as the experimental results (see Table 19). The deduced masses
of the baryons $\Xi $ and $\Omega $ are more than 98\% compatible with
experimental results.\ \ \ \ \ \ \ \ \ \ \ \ \ \ \ \ \ \ \ \ \ \ \ \ \ \ \ 

\ \ \ \ \ \ \ \ \ \ \ \ \ \ \ \ \ \ \ \ \ \ \ \ \ \ \ \ 

$%
\begin{tabular}{l}
\ \ \ \ \ \ \ \ \ \ \ \ \ \ \ \ \ \ Table 19. The Baryons $\Xi $ and the
Baryons $\Omega $ \\ 
\begin{tabular}{|l|l|l||l|l|l|}
\hline
Deduced & Experiment, $\Gamma $ & $\frac{\Delta \text{M}}{\text{M}}\%$ & 
Deduced & Experiment, $\Gamma $ & $\frac{\Delta \text{M}}{\text{M}}\%$ \\ 
\hline
$\Xi $(1299) & $\Xi $(1318) & 1.4 & $\Omega $(1659) & $\Omega $(1672) & 
\textbf{0.7} \\ \hline
$\Xi $(1479) & $\Xi $(1530), 9.9 & 3.3 & $\Omega $(2259) & $\Omega $(2252),
55 & 0.4 \\ \hline
$\Xi $(1659) & $\Xi $(1690), \TEXTsymbol{<}30 & 1.8 & $\Omega $(2439) & $%
\Omega (\overline{2427})^{\ast },\overline{48}$ & 0.5 \\ \hline
$\Xi $(1839) & $\Xi $(1823), 24 & 1.1 & $\Omega $(2979) & Prediction &  \\ 
\hline
$\Xi $(2019) & $\Xi $($\overline{\text{1988}}$)$^{\#}$,$\overline{40}$ & 1.6
& $\Omega $(3819) & Prediction &  \\ \hline
$\Xi $(2199) & $\Xi $($\overline{\text{2185}}$)$^{\$}$,33 & 0.64 &  &  &  \\ 
\hline
$\Xi $($\overline{\text{2369}}$) & $\Xi $(2370)$^{\ast }$,80 & 0.04 & $%
\Omega (\overline{\text{2427}})^{\ast }$ & =$\frac{1}{2}${\small [}$\Omega $%
{\small (2380)}$^{\ast }${\small +}$\Omega ${\small (2474)}$^{\ast }${\small %
]} &  \\ \hline
$\Xi $($\overline{\text{2549}}$) & $\Xi $(2500)$^{\ast }$,150 & 2.0 & $\Xi $(%
$\overline{\text{1988}}$)$^{\#}$ & =$\frac{1}{2}${\small [}$\Xi ${\small %
(1950)+}$\Xi ${\small (2025)]} &  \\ \hline
$\Xi $(2759) & Prediction &  & $\Xi $($\overline{\text{2185}}$)$^{\$}$ & =$%
\frac{1}{2}${\small [}$\Xi ${\small (}$\overline{2120}${\small )}$^{\ast }$%
{\small +}$\Xi ${\small (}$\overline{2250}${\small )}$^{\ast }${\small ]} & 
\\ \hline
\end{tabular}
\\ 
{\small *Evidences of existence for these baryons are only fair: they are
not listed } \\ 
\ {\small in the Baryon Summary Table \cite{Baryon04}.}%
\end{tabular}
\ \ ${\small .}$\ \ \ $

\ \ \ \ \ \ \ \ \ \ \ \ \ \ \ \ \ \ \ \ \ \ \ 

\ We compare the charmed $\Lambda _{c}^{+}$-baryon, $\mathbf{\Omega }_{C}$%
-baryon and $\Lambda _{b}^{0}$-baryon in Table 20. Their deduced intrinsic
quantum numbers are the same as the experimental results. \ The experimental
masses of the charmed baryons ($\Lambda _{c}^{+}$ and $\mathbf{\Omega }_{C}$%
) and bottom baryons ($\Lambda _{b}^{0}$) coincide more than 98\% with the
deduced results.\ 

\ \ \ \ \ \ \ \ \ \ \ \ \ \ \ \ \ \ \ \ \ \ \ \ \ \ 

\begin{tabular}{l}
\ \ \ \ \ \ Table 20. The\ $\Lambda _{c}^{+}$-Baryons, the $\Lambda _{b\text{
}}^{0}$-Baryons and the $\mathbf{\Omega }_{C}$-Baryons \\ 
\begin{tabular}{|l|l|l||l|l|l|}
\hline
Deduced & Experiment & $\frac{\Delta \text{M}}{\text{M}}\%$ & Deduced & 
Experiment & $\frac{\Delta \text{M}}{\text{M}}\%$ \\ \hline
$\Lambda _{c}^{+}$\textbf{(2279)} & $\ \Lambda _{c}^{+}$\textbf{(2285)} & 0.2
& $\Lambda _{b}^{0}$(5539) & $\Lambda _{b}^{0}$(5624) & 1.5 \\ \hline
$\Lambda _{C}^{+}$\textbf{(2609)} & 
\begin{tabular}{l}
$\Lambda _{C}^{+}$(2594) \\ 
$\Lambda _{C}^{+}$(2627) \\ 
$\Lambda _{c}^{+}$($\overline{\text{\textbf{2611}}}$)%
\end{tabular}
& 0.46 & $\Lambda _{b}^{0}$(9959) & Prediction &  \\ \hline
$\Lambda _{C}^{+}$\textbf{(2809)} & 
\begin{tabular}{l}
$\Lambda _{C}^{+}$(2765)$^{\ast }$ \\ 
$\Lambda _{C}^{+}$(2881)$^{\ast }$ \\ 
$\Lambda _{c}^{+}$($\overline{\text{\textbf{2823}}}$)$^{\ast }$%
\end{tabular}
& 0.85. &  &  &  \\ \hline
$\Lambda _{C}^{+}$\textbf{(3009)} & \ Prediction &  & $\mathbf{\Omega }_{C}$%
\textbf{(2709)} & $\mathbf{\Omega }_{C}$\textbf{(2698)} & 0.44 \\ \hline
$\Lambda _{C}^{+}$\textbf{(4019)} & \ Prediction &  & $\Omega _{C}$\textbf{%
(3089)} & Prediction &  \\ \hline
\end{tabular}
\\ 
{\small *Evidences of existence for these baryons are only fair;} \\ 
{\small they are not list in the Baryon Summary Table \cite{Baryon04}.}%
\end{tabular}

\ \ \ \ \ \ \ \ \ \ \ \ \ \ \ \ \ \ \ \ \ \ \ 

\ Finally we compare the deduced results with the experimental results for
the charmed strange baryons $\Xi _{c}$ and $\Sigma _{c}$ in Table 21.\ Their
intrinsic quantum numbers all match exactly and their masses also agree well
(more than 98\%).

\ \ 

\begin{tabular}{l}
\ \ \ \ \ \ \ \ \ \ \ \ \ \ \ \ Table 21. Charmed Strange Baryon $\Xi _{c}$
and $\Sigma _{c}$ \\ 
\begin{tabular}{|l|l|l||l|l|l|}
\hline
Deduced & Experiment & $\frac{\Delta \text{M}}{\text{M}}\%$ & Deduced & 
Experiment & $\frac{\Delta \text{M}}{\text{M}}\%$ \\ \hline
$\Xi _{c}$(\textbf{2499}) & 
\begin{tabular}{l}
$\Xi _{C}$(2469) \\ 
$\Xi _{C}$(2576) \\ 
$\Xi _{c}$($\overline{\text{\textbf{2523}}}$\textbf{)}%
\end{tabular}
& 1.0. & $\mathbf{\ \Sigma }_{c}$\textbf{(2479)} & 
\begin{tabular}{l}
$\Sigma _{c}$(2452)$\mathbf{\ }$ \\ 
$\Sigma _{c}$(2518) \\ 
$\mathbf{\Sigma }_{c}$\textbf{(}$\overline{\text{\textbf{2485}}}$\textbf{)}%
\end{tabular}
& 0.24 \\ \hline
$\Xi _{C}$\textbf{(2789)} & 
\begin{tabular}{l}
$\Xi _{C}$(2645) \\ 
$\Xi _{C}$(2790) \\ 
$\Xi _{C}$(2815) \\ 
$\Xi _{c}$($\overline{\text{\textbf{2750}}}$)%
\end{tabular}
& \textbf{1.4} & $\mathbf{\Sigma }_{c}\text{(\textbf{6799})}$ & Prediction & 
\\ \hline
$\Xi _{C}$\textbf{(2989)} & $\Pr $ediction &  &  &  &  \\ \hline
$\Xi _{C}$\textbf{(3819)} & Prediction &  &  &  &  \\ \hline
\end{tabular}
\\ 
\ \ {\small *Evidences of existence for these baryons are only fair; they
are not listed} \\ 
{\small \ \ \ \ \ \ in the Baryon Summary Table \cite{Baryon04}.}%
\end{tabular}

\ \ \ \ \ \ \ \ \ \ \ \ \ \ \ \ \ \ \ \ \ \ \ \ \ \ \ \ \ \ \ \ \ \ \ \ \ \
\ \ \ \ \ \ \ \ 

\ \ In summary, the phenomenological formulae explain all baryon
experimental intrinsic quantum numbers (100\%) and the rest masses (about
98\%). We explain virtually all experimentally-confirmed baryons in this
paper.

\section{The Meson Spectrum \ \ \ \ \ \ \ \ \ \ \ \ \ \ \ \ \ \ \ \ \ \ \ \
\ \ \ \ \ \ \ \ \ \ \ \ }

According to the Quark Model \cite{Quark Model}, a meson is made of a quark$%
\ $q$_{i}$ with one of the three colors and an antiquark $\overline{q_{j}}$
with the anticolor of the quark q$_{i}$. Thus, all mesons are colorless
particles. Because the mesons are colorless and the intrinsic quantum
numbers of the quarks are independent from the quarks' colors, we can omit
the colors when we deduce the intrinsic quantum numbers of mesons from the
quarks. Since we have already found the quark spectrum (Table 10 and Table
11), using sum laws, we can find the intrinsic quantum numbers (S, C, b, I
and Q) of the mesons (q$_{i}\overline{\text{q}_{j}}$): 
\begin{equation}
\begin{tabular}{l}
S$_{\text{M}}$= S$_{\text{q}_{i}}$+ S$_{\overline{q_{j}}}$, \\ 
C$_{\text{M}}$= C$_{\text{q}_{i}}$+ C$_{\overline{q_{j}}}$, \\ 
b$_{\text{M}}$= b$_{\text{q}_{i}}$+ b$_{\overline{q_{j}}}$, \\ 
Q$_{\text{M}}$= Q$_{\text{q}_{i}}$+ Q$_{\overline{q_{j}}}$, \\ 
$\overrightarrow{I}_{\text{M}}$= $\overrightarrow{I}_{\text{q}_{i}}$+ $%
\overrightarrow{I}_{\overline{q_{j}}}$.%
\end{tabular}
\label{Meson-Sum}
\end{equation}
We cannot, however, find the rest masses of mesons using a sum law because
of their binding energies.

\subsection{The Phenomenological Binding Energy Formula of Mesons}

There is not a theoretical formula for the binding energies; thus, we
propose a phenomenological formula. Because all quarks inside mesons are the
excited states of the elementary $\epsilon $-quarks, the binding energies
are roughly a constant (-$2\Delta $ - 337 Mev). Quarks and antiquarks all
have large rest masses. Although mesons are composed with large rest mass
quarks and antiquarks, the mesons themselves do not have large masses. Thus,
we use the (-2$\Delta $) to cancel the large part (2$\Delta $) of the quark
and antiquark masses. If the difference between the quark mass (m$_{i}$) and
the antiquark mass (m$_{j}$) in the quark pair (q$_{i}\overline{q_{j}}$) is
larger, the binding energy is smaller (100$\frac{\Delta \text{m}}{\text{m}%
_{g}}$). If [($\Delta $S)$_{i}$-($\Delta $S)$_{j}$] is larger, the binding
energy will be smaller [DS =$\left\vert \text{(}\Delta \text{S)}_{i}\text{- (%
}\Delta \text{S)}_{j}\right\vert $]. The pairs (q$_{i}\overline{q_{i}}$)
have larger binding energies than the pairs (q$_{i}\overline{q_{j}},i\neq j$%
), as shown in the following phenomenological formula for meson (q$_{i}%
\overline{q_{j}}$):

\begin{equation}
\text{E}_{B}\text{(q}_{i}\overline{q_{j}}\text{) = -2}\Delta \text{ -337
+100[}\frac{\Delta \text{m}}{\text{m}_{g}}\text{ +DS - }\widetilde{m}\text{
+ }\gamma (i,j)\text{ -0.78}\Delta \text{IS}_{i}\text{S}_{j}\text{ +(5.35}%
\Delta \text{I-2)I}_{i}\text{I}_{j}\text{]}  \label{M-Ebin}
\end{equation}
where $\Delta $ = $\frac{1}{3}\left\vert \text{E}_{bind}\right\vert $ (\ref%
{Dalta}) is $\frac{1}{3}$binding energy of a baryon (an unknown huge
constant, $\Delta $ \TEXTsymbol{>}\TEXTsymbol{>} m$_{\text{P}}$= 938 Mev); $%
\Delta $m = $\left\vert \text{m}_{i}\text{-m}_{j}\right\vert $, DS =$%
\left\vert \text{(}\Delta \text{S)}_{i}\text{- (}\Delta \text{S)}%
_{j}\right\vert $ and $\Delta $I = $\left\vert \text{I}_{i\text{ }}\text{-I}%
_{j}\right\vert $. m$_{g}$ = 939 (Mev) unless 
\begin{equation}
\begin{tabular}{|l|l|l|l|}
\hline
m$_{i}$(or m$_{j}$) equals & m$_{C}\geqslant $ 6073 & m$_{b}\geqslant $ 9333
& m$_{S}\geqslant $ 9613 \\ \hline
$\ \ \ \text{m}_{g}${\small \ will equal to} & 1753(Table 4) & 4913 (Table7)
& 3753(Table 4). \\ \hline
\end{tabular}
\label{m(g)}
\end{equation}

$\ \widetilde{m}$ = $\frac{m_{i}\times m_{j}}{\text{m}_{g_{i}}\times \text{m}%
_{g_{j}}}$ \ \ m$_{g_{i}}$ = m$_{g_{j}}^{\text{ \ }}$ = 939 (Mev) unless 
\begin{equation}
\begin{tabular}{|l|l|l|l|l|l|}
\hline
\ m$_{i}$(or m$_{j}$) & m$_{q_{_{N}}}$=313 & m$_{d_{s}}$=493 & m$%
_{u_{c}}\succeq $1753 & m$_{d_{S}}$\TEXTsymbol{>} 3753, & m$_{d_{b}}\succeq $
4913 \\ \hline
\ m$_{g_{j}}$ (or m$_{g_{j}}^{\text{ \ }})$ & 313 & 493 & 1753 & 3753, & 
4913. \\ \hline
\end{tabular}
\label{M(gi)}
\end{equation}
\ If\ q$_{i\text{ }}$and q$_{j}$ are both ground quarks, $\gamma $(i, j) =
0. If\ q$_{i\text{ }}$and q$_{j}$ are not both ground quarks, for q$_{i\text{
}}$= q$_{j}$, $\gamma $(i, j) = -$1$; for q$_{i\text{ }}\neq $ \ q$_{j}$,$\
\gamma $(i, j) = +1. S$_{i}$ (or S$_{j}$) is the strange number of the quark
q$_{i}$ (or q$_{j}$). I$_{i}$ (or I$_{j}$) is the isospin of the quark q$%
_{i} $ (or q$_{j}$).

We now have the sum laws (\ref{Meson-Sum}) and binding energy formulae (\ref%
{M-Ebin}). Using these formulae, we can deduce mesons from the quark
spectrum in Table 10 and Table 11.

\subsection{How to Deduce Mesons from Quarks Using these Formulae}

\ Three types of examples show how to deduce mesons using these formulae. We
will also deduce some important mesons at the same time.\ 

B1. The Mesons of the Ground Quarks and the Ground Antiquarks

We deduce the mesons that are composed of the five ground quarks ( u$^{0}$%
(313), d$^{0}$(313), d$_{S}^{1}$(493), u$_{C}^{1}$(1753) and d$_{b}^{1}$%
(4913)) and their antiquarks to show how to use the phenomenological
formulae (\ref{M-Ebin}) to deduce the rest masses of the mesons from the
quark spectrum. For the ground quarks, \ $\gamma $(i, j) = 0, $\widetilde{m}$
= 1 from (\ref{M(gi)}) and $\Delta $IS$_{i}$S$_{j}$ = $\Delta $I(I$_{i}$I$%
_{j})$ = 0; the formula (\ref{M-Ebin}) is simplified to

\begin{equation}
\text{{\Huge \ }E}_{B}\text{(q}_{i}\overline{q_{j}}\text{)= -2}\Delta \text{
- 437 + 100(}\frac{\Delta \text{m}}{\text{939}}\text{ + DS - 2I}_{i}\text{I}%
_{j}\text{).}  \label{GEbin}
\end{equation}
\ Using (\ref{GEbin}) we deduced the masses of the important mesons as shown
in Table 22:

\ \ \ \ \ \ \ \ \ \ \ \ \ \ \ 

\begin{tabular}{l}
$\ \ \ \ \ \ \ \ \ \ \ \ \text{Table 22.\ \ The Ground Mesons of the Ground
Quarks}$ \\ 
$%
\begin{tabular}{|l|l|l|l|l|l|l|}
\hline
$\text{q}_{i}^{\Delta S}\text{(m}_{i}\text{)}\ \overline{\text{q}%
_{j}^{\Delta S}\text{(m}_{j}\text{)}}$ & $\frac{100\Delta \text{m}}{\text{939%
}}$ & D{\small S} & $\text{2I}_{i}\text{I}_{j}$ & {\small E}$_{bind}$ & 
{\small Deduced} & {\small Experiment} \\ \hline
$\text{q}_{N}^{0}\text{(313+}\Delta \text{)}\overline{\text{q}_{N}^{0}\text{%
(313+}\Delta \text{)}}$ & {\small 0} & $0$ & $\frac{1}{2}$ & {\small - 487-}$%
2\Delta $ & $\pi ${\small (139)} & $\pi ${\small (138)} \\ \hline
$\text{q}_{N}^{0}\text{(313+}\Delta \text{)}\overline{\text{q}_{S}^{1}\text{%
(493+}\Delta \text{)}}$ & {\small 19} & {\small 1} & 0 & {\small - 318-}$%
2\Delta $ & {\small K(488)} & {\small K(494)} \\ \hline
$\text{q}_{S}^{1}\text{(493+}\Delta \text{)}\overline{\text{q}_{S}^{1}\text{%
(493+}\Delta \text{)}}$ & {\small 0} & {\small 0} & 0 & {\small - 437-}$%
2\Delta $ & $\eta ${\small (549)} & $\eta ${\small (548)} \\ \hline
$\text{q}_{C}^{1}\text{(1753+}\Delta \text{)}\overline{\text{q}_{N}^{0}\text{%
(313+}\Delta \text{)}}$ & {\small 153} & {\small 1} & 0 & {\small - 184-}$%
2\Delta $ & {\small D(1882)} & {\small D(1869)} \\ \hline
$\text{q}_{C}^{1}\text{(1753+}\Delta \text{)}\overline{\text{q}_{S}^{1}\text{%
(493+}\Delta \text{)}}$ & {\small 134} & {\small 0} & 0 & {\small - 303-}$%
2\Delta $ & {\small D}$_{S}${\small (1943)} & {\small D}$_{S}${\small (1969)}
\\ \hline
$\text{q}_{C}^{1}\text{(1753+}\Delta \text{)}\overline{\text{q}_{C}^{1}\text{%
(1753+}\Delta \text{)}}$ & {\small 0} & {\small 0} & 0 & {\small -437-}$%
2\Delta $ & {\small J/}$\psi ${\small (3069)} & {\small J/}$\psi ${\small %
(3097)} \\ \hline
$\text{q}_{N}^{0}\text{(313+}\Delta \text{)}\overline{\text{q}_{b}^{1}\text{%
(4913+}\Delta \text{)}}$ & {\small 490} & {\small 1} & 0 & \ {\small 153-}$%
2\Delta $ & {\small B(5379)} & {\small B(5279)} \\ \hline
$\text{q}_{S}^{1}\text{(493+}\Delta \text{)}\overline{\text{q}_{b}^{1}\text{%
(4913+}\Delta \text{)}}$ & {\small 471} & $0$ & 0 & \ {\small 34-}$2\Delta $
& {\small B}$_{S}${\small (5440)} & {\small B}$_{S}${\small (5370)} \\ \hline
$\text{q}_{C}^{1}\text{(1753+}\Delta \text{)}\overline{\text{q}_{b}^{1}\text{%
(4913+}\Delta \text{)}}$ & {\small 337} & {\small 0} & 0 & {\small -100-}$%
2\Delta $ & {\small B}$_{C}${\small (6566)} & {\small B}$_{C}${\small (6400)}
\\ \hline
$\text{q}_{b}^{1}\text{(4913+}\Delta \text{)}\overline{\text{q}_{b}^{1}\text{%
(4913+}\Delta \text{)}}$ & {\small 0} & {\small 0} & 0 & {\small -\ 437-}$%
2\Delta $ & $\Upsilon ${\small (9389)} & $\Upsilon ${\small (9460)} \\ \hline
\end{tabular}
\ $%
\end{tabular}

\ \ \ \ \ \ \ \ \ \ \ \ \ \ \ \ \ \ \ \ \ \ \ \ \ \ \ \ \ \ \ \ \ \ \ \ \ \
\ \ \ \ \ \ \ \ \ \ \ \ \ \ \ \ \ \ \ \ \ \ \ \ \ 

From Table 22, we can see that the terms $\Delta \ $of the quark and\
antiquark masses are always cancelled by term ($-2\Delta $) of the binding
energy.\ Thus we will omit the term $\Delta \ $in the quark masses and the
term $-2\Delta \ $in the binding energy from now on.\ \ \ \ \ \ \ \ \ \ \ \
\ \ \ \ \ \ \ \ \ \ \ \ \ \ \ \ \ \ \ \ \ \ \ \ \ \ \ \ 

B2. The Important Mesons of the Quark Pairs (q$_{i}$ = q$_{j}$)

For quark pairs, $\Delta $m = DS = $\Delta $I = 0, $\gamma $(i, j) = -1,
from (\ref{M-Ebin}), omitting -2$\Delta $, we have\qquad 
\begin{equation}
\text{E}_{B}\text{(q}_{i}\overline{q_{j}}\text{) = - 437 -100(}\widetilde{m}%
\text{ \ + 2I}_{i}\text{I}_{j}\text{)}.  \label{Ebin(Pair)}
\end{equation}
Using (\ref{Ebin(Pair)}) and the sum laws (\ref{Meson-Sum}), we deduced the
mesons that are shown in Table 23 from the quarks in Table 11:

\ \ \ \ \ \ \ \ \ \ \ \ \ \ \ \ \ \ \ 

\begin{tabular}{l}
$\ \ \ \ \ \ \ \ \ \ \ \ \ \ \ \ \ \ \ \ \ ${\small Table 23.\ \ The
Important Mesons of q(m)}$\overline{\text{q(m)}}${\small \ \ \ \ \ \ } \\ 
$%
\begin{tabular}{|l|l|l|l|l|}
\hline
\ \ q$_{i}^{\Delta s}$(m)$\overline{q_{j}^{\Delta s}\text{(m)}}$ & \ \ \ \
-100$\widetilde{m}$ & -200$\text{I}_{i}\text{I}_{j}$ & E$_{bind}$ & Deduced
\\ \hline
d$_{S}^{1}$(1933)$\overline{\text{{\small d}}_{S}^{1}\text{{\small (1933)}}}$
& -424$\leftarrow \frac{1933\times 1933}{939\times 939}$ & 0 & - 861 & $\eta 
$(3005) \\ \hline
d$_{S}^{\text{-1}}$(773)$\overline{\text{{\small d}}_{S}^{\text{-1}}\text{%
{\small (773)}}}$ & -68$\leftarrow \frac{773\times 773}{939\times 939}$ & 0
& -505 & $\eta $(1041) \\ \hline
$\text{d}_{S}^{\text{-1}}$(3753)$\overline{\text{d}_{S}^{\text{-1}}\text{%
{\small (3753)}}}$ & -1597$\leftarrow \frac{3753\times 3753}{939\times 939}$
& 0 & -2034 & $\eta $(5472) \\ \hline
{\small d}$_{\text{S}}^{0}${\small (1203})$\overline{d_{\text{S}}^{0}(1203)}$
& -164$\leftarrow \frac{1203\times 1203}{939\times 939}$ & 0 & -601 & $\eta $%
(1805) \\ \hline
{\small d}$_{\text{S}}^{0}${\small (1303})$\overline{d_{\text{S}}^{0}(1303)}$
& -193$\leftarrow \frac{1303\times 1303}{939\times 939}$ & 0 & -630 & $\eta $%
(1976) \\ \hline
{\small d}$_{S}^{0}${\small (1393)}$\overline{d_{S}^{0}(1393)}$ & -220$%
\leftarrow \frac{1393\times 1393}{939\times 939}$ & 0 & -657 & $\eta $(2129)
\\ \hline
{\small d}$_{\text{S}}^{0}${\small (2013})$\overline{d_{\text{S}}^{0}(2013)}$
& -460$\leftarrow \frac{2013\times 2013}{939\times 939}$ & 0 & -897 & $\eta $%
(3129) \\ \hline
$\text{d}_{S}$(2743)$\overline{\ \text{d}_{S}\text{{\small (2743)}}}$ & -853$%
\leftarrow \frac{2743\times 2743}{939\times 939}$ & 0 & -1290 & $\eta $(4196)
\\ \hline
{\small d}$_{S}^{1}${\small (1413)}$\overline{d_{S}^{1}(1413)}$ & -226$%
\leftarrow \frac{1413\times 1413}{939\times 939}$ & 0 & -663 & $\eta $(2163)
\\ \hline
{\small d}$_{S}^{1}${\small (1513)}$\overline{d_{S}^{1}(1513)}$ & -260$%
\leftarrow \frac{1513\times 1513}{939\times 939}$ & 0 & -697 & $\eta $(2329)
\\ \hline
{\small d}$_{\text{S}}^{-1}${\small (1503)}$\overline{d_{\text{S}}^{-1}(1503)%
}$ & -256$\leftarrow \frac{1503\times 1503}{939\times 939}$ & 0 & -693 & $%
\eta $(2313) \\ \hline
{\small d}$_{\text{S}}^{-1}${\small (1603)}$\overline{d_{\text{S}}^{-1}(1603)%
}$ & -291$\leftarrow \frac{1603\times 1603}{939\times 939}$ & 0 & -728 & $%
\eta $(2478) \\ \hline
q$_{\Sigma }^{0}$(583)$\overline{\text{{\small q}}_{\Sigma }^{0}\text{%
{\small (583)}}}$ & -39$\leftarrow \frac{583\times 583}{939\times 939}$ & 
-200 & -676 & $\eta $(490) \\ \hline
{\small q}$_{\text{N}}^{1,0}${\small (583)} $\overline{\text{{\small q}}_{%
\text{N}}^{1,0}\text{{\small (583)}}}$ & -39$\leftarrow \frac{583\times 583}{%
939\times 939}$ & -50 & -526 & $\eta $(640) \\ \hline
{\small q}$_{\text{N}}^{1}${\small (673)} $\overline{q_{\text{N}}^{1}\text{%
{\small (673)}}}$ & -51$\leftarrow \frac{673\times 673}{939\times 939}$ & -50
& -538 & $\eta $(808) \\ \hline
{\small q}$_{\text{N}}^{0}${\small (853)} $\overline{q_{\text{N}}^{0}\text{%
{\small (853)}}}$ & -83$\leftarrow \frac{853\times 853}{939\times 939}$ & -50
& -570 & $\eta $(1136) \\ \hline
{\small q}$_{\text{N}}^{0}${\small (1213)} $\overline{q_{\text{N}}^{0}\text{%
{\small (1213)}}}$ & -167$\leftarrow \frac{1213\times 1213}{939\times 939}$
& -50 & -654 & $\eta $(1772) \\ \hline
{\small q}$_{\text{N}}^{1,0}${\small (1303)} $\overline{q_{\text{N}}^{1,0}%
\text{{\small (1303)}}}$ & -193$\leftarrow \frac{1303\times 1303}{939\times
939}$ & -50 & -680 & $\eta $(1926) \\ \hline
{\small q}$_{\text{N}}^{1}${\small (1393)} $\overline{q_{\text{N}}^{1}\text{%
{\small (1393)}}}$ & -220$\leftarrow \frac{1393\times 1393}{939\times 939}$
& -50 & -707 & $\eta $(2079) \\ \hline
q$_{\text{N}}^{0}$(1573)$\overline{\text{{\small q}}_{\text{N}}^{0}\text{%
{\small (1573)}}}$ & -281$\leftarrow \frac{1573\times 1573}{939\times 939}$
& -50 & -768 & $\eta $(2378) \\ \hline
q$_{\text{N}}^{1,0}$(2023)$\overline{\text{{\small q}}_{\text{N}}^{1,0}\text{%
{\small (2023)}}}$ & -464$\leftarrow \frac{2023\times 20233}{939\times 939}$
& -50 & -951 & $\eta $(3095) \\ \hline
u$_{C}^{1}$(6073)$\overline{u_{C}^{1}\text{{\small (6073)}}}$ & -1200$%
\leftarrow \frac{6073\times 6073}{1753\times 1753}$ & 0 & -1637 & $\psi $%
(10509) \\ \hline
u$_{C}^{1}$(2133)$\overline{u_{C}^{1}\text{(2133)}}$ & -148$\leftarrow \frac{%
2133\times 2133}{1753\times 1753}$ & 0 & -585 & $\psi $(3681) \\ \hline
u$_{C}^{1}$(2333)$\overline{u_{C}^{1}\text{(2333)}}$ & -177$\leftarrow \frac{%
2333\times 2333}{1753\times 1753}$ & 0 & -614 & $\psi $(4052) \\ \hline
u$_{C}^{1}$(2533)$\overline{u_{C}^{1}\text{{\small (2533)}}}$ & -209$%
\leftarrow \frac{2533\times 2533}{1753\times 1753}$ & 0 & -646 & $\psi $%
(4420) \\ \hline
u$_{C}^{1}$(3543)$\overline{u_{C}^{1}\text{{\small (3543)}}}$ & -408$%
\leftarrow \frac{3543\times 3543}{1753\times 1753}$ & 0 & -845 & $\psi $%
(6241) \\ \hline
d$_{b}^{1}$(9333)$\overline{\text{{\small d}}_{b}^{1}\text{{\small (9333)}}}$
& -361$\leftarrow \frac{9333\times 9333}{4913\times 4913}$ & 0 & -798 & $%
\Upsilon $(17868) \\ \hline
\end{tabular}%
$%
\end{tabular}

\ \ \ \ \ \ \ \ \ \ \ \ \ \ \ \ \ \ \ \ \ \ \ \ \ \ \ \ \ \ \ \ \ \ \ \ \ \
\ \ \ \ \ \ \ 

B3. \ The Mesons of the Unpaired Quarks [q$_{i}\neq $ q$_{j}$]

For mesons (q$_{i}\neq $ q$_{j}$), $\gamma $(i, j) = 1, from (\ref{M-Ebin}),
omitting $-2\Delta ,$we have

\qquad 
\begin{equation}
\text{E}_{B}\text{(q}_{i}\overline{q_{j}}\text{) =\ -237+100[}\frac{\Delta 
\text{m}}{\text{m}_{g}}\text{ +DS -}\widetilde{m}\text{ -0.78}\Delta \text{IS%
}_{i}\text{S}_{j}\text{ +(5.35}\Delta \text{I-2)I}_{i}\text{I}_{j}\text{].}
\label{Un-P}
\end{equation}
Using the formula (\ref{Un-P}) and the sum laws (\ref{Meson-Sum}), we deduce
the mesons shown in Table 24A, Table 24B and Table 24C:

For mesons $\pi $(M) with S = C = b = 0 and I = 1, (\ref{Un-P}) can be
simplified to

\qquad E$_{B}$(q$_{i}\overline{q_{j}}$) =\ -237+100[$\frac{\Delta \text{m}}{%
\text{m}_{g}}$ +DS -$\widetilde{m}$ -0.78S$_{i}$S$_{j}$ +3.35I$_{i}$I$_{j}$%
]\ \ \ \ \ \ \ \ \ \ \ \ \ \ \ \ \ \ \ \ \ \ \ \ \ \ \ \ \ \ \ \ \ \ \ \ \ \
\ \ \ 

\ \ 
\begin{tabular}{l}
$\text{\ \ Table 24-A.\ }$The Light Unflavored Mesons (S = C = b = 0 and I =
1) \\ 
$%
\begin{tabular}{|l|l|l|l|l|l|l|l|}
\hline
\ \ q$_{i}$({\small m}$_{i}$)$\overline{q_{j}(m_{j})}$ & $\frac{100\Delta 
\text{m}}{\text{m}_{g}}$ & DS & 100$\widetilde{m}$ & S$_{i}$S$_{j}$ & I$_{i}$%
I$_{j}$ & {\small E}$_{bind}$ & {\small Deduced} \\ \hline
{\small d}$_{\Sigma }^{0}$(583)$\overline{\text{d}_{\text{S}}^{1}\text{(493)}%
}$ & 9.6 & 1 & 62.1 & 1 & 0 & -268 & $\pi $(808) \\ \hline
{\small d}$_{\Sigma }^{0}$(583)$\overline{\text{d}_{\text{S}}^{0}\text{(493)}%
}$ & 9.6 & 0 & 62.1 & 1 & 0 & -368 & $\pi $(708) \\ \hline
{\small d}$_{\Sigma }^{0}$(583)$\overline{\text{d}_{\text{S}}^{-1}\text{(493)%
}}$ & 9.6 & 1 & 62.1 & 1 & 0 & -268 & $\pi $(808) \\ \hline
\ q$_{N}^{0}$({\small 313})$\overline{q_{\Delta }^{0}(673)}$ & 38.3 & 0 & 
71.7 & 0 & $\frac{3}{4}$ & -19 & $\pi $($967$) \\ \hline
{\small d}$_{\Sigma }^{0}$(1033)$\overline{\text{d}_{\text{S}}^{1}\text{(493)%
}}$ & 57.0 & 1 & 110.0 & 1 & 0 & -268 & $\pi $(1258) \\ \hline
{\small d}$_{\Sigma }^{0}$(1033)$\overline{\text{d}_{\text{S}}^{0}\text{(493)%
}}$ & 57.0 & 0 & 110.0 & 1 & 0 & -368 & $\pi $(1158) \\ \hline
{\small d}$_{\Sigma }^{0}$(1033)$\overline{\text{d}_{\text{S}}^{-1}\text{%
(493)}}$ & 57.0 & 1 & 110.0 & 1 & 0 & -268 & $\pi $(1258) \\ \hline
q$_{N}^{0}$({\small 313})$\overline{q_{\Delta }^{0}(1033)}$ & 76.7 & 0 & 
110.0 & 0 & $\frac{3}{4}$ & -19 & $\pi ($1327$)$ \\ \hline
{\small d}$_{\Sigma }^{0}$(1303)$\overline{\text{d}_{\text{S}}^{1}\text{(493)%
}}$ & 86.3 & 1 & 138.8 & 1 & 0 & -268 & $\pi $(1528) \\ \hline
{\small d}$_{\Sigma }^{0}$(1303)$\overline{\text{d}_{\text{S}}^{0}\text{(493)%
}}$ & 86.3 & 0 & 138.8 & 1 & 0 & -368 & $\pi $(1428) \\ \hline
{\small d}$_{\Sigma }^{0}$(1303)$\overline{\text{d}_{\text{S}}^{-1}\text{%
(493)}}$ & 86.3 & 1 & 138.8 & 1 & 0 & -268 & $\pi $(1528) \\ \hline
q$_{N}^{0}$({\small 313)}$\overline{q_{\Delta }^{0}(1393)}$ {\small \ } & 
115.0 & 0 & 148.3 & 0 & $\frac{3}{4}$ & -19 & $\pi ($1687{\small )} \\ \hline
{\small d}$_{\Sigma }^{0}$(1753)$\overline{\text{d}_{\text{S}}^{1}\text{(493)%
}}$ & 134.2 & 1 & 186.7 & 1 & 0 & -268 & $\pi $(1978) \\ \hline
{\small d}$_{\Sigma }^{0}$(1753)$\overline{\text{d}_{\text{S}}^{0}\text{(493)%
}}$ & 134.2 & 0 & 186.7 & 1 & 0 & -168 & $\pi $(1878) \\ \hline
{\small d}$_{\Sigma }^{0}$(1753)$\overline{\text{d}_{\text{S}}^{-1}\text{%
(493)}}$ & 134.2 & 1 & 186.7 & 1 & 0 & -268 & $\pi $(1978) \\ \hline
q$_{N}^{0}$({\small 313)}$\overline{q_{\Delta }^{0}(1753)}$ & 153.4 & 0 & 
186.7 & 0 & $\frac{3}{4}$ & -19 & $\pi ($2077{\small )} \\ \hline
{\small d}$_{\Sigma }^{0}$(2023)$\overline{\text{d}_{\text{S}}^{1}\text{(493)%
}}$ & 163 & 1 & 215.4 & 1 & 0 & -268 & $\pi $(2248) \\ \hline
{\small d}$_{\Sigma }^{0}$(2023)$\overline{\text{d}_{\text{S}}^{1}\text{(493)%
}}$ & 163 & 0 & 215.4 & 1 & 0 & -368 & $\pi $(2148) \\ \hline
{\small d}$_{\Sigma }^{0}$(2023)$\overline{\text{d}_{\text{S}}^{1}\text{(493)%
}}$ & 163 & 1 & 215.4 & 1 & 0 & -268 & $\pi $(2248) \\ \hline
\end{tabular}%
\ $%
\end{tabular}

\newpage

\qquad For $\Delta $IS$_{i}$S$_{j}$ = $\Delta $I$\times $I$_{i}$I$_{j}$=0, E$%
_{B}$(q$_{i}\overline{q_{j}}$) =\ \ -\ 237+100[$\frac{\Delta \text{m}}{\text{%
m}_{g}}$ +DS -$\widetilde{m}$ -2I$_{i}$I$_{j}$]\ \ \ \ \ \ \ \ \ \ \ \ \ \ \
\ \ \ \ \ \ \ \ \ \ \ \ \ \ \ \ 

\begin{tabular}{l}
$\ \ \ \ \ \ \ \ \ \ \text{Table 24-B.\ }${\small The Mesons of the Unpaired
Quarks (}$\Delta $IS$_{i}$S$_{j}$ = $\Delta $I$\times $I$_{i}$I$_{j}$=0 ) \\ 
$%
\begin{tabular}{|l|l|l|l|l|l|l|}
\hline
$\ ${\small \ q(m}$_{i}${\small )}$\overline{\text{d}_{\text{S}}\text{(m}_{j}%
\text{)}}$ & {\small 100}$\frac{\Delta \text{m}}{\text{m}_{g}}$ & {\small DS}
& $\text{2I}_{i}\text{I}_{j}$ & {\small -100 }$\widetilde{m}$ & {\small E}$%
_{bind}$ & {\small Deduced} \\ \hline
{\small q}$_{\text{N}}${\small (313) }$\overline{\text{{\small q}}_{\text{N}%
}^{0}\text{{\small (583)}}}$ & {\small 29\ }$\leftarrow ${\small (}$\frac{270%
}{\text{939}})$ & {\small 0} & $\frac{1}{2}$ & {\small 62}$\leftarrow \frac{%
313\times 583}{313\times 939}$ & {\small -320} & $\eta (${\small 576)} \\ 
\hline
{\small q}$_{\text{N}}${\small (313)}$\overline{\text{{\small q}}_{\text{N}%
}^{1}\text{{\small (583)}}}$ & {\small 29\ }$\leftarrow \frac{270}{\text{939}%
}$ & {\small 1} & $\frac{1}{2}$ & {\small 62}$\leftarrow \frac{313\times 583%
}{313\times 939}$ & {\small -220} & $\eta ${\small (676)} \\ \hline
{\small q}$_{\text{N}}${\small (313)}$\overline{\text{{\small q}}_{\text{N}%
}^{1}\text{{\small (673)}}}$ & {\small 38\ .3}$\leftarrow \frac{\text{360}}{%
\text{939}}$ & {\small 1} & $\frac{1}{2}$ & {\small 71.7}$\leftarrow \frac{%
313\times 673}{313\times 939}$ & {\small -220} & $\eta ${\small (766)} \\ 
\hline
{\small q}$_{\text{S}}^{1}${\small (493)}$\overline{\text{{\small d}}%
_{S}^{-1}\text{{\small (773)}}}$ & {\small 30}$\leftarrow \frac{\text{280}}{%
\text{939}}$ & {\small 2} & {\small 0} & {\small 82.3}$\leftarrow \frac{%
493\times 773}{493\times 939}$ & {\small - 90} & $\eta ${\small (1177)} \\ 
\hline
{\small q}$_{\text{N}}${\small (313)}$\overline{\text{{\small q}}_{\text{N}%
}^{0}\text{{\small (1303)}}}$ & {\small 105.4\ }$\leftarrow \frac{\text{990}%
}{\text{939}}$ & {\small 0} & $\frac{1}{2}$ & {\small 138.8}$\leftarrow 
\frac{313\times 1303}{313\times 939}$ & {\small -320} & $\eta ${\small (1296)%
} \\ \hline
{\small q}$_{\text{N}}${\small (313)}$\overline{\text{{\small q}}_{\text{N}%
}^{1}\text{{\small (1303)}}}$ & {\small 105.4\ }$\leftarrow \frac{\text{990}%
}{\text{939}}$ & {\small 1} & $\frac{1}{2}$ & {\small 138.8}$\leftarrow 
\frac{313\times 1303}{313\times 939}$ & {\small -220} & $\eta ${\small (1396)%
} \\ \hline
{\small q}$_{\text{S}}^{1}${\small (493)}$\overline{\text{{\small d}}_{S}^{1}%
\text{{\small (1513)}}}$ & {\small 108.6}$\leftarrow \frac{\text{920}}{\text{%
939}}$ & {\small 0} & {\small 0} & {\small 161.1}$\leftarrow \frac{493\times
1513}{493\times 939}$ & {\small -290} & $\eta ${\small (1716)} \\ \hline
{\small q}$_{\text{S}}^{1}${\small (493)}$\overline{\text{{\small d}}%
_{S}^{-1}\text{{\small (1603)}}}$ & {\small 118.2}$\leftarrow \frac{\text{%
1110}}{\text{939}}$ & {\small 2} & {\small 0} & {\small 170.7}$\leftarrow 
\frac{493\times 1603}{493\times 939}$ & {\small -90} & $\eta ${\small (2006)}
\\ \hline
{\small q}$_{\text{S}}${\small (493)}$\overline{\text{{\small d}}_{S}\text{%
{\small (3753)}}}$ & {\small 347.2}$\leftarrow \frac{\text{3260}}{\text{939}}
$ & {\small 2} & {\small 0} & {\small 399.7}$\leftarrow \frac{493\times 3753%
}{493\times 939}$ & {\small -90} & $\eta ${\small (4156)} \\ \hline
{\small q}$_{\text{N}}${\small (313)}$\overline{\text{{\small d}}_{S}^{\pm }%
\text{{\small (773)}}}$ & {\small 49\ }$\leftarrow \frac{\text{460}}{\text{%
939}}$ & {\small 1} & {\small 0} & {\small 82.3}$\leftarrow \frac{313\times
773}{313\times 939}$ & {\small - 170} & {\small K(916)} \\ \hline
{\small q}$_{\text{N}}${\small (313)}$\overline{\text{{\small d}}_{S}^{0}%
\text{{\small (773)}}}$ & {\small 49\ }$\leftarrow \frac{\text{460}}{\text{%
939}}$ & {\small 0} & {\small 0} & {\small 82.3}$\leftarrow \frac{313\times
773}{313\times 939}$ & {\small -270} & {\small K(816)} \\ \hline
{\small q}$_{\text{N}}${\small (313)}$\overline{\text{{\small d}}_{S}^{0}%
\text{{\small (1203)}}}$ & {\small 94.8\ }$\leftarrow \frac{\text{890}}{%
\text{939}}$ & {\small 0} & {\small 0} & {\small 128.1}$\leftarrow \frac{%
313\times 1203}{313\times 939}$ & {\small -270} & {\small K(1246)} \\ \hline
{\small q}$_{\text{N}}${\small (313)}$\overline{\text{{\small d}}_{S}^{0}%
\text{{\small (1303})}}$ & {\small 105.4}$\leftarrow \frac{\text{990}}{\text{%
939}}$ & {\small 0} & {\small 0} & {\small 138.8}$\leftarrow \frac{313\times
1303}{313\times 939}$ & {\small -270} & {\small K(1346)} \\ \hline
{\small q}$_{\text{N}}${\small (313)}$\overline{\text{{\small d}}_{S}^{0}%
\text{{\small (1393)}}}$ & {\small 115\ }$\leftarrow \frac{\text{1080}}{%
\text{939}}$ & {\small 0} & {\small 0} & {\small 148.3}$\leftarrow \frac{%
313\times 1393}{313\times 939}$ & {\small -270} & {\small K(1436)} \\ \hline
{\small q}$_{\text{N}}${\small (313)}$\overline{\text{{\small d}}_{\text{S}%
}^{\text{-1}}\text{{\small (1513})}}$ & {\small 126.7}$\leftarrow \frac{%
\text{1190}}{\text{939}}$ & {\small 1} & {\small 0} & {\small 160}$%
\leftarrow \frac{313\times 1503}{313\times 939}$ & {\small -170} & {\small %
K(1646)} \\ \hline
{\small q}$_{\text{N}}${\small (313)}$\overline{\text{{\small d}}_{S}^{1}%
\text{{\small (1933)}}}$ & {\small 172.5}$\leftarrow \frac{\text{1620}}{%
\text{939}}$ & {\small 1} & {\small 0} & {\small 205.9}$\leftarrow \frac{%
313\times 1933}{313\times 939}$ & {\small -170} & {\small K(2076)} \\ \hline
{\small u}$_{C}${\small (6073)}$\overline{\text{{\small q}}_{\text{N}}\text{%
{\small (313)}}}$ & {\small 328,6}$\leftarrow \frac{\text{5760}}{\text{1753}}
$ & {\small 1} & {\small 0} & {\small 346.4}$\leftarrow \frac{313\times 6073%
}{313\times 1753}$ & {\small -155} & {\small D(6231)} \\ \hline
{\small u}$_{C}^{1}${\small (1753)}$\overline{\text{{\small d}}_{S}^{-1}%
\text{{\small (773)}}}$ & {\small 104.4}$\leftarrow \frac{\text{980}}{\text{%
939}}$ & {\small 2} & {\small 0} & {\small 82.3}$\leftarrow \frac{1753\times
773}{1753\times 939}$ & {\small -15} & {\small D}$_{S}${\small (2511)} \\ 
\hline
{\small u}$_{C}^{1}${\small (1753)}$\overline{\text{{\small d}}_{S}^{0}\text{%
{\small (773)}}}$ & {\small 104.4}$\leftarrow \frac{\text{980}}{\text{939}}$
& {\small 1} & {\small 0} & {\small 82.3}$\leftarrow \frac{1753\times 773}{%
1753\times 939}$ & {\small -115} & {\small D}$_{S}${\small (2411)} \\ \hline
{\small u}$_{C}^{1}${\small (1753)}$\overline{\text{{\small d}}_{S}^{1}\text{%
{\small (773)}}}$ & {\small 104.4}$\leftarrow \frac{\text{980}}{\text{939}}$
& {\small 0} & {\small 0} & {\small 82.3}$\leftarrow \frac{1753\times 773}{%
1753\times 939}$ & {\small -215} & {\small D}$_{S}${\small (2311)} \\ \hline
{\small q}$_{\text{N}}${\small (313)}$\overline{\text{{\small d}}_{b}^{1}%
\text{{\small (9333)}}}$ & {\small 183.6}$\leftarrow \frac{\text{9020}}{%
\text{4913}}$ & {\small 1} & {\small 0} & {\small 190}$\leftarrow \frac{%
313\times 9333}{313\times 4913}$ & {\small -143} & {\small B(9503)} \\ \hline
{\small d}$_{\text{S}}^{-1}${\small (493)}$\overline{\text{{\small d}}%
_{S}^{-1}\text{{\small (9613)}}}$ & {\small 243}$\leftarrow \frac{\text{9120}%
}{\text{3753}}$ & {\small 0} & {\small 0} & {\small 256.1}$\leftarrow \frac{%
493\times 9613}{493\times 3753}$ & {\small -250} & $\eta ${\small (9856)} \\ 
\hline
{\small d}$_{\text{S}}^{1}${\small (493)}$\overline{\text{{\small d}}%
_{S}^{-1}\text{{\small (9613)}}}$ & {\small 243}$\leftarrow \frac{\text{9120}%
}{\text{3753}}$ & {\small 2} & {\small 0} & {\small 256.1}$\leftarrow \frac{%
493\times 9613}{493\times 3753}$ & {\small -50} & $\eta ${\small (10056)} \\ 
\hline
{\small d}$_{\text{S}}^{-1}${\small (773)}$\overline{\text{{\small d}}%
_{S}^{-1}\text{{\small (9613)}}}$ & {\small 235.5}$\leftarrow \frac{\text{%
8840}}{\text{3753}}$ & {\small 0} & {\small 0} & {\small 210.9}$\leftarrow 
\frac{773\times 9613}{939\times 3753}$ & {\small -212} & $\eta ${\small %
(10174)} \\ \hline
{\small d}$_{\text{S}}^{0}${\small (1203)}$\overline{\text{{\small d}}%
_{S}^{-1}\text{{\small (9613)}}}$ & {\small 224}$\leftarrow \frac{\text{8410}%
}{\text{3753}}$ & {\small 1} & {\small 0} & {\small 328.2}$\leftarrow \frac{%
1203\times 9613}{939\times 3753}$ & {\small -241} & $\eta ${\small (10575)}
\\ \hline
{\small d}$_{\text{S}}^{1}${\small (1413)}$\overline{\text{{\small d}}%
_{S}^{-1}\text{{\small (9613)}}}$ & {\small 218.5}$\leftarrow \frac{\text{%
8200}}{\text{3753}}$ & {\small 2} & {\small 0} & {\small 385}$\leftarrow 
\frac{1413\times 9613}{939\times 3753}$ & {\small -204} & $\eta ${\small %
(10822)} \\ \hline
{\small d}$_{\text{S}}^{1}${\small (1513)}$\overline{\text{{\small d}}%
_{S}^{-1}\text{{\small (9613)}}}$ & {\small 215.8}$\leftarrow \frac{\text{%
8100}}{\text{3753}}$ & {\small 2} & {\small 0} & {\small 412.7}$\leftarrow 
\frac{1513\times 9613}{939\times 3753}$ & {\small -234} & $\eta ${\small %
(10892)} \\ \hline
{\small d}$_{\text{S}}^{0}${\small (1923)}$\overline{\text{{\small d}}%
_{S}^{-1}\text{{\small (9613)}}}$ & {\small 204.9}$\leftarrow \frac{\text{%
7690}}{\text{3753}}$ & {\small 1} & {\small 0} & {\small 524.5}$\leftarrow 
\frac{1923\times 9613}{939\times 3753}$ & {\small -456} & $\eta ${\small %
(11080)} \\ \hline
{\small d}$_{\text{S}}^{1}${\small (1933)}$\overline{\text{{\small d}}%
_{S}^{-1}\text{{\small (9613)}}}$ & {\small 204.6}$\leftarrow \frac{\text{%
7680}}{\text{3753}}$ & {\small 2} & {\small 0} & {\small 527.3}$\leftarrow 
\frac{1933\times 9613}{939\times 3753}$ & {\small -360} & $\eta ${\small %
(11186)} \\ \hline
\end{tabular}
\ \ $%
\end{tabular}

\ \ \ \ \newpage

\qquad For S$_{i}$S$_{j}$ = I$_{i}$I$_{j}$ = 0, E$_{B}$(q$_{i}\overline{q_{j}%
}$) =\ -237+100[$\frac{\Delta \text{m}}{\text{m}_{g}}$ +DS -$\widetilde{m}$
].

\begin{tabular}{l}
$\ \ \ \ \ \ \ \ \ \ \text{\ Table 24-C.\ }$The Mesons with S$_{i}$S$_{j}$ =I%
$_{i}$I$_{j}$ = 0 \\ 
$%
\begin{tabular}{|l|l|l|l|l|l|}
\hline
\ \ q$_{i}$({\small m}$_{i}$)$\overline{q_{j}(m_{j})}$ & $\frac{100\Delta 
\text{m}}{\text{m}_{g}}$ & DS & 100$\widetilde{m}$ & {\small E}$_{bind}$ & 
{\small Deduced} \\ \hline
u$_{C}^{1}$(1753)$\overline{\text{u}_{C}^{1}\text{(2133)}}$ & {\small 40.5}
& 0 & {\small 121.7} & -318 & $\psi $(3568) \\ \hline
u$_{C}^{1}$(1753)$\overline{\text{u}_{C}^{1}\text{(2333)}}$ & {\small 61.8}
& 0 & {\small 133.1} & -308 & $\psi $(3778) \\ \hline
u$_{C}^{1}$(1753)$\overline{\text{u}_{C}^{1}\text{(2533)}}$ & 83.1 & 0 & 
144.5 & -298 & $\psi $(3988) \\ \hline
q$_{N}^{0}$({\small 313)}$\overline{\text{u}_{C}^{1}\text{(2133)}}$ & 193.8
& 1 & 121.7 & -65 & D(2381) \\ \hline
q$_{N}^{0}$({\small 313)}$\overline{\text{u}_{C}^{1}\text{(2333)}}$ & 215.1
& 1 & 133.1 & -55 & D(2591) \\ \hline
q$_{N}^{0}$({\small 313)}$\overline{\text{u}_{C}^{1}\text{(2533)}}$ & 236.4
& 1 & 144.5 & -45 & D(2801) \\ \hline
u$_{C}^{1}$(2133)$\overline{\text{d}_{\text{S}}^{1}\text{(493)}}$ & 174.7 & 0
& 121.7 & -184 & D$_{S}$(2442) \\ \hline
u$_{C}^{1}$(2333)$\overline{\text{d}_{\text{S}}^{1}\text{(493)}}$ & 196.0 & 0
& 133.1 & -174 & D$_{S}$(2652) \\ \hline
u$_{C}^{1}$(2533)$\overline{\text{d}_{\text{S}}^{1}\text{(493)}}$ & 217.3 & 0
& 144.5 & -164 & D$_{S}$(2862) \\ \hline
$\text{q}_{N}^{0}\text{(583)}\overline{\text{q}_{b}^{1}\text{(4913)}}$ & 
461.1 & 1 & 62.1 & +262$^{\ast }$ & B(5758) \\ \hline
$\text{q}_{N}^{1}\text{(583)}\overline{\text{q}_{b}^{1}\text{(4913)}}$ & 
461.1 & 0 & 62.1 & +162$^{\ast }$ & B(5658) \\ \hline
$\text{q}_{N}^{0}\text{(853)}\overline{\text{q}_{b}^{1}\text{(4913)}}$ & 
432.4 & 1 & 90.8 & +205$^{\ast }$ & B(5971) \\ \hline
d$_{S}^{-1}\text{(773)}\overline{\text{q}_{b}^{1}\text{(4913)}}$ & 440.9 & 2
& 82.3 & +322$^{\ast }$ & B$_{S}$(6008) \\ \hline
d$_{S}^{0}\text{(773)}\overline{\text{q}_{b}^{1}\text{(4913)}}$ & 440.9 & 1
& 82.3 & +222$^{\ast }$ & B$_{S}$(5908) \\ \hline
d$_{S}^{1}\text{(773)}\overline{\text{q}_{b}^{1}\text{(4913)}}$ & 440.9 & 0
& 82.3 & +122$^{\ast }$ & B$_{S}$(5808) \\ \hline
\end{tabular}
\ \ \ $ \\ 
$\ \ ^{\ast }${\small \ }$\text{E}_{B}\text{(q}_{i}\overline{q_{j}}\text{) =
-2}\Delta \text{ -337+..., }\Delta >>${\small \ 337+.... Thus }$\text{E}_{B}%
\text{(q}_{i}\overline{q_{j}}\text{) \TEXTsymbol{<} 0.}$%
\end{tabular}
\ \ \ \ \ \ 

\ \ \ \ \ \ \ \ \ \ \ \ \ \ \ \ \ \ \ \ \ \ \ \ \ \ \ 

As in Table 22, Table 23 and Table 24, using the sum laws (\ref{Meson-Sum})
and the binding energy formula (\ref{M-Ebin}), we deduced a meson spectrum
from the quark spectrum (Table 10 and Table 11). Table 25-Table 31 will show
the results. At the same time, we will compare the deduced result with the
experimental results using Table 25-Table 31\ also in the following section.

\subsection{ Comparing with the Experimental Results\ of Mesons\ \ \ }

Using Table 25-Table 31, we compare the deduced meson spectrum with the
experimental results \cite{Meson04}. In the comparison, we do not take into
account the angular momenta of the experimental results. We assume that the
small differences of the masses in the same group of mesons with the same
intrinsic quantum numbers are from their different angular momenta. If we
ignore this effect, their masses would be essentially the same. In this
comparison, we use the meson name to represent the intrinsic quantum
numbers. If the names of the deduced meson and the experimental meson are
the same, this means that the intrinsic quantum numbers (I, S, C, b and Q)
are exactly the same. Since the unflavored mesons with the same intrinsic
quantum numbers but different angular momenta and parities have different
names, in order to compare the rest masses, we omit the differences of the
angular momenta and parities; we use meson $\eta $ to represent the mesons
with S = C = b = 0, I = Q = 0 ($\eta $, $\varpi $, $\phi $, h and f) (see
Table C5 and Table 27) and we use meson $\pi $ to represent the mesons with
S = C = b = 0, I = 1, Q = 1, 0, -1 ($\pi $, $\rho $, a and b) (see Table C6
and Table 28). Table 25-Table 31 show that the deduced intrinsic quantum
umbers (I, S, C, b and Q) exactly match with the experimental results.

Sometimes there are multiple possible mesons with the same intrinsic quantum
numbers and the same (or nearly same) rest masses, we will choose the meson
with the highest probability using (\ref{P(Qk)}). If there are several
possible mesons\ that have essentially the same probability, it is believed
that they form (superposition) one\ meson with the average mass of all the
mesons (see Table C3-Table C7).\ 

Using Table 25-Table 31, we compare the deduced mesons with the experimental
results. These tables show that although the names are not the same, the
intrinsic quantum numbers (I, S, C,b and Q) are the same for both deduced
meson and experimental meson. We do not repeat this conclusion for each
table. Their mass units are the same-``Mev''.

\ \ \ \ \newpage

\ \ We compare the heavy unflavored mesons in Table 25:\ 

\begin{tabular}{l}
$\ \text{Table 25. }$The Heavy Unflavored Mesons with\ S=C=b=I=0 \\ 
$%
\begin{tabular}{|l|l|l|l|}
\hline
d$_{S}^{-1}$(9613)$\overline{d_{S}(m)}$ & Deduced & Expert., $\Gamma $ (Mev)
& {\small R=}$\frac{\Delta \text{M}}{\text{M}}\%$ \\ \hline
d$_{b}$(4913)$\overline{\text{d}_{b}\text{(4913)}}$ & $\Upsilon $(9389) & $%
\Upsilon $(9460), 53 kev & 0.75 \\ \hline
d$_{S}^{-1}$(9613)$\overline{\text{d}_{S}^{0}\text{(493)}}$ & $\eta $($%
\overline{\text{9906}}$) & $\chi $($\overline{\text{9888}}$), \ \ ? & 0.18
\\ \hline
d$_{S}^{-1}$(9613)$\overline{\text{d}_{S}^{1}\text{(493)}}$ & $\eta $(10056)
& $\Upsilon $(10023), 43 kev & 0.13 \\ \hline
d$_{S}^{-1}$(9613)$\overline{\text{d}_{S}^{-1}\text{(773)}}$ & $\eta $($%
\overline{\text{10174}}$) & $\chi (\overline{\text{10251}}$), \ \ ? & 0.75
\\ \hline
u$_{C}$(6073)$\overline{\text{u}_{C}\text{(6073)}}$ & $\psi $(10509) & $%
\Upsilon $(10355), 26 kev & 1.5 \\ \hline
2d$_{S}^{-1}$(9613)$\overline{\text{d}_{S}^{0}\text{(1203)}_{F}}$\  & $\eta $%
($\overline{\text{10620}}$) & $\Upsilon $(10580), 20 & 0.66 \\ \hline
d$_{S}^{-1}$(9613)$\overline{\text{d}_{S}^{-1}\text{(1603)}_{F}}$\  & $\eta $%
($\overline{\text{10777}}$) & $\Upsilon $(10865), 110 & 0.73 \\ \hline
d$_{S}^{-1}$(9613)$\overline{\text{d}_{S}^{0}\text{(1923)}_{F}}$ & $\eta $%
(11079) & $\Upsilon $(11020), 79 & 0.54 \\ \hline
$\text{d}_{S}^{-1}\text{(9613)}\overline{\text{d}_{\text{S}}^{0}\text{(2013)}%
}$ & $\eta $(11151) & Prediction &  \\ \hline
d$_{S}^{-1}$(9613)$\overline{\text{d}_{S}^{-1}\text{(3753)}}$\  & $\eta $%
(12261) & Prediction &  \\ \hline
\end{tabular}
\ \ \ \ \ \ $%
\end{tabular}

\ \ \ \ \ \ \ \ \ \ \ \ \ \ \ \ \ \ \ \ \ \ \ \ \ \ \ \ \ \ \ \ \ \ \ \ \ \
\ \ \ \ \ \ \ \ 

{\small Table 25 shows that the masses of the deduced mesons are more than
98.5\% consistent with the experimental results.}

\ \ \ \ \ \ \ \ \ \ \ \ \ \ \ \ \ \ \ \ \ \ \ \ \ \ \ \ \ \ \ \ \ \ \ \ 

We compare the intermediate mass mesons in Table 26:

\begin{tabular}{l}
$\ \ \ \ $Table 26.\ \ The Intermediate Mass Mesons \ (S=C=b=I=0)\ \  \\ 
$%
\begin{tabular}{|l|l|l|l|}
\hline
\ \ q$_{i}$(m)$\overline{q_{i}\text{(m)}}$ & Deduced & Exper., $\Gamma $ 
{\small (Mev)} & {\small R=}$\frac{\Delta \text{M}}{\text{M}}\%$ \\ \hline
d$_{S}^{1}$(1933)$\overline{\text{d}_{S}^{1}\text{(1933)}}$ & $\eta $(3005)
& $\eta _{C}$(2980), 17 & 0.84 \\ \hline
$\text{q}_{C}^{1}\text{(1753)}\overline{\text{q}_{C}^{1}\text{({\small 1753})%
}}$ & J/$\psi $(3069) & J/$\psi $(3097), 91 kev & 0.91 \\ \hline
$\text{q}_{C}^{1}\text{(1753)}\overline{\text{u}_{C}^{1}\text{(2133)}}$ & $%
\psi $(3567) & $\chi $($\overline{\text{3494}}$), $\overline{\text{2}}$ & 
2.09 \\ \hline
u$_{C}^{1}$(2133)$\overline{\text{u}_{C}^{1}\text{(2133)}}$ & $\psi $(3681)
& $\psi $(3686), 281 kev & 0.14 \\ \hline
$\text{q}_{C}^{1}\text{(1753)}\overline{\text{u}_{C}^{1}\text{(2333)}}$ & $%
\psi $(3778) & $\psi $(3770), 23.6 & 0.21 \\ \hline
u$_{C}^{1}$(2333)$\overline{\text{u}_{C}^{1}\text{(2333)}}$ & $\psi $(4052)
& $\psi $(4040), 52 & 0.11 \\ \hline
d$_{S}^{-1}$(3753)$\overline{\text{d}_{S}^{1}\text{(493)}}$ & $\eta $(4156)
& $\psi $(4160), 78 & 0.10 \\ \hline
u$_{C}^{1}$(2533)$\overline{\text{u}_{\text{C}}^{1}\text{(2533)}}$ & $\psi $%
(4420) & $\psi $(4415), 43 & 0.30 \\ \hline
$\text{q}_{C}^{1}\text{(1753)}\overline{\text{u}_{C}^{1}\text{(3543)}}$ & $%
\psi $(4959) & Prediction &  \\ \hline
\end{tabular}
\ $%
\end{tabular}
\ \ \ \ \ \ \ \ \ \ \ \ \ \ \ \ \ \ \ \ \ \ \ \ \ \ \ \ \ \ \ \ \ \ 

{\small Table 26 shows that the masses of the deduced mesons are about 98\%
consistent with the experimental results.}

\ \ \ \ \ \ \ \ \ \ \ \ \ \ \ \ \ \ \ \ \ \ \ \ \ \ \ \ \ \ \ \ \ \ \ \ \ \
\ \ \ \ \ \ \ 

We compare the light unflavored mesons with I = 0 in Table 27:\ \ \ \ \ \ \
\ \ \ \ \ \ \ \ \ \ \ \ \ \ \ \ \ \ \ \ \ \ \ \ \ \ \ \ \ 

\bigskip 
\begin{tabular}{l}
Table 27. The Light Unflavored Mesons (S=C=b=0) I = 0 \\ 
$%
\begin{tabular}{|l|l|l|l|}
\hline
\ \ \ {\small q}$_{i}${\small (m}$_{i}${\small )}$\overline{\text{q}_{j}%
\text{(m}_{j}\text{)}}$ & Deduced & Exper., $\Gamma $ (Mev) & {\small R=}$%
\frac{\Delta \text{M}}{\text{M}}\%$ \\ \hline
$\ \text{q}_{S}^{1}\text{(493)}\overline{\text{q}_{S}^{1}\text{(493)}}$ & $%
\eta ${\small (549)} & $\eta ${\small (547), \ 1.18 Kev} & {\small 0.4} \\ 
\hline
\ {\small q}$_{N}^{0}${\small (313)}$\overline{\text{q}_{N}^{0}\text{(583)}}$
& $\eta $($\overline{\text{{\small 584}}}$) & $\eta ${\small (600)}$,$ 
{\small 800} & \ / \\ \hline
\ {\small q}$_{N}^{0}${\small (313)}$\overline{\text{q}_{N}^{1}\text{(673)}}$
& $\eta ${\small (}$\overline{\text{{\small 774}}}${\small )} & $\varpi $%
(782), 8.49 & {\small 1.15} \\ \hline
\ {\small d}$_{\text{S}}^{1}${\small (493)}$\overline{\text{{\small d}}_{S}^{%
\text{1}}\text{(773)}}$ & $\eta $(976) & $\eta (\overline{\text{969}}),%
\overline{\text{35}}$ & {\small 0.93} \\ \hline
\ {\small d}$_{S}^{-\text{1}}${\small (773)}$\overline{\text{{\small d}}%
_{S}^{-\text{1}}\text{{\small (773)}}}$ & $\eta ${\small (1041)} & $\phi $%
(1020), {\small 4.26} & 2.1 \\ \hline
\ {\small q}$_{N}^{0}${\small (313)}$\overline{\text{q}_{N}^{0}\text{(1213)}}
$ & $\eta $($\overline{\text{1166}}$) & {\small h}$_{1}${\small (1170), 360}
& {\small 0.26} \\ \hline
\ {\small q}$_{N}^{0}${\small (313)}$\overline{\text{q}_{N}^{0}\text{(1303)}}
$ & $\eta $(1296) & $\eta $($\overline{\text{1284}}$),$\overline{\text{88}}$
& {\small 0.94} \\ \hline
\ {\small q}$_{N}^{0}${\small (313)}$\overline{\text{q}_{N}^{1}\text{(1303)}}
$ & $\eta ${\small (1396 )} & $\eta $($\overline{\text{1403}}$){\small ,}$%
\overline{\text{{\small 147}}}$ & {\small 0.50} \\ \hline
\ {\small q}$_{N}^{0}${\small (313)}$\overline{\text{q}_{N}^{1}\text{(1393)}}
$ & $\eta ${\small (}$\overline{\text{1526}}${\small )} & $\eta $($\overline{%
\text{1503}}$), $\overline{\text{90}}$ & 1.53 \\ \hline
\begin{tabular}{l}
{\small d}$_{\text{S}}^{1}${\small (493)}$\overline{\text{{\small d}}_{S}^{0}%
\text{{\small (1303)}}}$ \\ 
{\small d}$_{\text{S}}^{1}${\small (493)}$\overline{\text{{\small d}}_{S}^{0}%
\text{(1393{\small )}}}$%
\end{tabular}
& $\eta $($\overline{\text{1666}}$) & $\eta $($\overline{\text{1670}}$%
{\small ),}$\overline{\text{{\small 191}}}$ & {\small 0.18} \\ \hline
\begin{tabular}{l}
{\small d}$_{\text{S}}^{1}${\small (493)}$\overline{\text{{\small d}}_{S}^{%
\text{1}}\text{{\small (1513)}}}$ \\ 
{\small d}$_{\text{S}}^{1}${\small (493)}$\overline{\text{{\small d}}_{S}^{%
\text{-1}}\text{{\small (1503)}}}$%
\end{tabular}
& $\eta (\overline{\text{{\small 1870}}}${\small )} & $\phi ${\small (}$%
\overline{\text{1899}}${\small ),}$\overline{\text{125}}$ & {\small 1.53} \\ 
\hline
\ {\small q}$_{N}^{0}${\small (313)}$\overline{\text{q}_{N}^{0}\text{(2023)}}
$ & $\eta (${\small 2016}$)$ & $\eta (\overline{\text{{\small 2011}}})$%
{\small , }$\overline{\text{259}}$ & 0.25 \\ \hline
\ \ 4{\small q}$_{N}^{0}${\small (313)}$\overline{\text{q}_{N}^{1}\text{%
(2023)}}$ & $\eta (\overline{\text{{\small 2116}}})$ & $\eta (\overline{%
\text{{\small 2130}}})^{\ast }$, $\overline{\text{187}}$ & 0.66 \\ \hline
\ {\small q}$_{N}^{0}${\small (313)}$\overline{\text{q}_{N}^{0}\text{(2293)}}
$ & $\eta (${\small 2216}$)$ & $\eta (\overline{\text{{\small 2218}}})^{\ast
}$, $\overline{\text{109}}$ & 0.09 \\ \hline
\ {\small d}$_{\text{S}}^{1}${\small (493)}$\overline{\text{{\small d}}%
_{S}^{0}\text{(2013{\small )}}}$ & $\eta $(2316) & $\eta $($\overline{\text{%
2318}}$)$^{\ast }$, $\overline{\text{234}}$ & 0.09 \\ \hline
\ {\small d}$_{\text{S}}^{-1}${\small (1603)}$\overline{\text{d}_{\text{S}%
}^{-1}\text{(1603)}}$ & $\eta (\overline{\text{2428}})$ & f$_{0}$(2465)$%
^{\ast }$, 255 & 1.5 \\ \hline
{\small q}$_{N}^{0}${\small (313)}$\overline{\text{q}_{N}^{0}\text{(2653)}}$
& $\eta $(2646) & Prediction &  \\ \hline
\ {\small q}$_{\text{S}}^{1}${\small (493)}$\overline{\text{{\small d}}%
_{S}^{0}\text{{\small (2743)}}}$ & $\eta ${\small (3046)} & Prediction &  \\ 
\hline
\end{tabular}
$%
\end{tabular}

\ \ \ \ \ \ \ \ \ \ \ \ \ \ \ \ \ \ \ \ \ \ \ \ 

{\small Table 27 shows that the masses of the deduced mesons are about 98\%
consistent with the experimental results.}

\ \ \ \ \ \ \ \ \ \ \ \newpage

We compare the light unflavored mesons\ with I = 1 in Table 28:

\ \ \ \ \ \ \ \ \ \ \ \ \ \ \ \ \ \ 

\begin{tabular}{l}
$\ \ \ \ \ \ \text{Table 28. }$The Light Unflavored Mesons (S=C=b=0) I = 1
\\ 
$%
\begin{tabular}{|l|l|l|l|l|}
\hline
q$_{N}$(313)$\overline{q_{j}(m_{j})}$ & E$_{bind}$ & Phenomen. & Exper., $%
\Gamma $ (Mev) & R=$\frac{\Delta \text{M}}{\text{M}}\%$ \\ \hline
q$_{N}^{0}$(313)$\overline{\text{q}_{N}^{0}\text{(313)}}$ & -487 & $\pi $%
(139) & $\pi $(138) & 0.72 \\ \hline
d$_{\text{S}}^{\pm 1,0}$(493)$\overline{\text{{\small d}}_{\Sigma }^{0}\text{%
(583)}}$ & $\overline{\text{-300}}$ & $\pi $($\overline{\text{775}}$) & $\pi
($776), 150 & 0.0 \\ \hline
q$_{N}^{0}$(313)$\overline{\text{q}_{\Delta }^{0}\text{(673)}}$ & -19 & $\pi 
$(967) & a$_{0}$(985), 75 & 1.8 \\ \hline
d$_{\text{S}}^{\pm 1,0}$(493)$\overline{\text{{\small d}}_{\Sigma }^{0}\text{%
(1033)}}$ & $\overline{\text{-300}}$ & $\pi $($\overline{\text{1225}}$) & $%
\pi $($\overline{\text{1230}}$),$\overline{\text{287}}$ & 0.24 \\ \hline
q$_{N}^{0}$(313)$\overline{\text{q}_{\Delta }\text{(1033)}}$ & -19 & $\pi $%
(1327) & $\pi $($\overline{\text{1331}}$), $\overline{\text{269}}$ & 0.30 \\ 
\hline
d$_{\text{S}}^{\pm 1,0}$(493)$\overline{\text{{\small d}}_{\Sigma }\text{%
(1303)}}$ & $\overline{\text{-300}}$ & $\pi $($\overline{\text{1495}}$) & $%
\pi $($\overline{\text{1470}}$),$\overline{\text{323}}$ & 1.8 \\ \hline
q$_{N}^{0}$(313)$\overline{\text{q}_{\Delta }\text{(1393)}}$\  & -19 & $\pi
( $1687) & $\pi $(1669), $\overline{\text{246}}$ & 1.1 \\ \hline
d$_{\text{S}}^{\pm 1,0}$(493)$\overline{\text{{\small d}}_{\Sigma }\text{%
(1753)}}$ & $\overline{\text{-300}}$ & $\pi $($\overline{\text{1825}}$) & $%
\pi $(1812),207 & 0.77 \\ \hline
q$_{N}^{0}$(313)$\overline{\text{q}_{\Delta }\text{(1753)}}$ & -19 & $\pi (%
\overline{\text{2001}}$) & $\pi $($\overline{\text{2010}}$), $\overline{%
\text{353}}$ & 0.40 \\ \hline
d$_{\text{S}}^{\pm 1,0}$(493)$\overline{\text{{\small d}}_{\Sigma }\text{%
(2023)}}$ & -268 & $\pi $($\overline{\text{2248}}$) & $\pi $($\overline{%
\text{2250}}$)$^{\ast },$ $\overline{\text{281}}$ & 0.09 \\ \hline
q$_{N}^{0}$(313)$\overline{\text{q}_{\Delta }\text{(2113)}}$ & -19 & $\pi ($%
2407) & $\pi $($\overline{\text{2450}}$)$^{\ast }$, $\overline{\text{400}}$
& 1.8 \\ \hline
d$_{\text{S}}^{\pm 1,0}$(493)$\overline{\text{{\small d}}_{\Sigma }\text{%
(2473)}}$ & $\overline{\text{-300}}$ & $\pi $(2666) & Prediction &  \\ \hline
\end{tabular}
\ $%
\end{tabular}

\ \ \ \ \ \ \ \ \ \ \ \ \ \ 

{\small Table 28 shows that the masses of the deduced mesons are more than
98\% consistent with the experimental results}.

\ \ \ \ \ \ \ \ \ \ \ \ \ \ 

\ We compare\ the mesons B and B$_{S}$ in Table 29: \ \ \ \ \ \ \ \ \ \ \ \
\ \ \ \ \ \ \ \ \ \ \ \ \ \ \ \ \ \ \ \ \ \ \ \ \ \ \ \ \ \ \ \ \ \ \ \ \ \
\ \ \ \ \ \ \ \ \ \ \ \ \ \ \ \ \ \ \ \ \ \ 

\begin{tabular}{l}
$\ \ \ \ \ \ \ \ \ \ $Table 29.\ \ The Mesons B(M) and B$_{S}$(M) \\ 
$%
\begin{tabular}{|l|l|l|l|}
\hline
\ \ d$_{b}$(m)$\overline{\text{q}_{\text{N}}\text{(m}_{j}\text{)}}$\  & 
Deduced & Exper., (Mev) & {\small R=}$\frac{\Delta \text{M}}{\text{M}}\%$ \\ 
\hline
q$_{\text{N}}^{0}$(313)$\overline{\text{q}_{b}^{1}\text{(4913)}}$\  & B(5378)
& B($\overline{\text{5302}}$)$^{\#}$ & 1.4 \\ \hline
$\text{q}_{N}^{1,0}\text{(583)}\overline{\text{q}_{b}^{1}\text{(4913)}}$ & 
B(5708) & B$_{j}^{\ast }$(5732) & 0.42 \\ \hline
$\text{q}_{N}^{0}\text{(853)}\overline{\text{q}_{b}^{1}\text{(4913)}}$ & 
B(5971) & Prediction &  \\ \hline
q$_{\text{N}}^{0}$(313)$\overline{\text{d}_{b}^{1}\text{(9333)}}$ & B(9503)
& Prediction &  \\ \hline
$\text{q}_{S}^{1}\text{(493)}\overline{\text{q}_{b}^{1}\text{(4913)}}$ & B$%
_{S}$(5440) & B$_{S}$(5370) & 1.3 \\ \hline
q$_{b}^{1}$(4913)$\overline{\text{d}_{\text{S}}\text{(773)}}$\  & B$_{S}$%
(5908) & B$_{SJ}^{\ast }$(5850) & 1.0 \\ \hline
$\text{q}_{S}\text{(493)}\overline{\text{q}_{b}\text{(9333)}}$ & B$_{S}$%
(9579) & Prediction &  \\ \hline
\end{tabular}
\ $ \\ 
$\ \ \ ${\small B(}$\overline{\text{{\small 5302}}}${\small )}$^{\#}${\small %
=}$\frac{1}{2}${\small [B(5279)+B(5325)]}%
\end{tabular}

\ \ \ \ \ \ \ \ \ \ \ \ \ \ \ \ \ \ \ \ \ \ \ \ \ \ \ \ \ \ 

{\small Table 29 shows that the masses of the deduced mesons are about 98\%
consistent with the experimental results.}

\ \ \ \ \ \ \ \ \ \ \ \ \ \ \ \ \ \ \ 

\begin{tabular}{l}
$\ \ \ \ \ $Table 30.\ \ The D-mesons and the D$_{S}$-Mesons \\ 
$%
\begin{tabular}{|l|l|l|l|}
\hline
\ u$_{C}$(m)$\overline{\text{q}_{\text{N}}^{0}\text{(313)}}$\  & Deduced & 
Exper.,(Mev) & R=$\frac{\Delta \text{M}}{\text{M}}\%$ \\ \hline
u$_{C}$(1753)$\overline{\text{q}_{\text{N}}^{0}\text{(313)}}$ & D(1882) & D($%
\overline{\text{1939}}$) & 2.9 \\ \hline
u$_{C}^{1}$(2133)$\overline{\text{q}_{\text{N}}^{0}\text{(313)}}$ & D(2381)
& D($\overline{\text{2440}}$) & 2.4 \\ \hline
$\text{u}_{\text{C}}^{1}\text{(2333)}\overline{\text{q}_{\text{N}}^{0}\text{%
(313)}}$ & D(2591) & D(2640)$^{\ast }$ & 1.9 \\ \hline
$\text{u}_{\text{C}}^{1}\text{(2533)}\overline{\text{q}_{\text{N}}^{0}\text{%
(313)}}$ & D(2801) & Prediction &  \\ \hline
$\text{u}_{\text{C}}^{1}\text{(6073)}\overline{\text{q}_{\text{N}}^{0}\text{%
(313)}}$ & D(6231) & Prediction &  \\ \hline
{*}*************** & ******* & ******** & *** \\ \hline
$\text{u}_{\text{C}}^{1}$(1753)$\overline{\text{d}_{\text{S}}^{1}\text{(493)}%
}$ & D$_{S}$(1943) & D$_{S}^{\pm }$(1968) & 1.3 \\ \hline
$\text{u}_{\text{C}}^{1}$(1753)$\overline{\text{d}_{\text{S}}\text{(493)}}$
& D$_{S}$($\overline{\text{2093}}$) & D$_{S}$($\overline{\text{2112}}$) & 
0.90 \\ \hline
$\text{u}_{\text{C}}^{1}\text{(2133)}\overline{\text{d}_{\text{S}}^{1}\text{%
(773)}}$ & D$_{S}$(2311) & D$_{S_{j}}$(2317) & 0.04 \\ \hline
$\text{u}_{\text{C}}^{1}\text{(2133)}\overline{\text{d}_{\text{S}}^{1}\text{%
(493)}}$ & D$_{S}$($\overline{\text{2455}}$) & D$_{S_{j}}$(2460) & 0.20 \\ 
\hline
u$_{C}^{1}$(2333)$\overline{\text{d}_{\text{S}}^{1}\text{(493)}}$ & D$_{S}$($%
\overline{\text{2597}}$) & D$_{S_{j}}$($\overline{\text{2555}}$) & 0.23 \\ 
\hline
$\text{u}_{\text{C}}^{1}\text{(2333)}\overline{\text{d}_{\text{S}}^{1}\text{%
(493)}}$ & D$_{S}$($\overline{\text{2761}}$) & Prediction &  \\ \hline
\end{tabular}
\ $%
\end{tabular}

\ \ \ \ \ \ \ \ \ \ \ \ \ \ \ \ \ \ \ \ \ \ \ \ 

\qquad\ \ \ \ \ \ \ \ \ \ \ \ \ \ \ \ \ \ \ \ \ \ \ \ \ \ \ \ \ \ \ \ \ \ \
\ \ \ \ \ \ \ \ \ \ \ \ \ \ \ \ \ \ \ \ \ \ \ \ \ \ \ \ \ \ \ \ \ \ \ \ \ \
\ \ \ \ \ \ \ \ \ \ \ \ \ \ \ \ \ \ \ \ \ \ \ \ \ \ \ \ \ \ \qquad

\begin{tabular}{l}
\ \ \ \ \ \ \ \ \ \ \ Table 31. The Strange Mesons (S = \ $\pm $1) \\ 
$%
\begin{tabular}{|l|l|l|l|}
\hline
\ q$_{\text{N}}$(313)$\ \overline{\ \text{d}_{S}\text{(m)}}$ & Deduced & 
Exper., $\Gamma $ (Mev) & R=$\frac{\Delta \text{M}}{\text{M}}\%$ \\ \hline
q$_{\text{N}}$(313)$\overline{\text{{\small d}}_{S}\text{(493)}}$ & K(488) & 
K(494) & 1.2 \\ \hline
q$_{\text{N}}$(313)$\overline{\text{{\small d}}_{S}\text{(773)}}$ & K($%
\overline{\text{883}}$) & K(892), 50 & 1.3 \\ \hline
q$_{\text{N}}$(313)$\overline{\text{{\small d}}_{S}^{0}\text{{\small (1203)}}%
}$ & K($\overline{\text{1279}}$) & K(1273), 90 & 0.47 \\ \hline
q$_{\text{N}}$(313)$\overline{\text{{\small d}}_{S}^{0}\text{(1393{\small )}}%
}$ & K($\overline{\text{1436}}$) & K($\overline{\text{1414}}$)$,\overline{%
\text{200}}$ & 1.6 \\ \hline
q$_{\text{N}}$(313)$\overline{\text{{\small d}}_{S}^{-1}\text{{\small (1503)}%
}}$ & K($\overline{\text{1606}}$) & K($\overline{\text{1615}}$)$^{\#}$, \ $%
\overline{\text{92}}$ & 0.50 \\ \hline
q$_{\text{N}}$(313)$\overline{\text{{\small d}}_{S}^{-1}\text{(1603{\small )}%
}}$ & K($\overline{\text{1766}}$) & K($\overline{\text{1771}}$), $\overline{%
\text{236}}$ & 0.28 \\ \hline
q$_{\text{N}}$(313)$\overline{\text{{\small d}}_{S}^{0}\text{(1923{\small )}}%
}$ & K(1966) & K($\overline{\text{1962}}$)$^{\#}$, $\overline{\text{287}}$ & 
0.20 \\ \hline
q$_{\text{N}}$(313)$\overline{\text{d}_{S}^{0}\text{(2013)}}$ & K(2056) & K$%
_{4}^{\ast }$(2045), 198 & 1.0 \\ \hline
q$_{N}^{0}$(2013)$\overline{\text{d}_{S}^{1}\text{(493)}}$ & K(2316) & K($%
\overline{\text{2317}}$)$^{\#}$, $\overline{\text{170}}$ & 0.04 \\ \hline
{\small q}$_{\text{N}}${\small (313)}$\overline{\text{d}_{S}^{0}\text{(2743)}%
}$ & K(2786) & Prediction & ? \\ \hline
\end{tabular}
\ \ \ \ \ \ $%
\end{tabular}

\ \ \ \ \ \ \ \ \ \ \ \ \ \ \ \ \ \ \ \ \ \ \ \ \ \ \ \ \ \ \ \ \ \ 

{\small Table 30 and Table 31 show that the deduced intrinsic quantum
numbers of the D-mesons, the D}$_{S}${\small -mesons and the K-mesons are
the same as the experimental results. They also show that the deduced rest
masses of the D-mesons, the D}$_{S}${\small -mesons and the K-mesons agree
(about 98\%) with the experimental results.}

In summary, the phenomenological formulae explain all meson experimental
intrinsic quantum numbers (100\%) and the rest masses (about 98\%). We
explain virtually all experimentally confirmed mesons in this paper and all
deduced low-mass mesons have already been discovered by experiments. \ \ \ \
\ \ \ \ \ \ \ \ \ \ \ \ \ 

\section{\ \ Evidence for the Deduced Quarks \ \ \ \ \ \ \ \ \ \ \ \ \ \ \ \ 
}

According to the Quark Model \cite{Quark Model}, we cannot see an individual
quark. We can only infer the existence of quarks from the existence of
baryons and mesons, which are made up of the quarks. Thus, if we find the
mesons that are made of certain quarks, it means that we also find the
quarks. For example, from meson J/$\Psi $(3097) = [u$_{C}^{1}$(1753)$%
\overline{\text{u}_{C}^{1}\text{(1753)}}$], we discovered the u$_{C}^{1}$%
(1753)-quark. Similarly the experimental baryon spectrum \cite{Baryon04} and
the meson spectrum \cite{Meson04} have already provided evidence of almost\
all of the quarks in Table 11. As examples, for the three
three-brother-quarks [d$_{S}^{\text{-1}}$(773), d$_{S}^{\text{-1}}$(3753)
and d$_{S}^{\text{-1}}$(9613)], [u(313), u$_{C}^{\text{1}}$(1753) and u$%
_{C}^{\text{1}}$(6073)] and [u$_{C}^{1}$(2133), u$_{C}^{1}$(2333) and u$%
_{C}^{1}$(2533)]\ and a four-brother-quark [d(313), d$_{S}^{\text{1}}$(493),
d$_{S}^{\text{1}}$(1933) and d$_{b}^{\text{1}}$(4913)], they all have strong
experimental evidence. \ \ \ \ \ \ \ \ \ \ 

\subsection{The Three Brother Quarks d$_{S}^{-1}$(773), d$_{S}^{-1}$(3753)
and d$_{S}^{-1}$(9613)}

There are three brother quarks, d$_{S}^{-1}$(773)$,$ d$_{S}^{-1}$(3753) and d%
$_{S}^{-1}$(9613). They are born on the single energy bands of the $\Delta $%
-axis at the symmetry point H from $\Delta $S = -1 (see Table 4). From Table
13, Table 25 and Table 26, we have:

\ \ \ \ \ \ \ \ \ \ \ \ \ \ \ \ \ \ \ \ \ \ \ \ \ \ \ \ \ \ \ \ \ \ \ \ \ \ 

\qquad A1. The d$_{S}^{-1}$(773)-quark (see Table 4)

The d$_{S}^{-1}$(773)-quark has Q = -$\frac{1}{3}$, n =(0, 0, 2). Table 32
shows its evidence:

\begin{tabular}{l}
\ \ \ \ \ \ \ \ \ \ \ \ \ \ \ \ \ \ \ \ \ \ \ \ \ Table 32. \ Evidence of
the $\text{d}_{S}^{-1}\text{(773)-quark}$ \\ 
\begin{tabular}{|l|l|l|l|}
\hline
where & The Quark Constitutions of the Hadrons & Exper. & $\frac{\Delta M}{M}%
\%$ \\ \hline
Table 17 & 
\begin{tabular}{l}
$\text{d}_{S}^{\text{-1}}\text{(773)u(313)d(313) = }\Lambda $(1399) \\ 
$\text{d}_{S}^{\text{-1}}\text{(773)u(313)u(313) = }\Sigma ^{+}$(1399) \\ 
$\text{d}_{S}^{\text{-1}}\text{(773)u(313)d(313) = }\Sigma ^{0}$(1399) \\ 
$\text{d}_{S}^{\text{-1}}\text{(773)d(313)d(313) = }\Sigma ^{-}$(1399)%
\end{tabular}
$\text{ }$ & 
\begin{tabular}{l}
$\Lambda $(1405) \\ 
$\Sigma ^{+}$(1383) \\ 
$\Sigma ^{0}$(1384) \\ 
$\Sigma ^{-}$(1387)%
\end{tabular}
& 
\begin{tabular}{l}
$0.35$ \\ 
$1.13$ \\ 
1.0 \\ 
0.87%
\end{tabular}
\\ \hline
Table 27 & $\ \text{d}_{S}^{\text{-1}}\text{(773)}\overline{\text{d}_{S}^{%
\text{-1}}\text{(773)}}\text{ = }\eta ${\small (1041)} & $\ \ \phi ${\small %
(1020)} & \ 2.1 \\ \hline
Table 31 & $\ \text{q}_{N}\text{(313)}\overline{\text{d}_{S}^{\text{-1}}%
\text{(773)}}\text{ =}${\small K(883)} & \ \ K(892) & \ 1.3 \\ \hline
Table 24B & \ d$_{\text{S}}$(493)$\overline{\text{d}_{S}^{\text{-1}}\text{%
(773)}}$ = $\eta $(1177) & \ \ h$_{1}${\small (1170)} & \ 0.1 \\ \hline
Table 24B & \ u$_{C}$(1753)$\overline{\text{d}_{S}^{\text{-1}}\text{(773)}}$
= D$_{S}$(2511) & \ D$_{S_{1}}$(2536) & 1.0 \\ \hline
\end{tabular}%
\end{tabular}
\ 

There is enough evidence to show that the d$_{S}^{-1}$(773)-quark really
does exist.\ \ \ \ \ \ \ \ \ \ \ \ \ \ \ \ \ \ \ \ \ \ \ \ \ \ \ \ 

\ \ \ \ \ \ \ \ \ \ \ \ \ \ \ \ \ \ \ \ \ \ \ \ \ \ \ \ \ \ \ \ \ \ \ \ \ \
\ \ \ \ \ \ \qquad

\qquad A2. The d$_{S}^{-1}$(3753)-quark (Table 4 and Table 26)

The d$_{S}^{-1}$(3753)-quark has Q = -$\frac{1}{3}$, n =(0, 0, 4). Table 33
show its evidence:

\begin{tabular}{l}
\ \ \ \ \ \ Table 33 \ Evidence of the d$_{S}^{-1}$(3753)$\text{-quark (see
Table 26)}$ \\ 
\begin{tabular}{|l|l|l|l|l|l|l|l|}
\hline
{\small \ \ q}$_{i}${\small (m)}$\overline{q_{i}\text{(m)}}$ & $DS$ & $\frac{%
100\Delta \text{m}}{939}$ & {\small -100}$\widetilde{m}$ & {\small E}$%
_{bind} $ & {\small Theory} & {\small Exper.} & $\frac{\Delta M}{M}\%$ \\ 
\hline
{\small d}$_{S}^{-1}${\small (3753)}$\overline{d_{S}^{1}(493)_{\Sigma }}$ & 
{\small 2} & {\small 347} & {\small -400} & {\small -90} & $\eta ${\small %
(4156)} & $\psi (4160$ & .093 \\ \hline
\end{tabular}%
\end{tabular}

\ \ \ \ \ \ \ \ \ \ \ \ \ \ \ \ \ \ \ \ \ \ \ \ \ \ \ \ \ \ \ \ \ 

\qquad A3. The d$_{S}^{-1}$(9613)-quark (Table 4 and Table 25)

The d$_{S}^{-1}$(9613)-quark has Q = -$\frac{1}{3}$, n =(0, 0, 6). Table 34
show its evidence:

\begin{tabular}{l}
$\ \ \ \ \ \ \text{\ }$Table 34. \ Evidence of the {\small d}$_{S}^{-1}$%
{\small (9613)}$\text{-quark (see Table 25)}$ \\ 
$%
\begin{tabular}{|l|l|l|l|l|}
\hline
{\small d}$_{S}^{-1}${\small (9613)}$\overline{d_{S}(m)}$ & {\small E}$%
_{bind}$ & {\small Theory} & {\small Exper., }$\Gamma $ (Mev) & R.=$\frac{%
\Delta m}{m}\%$ \\ \hline
{\small d}$_{S}^{-1}${\small (9613)}$\overline{d_{S}^{-1}(493)_{D}}$ & 
{\small -250} & $\eta ${\small (}$\overline{\text{{\small 9906}}}${\small )}
& $\chi ${\small (}$\overline{\text{{\small 9889}}}${\small ), \ \ ?} & 0.17
\\ \hline
{\small d}$_{S}^{-1}${\small (9613)}$\overline{d_{S}^{1}(493)_{\Sigma }}$ & 
{\small -50} & $\eta ${\small (10056)} & $\Upsilon ${\small (10023), 0.044}
& 0.13 \\ \hline
{\small d}$_{S}^{-1}${\small (9613)}$\overline{d_{S}^{-1}(773)_{\Delta }}$ & 
{\small - 212} & $\eta ${\small (}$\overline{\text{{\small 10174}}}${\small )%
} & $\chi (\overline{\text{{\small 10252}}}${\small ), \ \ ?} & 0.76 \\ 
\hline
{\small d}$_{S}^{-1}${\small (9613)}$\overline{d_{S}^{0}(1303)_{F}}${\small %
\ } & {\small - 247} & $\eta \overline{\text{{\small (10620)}}}$ & $\Upsilon 
${\small (10580), 14} & 0.66 \\ \hline
{\small d}$_{S}^{-1}${\small (9613)}$\overline{d_{S}^{-1}(1603)_{F}}${\small %
\ } & {\small - 347} & $\eta ${\small (}$\overline{\text{{\small 10777}}}$%
{\small )} & $\Upsilon ${\small (10860), 110} & 0.14 \\ \hline
{\small d}$_{S}^{-1}${\small (9613)}$\overline{d_{S}^{0}(1923)_{F}}$ & 
{\small -455} & $\eta ${\small (11079)} & $\Upsilon ${\small (11020), 79} & 
0.61 \\ \hline
\end{tabular}
\ \ \ \ $%
\end{tabular}

\ \ \ \ \ \ \ \ \ \ \ \ \ \ \ \ \ \ \ \ \ \ \ \ \ \ \ \ \ \ \ \ \ \ \ \ \ 

Thus, the three brother quarks [d$_{S}^{-1}$(773), d$_{S}^{-1}$(3753) and d$%
_{S}^{-1}$(9613)]\ really do exist.

\subsection{The Three Brother Quarks u$_{C}^{1}$(2133), u$_{C}^{1}$(2333)
and u$_{C}^{1}$(2533)}

There are three brother quarks u$_{C}^{1}$(2133), u$_{C}^{1}$(2333) and u$%
_{C}^{1}$(2533) with I = 0 and Q = $\frac{2}{3}$ on the D-axis. They are
born from $\Delta $S = +1 on the twofold bands with two inequivalent
n-values of the D-axis (see Table B5). From Table 14, Table 20, Table 26 and
Table 30, we can find:

\ \ \ \ \ \ \ \ \ \ \ \ \ \ \ \ 

\qquad B1. The u$_{C}^{1}$(2133)-quark (see Table B5, Table 26 and Table C7)

The u$_{C}$(2133)-quark\ has Q = $\frac{2}{3}$. Table 35 shows its
evidence:\ \ \ \ \ \ \ \ \ \ \ \ \ \ \ \ \ \ \ \ \ \ \ \ \ \ \ \ \ \ \ \ \ \
\ \ \ \ \ 

:\ 
\begin{tabular}{l}
\ \ \ \ \ Table 35. \ The Evidence of the u(313)-Quark \\ 
$%
\begin{tabular}{|l|l|l|}
\hline
\ Deduced Baryon or Mesons & experiment & $\frac{\Delta M}{M}\%$ \\ \hline
u$_{C}^{1}$(2133)u(313)d(313)= $\Lambda _{C}^{+}$(2609) & $\Lambda _{C}^{+}$(%
$\overline{\text{2611}}$) & 0.08 \\ \hline
u$_{C}^{1}$(2133)$\overline{\text{u}_{C}^{1}\text{(2133)}}$ = $\psi $(3681)
& $\psi $(3686), & 0.14 \\ \hline
q$_{N}$(313)$\overline{\text{u}_{C}^{1}\text{(2133) }}$= D(2381) & D($%
\overline{\text{2440}}$), $\ \ $ & 2.42 \\ \hline
u$_{C}^{1}$(2133)$\overline{\text{d}_{\text{S}}^{1}\text{(493)}}$ = D$_{S}$%
(2442) & D$_{S_{j}}$(2460) & 1.90 \\ \hline
\end{tabular}
\ \ \ \ \ $%
\end{tabular}

\ \ \ \ \ \ \ \ \ \ \ \ \ \ \ \ \ \ \ \ \ \ \ \ \ \ \ \ \ \ \ \ \ \ \ \ \ \
\ \ \ \ \ \ \ \ 

\qquad B2. The u$_{C}$(2333)-quark (Table B5 and Table 26)

The u$_{C}$(2333)-quark\ has Q = $\frac{2}{3}$. Table 36 shows its
evidence:\ \ \ \ \ \ \ \ \ \ \ \ \ \ \ \ \ \ \ \ \ \ \ \ \ \ \ \ \ \ \ \ \ \
\ \ \ \ \ 

: 
\begin{tabular}{l}
\ \ \ \ \ \ \ \ \ Table 36. \ The Evidence of the u(313)-Quark \\ 
$%
\begin{tabular}{|l|l|l|}
\hline
\ \ Deduced Baryons or Mesons & Experiments & $\frac{\Delta M}{M}\%$ \\ 
\hline
u$_{C}^{1}$(2333)u(313)d(313)= $\Lambda _{C}^{+}$(2809) & $\Lambda _{C}^{+}$(%
$\overline{\text{2823}}$) & 0.50 \\ \hline
u$_{C}^{1}$(2333)$\overline{\text{u}_{C}^{1}\text{(2333)}}$ = $\psi $(4052)
& $\psi $(4040)) & 0.30 \\ \hline
q$_{N}$(313)$\overline{\text{u}_{C}^{1}\text{(2333) }}$= D(2591) & D(2640)$%
^{\ast }$ $\ \ $ & 1.90 \\ \hline
\end{tabular}
\ \ \ \ $%
\end{tabular}

\ \ \ \ \ \ \ \ \ \ \ \ \ \ \ \ \ \ \ \ \ \ \ \ \ 

\qquad B3. The u$_{C}^{1}$(2533)-quark (Table B5 and Table 26)

The u$_{C}^{1}$(2533)-quark\ has Q = $\frac{2}{3}$. Table 37 shows its
evidence:\ \ \ \ \ \ \ \ \ \ \ \ \ \ \ \ \ \ \ \ \ \ \ \ \ \ \ \ \ \ \ \ \ \
\ \ \ \ \ 

:\ 
\begin{tabular}{l}
\ \ \ \ \ Table 37. \ The Evidence of the u(313)-Quark \\ 
$%
\begin{tabular}{|l|l|l|}
\hline
Deduced Baryon or Mesons & \ \ \ Experiments & $\frac{\Delta M}{M}\%$ \\ 
\hline
u$_{C}^{1}$(2533)$\overline{\text{u}_{C}^{1}\text{(2533)}}$ = $\psi $(4420)
& \ $\psi $(4415), $\Gamma $ = \ 43 & 0.11 \\ \hline
\end{tabular}
\ \ \ \ $%
\end{tabular}

\ \ \ \ \ \ \ \ \ \ \ \ \ \ \ \ \ \ 

Table 35-Table 37 show that the three brother quarks u$_{C}$(2133), u$_{C}$%
(2333) and u$_{\text{C}}$(2533) do exist.\ \ \ 

\subsection{The Three Brother Quarks u(313), u$_{C}^{1}$(1753) and u$%
_{C}^{1} $(6073)}

There are three brother quarks u(313),\ u$_{C}^{1}$(1753) and u$_{C}^{1}$%
(6073) also.\ They are born at the $\Gamma $-point of the single energy
bands of the $\Delta $-axis (table 4). With Table 16, Table 22, Table 25\
and Table 26, we get:\ \ \ \ \ \ \ \ \ \ \ \ \ \ \ \ \ \ \ \ \ \ \ \ \ \ \ \
\ \ \ \ \ \ 

\qquad C1. The u(313)-quark (Table 4)

The u(313)-quark\ has Q = $\frac{2}{3}$, $\overrightarrow{n}$ = (0, 0, 0).
Table 38 shows its evidence:

\begin{tabular}{l}
\ \ \ \ \ \ \ \ \ \ \ \ \ \ \ \ \ \ \ Table 38. \ The Evidence of the
u(313)-Quark \\ 
$%
\begin{tabular}{|l|l|}
\hline
Deduced Baryon or Mesons & Experiments \\ \hline
u(313)u(313)d(313)=p(939) & \ P(938), $\tau $= 2.1$\times $10$^{\text{29}}$%
years \\ \hline
u(313)$\overline{\text{u(313)}}$=$\pi ^{0}$(139) & \ $\pi ^{0}$(135), $\tau $%
= (8.4 $\pm $ 0.6) $\times $ 10$^{-17}$ s \\ \hline
q$_{N}$(313)$\overline{\text{u(313)}}$= $\mathfrak{\langle }$%
\begin{tabular}{l}
$\pi ^{0}$(139) \\ 
$\pi ^{-}$(139)%
\end{tabular}
& 
\begin{tabular}{l}
$\pi ^{0}$(135), $\tau $= (8.4 $\pm $ 0.6) $\times $ 10$^{-17}$ s \\ 
$\pi ^{-}$(140), $\tau $= (2.6033 $\pm $ 0.0005) $\times $ 10$^{-8}$ s%
\end{tabular}
\\ \hline
$\text{u(313)}\overline{\text{d}_{S}^{1}\text{(493)}}$= K$^{+}$(488) & \ \ K$%
^{+}$(494), (1.2384 $\pm $ 0.0024) $\times $\ 10$^{-8}$ s \\ \hline
\end{tabular}
$%
\end{tabular}
{\LARGE \ \ \ }\ \ \ \ \ \ \ \ \ \ \ \ \ \ \ \ \ \ \ \ \ \ \ \ \ \ \ \ 

\ \ \ \ \ \ \ \ \ \ \ \ 

\qquad C2. The u$_{C}^{1}$(1753)-quark (Table 4)

The u$_{C}^{1}$(1753)-quark\ has Q = $\frac{2}{3}$, $\overrightarrow{n}$ =
(0, 0, -2). Table 39 shows its evidence:\ \ \ \ \ \ \ \ \ \ \ \ \ \ \ \ \ \
\ \ \ \ \ \ \ \ \ \ \ \ \ \ \ \ \ \ \ \ \ \ 
\begin{tabular}{l}
\ \ \ \ \ \ \ \ \ \ \ \ \ \ \ \ \ \ \ \ Table 39. \ The Evidence of the u$%
_{C}^{1}$(1753)-Quark \\ 
$%
\begin{tabular}{|l|l|}
\hline
Deduced Baryon or Mesons & Experiments \\ \hline
u$_{C}^{1}$(1753)u(313)d(313)= $\Lambda _{C}^{+}$(2279) & $\Lambda _{C}^{+}$%
(2285), $\tau $= (200 $\pm $ 6) $\times $ 10$^{-15}$ s \\ \hline
u$_{C}^{1}$(1753)$\overline{\text{u}_{C}^{1}\text{(1753)}}$ = $\psi $(3069)
& J/$\psi $(3097), $\Gamma $ = \ 91 $\pm $ 3.2 kev \\ \hline
q$_{N}$(313)$\overline{\text{u}_{C}^{1}\text{(1753)}}$= D(1882) & D(1869), $%
\ \ \tau $= (1040 $\pm $ 7) $\times $ 10$^{-15}$ s \\ \hline
u$_{C}^{1}$(1753)$\overline{\text{d}_{\text{S}}\text{(493)}}$= D$_{S}$(1943)
& D$_{S}$(1968), $\tau $= (490 $\pm $ 9) $\times $ 10$^{-15}$ s \\ \hline
\end{tabular}
\ \ \ \ \ $%
\end{tabular}

\ \ \ \ \ \ \ \ \ \ \ \ \ \ \ \ \ \ \ \ \ 

\qquad C3. The u$_{C}^{1}$(6073)-quark\ (see Table4 and Table 25)\ \ \ \ \ \
\ \ \ \ \ \ 

The u$_{C}^{1}$(6073)-quark\ has Q = $\frac{2}{3}$, $\overrightarrow{n}$ =
(0, 0, -4). Table 40 shows its evidence:

\begin{tabular}{l}
\ \ \ \ \ \ \ \ \ \ \ \ \ \ \ \ \ \ \ Table 40. The Evidence of the
u(6073)-quark \\ 
\begin{tabular}{|l|l|l|l|l|l|}
\hline
\ \ \ \ {\small q(m)}$\overline{q(m)}$ & \ \ \ \ -100$\widetilde{m}$ & $%
\text{I}_{i}\text{I}_{j}$ &  & {\small Deduced} & {\small Exper., }$\Gamma $
(kev) \\ \hline
u$_{C}^{1}${\small (6073)}$\overline{\text{u}_{C}^{1}\text{{\small (6073)}}}$
& -1200$\leftarrow \frac{6073\times 6073}{1753\times 1753}$ & 0 & -1637 & $%
\psi ${\small (10509)} & $\Upsilon ${\small (10355), 26} \\ \hline
\end{tabular}%
\end{tabular}

\ \ \ \ \ \ \ \ \ \ \ \ 

Table 40 shows very strong evidence of the u$_{C}^{1}$(6073)-quark: 1) $\psi 
$(10509) and $\Upsilon $(10355) have exactly the same intrinsic quantum
numbers ( I = S = C = b = Q = 0), 2) they have essentially the same rest
masses, and 3) the width $\Gamma $ = 26 kev of $\Upsilon $(10355) 
\TEXTsymbol{<} the width $\Gamma $ = 87 kev of J/$\psi $(3067). It is well
known that the discovery of J/$\psi $(3067) is also the discovery of the u$%
_{C}^{1}$(1753)-quark. Thus, the discovery of $\Upsilon $(10355) [$\psi $%
(10509)] is also the discovery of the u$_{C}^{1}$(6073)-quark.

Table 38-Table 40 shows that the three brother quarks really do exist.\ \ \
\ 

\subsection{The Four Brother Quarks d(313), d$_{S}^{1}$(493), d$_{S}^{1}$%
(1933) and d$_{b}^{1}$(4913)}

From Table 7, we see the following four brother quarks: at E$_{\Gamma }$= 0, 
$\overrightarrow{n}=$(0,0,0), d(313); at E$_{N}$=1/2,\ $\overrightarrow{n}$
= (1,1,0), d$_{S}^{1}$(493); at E$_{N}$=9/2, $\overrightarrow{n}$=(2,2,0), d$%
_{S}^{1}$(1933); and at E$_{N}$=25/2, $\overrightarrow{n}$=(3,3,0), d$%
_{b}^{1}$(4913). They are born on the single energy bands of the $\Sigma $%
-axis. The four \textquotedblleft brothers\textquotedblright\ have the same
electric charge (Q = -1/3) and the same generalized strange number (S$_{G}$
= S + C + b = -1). Using Table 7, Table 22, Table 26, Table 28 and Table 30,
we find:

\ \ \ \ \ \ \ \ \ \ \ \ \ \ \ \ \ \ \ \ \ 

\qquad D1. The d(313)-Quark (Table 7)

The d(313)-Quark has Q = -$\frac{1}{3}$, $\overrightarrow{\text{n}}$= (0, 0,
0). Table 41 show its evidence:

\begin{tabular}{l}
\ \ \ \ \ \ \ \ \ \ \ \ \ \ \ \ \ \ \ \ Table 41. \ The Evidence of the
d(313)-Quark \\ 
$%
\begin{tabular}{|l|l|}
\hline
d(313)u(313)d(313)=$n$(939) & \ n(940), $\tau $=(885.7$\pm $\ 0.8)s \\ \hline
d(313)$\overline{\text{d(313)}}$=$\pi ^{0}$(139) & \ $\ \pi ^{0}$(135), $%
\tau $=(2.6033 $\pm $\ \ \ 0.0005)$\times $\ \ 10$^{-8}$s \\ \hline
q$_{N}$(313)$\overline{\text{d(313)}}$= $\text{ }\langle $ 
\begin{tabular}{l}
$\pi ^{+}$(139) \\ 
$\pi ^{0}$(139)%
\end{tabular}
& 
\begin{tabular}{l}
$\pi ^{+}$(139), $\tau $=(2.6033 $\pm $\ \ \ 0.0005)$\times $\ \ 10$^{-8}$s
\\ 
$\pi ^{0}$(135), $\tau $= (8.4 $\pm $ 0.6) $\times $ 10$^{-17}$ s%
\end{tabular}
\\ \hline
d$_{S}^{1}$(493)$\overline{\text{d(313)}}$= K$^{0}$(488) & \ K$^{0}$(494), $%
\ \tau $= (1.2384 $\pm $ 0.0024) $\times $ 10$^{-8}$ s \\ \hline
\end{tabular}
\ \ \ \ \ $%
\end{tabular}

\ \ \ \ \ \ \ \ \ \ \ \ \ \ \ \ \ \ \ \ \ \ \ \ \ \ \ \ \ \ \ \ \ \ \ \ \ \
\ 

\qquad D2. The d$_{S}^{1}$(493)-quark (Table 7)

The d$_{S}^{1}$(493)-Quark has Q = -$\frac{1}{3}$, $\overrightarrow{\text{n}}%
=$(1, 1, 0). Table 42 shows its evidence:

\begin{tabular}{l}
\ \ \ \ \ \ \ \ \ \ \ \ \ \ \ \ \ \ \ \ Table 42. \ Evidence of the d$%
_{S}^{1}$(493)-quark \\ 
\begin{tabular}{|l|l|}
\hline
d$_{S}^{1}$(493)u(313)d(313)=$\Lambda $(1119) & $\Lambda $(1116), $\tau $%
=(2.632$\pm $ 0.020)$\times $ \ 10$^{-10}$s \\ \hline
d$_{S}^{1}$(493)$\overline{\text{d}_{S}^{1}\text{(493)}}$=$\eta $(549) & $%
\eta $(548), $\Gamma $=(1.29 $\pm $ 0.11) kev \\ \hline
q$_{N}$(313)$\overline{\text{d}_{S}^{1}\text{(493)}}$= K(488) & K(494), $%
\Gamma =$ (1.2384 $\pm $ 0.0024) $\times $ \ 10$^{-8}$s \\ \hline
\end{tabular}%
\end{tabular}

\ \ \ \ \ \ \ \ \ \ \ \ \ \ \ \ \ \ \ \ \ 

\qquad D3. The d$_{S}^{1}$(1933)-quark (Table 7)

The d$_{S}^{1}$(1933)-quark has Q = -$\frac{1}{3}$, $\overrightarrow{n}=$(2,
2, 0). Table 43 shows its evidence:

\begin{tabular}{l}
\ \ \ \ \ \ \ \ \ \ Table 43. \ The Evidence of the d$_{S}$(1933)-Quarks \\ 
$%
\begin{tabular}{|l|l|}
\hline
d$_{S}$(1933)u(313)d(313)=$\Lambda $(2559) & $\Lambda $(2585), $\tau $=?; \\ 
\hline
d$_{S}^{1}$(1933)$\overline{\text{d}_{S}^{1}\text{(1933)}}$= $\eta $(3005) & 
$\eta _{C}$(2980), $\Gamma $=17.3 Mev \\ \hline
q$_{N}$(313)$\overline{\text{d}_{S}^{1}\text{(1933)}}$=K(2076) & K$%
_{4}^{\ast }$(2045), $\Gamma $ = 198 $\pm $ 30 Mev \\ \hline
d$_{S}^{1}$(493)$\overline{\text{d}_{S}^{1}\text{(1933)}}$=$\eta $(2127) & f$%
_{2}$(2150), $\Gamma $= 167$\pm 30$ Me \\ \hline
\end{tabular}
\ \ \ $%
\end{tabular}
\ \ 

\ \ \ \ \ \ \ \ \ \ \ \ \ \ \ \ \ \ \ \ \ \ \ \ \ \ \ 

\qquad D4. The d$_{b}^{1}$(4913)-quark (Table 7)

The d$_{b}^{1}$(4913)-quark has Q = -$\frac{1}{3}$, $\overrightarrow{n}=$(3,
3, 0). Table 44 shows its evidence:\ \ \ \ \ \ \ \ \ \ \ \ \ \ \ \ \ 
\begin{tabular}{l}
\ \ \ \ \ \ \ \ \ \ \ \ \ \ \ \ \ \ \ \ Table 44. \ The Evidence of the%
{\small \ }d$_{S}^{1}$(4913)-$\text{quark}$ \\ 
$%
\begin{tabular}{|l|l|}
\hline
d$_{b}^{1}$(4913)u(313)d(313)=$\Lambda _{b}$(5539) & $\Lambda _{b}$(5624), $%
\tau $=(1.229$\pm $.080)$\times $10$^{-12}$s; \\ \hline
d$_{b}^{1}$(4913)$\overline{\text{d}_{b}^{1}\text{(4913)}}$=$\Upsilon $%
(1S)(9389) & $\Upsilon $(1S)(9460), $\Gamma $=(53.0 $\pm $1.5) kev \\ \hline
q$_{N}$(313)d$_{b}^{1}$(4913)= B(5378) & B(5279), $\tau $=(1.671$\pm $.018)$%
\times $10$^{-12}$s; \\ \hline
d$_{b}^{1}$(493)$\overline{\text{d}_{b}^{1}\text{(4913)}}$= B$_{S}$(5344) & B%
$_{S}$(5370), $\Gamma $= (1.461 $\pm $\ 0.057) $\times $10$^{-12}$s] \\ 
\hline
\end{tabular}
\ $%
\end{tabular}

\ \ \ \ \ \ \ \ \ \ \ \ \ \ \ \ \ \ \ \ \ \ \ \ 

In Table 41-Table 44, we can see that the four brother quarks [d(313), d$%
_{S}^{1}$(493), d$_{S}^{1}$(1933) and d$_{b}^{1}$(4913)] really do exist.\ \
\ 

Please pay attention to $\overrightarrow{n}$ = (n$_{1}$, n$_{2}$, n$_{3}$)\
values of the brothers of the three families: for the first one, n =(0, 0,
2), n =(0, 0, 4), n =(0, 0, 6); for the second one, $\overrightarrow{\text{n}%
}$= (0, 0, 0), $\overrightarrow{\text{n}}$= (1, 1, 0), $\overrightarrow{%
\text{n}}$= (2, 2, 0), $\overrightarrow{\text{n}}$= (3, 3, 0); for the third
one $\overrightarrow{\text{n}}$ =(0, 0, 0), $\overrightarrow{\text{n}}$ =(0,
0, -2), $\overrightarrow{\text{n}}$ =(0, 0, -4). These values are very
interesting. They might be hitting a law or a solution of an equation. Since
each $\overrightarrow{\text{n}}$ represents a discovered quark, they might
be implicating a physical law or a fundamental physical equation.

As with Table 32-Table 44, we can show that most of the deduced quarks in
Table 11 have already been discovered by experiments.\ 

\subsection{Experiments Have Already Discovered Most of the Deduced Quarks
in Table 11}

Using the phenomenological formulae, this paper deduces an excited quark
spectrum, shown in Table 10 and Table 11. Experiments have already
discovered most of these quarks inside baryons and mesons. The q$_{N}$%
(m)-quarks are inside the N-baryons (Table 18), the light unflavored mesons
with I = 0 (Table 27) and the strange \ K(M)-mesons (Table 31). The q$%
_{\Delta }$(m)-quarks are inside the $\Delta $-baryons (Table 18) and the
light unflavored mesons with I = 1 (Table 28). The d$_{S}$(m)-quarks are
inside the $\Lambda $-baryons (Table 17), the light unflavored mesons with I
= 0 (Table 27) and I = 1 (Table 28), the strange K-mesons (Table 31) and the
heavy unflavored mesons (Table 25). The q$_{\Sigma }$(m)-quarks are inside
the $\Sigma $-baryons (Table 17) and the light unflavored mesons with I = 1
(Table 28). The u$_{C}$(m)-quarks are inside the $\Lambda _{C}^{+}$-baryons
(Table 14 and Table 20), the D-mesons (Table 30) and the intermediate mass $%
\psi $-mesons (Table 26). The d$_{b}$(m)-quarks are inside the $\Lambda
_{b}^{0}$-baryon (Table 20), the B(M)-mesons (Table 29) and the $\Upsilon $%
-meson (Table 25). The q$_{\Xi _{C}}$(m)-quarks are inside the $\Xi _{C}$%
-baryons (Table 21). The $\Omega _{C}$(2133)-quark is inside the $\Omega
_{C} $(2907)-baryon (Table B7, Table 14 and Table 20). The u$_{C}$%
(6023)-quark (Table 4) is inside the meson $\Upsilon $(10355) (Table 25).

These experimental results are enough to support the deduced quark spectrum
and the phenomenological formulae.\ \ \ \ \ \ \ \ \ \ \ \ 

\ It is very important to pay attention to the $\Upsilon $(3S)-meson (mass m
= 10.3052 $\pm $ 0.00004 Gev, full width $\Gamma $\ = 26.3 $\pm $ 3.5 kev). $%
\Upsilon $(3S) has more than three times larger mass than J/$\psi $(1S)\ (m
= 3096.87 $\pm $ 0.04 Mev)\ and more than three times longer lifetime than J/%
$\psi $(1S)\ (full width $\Gamma $ = 87 $\pm $ 5kev). It is well known that
the discovery\ of J/$\psi $(1S)\ is also the discovery of charmed\ quark c (u%
$_{c}$(1753)). Similarly the discovery of $\Upsilon $(3S) will be the
discovery of a very important new quark---the u$_{C}$(6073)-quark.

\section{Predictions}

\subsection{New Quarks, Baryons and Mesons}

This paper\ predicts many quarks, baryons and mesons with high rest masses.
Since these particles have high rest masses, they are difficult to discover.
The following quarks, baryons and mesons in Table 45 have a possibility of
being discovered in the not too distant future:\ \ \ \ \ \ \ \ \ \ \ \ \ \ \
\ \ \ \ \ \ \ \ \ 

\begin{tabular}{l}
\ \ \ \ \ Table 45. The Important Predictive Quarks, Baryons and Mesons \\ 
\begin{tabular}{|l|l|l|l|l|l|l|}
\hline
& {\small n}$_{1}${\small ,n}$_{2}${\small ,n}$_{3}$ & {\small Quark} & 
{\small Baryon} & {\small q(m)}$\overline{\text{q(m)}}$ & {\small q}$_{N}$%
{\small (313)}$\overline{\text{q}}$ & {\small d}$_{S}(493)\overline{\text{q}}
$ \\ \hline
{\small E}$_{\text{P}}${\small =}$\frac{\text{27}}{\text{4}}$ & {\small (2,
2, 2)} & $\text{d}_{S}^{0}\text{({\small 2743})}$ & $\Lambda ${\small (3369)}
& $\eta ${\small (4196)} & {\small K(2786)} & $\eta ${\small (3047)} \\ 
\hline
$\text{E}_{H}\text{= }${\small 9} & {\small (0, 0, 4)} & $\text{d}_{S}^{%
\text{-1}}\text{({\small 3753})}$ & $\Lambda ${\small (4379)} & $\eta $%
{\small (5472)} & {\small K(3896)} & $\eta ${\small (4156)}$^{\ast }$ \\ 
\hline
$\text{E}_{H}\text{= }${\small 25} & {\small (0, 0, 6)} & $\text{d}_{S}^{-1}%
\text{({\small 9613})}$ & $\Lambda ${\small (10239)} & $\eta ${\small (18133)%
} & {\small K(9781)} & $\eta ${\small (10056)}$^{\ast }$ \\ \hline
$\text{E}_{\Gamma }\text{=}${\small 16} & {\small (0, 0,}$\overline{\text{%
{\small 4}}}${\small )} & {\small u}$_{C}^{1}${\small (6073)} & $\Lambda
_{C} ${\small (6599)} & $\psi ${\small (10509)}$^{\ast }$ & {\small D(6231)}
& {\small D}$_{S}${\small (6312)} \\ \hline
{\small E}$_{N}${\small =}$\frac{\text{49}}{\text{2}}$ & {\small (4, 4, 0)}
& {\small d}$_{b}^{1}${\small (9333)} & $\Lambda _{b}${\small (9959)} & $%
\Upsilon ${\small (17868)} & {\small B(9503)} & {\small B}$_{S}${\small %
(9579)} \\ \hline
\end{tabular}
\\ 
\begin{tabular}{l}
{\small {*} Some mesons have been discovered: u}$_{C}${\small (6073)}$%
\overline{\text{{\small u}}_{C}\text{{\small (6073)}}}${\small =}$\psi $%
{\small (10509)}$^{\ast }${\small [}$\Upsilon ${\small (10355)]}$;$ \\ 
$\text{d}_{S}\text{(9613)}\overline{\text{d}_{S}\text{(493)}}${\small =}$%
\eta ${\small (10056)}$^{\ast }${\small [}$\Upsilon ${\small (10023)] and }$%
\text{d}_{S}^{\text{-1}}${\small (3753)}$\overline{\text{d}_{S}\text{{\small %
(493)}}}${\small =}$\eta ${\small (4156)}$^{\ast }${\small [}$\psi ${\small %
(4160)].}%
\end{tabular}%
\end{tabular}

\ \ \ \ \ \ \ \ \ \ \ \ \ \ \ \ \ \ \ \ \ \ \ \ \ \ \ \ \ \ \ \ \ \ \ \ \ \
\ \ \ \ \ \ \ \ \ \ \ \ \ \ \ \ \ 

The $\Lambda _{C}$(6599)-baryon, the D(6231)-meson, the K(2786)-meson and
the $\Lambda _{b}$(9959)-baryon have a better chance of being discovered. In
order to confirm the u$_{C}$(6073)-quark, we need to find the $\Lambda _{C}$%
(6599)-baryon first.

\subsection{The ``fine structure''\ of Particle Physics}

The large full widths of some mesons and baryons might indicate a ``fine
structure''\ of particle physics. There are several mesons (baryons) that
have the same intrinsic quantum numbers (I, S, C, b and Q), angular
momentums and parities but different rest masses. They are different mesons
(baryons). Since there are only five quarks in current Quark Model,
physicists think that there is only one meson (baryon) with large full
width. For example, f$_{0}$(600) might be q$_{N}$(313)$\overline{\text{q}_{N}%
\text{{\small (583)}}}$ + q$_{N}$(583)$\overline{\text{q}_{N}\text{{\small %
(583)}}}$ + q$_{\Sigma }$(583)$\overline{\text{{\small q}}_{\Sigma }\text{%
{\small (583)}}}$ + q$_{\Delta }$(673)$\overline{\text{{\small d}}_{\Delta }%
\text{{\small (673)}}}$ + q$_{\Xi }$(673)$\overline{\text{{\small d}}_{\Xi }%
\text{{\small (673)}}}$, (see Table C5), shown in Table 46 .

\ \ \ \ \ \ \ \ \ \ \ \ \ \ \ \ \ \ \ \ \ \ \ \ \ \ \ \ 

\begin{tabular}{l}
\ \ \ \ Table 46. \ The \textquotedblleft Fine Structure\textquotedblright\
of f$_{0}$(600) \cite{Meson04} \\ 
\begin{tabular}{|l|l|l|l|}
\hline
\begin{tabular}{l}
q$_{N}$(313)$\overline{\text{q}_{N}\text{{\small (583)}}}$ \\ 
q$_{N}$(583)$\overline{\text{q}_{N}\text{{\small (583)}}}$ \\ 
q$_{\Sigma }$(583)$\overline{\text{{\small q}}_{\Sigma }\text{{\small (583)}}%
}$ \\ 
q$_{\Delta }$(673)$\overline{\text{{\small d}}_{\Delta }\text{{\small (673)}}%
}$ \\ 
q$_{\Xi }$(673)$\overline{\text{{\small d}}_{\Xi }\text{{\small (673)}}}$%
\end{tabular}
& 
\begin{tabular}{l}
-320 \\ 
-528 \\ 
-676 \\ 
-938 \\ 
-538%
\end{tabular}
& 
\begin{tabular}{l}
$\eta $(576) \\ 
$\eta $(638) \\ 
$\eta $(676) \\ 
$\eta $(408) \\ 
$\eta \left( \text{{\small 808}}\right) $%
\end{tabular}
& 
\begin{tabular}{l}
f$_{0}$(600) \\ 
$\Gamma $= 600-1000%
\end{tabular}
\\ \hline
\end{tabular}%
\end{tabular}

\ \ \ \ \ \ \ \ \ \ \ \ \ \ \ \ \ \ \ \ \ \ \ \ \ \ \ 

Thus, we propose that experimental physicists separate both the mesons and
the baryons with large widths using experiments. First, we shall separate
the f$_{0}$(600)-meson (with $\Gamma $= 600-1000 Mev). The experimental
investigation of the predicted ``fine structure''\ for both mesons and
baryons with large widths provides a crucial test for our phenomenological
formulae.

\section{Discussion\ \ \ \ \ \ \ \ \ \ \ \ \ \ \ \ \ \ \ \ \ \ \ \ \ \ \ \ \
\ \ \ \ \ \ \ }

1. These phenomenological formulae not only decrease the number of
elementary quarks, but also increase the number of excited quarks that
compose new baryons and mesons. The decrease in the number of elementary
quarks will decrease the number of parameters and may simplify the
calculations. The increase in the number of excited quarks provides the
explanation for many newly discovered high-mass baryons and mesons.

2. Since individual free quarks cannot be found, we cannot directly measure
the binding energies of the baryons and mesons. Although we cannot measure
these binding energies, they do exist. The fact that physicists have not
found individual free quarks shows that the binding energies of the baryons
and the mesons are huge. The binding energies (-3$\Delta $) (\ref{Baryon M})
of the baryons [or mesons (-$2\Delta $) (\ref{M-Ebin})] are always canceled
by the corresponding parts (3$\Delta $) of three quarks [or\ two quarks (2$%
\Delta )$] (see Table 11). Thus, we can omit the binding energies [baryons
(-3$\Delta $) and mesons (-$2\Delta $)] and the corresponding part ($\Delta $%
) (see Table 11) of the rest masses of the quarks when we calculate the rest
masses of the baryons and the mesons. The effect looks as if there is not
binding energy in baryons and mesons.

3. The phenomenological formulae only apply to strong interaction static
properties of quarks, baryons and mesons. They do not apply to
electromagnetic and weak interactions. They cannot work for strong
interaction scattering processes.

4. These formulae do not break any principles of the Quark Model, such as
SU(n) symmetries, the sum laws, the``qqq''\ baryon model and the $``$q$%
\overline{q}"$ meson model. In fact, they preserve and improve these
principles. They show that all quarks are the excited state of the same
elementary quark $\epsilon $ to provide a physical foundation of SU(n)
symmetries. They deduce many new quarks to substantiate the``qqq''\ baryon
model and the ``q$\overline{\text{q}}$''\ meson model. They deduce the rest
masses and the intrinsic quantum numbers of quarks, baryons and mesons to
consolidate\ the Quark Model. They infer the large binding energies to
strengthen the foundation of the confinement of the quarks.

\section{Conclusions\ \ \ \ \ \ \ \ \ \ \ \ \ \ \ \ \ \ \ \ \ \ \ \ \ \ \ \
\ \ \ \ \ \ }

1. There is only one unflavored elementary quark family $\epsilon $ with
three colors that have two isospin states ($\epsilon _{u}$ with I$_{Z}$ = $%
\frac{1}{2}$ and Q = +$\frac{2}{3}$, $\epsilon _{d}$ with I$_{Z}$ = $\frac{-1%
}{2}$ and Q = -$\frac{1}{3}$) for each color. Thus there are six Fermi (s = $%
\frac{1}{2}$) elementary quarks with S = C = b = 0 in the vacuum.

2. When an elementary quark ($\epsilon $) with one color (red, yellow or
blue) and a charge Q (= $\frac{2}{3}$ or -$\frac{1}{3}$) is excited from the
vacuum, its color and electric charge will remain unchanged. At the same
time, it may\ go into an energy band of (\ref{E(n,k)}) to get the rest mass
and intrinsic quantum numbers (I, S, C, and b) and become an excited quark.

3. Using the phenomenological formulae, we deduced an excited quark spectrum
of the $\epsilon $-quarks (see Table 10 and Table 11) from the elementary $%
\epsilon $-quarks. Experiments have already provided evidence for almost all
of the deduced quarks inside the baryons (see Table 16-Table 21) and mesons
(see Table 25-Table 31). All quarks inside the baryons \cite{Baryon04} and
the mesons \cite{Meson04} are the excited states of the elementary $\epsilon 
$-quarks.

4. The five quarks of the current Quark Model correspond to the five deduced
ground quarks [ u$\leftrightarrow $u(313), d$\leftrightarrow $d(313), s$%
\leftrightarrow $d$_{s}$(493), c$\leftrightarrow $u$_{c}$(1653) and b$%
\leftrightarrow $d$_{b}$(4913)] (see Table 11). The current Quark Model uses
only the five independent ``elementary''\ quarks to explain the baryons and
mesons, there is no excited quark spectrum and no phenomenological formulae
in the model. Thus, the current Quark Model is the five ground quark
approximation of the new Quark Model with the phenomenological formulae.

5. With the phenomenological formulae, we deduce the rest masses and the
intrinsic quantum numbers (I, S, C, b and Q) of the quarks, baryons and
mesons. These deduced intrinsic quantum numbers match exactly with the
experimental results. The deduced rest masses are consistent with
experimental results at a 98\% level of confidence. Without the
phenomenological formulae, the current Quark Model cannot deduce the rest
masses of the quarks, baryons and mesons. With the formulae, the Quark Model
is much more powerful.

6. This paper has deduced the intrinsic quantum numbers of the quarks using
the following formulae: a). the strange numbers are from S = R - 4 (\ref%
{S-Number})\textbf{;} b). the isospins are from deg = 2I + 1 (\ref{IsoSpin}%
); c). the charmed numbers are from (\ref{Charmed}); d). the bottom numbers
are from (\ref{Battom}); e). the electric charges are from the elementary
quarks (\ref{2/3}) and (\ref{-1/3}). From the deduced intrinsic quantum
numbers of the quarks, using the sum laws, this paper has deduced the
intrinsic quantum numbers of the baryons and mesons.

7. Using the phenomenological mass formula (\ref{Rest Mass}), this paper has
deduced the rest masses of the quarks (see Table 11). From these deduced
rest masses of the quarks, using the sum laws and the phenomenological
binding energy formula (\ref{CB-Eb}) and (\ref{M-Ebin}), this paper has
deduced the baryon mass spectrum (see Table 16-Table 21) and the meson mass
spectrum (see Table 25-Table 31) respectively. The deduced rest masses of
the baryons and mesons are consistent with the experimental results (with a
98\% level of confidence).

8. The SU(3) (based on u(313), d(313) and d$_{S}$(493)), SU(4), SU(5), ...
SU(n), ... are natural extensions of the SU(2) (based on $\epsilon _{u}$ and 
$\epsilon _{d}$). These phenomenological formulae provide a physical
foundation for SU(3)... ,SU(n), ....

9. For baryons, the binding energy is roughly a constant (\ref{CB-Eb}). For
mesons, the binding energy is roughly a constant\ (\ref{M-Ebin}) too. These
binding energies are the phenomenological approximations of the color's
strong interaction energies. Although we do not know their exact values, we
believe that the binding energies of the baryons and mesons are very large.
These large binding energies provide a possible foundation for the
confinement of the quarks inside the hadrons.

10. The experimental investigation of the \textquotedblleft fine
structure\textquotedblright\ for both the mesons and baryons with large
widths (for example f$_{0}$(600) with $\Gamma $ = 600-1000 \cite{Meson04}),
provides a crucial test for our phenomenological formulae.

\begin{center}
\bigskip \textbf{Acknowledgments}
\end{center}

I sincerely thank Professor Robert L. Anderson for his valuable advice. I
acknowledge\textbf{\ }my indebtedness to Professor D. P. Landau for his help
also. I would like to express my heartfelt gratitude to Dr. Xin Yu for
checking the calculations. I sincerely thank Professor Yong-Shi Wu for his
important advice and help. I thank Professor Wei-Kun Ge for his support and
help. I sincerely thank Professor Kang-Jie Shi for his advice. There are
many friends who have already given advice and help to me; although I cannot
mention all of their names here, I really thank them very much.

\ \ \ \ 

\bigskip \newpage

{\Large Appendix A. The Regular Rhombic Dodecahedron and }$\overrightarrow{n}
${\Large \ Values\ }

There are many symmetry points ($\Gamma $, H, P, N, M, ...) and symmetry
axes ($\Delta $, $\Lambda $, $\Sigma $, D, F, G, ...) in the regular rhombic
dodecahedron (see Fig. 1).

We give a definition of the equivalent symmetry points: taking $%
\overrightarrow{n}$ (n$_{1}$, n$_{2}$, n$_{3}$) = 0 in (\ref{E(n,k)}), if
for different points $\overrightarrow{k}$ = ($\xi $, $\eta $, $\varsigma $),
we can get the same E($\vec{k}$,$\vec{n}$) value. We call these points the
\textquotedblleft equivalent\textquotedblright\ points. Table A1 shows the
equivalent points\ of the regular rhombic dodecahedron:

\ \ \ \ \ \ \ \ \ \ \ \ \ \ \ \ \ \ \ \ \ \ \ \ \ \ \ \ \ \ \ \ 

\begin{tabular}{l}
\ \ \ \ \ \ \ \ \ \ \ \ \ \ \ \ \ \ \ \ \ \ \ Table A1. The Equivalent Points
\\ 
$%
\begin{tabular}{|l|l|}
\hline
Sym. Point ($\xi $, $\eta $, $\varsigma $) & The Equivalent Points ($\xi $, $%
\eta $, $\varsigma $) \\ \hline
$\Gamma \text{ = (0, 0, 0)}$ & $\text{(0, 0, 0)}$ \\ \hline
P = ($\frac{1}{2}$, $\frac{1}{2}$, $\frac{1}{2}$) & ($\pm \frac{1}{2}$, $\pm 
\frac{1}{2}$, $\pm \frac{1}{2}$) \\ \hline
H = (0, 0, 1) & (0, 0, $\pm $1), (0, $\pm $1, 0), ($\pm $1, 0, 0) \\ \hline
N = ($\frac{1}{2}$, $\frac{1}{2}$, 0) & ($\pm \frac{1}{2}$, $\pm \frac{1}{2}$%
, 0), ($\pm \frac{1}{2}$, 0, $\pm \frac{1}{2}$), (0, $\pm \frac{1}{2}$, $\pm 
\frac{1}{2}$) \\ \hline
\end{tabular}
\ $%
\end{tabular}

\ \ \ \ \ \ \ \ \ \ \ \ \ \ \ \ \ \ \ \ \ \ \ \ \ \ \ \ \ \ \ \ \ \ \ \ \ 

Similarly, we can give the definition of the equivalent symmetry axes. Table
A2 shows the equivalent axes inside the regular rhombic dodecahedron:

\ \ \ \ \ \ \ \ \ \ \ \ \ \ \ \ \ \ \ \ \ \ \ \ \ \ \ \ \ \ \ \ 

$%
\begin{tabular}{l}
\ \ \ \ \ \ \ \ Table A2. \ The\ Equivalent Symmetry Axes \\ 
:$%
\begin{tabular}{|l|l|}
\hline
Symmetry Axis & Equivalent Symmetry Axes \\ \hline
$\Gamma \text{(000)-H(001)}$ & $\Gamma \text{-(0,}\pm \text{1,0), }\Gamma 
\text{-M(}\pm \text{1,0,0), }\Gamma \text{-H'(}0\text{,0,-1)}$ \\ \hline
$\Gamma \text{(000)-P(}\frac{1}{2}\frac{1}{2}\frac{1}{2}\text{)}$ & $\Gamma 
\text{-P'(}\pm \frac{1}{2},\pm \frac{1}{2},\pm \frac{1}{2}\text{)}$ \\ \hline
$\Gamma \text{(000)-N(}\frac{1}{2}\frac{1}{2}0\text{)}$ & $\Gamma \text{-N'(}%
\pm \frac{1}{2},\pm \frac{1}{2},0\text{), }\Gamma \text{-N'(}\pm \frac{1}{2}%
,0,\pm \frac{1}{2}\text{), }\Gamma \text{-N'(}0,\pm \frac{1}{2},\pm \frac{1}{%
2}\text{)}$ \\ \hline
\end{tabular}
$%
\end{tabular}
$

\ \ \ \ \ \ \ \ \ \ \ \ \ \ \ \ \ \ \ \ \ \ 

In order to distinguish the axes inside the regular rhombic dodecahedron
from the axes on its surfaces, we define an in-out number $\Theta $ (for $%
\Theta $ = 0, the axis is inside; for $\Theta $=1, the axis is on the
surface):

\begin{equation}
\Theta \text{ = }\overrightarrow{\text{k}_{Start}}\bullet \overrightarrow{%
\text{k}_{End}}\text{ = 2(}\xi _{Start}\times \xi _{End}\text{+}\eta
_{Start}\times \eta _{End}\text{+}\varsigma _{Start}\times \varsigma _{End}%
\text{).}  \label{f(se)}
\end{equation}%
For the axes inside it, $\Theta $ =0; for the axes on its surfaces, $\theta $
= 1. Table A3 shows the in-out number $\Theta $, the symmetry rotary fold R,
the strange number S, the symmetry operation P of the symmetry axes and the
first (second) division K$_{fir}$= 0 (K$_{Sec}$=1):

\ \ \ \ \ \ \ \ \ \ \ \ \ \ \ \ \ \ \ \ \ \ \ \ \ \ \ \ \ 

\begin{tabular}{l}
\ Table A3. The $\Theta ${\small , R, S and Symmetry Operation(p) Values} \\ 
\begin{tabular}{|l|l|l|l|l|l|l|l|l|}
\hline
{\small Axis} & {\small Start Point} & {\small End Point} & $\Theta $ & 
{\small R} & {\small S} & {\small P} & K$_{first}$ & K$_{Secend}$ \\ \hline
$\Delta ${\small (}$\Gamma ${\small -H)} & {\small (0, 0, 0)} & {\small (0,
0, 1)} & {\small 0} & {\small 4} & {\small 0} & {\small 8} & 0 & \ \ \ / \\ 
\hline
$\Lambda ${\small (}$\Gamma ${\small -P)} & {\small (0, 0, 0)} & {\small (}$%
\frac{1}{2}${\small ,}$\frac{1}{2}${\small ,}$\frac{1}{2}${\small )} & 
{\small 0} & {\small 3} & {\small -1} & {\small 6} & 0 & \ \ \ / \\ \hline
$\Sigma ${\small (}$\Gamma ${\small -N)} & {\small (0, 0, 0)} & {\small (}$%
\frac{1}{2}${\small ,}$\frac{1}{2}${\small ,0)} & {\small 0} & {\small 2} & 
{\small -2} & {\small 4} & 0 & \ \ \ / \\ \hline
{\small D(P-N)} & {\small (}$\frac{1}{2}${\small ,}$\frac{1}{2}${\small ,}$%
\frac{1}{2}${\small )} & {\small (}$\frac{1}{2}${\small ,}$\frac{1}{2}$%
{\small ,0)} & {\small 1} & {\small 2} & {\small 0} & {\small 4} & 0 & \ \ \
/ \\ \hline
{\small F(P-H)} & {\small (}$\frac{1}{2}${\small ,}$\frac{1}{2}${\small ,}$%
\frac{1}{2}${\small )} & {\small (0, 0, 1)} & {\small 1} & {\small 3} & 
{\small -1} & {\small 6} & 0 & \ \ \ 1$^{\#}$ \\ \hline
{\small G(M-N)} & {\small (1, 0, 0)} & {\small (}$\frac{1}{2}${\small ,}$%
\frac{1}{2}${\small ,0)} & {\small 1} & {\small 2} & {\small -2} & {\small 4}
& 0 & \ \ \ 1$^{\#}$ \\ \hline
\end{tabular}
\\ 
$^{\ \#}${\small K=1 corresponds to the second divisions in sixfold bands.}%
\end{tabular}

\ \ \ \ \ \ \ \ \ \ \ \ \ \ \ \ \ 

From (\ref{E(n,k)}) we can also give a definition of the equivalent $%
\overrightarrow{n}$: for $\xi $ = $\eta $ = $\varsigma $ = 0, all $%
\overrightarrow{n}$ values that yield a same E($\overrightarrow{n}$,0) value
are equivalent n-values. We show the low level equivalent $\overrightarrow{n}
$-values that satisfy conditions (\ref{l-n}) in the following list (\ref%
{Equ-N}) (note $\overline{\text{n}_{i}}$ = - n$_{i}$): \ \ \ \ \ \ \ \ \ \ \
\ \ \ \ 
\begin{equation}
\begin{tabular}{|l|}
\hline
{\small E(}$\overrightarrow{n}${\small ,0) = 0\ : (0, 0, 0)} \\ \hline
{\small E(}$\overrightarrow{n}${\small ,0) = 2\ :}$\ ${\small (101, }$%
\overline{\text{{\small 1}}}${\small 01, 011, 0}$\overline{\text{{\small 1}}}
${\small 1, 110, 1}$\overline{\text{{\small 1}}}${\small 0, }$\overline{%
\text{{\small 1}}}${\small 10, }$\overline{\text{{\small 1}}}\overline{\text{%
{\small 1}}}${\small 0, 10}$\overline{\text{{\small 1}}}${\small , }$%
\overline{\text{{\small 1}}}${\small 0}$\overline{\text{{\small 1}}}${\small %
, 01}$\overline{\text{{\small 1}}}${\small , 0}$\overline{\text{{\small 1}}}%
\overline{\text{{\small 1}}}${\small )} \\ \hline
{\small E(}$\overrightarrow{n}${\small ,0) = 4\ :\ \ (002, 200}, {\small 200}%
$\text{, }\overline{\text{{\small 2}}}\text{{\small 00}, }${\small 0}$%
\overline{\text{{\small 2}}}${\small 0, 00}$\overline{\text{{\small 2}}}$%
{\small )} \\ \hline
{\small E(}$\overrightarrow{n}${\small ,0) = 6:} 
\begin{tabular}{l}
{\small 112, 211, 121, }$\overline{\text{{\small 1}}}${\small 21,}$\overline{%
\text{{\small 1}}}${\small 12, 2}$\overline{\text{{\small 1}}}${\small 1}, 
{\small 1}$\overline{\text{{\small 1}}}${\small 2, 21}$\overline{\text{%
{\small 1}}}${\small ,12}$\overline{\text{{\small 1}}}${\small ,}$\overline{%
\text{{\small 2}}}${\small 11, 1}$\overline{\text{{\small 2}}}${\small 1, 11}%
$\overline{\text{{\small 2}}}$,{\small \ } \\ 
$\overline{\text{{\small 11}}}${\small 2, }$\overline{\text{{\small 1}}}$%
{\small 2}$\overline{\text{{\small 1}}}$,{\small \ 2}$\overline{\text{%
{\small 11}}}${\small , }$\overline{\text{{\small 21}}}${\small 1}, $%
\overline{\text{{\small 12}}}${\small 1, 1}$\overline{\text{{\small 12}}}$%
{\small , 1}$\overline{\text{{\small 21}}}${\small ,}$\overline{\text{%
{\small 1}}}${\small 1}$\overline{\text{{\small 2}}}${\small ,}$\overline{%
\text{{\small 2}}}$1$\overline{\text{{\small 1}}}$,{\small \ }$\overline{%
\text{{\small 211}}}${\small , }$\overline{\text{{\small 121}}},${\small \ }$%
\overline{\text{{\small 112}}},$%
\end{tabular}
\\ \hline
{\small E(}$\overrightarrow{n}${\small ,0) = 8:}$\ ${\small (220, 2}$%
\overline{\text{{\small 2}}}${\small 0, }$\overline{\text{{\small 2}}}$%
{\small 20, }$\overline{\text{{\small 2}}}\overline{\text{{\small 2}}}$%
{\small 0, 202, 20}$\overline{\text{{\small 2}}}${\small , }$\overline{\text{%
{\small 2}}}${\small 02, }$\overline{\text{{\small 2}}}${\small 0}$\overline{%
\text{{\small 2}}}${\small , 022, 02}$\overline{\text{{\small 2}}}${\small ,
0}$\overline{\text{{\small 2}}}${\small 2, 0}$\overline{\text{{\small 2}}}%
\overline{\text{{\small 2}}}${\small )} \\ \hline
\end{tabular}
\label{Equ-N}
\end{equation}

\ \ \ \ \ \ \ \ \ \ \ \ \ \ \ \ \ \ \ \ \ \ \ \ \ \ \ \ \ \ \ \ \ \ \ \ \ \
\ \ \ \ \ \ \ \ \ \ \ \ \ \ \ \ \ \ \ \ \ \ \ 

\begin{tabular}{l}
Table A4. \ The Energy Bands with $\Delta $S = 0 \\ 
\begin{tabular}{|l|l|l|l|l|l|l|l|}
\hline
Axis & $\Delta $ & $\Sigma $ & $\Lambda $ & $\Lambda $ & D & F & G \\ \hline
R & 4 & 2 & 3 & 3 & 2 & 3 & 2 \\ \hline
S$_{Ax}$ & 0 & -2 & -1 & -1 & 0 & -1 & -2 \\ \hline
deg & 4 & 2 & 3 & 1 & 2 & 1 & 2 \\ \hline
$\Delta S$ & 0 & 0 & 0 & 0 & 0 & 0 & 0 \\ \hline
q$_{Name}$ & q$_{\Delta }$ & q$_{\Xi }$ & q$_{\Sigma }$ & d$_{S}$ & q$_{%
\text{N}}$ & d$_{S}$ & q$_{\Xi }$ \\ \hline
\end{tabular}%
\end{tabular}
\ \ \ \ \ \ \ \ 

\ \ \ \ \ \ \ \ \ \ \ \ \ \ \ \ \ \ \ \ \ \ \ \ \ 

\ \ \ \ \ \ \ \ \ \ \ \ \ \ \ \ \ \ \ \ \ \ \ \ \ \ \ \ \ \ \ \ \ \ \ \ \ \
\ \ \ \ \ \ \ \ \ \ \ \ \ \ \ \ \ \ \ \ \ \ \ \ \ \ \ 

\ 
\begin{tabular}{l}
\ \ \ Table A5. The Energy Bands with the Fluctuation $\Delta $S$\neq $0 \\ 
\begin{tabular}{|l|l|l|l|l|l|l|l|l|l|l|l|l|}
\hline
Axis & $\Delta $ & $\Delta $ & $\Sigma $ & $\Sigma $ & $\Sigma $ & D & D & F
& F & F & G & G \\ \hline
R & 4 & 4 & 2 & 2 & 2 & 2 & 2 & 3 & 3 & 3 & 2 & 2 \\ \hline
S$_{Ax}$ & 0 & 0 & -2 & -2 & -2 & 0 & 0 & -1 & -1 & -1 & -2 & -2 \\ \hline
deg & 1 & 1 & 1 & 1 & 1 & 1 & 1 & 2 & 2 & 2 & 1 & 1 \\ \hline
$\Delta S$ & 1 & -1 & 1 & 1 & -1 & 1 & -1 & 1 & 1 & -1 & 1 & -1 \\ \hline
S & 0 & -1 & -1 & 0 & -3 & 0 & -1 & 0 & -1 & -2 & -2 & -3 \\ \hline
C & 1 & 0 & 0 & 0 & 0 & 1 & 0 & 0 & 1 & 0 & 1 & 0 \\ \hline
b & 0 & 0 & 0 & -1 & 0 & 0 & 0 & 0 & 0 & 0 & 0 & 0 \\ \hline
quark & u$_{C}$ & d$_{S}$ & d$_{S}^{\ast }$ & d$_{b}^{\#}$ & d$_{\Omega }$ & 
u$_{C}$ & d$_{S}$ & q$_{N}$ & q$_{\Xi _{C}}$ & q$_{\Xi }$ & d$_{\Omega _{C}}$
& d$_{\Omega }$ \\ \hline
\end{tabular}
\\ 
$\ \ \ ^{\ast }$If $\Delta $E = 0, the quark is d$_{S}$; $^{\#}$If $\Delta $%
E $\neq $ 0, the quark is d$_{b}$.%
\end{tabular}

\ \ \ \ \ \ \ \ \ \ \ \ \ \ \ \ \ \ \ \ \ \ \ \ \ \ \ \ \ \ \ \ \ \ \ \ \ \
\ \ \ \ \ \ \ \ \ \ \ \ \ \ \ \ \ \ \ \ \ \ \ \ \ \ \ 

\ \ \ \ \ \ \ \ \ \ \ \ \ \ \ \ \ \ \ \ \ \ \ \ \ \ \ \ \ \ \ \ \ 

\begin{tabular}{l}
\ A6 The Possible Maximum Isospin of baryons at symmetry points \\ 
\begin{tabular}{|l|l|l|l|l|l|l|l|l|l|l|l|l|}
\hline
Symmetry Axis & $\Delta $ & $\Delta $ & $\Sigma $ & $\Sigma $ & $\Lambda $ & 
$\Lambda $ & D & D & F & F & G & G \\ \hline
Symmetry Point & $\Gamma $ & H & $\Gamma $ & N & $\Gamma $ & P & P & N & P & 
H & M & N \\ \hline
Symmetry rotatory R & 4 & 4 & 2 & 2 & 3 & 3 & 2 & 2 & 3 & 3 & 2 & 2 \\ \hline
Maximum Equ-n Value & 8 & 8 & 4 & 4 & 6 & 6 & 2 & 2 & 2 & 2 & 2 & 2 \\ \hline
Maximum I of Quarks & $\frac{3}{2}$ & $\frac{3}{2}$ & $\frac{1}{2}$ & $\frac{%
1}{2}$ & 1 & 1 & $\frac{1}{2}$ & $\frac{1}{2}$ & $\frac{1}{2}$ & $\frac{1}{2}
$ & $\frac{1}{2}$ & $\frac{1}{2}$ \\ \hline
Possibility Large I$_{Baryon}$ & $\frac{3}{2}$ & $\frac{3}{2}$ & $\frac{1}{2}
$ & $\frac{1}{2}$ & 1 & 1 & $\frac{3}{2}$ & $\frac{1}{2}$ & $\frac{3}{2}$ & $%
\frac{3}{2}$ & $\frac{1}{2}$ & $\frac{1}{2}$ \\ \hline
\end{tabular}%
\end{tabular}

\ \ \ \ \ \ \ \ \ \ \ \ \ \ \ \ \ \ \ \ \ \ \ \ \ \ \ \ \ \ \ \ \ \ \ \ \ \
\ \ \ \ \ \ \ \ \ \ \ \ \ \ \ \ \ \ \ 

\newpage

\ \ \ \ \ \ \ \ \ \ \ \ \ \ 

{\LARGE Appendix B. Energy Bands and Corresponding Quarks}

From Table 2, Table 3 and Table 4, omitting $\Delta $ part mass of quarks,
we have the quarks (q$_{\Delta }^{0}$,\ d$_{S}^{\text{-1}}$ and u$_{C}^{1}$)
shown in Table B1 (note $\overline{\text{n}_{i}}$ = - n$_{i}$):\ \ \ \ \ \ \
\ \ \ 

\ \ \ \ \ \ \ \ \ \ \ Table B1. The Energy Bands and the Quarks of the $%
\Delta $-Axis (S = 0)

$%
\begin{tabular}{|l|l|l|l|l|l|l|l|l|l|}
\hline
$\text{E}_{Start}$ & {\small E(}$\vec{k}${\small ,}$\vec{n}${\small )} & 
{\small \ \ d} & $\Delta $S & J & $\text{(n}_{1}\text{n}_{2}\text{n}_{3}$%
{\small , }$\text{...); }$ J & {\small S} & {\small C} & $\Delta $E & 
{\small \ }q(m{\small \ }$_{\text{{\small (Mev)}}}$) \\ \hline
$\text{E}_{\Gamma }\text{=0}$ & {\small 313} & {\small \ \ 1} & 0 & 0 & 
{\small (000): } $\text{J}_{\Gamma }\text{ = 0; }$ & {\small 0} & {\small 0}
& {\small 0} & {\small \ \ u(313)} \\ \hline
$\text{E}_{H}\text{=1}$ & {\small 673} & {\small \ \ 4} & 0 & 0 & $\text{%
(101,}\overline{\text{1}}\text{01,011,0}\overline{\text{1}}\text{1)}$ & 
{\small 0} & {\small 0} & {\small 0} & {\small \ \ q}$_{\Delta }${\small %
(673)} \\ \hline
$\text{E}_{H}\text{=1}$ & {\small 673} & {\small \ \ 1} & -1 & 1 & {\small %
(002); }$\text{ }$ $\text{J}_{\text{S,H}}\text{ = 1}$ & {\small -1} & 
{\small 0} & {\small 100} & {\small \ \ d}$_{S}^{-1}${\small (773)} \\ \hline
$\text{E}_{\Gamma }\text{=2}$ & {\small 1033} & {\small \ \ 4} & 0 & 0 & $%
\text{(110,1}\overline{\text{1}}\text{0,}\overline{\text{1}}\text{10,}%
\overline{\text{1}}\overline{\text{1}}\text{0)}$ & {\small 0} & {\small 0} & 
{\small 0} & {\small \ \ q}$_{\Delta }${\small (1033)} \\ \hline
$\text{E}_{\Gamma }\text{=2}$ & {\small 1033} & {\small \ \ 4} & 0 & 0 & $%
\text{(10}\overline{\text{1}}\text{,}\overline{\text{1}}\text{0}\overline{%
\text{1}}\text{,01}\overline{\text{1}}\text{,0}\overline{\text{11}}\text{)}$
& {\small 0} & {\small 0} & {\small 0} & {\small \ \ q}$_{\Delta }${\small %
(1033)} \\ \hline
$\text{E}_{H}\text{=3}$ & {\small 1393} & {\small \ \ 4} & 0 & 0 & $\text{%
(112,1}\overline{\text{1}}\text{2,}\overline{\text{1}}\text{12,}\overline{%
\text{1}}\overline{\text{1}}\text{2)}$ & {\small 0} & {\small 0} & {\small 0}
& {\small \ \ q}$_{\Delta }${\small (1393)} \\ \hline
$\text{E}_{\Gamma }\text{=4}$ & {\small 1753} & {\small \ \ 4} & 0 & 0 & $%
\text{(200,}\overline{\text{2}}\text{00,020,0}\overline{\text{2}}\text{0)}$
& {\small 0} & {\small 0} & {\small 0} & {\small \ \ q}$_{\Delta }${\small %
(1753)} \\ \hline
$\text{E}_{\Gamma }\text{=4}$ & {\small 1753} & {\small \ \ 1} & 1 & 1 & 
{\small (00}$\overline{2}${\small ); }$\text{J}_{\text{C,}\Gamma }\text{ = 1}
$ & {\small 0} & {\small 1} & {\small 0} & {\small \ \ u}$_{C}^{1}${\small %
(1753)} \\ \hline
$\text{E}_{H}\text{=5}$ & {\small 2113} & 
\begin{tabular}{l}
{\small 4} \\ 
{\small 4}%
\end{tabular}
& 0 & 0 & $%
\begin{array}{l}
\text{(121,1}\overline{\text{2}}\text{1,}\overline{\text{1}}\text{21,}%
\overline{\text{1}}\overline{\text{2}}\text{1)} \\ 
\text{(211,2}\overline{\text{1}}\text{1,}\overline{\text{2}}\text{11,}%
\overline{\text{21}}\text{1)}%
\end{array}
$ & {\small 0} & {\small 0} & {\small 0} & 
\begin{tabular}{l}
{\small q}$_{\Delta }${\small (2113)} \\ 
{\small q}$_{\Delta }${\small (2113)}%
\end{tabular}
\\ \hline
$\text{E}_{H}\text{=5}$ & {\small 2113} & {\small \ \ 4} & 0 & 0 & {\small %
(202,}$\overline{\text{2}}${\small 02,022,0}$\overline{\text{2}}${\small 2)}
& {\small 0} & {\small 0} & {\small 0} & {\small \ q}$_{\Delta }${\small %
(2113)} \\ \hline
$\text{E}_{H}\text{=5}$ & {\small 2113} & {\small \ \ 4} & 0 & 0 & {\small %
(013,0}$\overline{\text{1}}${\small 3,103,}$\overline{\text{1}}${\small 03)}
& {\small 0} & {\small 0} & {\small 0} & {\small \ q}$_{\Delta }${\small %
(2113)} \\ \hline
$\text{E}_{\Gamma }\text{=6}$ & {\small 2473} & 
\begin{tabular}{l}
{\small 4} \\ 
{\small 4}%
\end{tabular}
& 0 & 0 & $%
\begin{array}{l}
\text{(12}\overline{\text{{\small 1}}}\text{,1}\overline{\text{{\small 21}}}%
\text{,}\overline{\text{{\small 1}}}\text{2}\overline{\text{{\small 1}}}%
\text{,}\overline{\text{{\small 121}}}\text{)} \\ 
\text{(21}\overline{\text{{\small 1}}}\text{,2}\overline{\text{{\small 11}}}%
\text{,}\overline{\text{{\small 2}}}\text{1}\overline{\text{{\small 1}}}%
\text{,}\overline{\text{{\small 211}}}\text{)}%
\end{array}
$ & {\small 0} & {\small 0} & {\small 0} & 
\begin{tabular}{l}
{\small q}$_{\Delta }${\small (2473)} \\ 
{\small q}$_{\Delta }${\small (2473)}%
\end{tabular}
\\ \hline
$\text{E}_{\Gamma }\text{=6}$ & {\small 2473} & {\small \ 4} & 0 & 0 & 
{\small (11}$\overline{\text{2}}${\small ,1}$\overline{\text{12}}${\small ,}$%
\overline{\text{1}}${\small 1}$\overline{\text{2}}${\small ,}$\overline{%
\text{112}}${\small )} & {\small 0} & {\small 0} & {\small 0} & {\small \ q}$%
_{\Delta }${\small (2473)} \\ \hline
$\text{E}_{H}\text{=9}$ & {\small 3553} & {\small \ \ 1} & -1 & 2 & {\small %
(004); }$\text{ }$ $\text{J}_{\text{S,H}}\text{ = 2}$ & {\small -1} & 
{\small 0} & {\small 200} & {\small \ d}$_{S}^{-1}${\small (3753)} \\ \hline
$\text{E}_{\Gamma }\text{=16}$ & {\small 6073} & {\small \ \ 1} & 1 & 2 & 
{\small (00}$\overline{\text{{\small 4}}}${\small ); } $\text{J}_{\text{C,}%
\Gamma }\text{ = 2}$ & {\small 0} & {\small 1} & {\small 0} & {\small \ u}$%
_{C}^{1}${\small (6073)} \\ \hline
$\text{E}_{H}\text{=25}$ & {\small 9313} & {\small \ \ 1} & -1 & 3 & {\small %
(}$\text{006); }$ $\text{J}_{\text{S,H}}\text{ = 3}$ & {\small -1} & {\small %
0} & {\small 300} & {\small \ d}$_{S}^{-1}${\small (9613)} \\ \hline
$\text{E}_{\Gamma }\text{=36}$ & {\small 13273} & {\small \ \ 1} & 1 & 3 & 
{\small (00}$\overline{\text{{\small 6}}}${\small ); } $\text{J}_{\text{C,}%
\Gamma }\text{ = 3}$ & {\small 0} & {\small 1} & {\small 0} & {\small \ \ u}$%
_{C}^{1}${\small (13273)} \\ \hline
$\text{E}_{H}\text{=49}$ & {\small 17953} & {\small \ \ 1} & -1 & 4 & 
{\small (}$\text{008); }\ \text{J}_{\text{S,H}}\text{ = 4}$ & {\small -1} & 
{\small 0} & {\small 400} & {\small \ d}$_{S}^{-1}${\small (18353)} \\ \hline
... & ... & ... & ... & ... & ... & ... & ... & ... & .... \\ \hline
\end{tabular}
\ \ \ \ $

E$_{\text{Start}}$ = [(n$_{1}$-$\xi $)$^{2}$+(n$_{2}$-$\eta $)$^{2}$+(n$_{3}$%
-$\zeta $)$^{2}$], {\small E(}$\vec{k}${\small ,}$\vec{n}${\small )} is the
starting and minimum energy of the energy band; $\overline{\text{n}_{i}}$ =
-n$_{i}$; $\overline{\text{n}_{1}\text{n}_{2}\text{n}_{3}}$ = -n$_{1}$, -n$%
_{2}$, -n$_{3}$ [see Eq. (\ref{Equ-N})].\ 

\newpage

From Table 2, Table 5, Table 6 and Table 7, omitting $\Delta $ part mass of
quarks, we have the quarks (q$_{\Xi }^{0}$,\ d$_{S}^{1}$ , d$_{b}^{1}$ and 
{\small d}$_{\Omega }^{-1}$) shown in Table B2 (note $\overline{\text{n}_{i}}
$ = - n$_{i}$):\ \ \ \ \ \ \ \ \ \ \ \ \ \ \ \ \ \ \ \ \ \ \ 

\qquad\ \ \ \ \ 

\ \ \ \ \ \ \ \ \ Table B2. The Energy Bands and the Quarks of the $\Sigma $%
-Axis (S = -2)

\begin{tabular}{|l|l|l|l|l|l|l|l|l|l|}
\hline
$\text{E}_{\text{Start}}$ & {\small E(}$\vec{k}${\small ,}$\vec{n}${\small )}
& {\small \ \ d} & $\Delta ${\small S} & J & $\text{(n}_{1}\text{n}_{2}\text{%
n}_{3}$); J & {\small S} & {\small b} & $\Delta $E & q(m{\small \ }$_{\text{%
{\small (Mev)}}}$) \\ \hline
$\text{E}_{\Gamma }\text{= 0}$ & {\small 313} & {\small \ \ 1} & -2 & 0 & $%
\text{(000); }${\small \ }$\text{J}_{\Gamma }${\small =0} & {\small 0} & 
{\small 0} & {\small \ \ \ 0} & {\small \ \ d(313)} \\ \hline
$\text{E}_{N}\text{=1/2}$ & {\small 493} & {\small \ \ 2} & 1 & 1 & ($\text{%
110}${\small ); }$\text{J}_{\text{N}}${\small =1;} & {\small -1} & {\small 0}
& {\small \ \ \ 0} & {\small \ \ d}$_{S}^{1}${\small (493)} \\ \hline
$\text{E}_{N}\text{=3/2}$ & {\small 853} & 
\begin{tabular}{l}
{\small 2} \\ 
{\small 2}%
\end{tabular}
& \ \ 0 & 0 & $%
\begin{tabular}{l}
($\text{101,10}\overline{\text{1}}$) \\ 
($\text{011,01}\overline{\text{1}}$)%
\end{tabular}%
\ \ \ \ \ \ \ $ & {\small -2} & {\small 0} & {\small \ \ \ 0} & 
\begin{tabular}{l}
{\small q}$_{\Xi }^{0}${\small (853)} \\ 
{\small q}$_{\Xi }^{0}${\small (853)}%
\end{tabular}
\\ \hline
$\text{E}_{\Gamma }\text{= 2}$ & {\small 1033} & {\small \ \ 2} & \ \ 0 & 0
& {\small (}$\text{1}\overline{\text{1}}\text{0,}\overline{\text{1}}\text{10)%
}$ & {\small -2} & {\small 0} & {\small \ \ \ 0} & {\small \ \ q}$_{\Xi }^{0}
${\small (1033)} \\ \hline
$\text{E}_{\Gamma }\text{= 2}$ & {\small 1033} & 
\begin{tabular}{l}
{\small 2} \\ 
{\small 2}%
\end{tabular}
& \ \ 0 & 0 & 
\begin{tabular}{l}
{\small (}$\overline{\text{1}}${\small 01,}$\overline{\text{1}}${\small 0}$%
\overline{\text{1}}${\small )} \\ 
{\small (0}$\overline{\text{1}}${\small 1,0}$\overline{\text{1}}\overline{%
\text{1}}${\small )}%
\end{tabular}
& {\small -2} & {\small 0} & {\small \ \ \ 0} & 
\begin{tabular}{l}
{\small q}$_{\Xi }^{0}${\small (1033)} \\ 
{\small q}$_{\Xi }^{0}${\small (1033)}%
\end{tabular}
\\ \hline
$\text{E}_{\Gamma }\text{= 2}$ & {\small 1033} & {\small \ \ 2} & -1 & 1 & 
{\small (}$\overline{\text{{\small 11}}}${\small 0); \ }$\text{J}_{\Gamma }$%
=1 & {\small -3} & {\small 0} & {\small \ \ \ 0} & {\small \ \ d}$_{\Omega
}^{-1}${\small (1033)} \\ \hline
$\text{E}_{N}\text{=5/2}$ & {\small 1213} & {\small \ \ 2} & \ \ 0 & 0 & $%
\text{(200,020)}$ & {\small -2} & {\small 0} & {\small \ \ \ 0} & {\small \
\ q}$_{\Xi }^{0}${\small (1213)} \\ \hline
$\text{E}_{N}\text{=7/2}$ & {\small 1573} & 
\begin{tabular}{l}
{\small 2} \\ 
{\small 2}%
\end{tabular}
& 
\begin{tabular}{l}
{\small 0} \\ 
{\small 0}%
\end{tabular}
& 0 & 
\begin{tabular}{l}
{\small (121,12}$\overline{\text{{\small 1}}}${\small )} \\ 
{\small (211,21}$\overline{\text{{\small 1}}}${\small )}%
\end{tabular}
& {\small -2} & {\small 0} & {\small \ \ \ 0} & 
\begin{tabular}{l}
{\small q}$_{\Xi }^{0}${\small (1573)} \\ 
{\small q}$_{\Xi }^{0}${\small (1573)}%
\end{tabular}
\\ \hline
$\text{E}_{\Gamma }\text{= 4}$ & {\small 1753} & {\small \ \ 2} & \ \ 0 & 0
& $\text{(002,00}\overline{\text{2}}\text{)}$ & {\small -2} & {\small 0} & 
{\small \ \ \ 0} & {\small \ \ q}$_{\Xi }^{0}${\small (1753)} \\ \hline
$\text{E}_{\Gamma }\text{= 4}$ & {\small 1753} & {\small \ \ 2} & \ \ 0 & 0
& $\text{(}\overline{\text{2}}\text{00,0}\overline{\text{2}}\text{0)}$ & 
{\small -2} & {\small 0} & {\small \ \ \ 0} & {\small \ \ q}$_{\Xi }^{0}$%
{\small (1753)} \\ \hline
$\text{E}_{N}\text{=9/2}$ & {\small 1933} & {\small \ \ 2} & \ \ 0 & 0 & $%
\text{(112,11}\overline{\text{2}}\text{)}$ & {\small -2} & {\small 0} & 
{\small \ \ \ 0} & {\small \ \ q}$_{\Xi }^{0}${\small (1933)} \\ \hline
$\text{E}_{N}\text{=9/2}$ & {\small 1933} & {\small \ \ 1} & \ 1 & 2 & 
{\small (220); }$\text{J}_{\text{N}}$=2 & {\small -1} & {\small 0} & {\small %
\ \ \ 0} & {\small \ \ d}$_{S}^{1}${\small (1933)} \\ \hline
$\text{E}_{N}\text{=}\frac{\text{11}}{2}$ & {\small 2293} & \ \ {\small 1} & 
\ \ 0 & 0 & 
\begin{tabular}{l}
{\small (}$\overline{\text{{\small 1}}}${\small 21,}$\overline{\text{{\small %
1}}}${\small 2}$\overline{\text{{\small 1}}}${\small ,} \\ 
{\small 2}$\overline{\text{{\small 1}}}${\small 1,2}$\overline{\text{{\small %
11}}})$%
\end{tabular}
& {\small -2} & {\small 0} & {\small \ \ \ 0} & 
\begin{tabular}{l}
{\small q}$_{\Xi }^{0}${\small (2293)} \\ 
{\small q}$_{\Xi }^{0}${\small (2293)}%
\end{tabular}
\\ \hline
$\text{E}_{\Gamma }\text{= 8}$ & {\small 3193} & {\small \ \ 1} & -1 & 2 & 
{\small (}$\overline{\text{{\small 22}}}${\small 0); }$\text{J}_{\Gamma }$%
{\small =2; } & {\small -3} & {\small 0} & {\small \ \ \ 0} & {\small \ \ d}$%
_{\Omega }^{-1}${\small (3193)} \\ \hline
$\text{E}_{N}\text{=25/2}$ & {\small 4813} & {\small \ \ 1} & \ 1 & 3 & 
{\small (330); }$\text{J}_{\text{N}}${\small =3;} & {\small 0} & {\small -1}
& {\small 100} & {\small \ \ d}$_{b}^{1}${\small (4913)} \\ \hline
$\text{E}_{\Gamma }\text{= 18}$ & {\small 6793} & {\small \ \ 1} & -1 & 3 & 
{\small (}$\overline{\text{{\small 33}}}${\small 0); }$\text{J}_{\Gamma }$%
{\small =3;} & {\small -3} & {\small 0} & {\small -300} & {\small \ \ d}$%
_{\Omega }^{-1}${\small (6493)} \\ \hline
$\text{E}_{N}\text{=49/2}$ & {\small 9133} & {\small \ \ 1} & \ 1 & 4 & 
{\small (440); \ }$\text{J}_{\text{N}}${\small =4;} & {\small 0} & {\small -1%
} & {\small 200} & {\small \ \ d}$_{b}^{1}${\small (9333)} \\ \hline
$\text{E}_{\Gamma }\text{= 32}$ & {\small 11833} & {\small \ \ 1} & -1 & 4 & 
{\small (}$\overline{\text{{\small 44}}}${\small 0); }$\text{J}_{\Gamma }$%
{\small =4;} & {\small -3} & {\small 0} & {\small -600} & {\small \ \ d}$%
_{\Omega }^{-1}${\small (11233)} \\ \hline
$\text{E}_{N}\text{=81/2}$ & {\small 14893} & {\small \ \ 1} & \ 1 & 5 & 
{\small (550); \ }$\text{J}_{\text{N}}${\small =5; } & {\small 0} & {\small %
-1} & {\small 300} & {\small \ \ d}$_{b}^{1}${\small (15193)} \\ \hline
{\small ...} & {\small ...} & {\small ...} & ... & ... & {\small ...} & ...
& ... & ... & .... \\ \hline
\end{tabular}

E$_{\text{Start}}$ = [(n$_{1}$-$\xi $)$^{2}$+(n$_{2}$-$\eta $)$^{2}$+(n$_{3}$%
-$\zeta $)$^{2}$], {\small E(}$\vec{k}${\small ,}$\vec{n}${\small )} is the
starting and minimum energy of the energy band; $\overline{\text{n}_{i}}$ =
-n; $\overline{\text{n}_{1}\text{n}_{2}\text{n}_{3}}$ = -n$_{1}$, -n$_{2}$,
-n$_{3}$ [see Eq. (\ref{Equ-N})].

From Table 2, Table 8 and Table 9, omitting $\Delta $ part mass of quarks,
we have the quarks (q$_{\Sigma }^{0}$ and\ d$_{S}^{0}$)\ shown in Table B3
(note $\overline{\text{n}_{i}}$ = - n$_{i}$);\qquad\ \ \ \qquad \qquad

\ \ \ \ \ \ Table B3. The Energy Bands and the Quarks of the $\Lambda $-Axis
(S = -1)

\begin{tabular}{|l|l|l|l|l|l|l|l|}
\hline
$\text{E}_{\text{Start}}$ & {\small E(}$\vec{k}${\small ,}$\vec{n}${\small )}
& \ \ d & $\Delta $S & J & $\text{ (n}_{1}\text{n}_{2}\text{n}_{3}\text{,
...)}$ & S & q(m{\small \ }$_{\text{{\small (Mev)}}}$) \\ \hline
$\text{E}_{\Gamma }\text{=0}$ & {\small 313} & {\small \ \ 1} & -1 & 0 & $\ 
\text{(000) }$ & {\small \ 0} & {\small \ \ d(313)} \\ \hline
$\text{E}_{\text{P}}\text{=3/4 \ }$ & {\small 583} & {\small \ \ 3} & 0 & 0
& $\ \text{(101,011,110)}$ & {\small -1} & {\small \ \ q}$_{\Sigma }^{0}$%
{\small (583)} \\ \hline
$\text{E}_{\Gamma }\text{= 2 \ \ }$ & {\small 1033} & 
\begin{tabular}{l}
{\small 3} \\ 
{\small 3}%
\end{tabular}
& 0 & 0 & $\text{%
\begin{tabular}{l}
$\text{(1}\overline{\text{1}}\text{0,}\overline{\text{1}}\text{10,01}%
\overline{\text{1}}$) \\ 
($\text{0}\overline{\text{1}}\text{1,10}\overline{\text{1}}\text{,}\overline{%
\text{1}}\text{01)}$%
\end{tabular}
}$ & {\small -1} & 
\begin{tabular}{l}
{\small q}$_{\Sigma }^{0}${\small (1033)} \\ 
{\small q}$_{\Sigma }^{0}${\small (1033)}%
\end{tabular}
\\ \hline
$\text{E}_{\Gamma }\text{= 2 \ \ }$ & {\small 1033} & {\small \ \ 3} & 0 & 0
& $\ \text{(}\overline{\text{1}}\text{0}\overline{\text{1}}\text{,0}%
\overline{\text{1}}\overline{\text{1}}\text{,}\overline{\text{1}}\overline{%
\text{1}}\text{0)}$ & {\small -1} & {\small \ \ q}$_{\Sigma }^{0}${\small %
(1033)} \\ \hline
$\text{E}_{\text{P}}\text{=11/4}$ & {\small 1303} & {\small \ \ 3} & 0 & 0 & 
$\ \text{(020,002,200)}$ & {\small -1} & {\small \ \ q}$_{\Sigma }^{0}$%
{\small (1303)} \\ \hline
$\text{E}_{\text{P}}\text{=11/4}$ & {\small 1303} & {\small \ \ 3} & 0 & 0 & 
$\ \text{(121,211,112)}$ & {\small -1} & {\small \ \ q}$_{\Sigma }^{0}$%
{\small (1303)} \\ \hline
$\text{E}_{\Gamma }\text{= 4 \ \ }$ & {\small 1753} & {\small \ \ 3} & 0 & 0
& \ {\small (0}$\overline{2}${\small 0,}$\overline{2}${\small 00,00}$%
\overline{2}${\small )} & {\small -1} & {\small \ \ q}$_{\Sigma }^{0}$%
{\small (1753)} \\ \hline
$\text{E}_{\text{P}}\text{=19/4}$ & {\small 2023} & 
\begin{tabular}{l}
{\small 3} \\ 
{\small 3}%
\end{tabular}
& 0 & 0 & $\text{%
\begin{tabular}{l}
$\text{(1}\overline{\text{1}}\text{2,}\overline{\text{1}}\text{12,21}%
\overline{\text{1}}$) \\ 
($\text{2}\overline{\text{1}}\text{1,12}\overline{\text{1}}\text{,}\overline{%
\text{1}}\text{21)}$%
\end{tabular}
}$ & {\small -1} & 
\begin{tabular}{l}
{\small q}$_{\Sigma }^{0}${\small (2023)} \\ 
{\small q}$_{\Sigma }^{0}${\small (2023)}%
\end{tabular}
\\ \hline
$\text{E}_{\text{P}}\text{=19/4}$ & {\small 2023} & \ {\small 3} & 0 & 0 & $%
\ \text{(202,022,220)}$ & {\small -1} & {\small \ q}$_{\Sigma }^{0}${\small %
(2023)} \\ \hline
$\text{E}_{\Gamma }\text{= 6 \ \ }$ & {\small 2473} & 
\begin{tabular}{l}
{\small 3} \\ 
{\small 3}%
\end{tabular}
& 0 & 0 & $\text{%
\begin{tabular}{l}
$\text{(}\overline{2}\text{11,2}\overline{\text{1}}\overline{\text{1}}\text{,%
}\overline{\text{1}}\overline{\text{1}}\text{2}$) \\ 
($\text{11}\overline{2}\text{,}\overline{\text{1}}\text{2}\overline{\text{1}}%
\text{,1}\overline{2}\text{1)}$%
\end{tabular}
}$ & {\small -1} & 
\begin{tabular}{l}
{\small q}$_{\Sigma }^{0}${\small (2473)} \\ 
{\small q}$_{\Sigma }^{0}${\small (2473)}%
\end{tabular}
\\ \hline
$\text{E}_{\Gamma }\text{= 6 \ \ }$ & {\small 2473} & 
\begin{tabular}{l}
{\small 3} \\ 
{\small 3}%
\end{tabular}
& 0 & 0 & $\text{%
\begin{tabular}{l}
$\text{(}\overline{\text{1}}\overline{2}\text{1,1}\overline{2}\overline{%
\text{1}}\text{,}\overline{\text{1}}\text{1}\overline{2}$) \\ 
$\text{ (1}\overline{\text{1}}\overline{2}\text{,}\overline{2}\text{1}%
\overline{\text{1}}\text{,}\overline{2}\overline{\text{1}}\text{1)}$%
\end{tabular}
}$ & {\small -1} & 
\begin{tabular}{l}
{\small q}$_{\Sigma }^{0}${\small (2473)} \\ 
{\small q}$_{\Sigma }^{0}${\small (2473)}%
\end{tabular}
\\ \hline
$\text{E}_{\Gamma }\text{= 6 \ \ }$ & {\small 2473} & {\small \ \ 3} & 0 & 0
& $\text{(}\overline{\text{1}}\overline{2}\overline{\text{1}}\text{,}%
\overline{\text{1}}\overline{\text{1}}\overline{2}\text{,}\overline{2}%
\overline{\text{1}}\overline{\text{1}}\text{)}$ & {\small -1} & {\small \ \ q%
}$_{\Sigma }^{0}${\small (2473)} \\ \hline
$\text{E}_{\text{P}}\text{=27/4}$ & {\small 2743} & 
\begin{tabular}{l}
{\small 3} \\ 
{\small 3}%
\end{tabular}
& 0 & 0 & 
\begin{tabular}{l}
{\small (013,031,310)} \\ 
{\small (130,301,103)}%
\end{tabular}
& {\small -1} & 
\begin{tabular}{l}
{\small q}$_{\Sigma }^{0}${\small (2743)} \\ 
{\small q}$_{\Sigma }^{0}${\small (2743)}%
\end{tabular}
\\ \hline
& {\small 2743} & {\small \ \ 1} & 0 & 0 & $\text{(222)}$ & {\small -1} & 
{\small \ \ d}$_{S}^{0}${\small (2743)} \\ \hline
\end{tabular}

\ \ \ \ \ \ \ \ \ \ \ \ \ \ \ \ \ \ \ \ \ \ \ \ \ \ \ \ \ \ \ \ \ \ \ \ \ \
\ \ \ \ \ \ \ \ \ \ \ \ \ \ \ \ \ \ \ \ \ \ \ \ \ \ \ \ \ \ \ \ \ \ \ \ \ 

For the D-, F- and G-axes on the surfaces of the regular rhombic
dodecahedron (see Fig. 1), the energy bands with the same energy might not
have all equivalent $\overrightarrow{n}$ values. Using (\ref{deg > R}), (\ref%
{Subdeg}) and (\ref{IsoSpin}) [the first division, K = 0], we can get
isospin and other intrinsic quantum numbers.

For sixfold energy bands of the F-axis and the G-axis, we need \textbf{a
second division, K = 1}. Using (\ref{Kersa-C}) and (\ref{Omiga-C}), we can
obtain the q$_{\Xi _{C}}$-quark and the q$_{\Omega _{C}}$-quark, shown in
Table B6 and Table B7.\ 

There are three energy bands ($\overrightarrow{n}$ = (000), $\overrightarrow{%
n}$ = (100) and $\overrightarrow{n}$ = (200)) that have already been
recognized on the three axes $\Gamma $-H, $\Gamma $-P and $\Gamma $-N. The
bands on the surfaces of the regular rhombic dodecahedron are the same
quarks as those inside:\ \ \ \ \ \ \ \ \ \ \ \ \ \ \ \ \ \ \ \ 

\begin{tabular}{l}
\ \ \ \ \ \ \ \ \ \ \ \ \ \ \ \ \ \ \ \ \ \ \ \ \ \ \ \ \ \ \ \ Table B4.
The Special Energy bands \\ 
$%
\begin{tabular}{|l|l|l|l|}
\hline
$\ \ \overrightarrow{n}$ & {\small Bands (Inside dodecahedron)} & \ \ \ \ \
\ {\small Bands (on Surface )} & {\small Quark} \\ \hline
\ {\small (000)} & 
\begin{tabular}{l}
{\small E}$_{\Gamma }${\small (0)}$\rightarrow ${\small E}$_{N}${\small (}$%
\frac{1}{2}${\small )} \\ 
{\small E}$_{\Gamma }${\small (0)}$\rightarrow ${\small E}$_{p}${\small (}$%
\frac{3}{4}${\small )} \\ 
{\small E}$_{\Gamma }${\small (0)}$\rightarrow ${\small E}$_{H}${\small (1)}%
\end{tabular}
& 
\begin{tabular}{l}
{\small E}$_{N}${\small (}$\frac{1}{2}${\small )}$\rightarrow ${\small E}$%
_{p}${\small (}$\frac{3}{4}${\small )} \\ 
{\small E}$_{p}${\small (}$\frac{3}{4}${\small )}$\rightarrow ${\small E}$%
_{H}${\small (1)} \\ 
{\small E}$_{N}${\small (}$\frac{1}{2}${\small )}$\rightarrow ${\small E}$%
_{M}${\small (1)}%
\end{tabular}
& {\small q}$_{N}${\small (313)} \\ \hline
\begin{tabular}{l}
{\small (110)} \\ 
{\small (1}$\overline{\text{{\small 1}}}${\small 0)}%
\end{tabular}
& \ \ \ {\small E}$_{N}${\small (}$\frac{1}{2}${\small )}$\rightarrow $%
{\small E}$_{\Gamma }${\small (2)} & 
\begin{tabular}{l}
{\small E}$_{N}${\small (}$\frac{1}{2}${\small )}$\rightarrow ${\small E}$%
_{p}${\small (}$\frac{3}{4}${\small )} \\ 
{\small E}$_{p}${\small (}$\frac{3}{4}${\small )$\rightarrow $E}$_{H}$%
{\small (3)} \\ 
{\small E}$_{N}${\small (}$\frac{1}{2}${\small )$\rightarrow $E}$_{M}$%
{\small (1)} \\ 
{\small E}$_{M}${\small (1)}$\leftarrow ${\small E}$_{N}${\small (}$\frac{5}{%
2}$)%
\end{tabular}
& {\small d}$_{S}${\small (493)} \\ \hline
{\small (200)} & \ \ {\small E}$_{H}${\small (1)}$\rightarrow ${\small E}$%
_{\Gamma }${\small (2)} & \ \ {\small E}$_{M}${\small (1)}$\rightarrow $%
{\small E}$_{N}${\small (}$\frac{5}{2}${\small ),} & {\small d}$_{S}${\small %
(773)} \\ \hline
\ {\small (002)} &  & {\small \ \ E}$_{H}${\small (1)}$\rightarrow ${\small E%
}$_{p}${\small (}$\frac{11}{4}${\small )}$\rightarrow ${\small E}$_{N}$%
{\small (}$\frac{9}{2}${\small )} & {\small d}$_{S}${\small (773)} \\ \hline
\end{tabular}
\ \ \ \ $%
\end{tabular}
$\ \ $

For each symmetry axis, from (\ref{Dalta-E}), we can get simple binding
energy formulae:

\begin{equation}
\begin{tabular}{l}
For the D-axis, $\Theta $ = 1, CI = CS = K = b = S$_{Ax}$ = 0, from(\ref%
{Dalta-E}): \\ 
twofold, 
\begin{tabular}{l}
$\Delta $E =100(2CJ$_{C}\text{+ }$SJ$_{S}\Delta $S$)$\ \ \ J$_{C}$ = 1, 2,
3, ... ,\ J$_{S,2}$= 2, 3, .... ; \\ 
$\Delta $E = 0, J$_{S,2}$= 1;\ \ \ \ \ \ \ \ 
\end{tabular}
\\ 
fourfold, C = $\Delta $S = 0, $\Delta $E = 0.%
\end{tabular}
\label{D-E}
\end{equation}

\begin{equation}
\begin{tabular}{l}
$\text{For the F-Axis, }\Theta \text{ }\text{= 1, b = 0, S}_{Ax}\text{= -1, }
$from(\ref{Dalta-E}), we have: \ \ \  \\ 
threefold, K = C = (1+S$_{Ax}$) = 0, $\Delta $E = 0; \\ 
sixfold, \ \ \ K=1, $\Delta \text{E }\text{= 100[C(2J}_{C}\text{-I)-CS+S],\
\ J}_{C}\text{ = 1, 2, 3, ...;\ }$%
\end{tabular}
\label{F-DE}
\end{equation}%
$\ $

\begin{equation}
\begin{tabular}{l}
For the G-Axis, $\Theta $ = 1,\ b = 0, S$_{A}$ = -2. From (\ref{Dalta-E}),
we have: \\ 
twofold C =K = 0, $\Delta $E = (-S)(J$_{S,2}$-2)$\Delta $S \ \ J$_{S,2}$ =
4, 5 6, ....; $\Delta $E = 0$\text{ \ J}_{S,2}$, $\leq $ 3 ; \\ 
fourfold C= K =$\Delta $S = 0, \ $\Delta $E = 0; \ \ \ \ \ \ \ \ \ \ \ \ \ \ 
\\ 
sixfold K = 1, $\left\{ 
\begin{tabular}{l}
$\Delta \text{E = 100\{C(2J}_{C}\text{-S)+S -S(J}_{S,6}\text{-2)}\Delta 
\text{S\} J}_{C}\text{=1, 2, ...; \ }$ \\ 
$\text{J}_{S,6}\text{= 4, 5, ... .\ \ }\Delta \text{E = 0, J}_{S,6}$ 
\TEXTsymbol{<}$\text{ 4.}$%
\end{tabular}
\ \right\} $%
\end{tabular}
\ \ \ \ \ \ \text{\ \ }  \label{E(G-6)}
\end{equation}

For the D-axis, from Table 2, (\ref{E(n,k)}), (\ref{Equ-N}), (\ref{DaltaS}),
(\ref{Charmed}), (\ref{D-E}), (\ref{Rest Mass}) and Table B4, omitting $%
\Delta $ part energy of quarks, we have:

\ \ \ \ \ \ Table B5. The Energy Bands and the Quarks of the D-Axis (S = 0)

$%
\begin{tabular}{|l|l|l|l|l|l|l|l|l|}
\hline
$\text{E}_{Start}$ & {\small E} & (n$_{1}$n$_{2}$n$_{3}${\small ),...} & $%
\Delta $S & J & {\small S} & {\small C} & $\Delta $E & q(m{\small \ }$_{%
\text{{\small (Mev)}}}$) \\ \hline
{\small E}$_{\text{N}}${\small \ = }$\frac{1}{2}$ & {\small 493} & 
\begin{tabular}{l}
{\small (000)} \\ 
({\small 110)}%
\end{tabular}
{\small \ } & 
\begin{tabular}{l}
{\small 0} \\ 
{\small -1}%
\end{tabular}
& 
\begin{tabular}{l}
{\small 0} \\ 
J$_{S,2}$=1%
\end{tabular}
& 
\begin{tabular}{l}
{\small 0} \\ 
{\small -1}%
\end{tabular}
& 
\begin{tabular}{l}
{\small 0} \\ 
0%
\end{tabular}
& {\small \ \ \ 0} & 
\begin{tabular}{l}
{\small u(313)} \\ 
{\small d}$_{S}^{-1}${\small (493)}%
\end{tabular}
\\ \hline
{\small E}$_{\text{P}}${\small \ = }$\frac{3}{4}$ & {\small 583} & \ {\small %
(101, 011)} & {\small \ \ 0} & \ \ 0 & {\small \ \ 0} & {\small \ \ 0} & 
{\small \ \ \ 0} & {\small \ \ q}$_{N}^{0}${\small (583)} \\ \hline
{\small E}$_{\text{N}}${\small \ = }$\frac{3}{2}$ & {\small 853} & \ {\small %
(10}$\overline{\text{{\small 1}}}${\small , 01}$\overline{1}${\small )} & 
{\small \ \ 0} & \ \ 0 & {\small \ \ 0} & {\small \ \ 0} & {\small \ \ \ 0}
& {\small \ \ q}$_{N}^{0}${\small (853)} \\ \hline
{\small E}$_{\text{N}}${\small \ = }$\frac{5}{2}$ & {\small 1213} & 
\begin{tabular}{l}
{\small (1}$\overline{\text{{\small 1}}}${\small 0, }$\overline{\text{%
{\small 1}}}${\small 10)} \\ 
{\small (020, 200)}%
\end{tabular}
& 
\begin{tabular}{l}
{\small 0} \\ 
{\small 0}%
\end{tabular}
& 
\begin{tabular}{l}
{\small 0} \\ 
{\small 0}%
\end{tabular}
& 
\begin{tabular}{l}
{\small 0} \\ 
{\small 0}%
\end{tabular}
& 
\begin{tabular}{l}
{\small 0} \\ 
{\small 0}%
\end{tabular}
& {\small \ \ \ 0} & 
\begin{tabular}{l}
{\small q}$_{N}^{0}${\small (1213)} \\ 
{\small q}$_{N}^{0}${\small (1213)}%
\end{tabular}
\\ \hline
{\small E}$_{\text{P}}${\small \ = }$\frac{11}{4}$ & {\small 1303} & 
\begin{tabular}{l}
{\small (}$\overline{\text{{\small 1}}}${\small 01, 0}$\overline{1}${\small %
1,} \\ 
{\small (211, 121)}%
\end{tabular}
& 
\begin{tabular}{l}
{\small 0} \\ 
{\small 0}%
\end{tabular}
& 
\begin{tabular}{l}
{\small 0} \\ 
{\small 0}%
\end{tabular}
& 
\begin{tabular}{l}
{\small 0} \\ 
{\small 0}%
\end{tabular}
& 
\begin{tabular}{l}
{\small 0} \\ 
{\small 0}%
\end{tabular}
& {\small \ \ \ 0} & 
\begin{tabular}{l}
{\small q}$_{N}^{0}${\small (1303)} \\ 
{\small q}$_{N}^{0}${\small (1303)}%
\end{tabular}
\\ \hline
{\small E}$_{\text{P}}${\small \ = }$\frac{11}{4}$ & {\small 1303} & 
\begin{tabular}{l}
{\small (002)} \\ 
{\small (112)}%
\end{tabular}
& 
\begin{tabular}{l}
{\small -1} \\ 
{\small -1}%
\end{tabular}
& 
\begin{tabular}{l}
J$_{S,2}$={\small 2} \\ 
J$_{S,2}$={\small 3}%
\end{tabular}
& 
\begin{tabular}{l}
{\small -1} \\ 
{\small -1}%
\end{tabular}
& 
\begin{tabular}{l}
{\small 0} \\ 
{\small 0}%
\end{tabular}
& 
\begin{tabular}{l}
{\small 200} \\ 
{\small 300}%
\end{tabular}
& 
\begin{tabular}{l}
{\small d}$_{S}^{-1}${\small (1503)} \\ 
{\small d}$_{S}^{-1}${\small (1603)}%
\end{tabular}
\\ \hline
{\small E}$_{\text{N}}${\small \ = }$\frac{7}{2}$ & {\small 1573} & 
\begin{tabular}{l}
{\small (12}$\overline{\text{{\small 1}}}${\small , 21}$\overline{\text{%
{\small 1}}}${\small )} \\ 
{\small (}$\overline{1}${\small 0}$\overline{1}${\small , 0}$\overline{11})$%
\end{tabular}
& 
\begin{tabular}{l}
{\small 0} \\ 
{\small 0}%
\end{tabular}
& 
\begin{tabular}{l}
{\small 0} \\ 
{\small 0}%
\end{tabular}
& 
\begin{tabular}{l}
{\small 0} \\ 
{\small 0}%
\end{tabular}
& 
\begin{tabular}{l}
{\small 0} \\ 
{\small 0}%
\end{tabular}
& {\small \ \ \ 0} & 
\begin{tabular}{l}
{\small q}$_{N}^{0}${\small (1573)} \\ 
{\small q}$_{N}^{0}${\small (1573)}%
\end{tabular}
\\ \hline
{\small E}$_{\text{N}}${\small \ =}$\frac{\ 9}{2}$ & {\small 1933} & 
\begin{tabular}{l}
{\small (220,} \\ 
$\overline{1}\overline{1}${\small 0)}%
\end{tabular}
& 
\begin{tabular}{l}
{\small -1} \\ 
{\small 1}%
\end{tabular}
& 
\begin{tabular}{l}
J$_{S,2}$=4 \\ 
J$_{\text{C}}$= 1%
\end{tabular}
& 
\begin{tabular}{l}
{\small -1} \\ 
{\small 0}%
\end{tabular}
& 
\begin{tabular}{l}
{\small 0} \\ 
{\small 1}%
\end{tabular}
& 
\begin{tabular}{l}
{\small 400} \\ 
{\small 200}%
\end{tabular}
& 
\begin{tabular}{l}
{\small d}$_{S}^{-1}${\small (2333)} \\ 
{\small u}$_{C}${\small (2133)}%
\end{tabular}
\\ \hline
{\small E}$_{\text{N}}${\small \ = }$\frac{9}{2}$ & {\small 1933} & 
\begin{tabular}{l}
{\small (11}$\overline{2}${\small )} \\ 
{\small (00}$\overline{2}${\small )}%
\end{tabular}
& 
\begin{tabular}{l}
{\small 1} \\ 
{\small 1}%
\end{tabular}
& 
\begin{tabular}{l}
J$_{\text{C}}$=2 \\ 
J$_{\text{C}}$=3%
\end{tabular}
& 
\begin{tabular}{l}
{\small 0} \\ 
{\small 0}%
\end{tabular}
& 
\begin{tabular}{l}
{\small 1} \\ 
{\small 1}%
\end{tabular}
& 
\begin{tabular}{l}
{\small 400} \\ 
{\small 600}%
\end{tabular}
& 
\begin{tabular}{l}
{\small u}$_{C}${\small (2333)} \\ 
{\small u}$_{C}${\small (2533)}%
\end{tabular}
\\ \hline
{\small E}$_{\text{p}}${\small \ =}$\frac{19}{4}$ & {\small 2023} & \ 
{\small (}$\overline{1}${\small 21,2}$\overline{1}${\small 1)} & {\small \ \
0} & \ \ 0 & {\small \ \ 0} & {\small \ \ 0} & {\small \ \ \ 0} & {\small \
\ q}$_{N}^{0}${\small (2023)} \\ \hline
{\small E}$_{\text{P}}${\small \ =}$\frac{19}{4}$ & {\small 2023} & 
\begin{tabular}{l}
{\small (}$\overline{1}${\small 12,1}$\overline{1}${\small 2)} \\ 
{\small (202,022)}%
\end{tabular}
& 
\begin{tabular}{l}
{\small 0} \\ 
{\small 0}%
\end{tabular}
& 
\begin{tabular}{l}
{\small 0} \\ 
{\small 0}%
\end{tabular}
& 
\begin{tabular}{l}
{\small 0} \\ 
{\small 0}%
\end{tabular}
& 
\begin{tabular}{l}
{\small 0} \\ 
{\small 0}%
\end{tabular}
& {\small \ \ \ 0} & 
\begin{tabular}{l}
{\small q}$_{N}^{0}${\small (2023)} \\ 
{\small q}$_{N}^{0}${\small (2023)}%
\end{tabular}
\\ \hline
{\small E}$_{\text{N}}${\small \ =}$\frac{11}{2}$ & {\small 2293} & {\small %
\ \ (2}$\overline{11}${\small ,}$\overline{1}${\small 2}$\overline{1}$%
{\small )} & {\small \ \ 0} & \ \ 0 & {\small \ \ 0} & {\small \ \ 0} & 
{\small \ \ \ 0} & {\small \ \ q}$_{N}^{0}${\small (2293)} \\ \hline
{\small E}$_{\text{N}}${\small \ =}$\frac{13}{2}$ & {\small 2653} & 
\begin{tabular}{l}
{\small (310, 130)} \\ 
{\small (}$\overline{\text{{\small 2}}}${\small 00, 0}$\overline{2}${\small %
0)}%
\end{tabular}
& 
\begin{tabular}{l}
{\small 0} \\ 
{\small 0}%
\end{tabular}
& 
\begin{tabular}{l}
{\small 0} \\ 
{\small 0}%
\end{tabular}
& 
\begin{tabular}{l}
{\small 0} \\ 
{\small 0}%
\end{tabular}
& 
\begin{tabular}{l}
{\small 0} \\ 
{\small 0}%
\end{tabular}
& {\small \ \ \ 0} & 
\begin{tabular}{l}
{\small q}$_{N}^{0}${\small (2653)} \\ 
{\small q}$_{N}^{0}${\small (2653)}%
\end{tabular}
\\ \hline
{\small E}$_{\text{P}}${\small \ =}$\frac{27}{4}$ & {\small 2743} & {\small %
\ \ (222)} & {\small \ \ \ \ 1} & J$_{\text{C}}$=4 & {\small \ \ 0} & 
{\small \ \ 1} & {\small \ 800} & {\small \ \ u}$_{C}${\small (3543)} \\ 
\hline
\end{tabular}%
\ \ \ \ \ \ $

$\ $E$_{\text{Start}}$ = [(n$_{1}$-$\xi $)$^{2}$+(n$_{2}$-$\eta $)$^{2}$+(n$%
_{3}$-$\zeta $)$^{2}$], {\small E} is the starting and minimum energy of the
energy band; $\overline{\text{n}_{i}}$ = -n; $\overline{\text{n}_{1}\text{n}%
_{2}\text{n}_{3}}$ = -n$_{1}$, -n$_{2}$, -n$_{3}$ [see Eq. (\ref{l-n})].

For the F-axis, from Table 2, (\ref{E(n,k)}), (\ref{Equ-N}), (\ref{DaltaS}),
(\ref{Kersa-C}), (\ref{F-DE}), (\ref{Rest Mass}) and Table B4, omitting $%
\Delta $ part energy of quarks, we have:

\newpage

\ \ \ \ \ \ \ Table B6. The Energy Bands and the Quarks of the F-Axis\ (S =
-1)\ \ \ \ \ \ \ \ \ \ \ \ \ \ \ \ \ \ \ \ \ \ \ \ \ \ 

\begin{tabular}{|l|l|l|l|l|l|l|l|l|}
\hline
$\text{E}_{Start}$ & {\small E} & $\ \text{E-Band }$ & $\Delta S$ & \ J & \
\ {\small S} & {\small C} & $\ \ \Delta $E & \ q(m{\small \ }$_{\text{%
{\small (Mev)}}}$) \\ \hline
{\small E}$_{\text{P}}${\small =}$\frac{3}{4}$ & {\small 583} & 
\begin{tabular}{l}
{\small (000)} \\ 
{\small (011,101)}%
\end{tabular}
& 
\begin{tabular}{l}
{\small 1} \\ 
{\small 1}%
\end{tabular}
& 
\begin{tabular}{l}
{\small 0} \\ 
{\small 1}%
\end{tabular}
& 
\begin{tabular}{l}
{\small 0} \\ 
{\small 0}%
\end{tabular}
& 
\begin{tabular}{l}
{\small 0} \\ 
{\small 0}%
\end{tabular}
&  \ 
\begin{tabular}{l}
{\small 0} \\ 
{\small 0}%
\end{tabular}
& 
\begin{tabular}{l}
{\small d(313)} \\ 
{\small q}$_{\text{N}}^{1}${\small (583)}%
\end{tabular}
\\ \hline
{\small E}$_{\text{P}}${\small =}$\frac{3}{4}$ & {\small 583} & \ {\small %
(110)} & {\small \ \ 0} & \ \ 0 & {\small \ \ -1} & {\small \ \ 0} & \ \ \ 
{\small 0} & {\small \ \ d}$_{S}^{0}${\small (493)} \\ \hline
{\small E}$_{\text{H}}${\small = 1} & {\small 673} & 
\begin{tabular}{l}
{\small (002)} \\ 
{\small (}$\overline{\text{{\small 1}}}${\small 01,0}$\overline{\text{%
{\small 1}}}${\small 1)}%
\end{tabular}
& 
\begin{tabular}{l}
{\small 0} \\ 
{\small 1}%
\end{tabular}
& 
\begin{tabular}{l}
{\small 0} \\ 
2%
\end{tabular}
& 
\begin{tabular}{l}
{\small -1} \\ 
{\small \ 0}%
\end{tabular}
& 
\begin{tabular}{l}
{\small 0} \\ 
{\small 0}%
\end{tabular}
& {\small \ \ \ \ 0} & 
\begin{tabular}{l}
{\small d}$_{S}^{0}${\small (773)} \\ 
{\small q}$_{\text{N}}^{1}${\small (673)}%
\end{tabular}
\\ \hline
{\small E}$_{\text{P}}${\small =}$\frac{11}{2}$ & {\small 1303} & 
\begin{tabular}{l}
{\small (112)} \\ 
{\small (1}$\overline{\text{{\small 1}}}${\small 0,}$\overline{\text{{\small %
1}}}${\small 10)}%
\end{tabular}
& 
\begin{tabular}{l}
{\small 0} \\ 
{\small 1}%
\end{tabular}
& 
\begin{tabular}{l}
{\small 0} \\ 
{\small 3}%
\end{tabular}
& 
\begin{tabular}{l}
{\small -1} \\ 
{\small \ 0}%
\end{tabular}
& 
\begin{tabular}{l}
{\small 0} \\ 
{\small 0}%
\end{tabular}
& {\small \ \ \ \ 0} & 
\begin{tabular}{l}
{\small d}$_{S}^{0}${\small (1303)} \\ 
{\small q}$_{\text{N}}^{1}${\small (1303)}%
\end{tabular}
\\ \hline
{\small E}$_{\text{P}}${\small =}$\frac{11}{2}$ & {\small 1303} & 
\begin{tabular}{l}
{\small (01}$\overline{\text{{\small 1}}}${\small ,10}$\overline{\text{%
{\small 1}}}${\small )} \\ 
{\small (020)} \\ 
{\small (200)} \\ 
{\small (121,211)}%
\end{tabular}
& 
\begin{tabular}{l}
{\small 1} \\ 
{\small 0} \\ 
{\small 0} \\ 
{\small 1}%
\end{tabular}
& 
\begin{tabular}{l}
{\small 4} \\ 
0 \\ 
0 \\ 
{\small 5}%
\end{tabular}
& 
\begin{tabular}{l}
{\small \ 0} \\ 
{\small -1} \\ 
{\small -1} \\ 
{\small \ 0}%
\end{tabular}
& 
\begin{tabular}{l}
{\small 0} \\ 
{\small 0} \\ 
{\small 0} \\ 
{\small 0}%
\end{tabular}
& 
\begin{tabular}{l}
{\small \ \ 0} \\ 
{\small -100} \\ 
{\small -100} \\ 
{\small \ \ 0}%
\end{tabular}
& 
\begin{tabular}{l}
{\small q}$_{\text{N}}^{1}${\small (1303)} \\ 
{\small d}$_{S}^{0}${\small (1203)} \\ 
{\small d}$_{S}^{0}${\small (1203)} \\ 
{\small q}$_{\text{N}}^{1}${\small (1303)}%
\end{tabular}
\\ \hline
{\small E}$_{\text{H}}${\small = 3} & {\small 1393} & 
\begin{tabular}{l}
{\small (}$\overline{\text{{\small 1}}}\overline{\text{{\small 1}}}${\small %
0)} \\ 
{\small (}$\overline{\text{{\small 1}}}${\small 12,1}$\overline{\text{%
{\small 1}}}${\small 2)}%
\end{tabular}
& 
\begin{tabular}{l}
{\small 0} \\ 
{\small 1}%
\end{tabular}
& 
\begin{tabular}{l}
{\small 0} \\ 
{\small 6}%
\end{tabular}
& 
\begin{tabular}{l}
{\small -1} \\ 
{\small \ 0}%
\end{tabular}
& 
\begin{tabular}{l}
{\small 0} \\ 
{\small 0}%
\end{tabular}
& \ \ {\small 0} & 
\begin{tabular}{l}
{\small d}$_{S}^{0}${\small (1393)} \\ 
{\small q}$_{\text{N}}^{1}${\small (1393)}%
\end{tabular}
\\ \hline
{\small E}$_{\text{H}}${\small = 3} & {\small 1393} & \ {\small (}$\overline{%
\text{{\small 1}}}\overline{\text{{\small 1}}}${\small 2)} & {\small \ \ 0}
& \ \ 0 & {\small \ \ -1} & {\small \ \ 0} & {\small \ \ \ \ 0} & {\small \
\ \ d}$_{S}^{0}${\small (1393)} \\ \hline
{\small E}$_{\text{P}}${\small =}$\frac{19}{4}$ & {\small 2023} & 
\begin{tabular}{l}
{\small (0}$\overline{\text{{\small 11}}}${\small ,}$\overline{\text{{\small %
1}}}${\small 0}$\overline{\text{{\small 1}}})$ \\ 
{\small (}$\overline{\text{{\small 1}}}${\small 21)} \\ 
{\small (2}$\overline{\text{{\small 1}}}${\small 1)} \\ 
{\small (202, 022)}%
\end{tabular}
& 
\begin{tabular}{l}
{\small -1} \\ 
{\small 0} \\ 
{\small 0} \\ 
{\small 1}%
\end{tabular}
& 
\begin{tabular}{l}
J$_{\text{S,6}}$= 1 \\ 
0 \\ 
0 \\ 
J$_{C}$=1%
\end{tabular}
& 
\begin{tabular}{l}
{\small -2} \\ 
{\small -1} \\ 
{\small -1} \\ 
{\small -1}%
\end{tabular}
& 
\begin{tabular}{l}
{\small 0} \\ 
{\small 0} \\ 
{\small 0} \\ 
{\small 1}%
\end{tabular}
& 
\begin{tabular}{l}
{\small -200} \\ 
{\small -100} \\ 
{\small -100} \\ 
{\small -150}%
\end{tabular}
& 
\begin{tabular}{l}
{\small q}$_{\Xi }^{-1}${\small (1823)} \\ 
{\small d}$_{S}^{0}${\small (1923)} \\ 
{\small d}$_{S}^{0}${\small (1923)} \\ 
{\small q}$_{\Xi _{C}}${\small (1873)}%
\end{tabular}
\\ \hline
{\small E}$_{\text{P}}${\small =}$\frac{19}{4}$ & {\small 2023} & 
\begin{tabular}{l}
{\small (220)} \\ 
{\small (21}$\overline{\text{{\small 1}}}${\small ,12}$\overline{\text{%
{\small 1}}}${\small )}%
\end{tabular}
& 
\begin{tabular}{l}
{\small 0} \\ 
{\small 1}%
\end{tabular}
& 
\begin{tabular}{l}
{\small 0} \\ 
{\small 7}%
\end{tabular}
& 
\begin{tabular}{l}
{\small -1} \\ 
{\small 0}%
\end{tabular}
& 
\begin{tabular}{l}
{\small 0} \\ 
{\small 0}%
\end{tabular}
& {\small \ \ \ 0} & 
\begin{tabular}{l}
{\small d}$_{S}^{0}${\small (2023)} \\ 
{\small q}$_{\text{N}}^{1}${\small (2023)}%
\end{tabular}
\\ \hline
{\small E}$_{\text{H}}${\small = 5} & {\small 2113} & 
\begin{tabular}{l}
{\small (0}$\overline{\text{{\small 2}}}${\small 0,}$\overline{\text{{\small %
2}}}${\small 00)} \\ 
{\small (}$\overline{\text{{\small 2}}}${\small 11)} \\ 
{\small (1}$\overline{\text{{\small 2}}}${\small 1)} \\ 
{\small (013,103)}%
\end{tabular}
& 
\begin{tabular}{l}
{\small -1} \\ 
{\small 0} \\ 
{\small 0} \\ 
{\small 1}%
\end{tabular}
& 
\begin{tabular}{l}
J$_{\text{S,6}}$= 2 \\ 
0 \\ 
0 \\ 
J$_{C}$=2%
\end{tabular}
& 
\begin{tabular}{l}
{\small -2} \\ 
{\small -1} \\ 
{\small -1} \\ 
{\small -1}%
\end{tabular}
& 
\begin{tabular}{l}
{\small 0} \\ 
{\small 0} \\ 
{\small 0} \\ 
{\small 1}%
\end{tabular}
& 
\begin{tabular}{l}
{\small -200} \\ 
{\small -100} \\ 
{\small -100} \\ 
{\small \ 50}%
\end{tabular}
& 
\begin{tabular}{l}
{\small q}$_{\Xi }^{-1}${\small (1913)} \\ 
{\small d}$_{S}^{0}${\small (2013)} \\ 
{\small d}$_{S}^{0}${\small (2013)} \\ 
{\small q}$_{\Xi _{C}}${\small (2163)}%
\end{tabular}
\\ \hline
{\small E}$_{\text{H}}${\small = 5} & {\small 2113} & 
\begin{tabular}{l}
{\small (}$\overline{\text{{\small 21}}}${\small 1, }$\overline{\text{%
{\small 12}}}${\small 1)} \\ 
{\small (0}$\overline{\text{{\small 2}}}${\small 2)} \\ 
{\small (}$\overline{\text{{\small 2}}}${\small 02)} \\ 
{\small (0}$\overline{\text{{\small 1}}}${\small 3,}$\overline{1}${\small 03)%
}%
\end{tabular}
& 
\begin{tabular}{l}
{\small -1} \\ 
{\small 0} \\ 
{\small 0} \\ 
{\small 1}%
\end{tabular}
& 
\begin{tabular}{l}
J$_{\text{S,6}}$= 3 \\ 
0 \\ 
0 \\ 
J$_{C}$=3%
\end{tabular}
& 
\begin{tabular}{l}
{\small -2} \\ 
{\small -1} \\ 
{\small -1} \\ 
{\small -1}%
\end{tabular}
& 
\begin{tabular}{l}
{\small 0} \\ 
{\small 0} \\ 
{\small 0} \\ 
{\small 1}%
\end{tabular}
& 
\begin{tabular}{l}
{\small -200} \\ 
{\small -100} \\ 
{\small -100} \\ 
{\small \ 250}%
\end{tabular}
& 
\begin{tabular}{l}
{\small q}$_{\Xi }^{-1}${\small (1913)} \\ 
{\small d}$_{S}^{0}${\small (2013)} \\ 
{\small d}$_{S}^{0}${\small (2013)} \\ 
{\small q}$_{\Xi _{C}}${\small (2363)}%
\end{tabular}
\\ \hline
{\small E}$_{\text{P}}${\small =}$\frac{27}{4}$ & {\small 2743} & 
\begin{tabular}{l}
{\small (2}$\overline{\text{{\small 11}}}$,$\overline{\text{{\small 1}}}$%
{\small 2}$\overline{\text{{\small 1}}}${\small )}%
\end{tabular}
& 
\begin{tabular}{l}
{\small 1}%
\end{tabular}
& \ J$_{C}$=4 & 
\begin{tabular}{l}
{\small -1}%
\end{tabular}
& 
\begin{tabular}{l}
{\small 1}%
\end{tabular}
&  
\begin{tabular}{l}
{\small 450}%
\end{tabular}
& 
\begin{tabular}{l}
{\small q}$_{\Xi _{C}}${\small (3193)}%
\end{tabular}
\\ \hline
\end{tabular}

E$_{\text{Start}}$ = [(n$_{1}$-$\xi $)$^{2}$+(n$_{2}$-$\eta $)$^{2}$+(n$_{3}$%
-$\zeta $)$^{2}$], {\small E} is the starting and minimum energy of the
energy band; $\overline{\text{n}_{i}}$ = -n; $\overline{\text{n}_{1}\text{n}%
_{2}\text{n}_{3}}$ = -n$_{1}$, -n$_{2}$, -n$_{3}$ [see Eq. (\ref{l-n})].

\ For the G-axis, from Table 2,\ (\ref{E(n,k)}), (\ref{Equ-N}), (\ref{DaltaS}%
), (\ref{Omiga-C}),(\ref{E(G-6)}) and (\ref{Rest Mass}), omitting $\Delta $
part energy of quarks, we have:\ \ $\ $\ \ \ \ \ \ \ \ \ \ \ \ \ \ \ \ \ \ \
\ \ \ \ \ \ \ \ \ \ \ \ \ \ \ \ \ \ \ \ \ \ \ \ \ \ \ \ \ \ \ \ \ \ \ \ \ \
\ \ \ \ \ \ \ \ \ \ \ \ \ \ \ \ \ \ \ \ 

\ \ \ \ \ Table B7. The Energy Bands and the Quarks of the G-Axis (S = -2)\ 

$%
\begin{tabular}{|l|l|l|l|l|l|l|l|l|}
\hline
$\text{E}_{Start}$ & {\small E} & $\text{ (n}_{1}\text{n}_{2}\text{n}_{3}%
\text{, ...)}$ & $\Delta $S & \ J & \ \ {\small S} & \ {\small C} & $\
\Delta $E & q(m{\small \ }$_{\text{{\small (Mev)}}}$) \\ \hline
{\small E}$_{\text{N}}${\small =}$\frac{1}{2}$ & {\small 493} & 
\begin{tabular}{l}
{\small (000)} \\ 
({\small 110)}%
\end{tabular}
& 
\begin{tabular}{l}
{\small 0} \\ 
{\small 1}%
\end{tabular}
& 
\begin{tabular}{l}
{\small 0} \\ 
J$_{S,2}$={\small 1}%
\end{tabular}
& 
\begin{tabular}{l}
{\small 0} \\ 
{\small -1}%
\end{tabular}
& 
\begin{tabular}{l}
{\small 0} \\ 
{\small 0}%
\end{tabular}
& {\small \ } 
\begin{tabular}{l}
{\small 0} \\ 
{\small 0}%
\end{tabular}
& 
\begin{tabular}{l}
{\small d(313)} \\ 
{\small d}$_{S}^{1}${\small (493)}%
\end{tabular}
\\ \hline
{\small E}$_{\text{M}}${\small = 1} & {\small 673} & \ {\small (101, 10}$%
\overline{1}${\small )} & {\small \ \ 0} & \ \ 0 & {\small \ \ -2} & {\small %
\ \ 0} & {\small \ \ \ \ 0} & {\small \ \ q}$_{\Xi }^{0}${\small (673)} \\ 
\hline
{\small E}$_{\text{M}}${\small = 1} & {\small 673} & 
\begin{tabular}{l}
{\small 1}$\overline{1}${\small 0)} \\ 
{\small (200)}%
\end{tabular}
& 
\begin{tabular}{l}
{\small 1} \\ 
{\small 1}%
\end{tabular}
& 
\begin{tabular}{l}
J$_{S,2}$=2 \\ 
J$_{S,2}$=3%
\end{tabular}
& 
\begin{tabular}{l}
{\small -1} \\ 
{\small -1}%
\end{tabular}
& 
\begin{tabular}{l}
{\small 0} \\ 
{\small 0}%
\end{tabular}
& 
\begin{tabular}{l}
{\small \ \ 0} \\ 
{\small 100}%
\end{tabular}
& 
\begin{tabular}{l}
{\small d}$_{S}^{1}${\small (493)} \\ 
{\small d}$_{S}^{1}${\small (773)}%
\end{tabular}
\\ \hline
{\small E}$_{\text{N}}${\small =}$\frac{3}{2}$ & {\small 853} & \ {\small %
(011, 01}$\overline{1}${\small )} & {\small \ \ 0} & \ \ 0 & {\small \ \ -2}
& {\small \ \ 0} & {\small \ \ \ \ \ 0} & {\small \ \ q}$_{\Xi }^{0}${\small %
(853)} \\ \hline
{\small E}$_{\text{N}}${\small =}$\frac{5}{2}$ & {\small 1213} & 
\begin{tabular}{l}
{\small (020) } \\ 
($\overline{1}${\small 10)}%
\end{tabular}
& 
\begin{tabular}{l}
{\small 1} \\ 
{\small 1}%
\end{tabular}
& 
\begin{tabular}{l}
J$_{S,2}$=4 \\ 
J$_{S,2}$=5%
\end{tabular}
& 
\begin{tabular}{l}
{\small -1} \\ 
{\small -1}%
\end{tabular}
& 
\begin{tabular}{l}
{\small 0} \\ 
{\small 0}%
\end{tabular}
& 
\begin{tabular}{l}
{\small -200} \\ 
{\small -300}%
\end{tabular}
& 
\begin{tabular}{l}
{\small d}$_{S}^{1}${\small (1413)} \\ 
{\small d}$_{S}^{1}${\small (1513)}%
\end{tabular}
\\ \hline
{\small E}$_{\text{M}}${\small = 3} & {\small 1393} & 
\begin{tabular}{l}
({\small 0}$\overline{1}${\small 1, 0}$\overline{11})$ \\ 
({\small 211, 21}$\overline{1})$%
\end{tabular}
& 
\begin{tabular}{l}
{\small 0} \\ 
{\small 0}%
\end{tabular}
& 
\begin{tabular}{l}
{\small 0} \\ 
{\small 0}%
\end{tabular}
& 
\begin{tabular}{l}
{\small -2} \\ 
{\small -2}%
\end{tabular}
& 
\begin{tabular}{l}
{\small 0} \\ 
{\small 0}%
\end{tabular}
& {\small \ \ } 
\begin{tabular}{l}
{\small 0} \\ 
{\small 0}%
\end{tabular}
& 
\begin{tabular}{l}
{\small q}$_{\Xi }^{0}${\small (1393)} \\ 
{\small q}$_{\Xi }^{0}${\small (1393)}%
\end{tabular}
\\ \hline
{\small E}$_{\text{M}}${\small = 3} & {\small 1393} & {\small \ \ (2}$%
\overline{1}${\small 1, 2}$\overline{11})$ & {\small \ \ 0} & \ 0 & {\small %
\ \ -2} & {\small \ \ 0} & {\small \ \ \ \ \ 0} & {\small \ \ q}$_{\Xi }^{0}$%
{\small (1393)} \\ \hline
{\small E}$_{\text{N}}${\small =}$\frac{7}{2}$ & {\small 1573} & 
\begin{tabular}{l}
{\small (}$\overline{1}${\small 01,}$\overline{101})$ \\ 
{\small (121, 12}$\overline{1})$%
\end{tabular}
& 
\begin{tabular}{l}
{\small 0} \\ 
{\small 0}%
\end{tabular}
& 
\begin{tabular}{l}
{\small 0} \\ 
{\small 0}%
\end{tabular}
& 
\begin{tabular}{l}
{\small -2} \\ 
{\small -2}%
\end{tabular}
& 
\begin{tabular}{l}
{\small 0} \\ 
{\small 0}%
\end{tabular}
& {\small \ \ } 
\begin{tabular}{l}
{\small 0} \\ 
{\small 0}%
\end{tabular}
& 
\begin{tabular}{l}
{\small q}$_{\Xi }^{0}${\small (1573)} \\ 
{\small q}$_{\Xi }^{0}${\small (1573)}%
\end{tabular}
\\ \hline
{\small E}$_{\text{N}}${\small =}$\frac{9}{2}$ & {\small 1933} & 
\begin{tabular}{l}
{\small (112, 11}$\overline{2})$ \\ 
{\small (002,\ 00}$\overline{2})$ \\ 
{\small (220)} \\ 
{\small (}$\overline{1}\overline{1}${\small 0)}%
\end{tabular}
& 
\begin{tabular}{l}
{\small 0} \\ 
{\small 0} \\ 
{\small 1} \\ 
{\small -1}%
\end{tabular}
& 
\begin{tabular}{l}
{\small 0} \\ 
{\small 0} \\ 
J$_{C}$=1 \\ 
J$_{S,6}$=1%
\end{tabular}
& 
\begin{tabular}{l}
{\small -2} \\ 
{\small -2} \\ 
{\small -2} \\ 
{\small -3}%
\end{tabular}
& 
\begin{tabular}{l}
{\small 0} \\ 
{\small 0} \\ 
{\small 1} \\ 
{\small 0}%
\end{tabular}
& 
\begin{tabular}{l}
{\small -200} \\ 
{\small -200} \\ 
{\small \ 200} \\ 
{\small -300}%
\end{tabular}
& 
\begin{tabular}{l}
{\small q}$_{\Xi }^{0}${\small (1733)} \\ 
{\small q}$_{\Xi }^{0}${\small (1733)} \\ 
{\small d}$_{\Omega _{C}}${\small (2133)} \\ 
{\small d}$_{\Omega }${\small (1633)}%
\end{tabular}
\\ \hline
{\small E}$_{\text{M}}${\small = 5} & {\small 2113} & 
\begin{tabular}{l}
{\small (202, 20}$\overline{2}${\small )} \\ 
{\small (1}$\overline{1}${\small 2, 1}$\overline{12}${\small )} \\ 
{\small (310) } \\ 
{\small (0}$\overline{2}${\small 0)}%
\end{tabular}
& 
\begin{tabular}{l}
{\small 0} \\ 
{\small 0} \\ 
{\small 1} \\ 
{\small -1}%
\end{tabular}
& 
\begin{tabular}{l}
{\small 0} \\ 
{\small 0} \\ 
J$_{C}$=2 \\ 
J$_{S,6}$=2%
\end{tabular}
& 
\begin{tabular}{l}
{\small -2} \\ 
{\small -2} \\ 
{\small -2} \\ 
{\small -3}%
\end{tabular}
& 
\begin{tabular}{l}
{\small 0} \\ 
{\small 0} \\ 
{\small 1} \\ 
{\small 0}%
\end{tabular}
& 
\begin{tabular}{l}
{\small -200} \\ 
{\small -200} \\ 
{\small \ 400} \\ 
{\small -300}%
\end{tabular}
& 
\begin{tabular}{l}
{\small q}$_{\Xi }^{0}${\small (1913)} \\ 
{\small q}$_{\Xi }^{0}${\small (1913)} \\ 
{\small q}$_{\Omega _{C}}${\small (2513)} \\ 
{\small q}$_{\Omega }${\small (1813)}%
\end{tabular}
\\ \hline
{\small E}$_{\text{M}}${\small = 5} & {\small 2113} & 
\begin{tabular}{l}
{\small (301,30}$\overline{1}${\small )} \\ 
{\small (1}$\overline{2}${\small 1,1}$\overline{21})$%
\end{tabular}
& 
\begin{tabular}{l}
{\small 0} \\ 
{\small 0}%
\end{tabular}
& 
\begin{tabular}{l}
{\small 0} \\ 
{\small 0}%
\end{tabular}
& 
\begin{tabular}{l}
{\small -2} \\ 
{\small -2}%
\end{tabular}
& 
\begin{tabular}{l}
{\small 0} \\ 
{\small 0}%
\end{tabular}
& {\small \ \ } 
\begin{tabular}{l}
{\small 0} \\ 
{\small 0}%
\end{tabular}
& 
\begin{tabular}{l}
{\small q}$_{\Xi }^{0}${\small (2113)} \\ 
{\small q}$_{\Xi }^{0}${\small (2113)}%
\end{tabular}
\\ \hline
{\small E}$_{\text{M}}${\small = 5} & {\small 2113} & 
\begin{tabular}{l}
{\small (3}$\overline{\text{{\small 1}}}${\small 0)} \\ 
{\small (2}$\overline{2}${\small 0)}%
\end{tabular}
& 
\begin{tabular}{l}
{\small 1} \\ 
{\small 1}%
\end{tabular}
& 
\begin{tabular}{l}
J$_{S,2}$=6 \\ 
J$_{S,2}$=7%
\end{tabular}
& 
\begin{tabular}{l}
{\small -1} \\ 
{\small -1}%
\end{tabular}
& 
\begin{tabular}{l}
{\small 0} \\ 
{\small 0}%
\end{tabular}
& 
\begin{tabular}{l}
{\small 400} \\ 
{\small 500}%
\end{tabular}
& 
\begin{tabular}{l}
{\small d}$_{S}^{1}${\small (2513)} \\ 
{\small d}$_{S}^{1}${\small (2613)}%
\end{tabular}
\\ \hline
{\small E}$_{\text{N}}${\small =}$\frac{11}{2}$ & {\small 2293} & \ {\small (%
}$\overline{1}${\small 21,}$\overline{1}${\small 2}$\overline{1})$ & {\small %
\ \ 0} & \ 0 & {\small \ \ -2} & {\small \ \ 0} & {\small \ \ \ \ 0} & 
{\small \ \ q}$_{\Xi }^{0}${\small (2293)} \\ \hline
\end{tabular}
\ $

E$_{\text{Start}}$ = [(n$_{1}$-$\xi $)$^{2}$+(n$_{2}$-$\eta $)$^{2}$+(n$_{3}$%
-$\zeta $)$^{2}$], {\small E} is the starting and minimum energy of the
energy band; $\overline{\text{n}_{i}}$ = -n; $\overline{\text{n}_{1}\text{n}%
_{2}\text{n}_{3}}$ = -n$_{1}$, -n$_{2}$, -n$_{3}$ [see Eq. (\ref{l-n})].

{\LARGE Appendix C. Omitting Angular Momenta, Dividing Groups and Obtaining
a Representative Particle of the Group\ \ }

We show how to omit angular momenta and parities of the baryons or mesons
and how\ to divide experimental baryons and mesons into groups using some
examples in Table C1-- C7. There might be many members in each group for
experimental and deduced results. For a group with many members, we find a
representative particle. The representative particle has the same intrinsic
quantum numbers (I, S, C, b and Q) with the same name and the average rest
mass of the members in the group. In Table C1, we show the representative
particles of the deduced and experimental results. Similarly, we can get all
representative baryons and mesons shown in Table C2 - Table C7.

The unflavored mesons with the same intrinsic quantum numbers but different
angular momenta or parities have different names. In order to compare their
rest masses, omitting the differences of the angular momenta and parities,
we use meson $\eta $ to represent the mesons with S = C = b = 0, I = Q = 0 ($%
\eta $, $\varpi $, $\phi $, h and f) (see Table C5) and we use meson $\pi $
to represent the mesons with S = C = b = 0, I = 1, Q = 1, 0, -1 ($\pi $, $%
\rho $, a and b) (see Table C6).

For example, the number ($\overline{\text{1498}}$) inside N($\overline{\text{%
1498}}$) is the rest mass of the baryon N($\overline{\text{1498}}$). The top
line of the number ($\overline{\text{1498}}$) means that the number down the
line is the average rest mass of the group. $\Gamma $ is the full width of
the baryon (or meson) in unit Mev, $\overline{\Gamma }$ means the average of
the member full widths ($\Gamma s$) in the group.

\ \ \ 

\bigskip

\ 
\begin{tabular}{l}
\ \ \ \ \ \ \ \ \ \ \ Table C1. The Unflavored Baryons $N$ and $\Delta $ ($S$%
= $C$=$b$ = 0) \\ 
\begin{tabular}{|l|l||l|l|}
\hline
\ Deduced & \ Experiment & \ Deduced & Experiment \\ \hline
\begin{tabular}{l}
{\small 2}$N(1209)$ \\ 
2$N(1299)$ \\ 
N($\overline{1254}$)%
\end{tabular}
&  & $%
\begin{array}{c}
2\Delta (1209) \\ 
2\Delta (1299) \\ 
\Delta \mathbf{(}\overline{\mathbf{1254}}\mathbf{)}%
\end{array}
$ & $\ \Delta (1232),120$ \\ \hline
\ \textbf{N(1479)} & 
\begin{tabular}{l}
$N(1440),350$ \\ 
$N(1520),120$ \\ 
$N(1535),150$ \\ 
$\mathbf{N(}\overline{\mathbf{1498}}\mathbf{),}\overline{\mathbf{207}}$%
\end{tabular}
&  &  \\ \hline
\begin{tabular}{l}
$N(1659)$ \\ 
$N(1659)$ \\ 
$\mathbf{N(}\overline{\mathbf{1650}}\mathbf{)}$%
\end{tabular}
& 
\begin{tabular}{l}
$N(1650),150$ \\ 
$N(1675),150$ \\ 
$N(1680),130$ \\ 
$N(1700),100$ \\ 
$N(1710),100$ \\ 
$N(1720),150$ \\ 
$\mathbf{N(}\overline{\mathbf{1689}}\mathbf{),}\overline{\mathbf{130}}%
\mathbf{)}$%
\end{tabular}
& 
\begin{tabular}{l}
$\Delta (1659)$ \\ 
$\Delta (1659)$ \\ 
$\mathbf{\Delta }\mathbf{(}\overline{\mathbf{1659}}\mathbf{)}$%
\end{tabular}
& 
\begin{tabular}{l}
$\Delta (1600),350$ \\ 
$\Delta (1620),150$ \\ 
$\Delta (1700),300$ \\ 
$\mathbf{\Delta (}\overline{\mathbf{1640}}\mathbf{),}\overline{\mathbf{267}}$%
\end{tabular}
\\ \hline
\begin{tabular}{l}
2$N(1839)$ \\ 
5$N(1929)$ \\ 
2$N(2019)$ \\ 
$\mathbf{N(}\overline{\mathbf{1929}}\mathbf{)}$%
\end{tabular}
& 
\begin{tabular}{l}
$N(1900)^{\ast }$, 500 \\ 
$N(1990)^{\ast }$, 500 \\ 
$N(2000)^{\ast }$, 400 \\ 
$N(2080)^{\ast }$, 400 \\ 
$N(2090)^{\ast }$, 400 \\ 
$\mathbf{N(}\overline{\mathbf{1912}}\mathbf{)}^{\ast }$, $\overline{440}$%
\end{tabular}
& 
\begin{tabular}{l}
2$\Delta (1929)$ \\ 
3$\Delta ${\small (1929)} \\ 
2$\Delta ${\small (2019)} \\ 
$\mathbf{\Delta }\mathbf{(}\overline{\mathbf{1955}}\mathbf{)}$%
\end{tabular}
& 
\begin{tabular}{l}
$\Delta (1905),350$ \\ 
$\Delta (1910),250$ \\ 
$\Delta (1920),200$ \\ 
$\Delta (1930),350$ \\ 
$\Delta (1950),300$ \\ 
$\mathbf{\Delta (}\overline{\mathbf{1923}}\mathbf{),}\overline{\mathbf{264}}$%
\end{tabular}
\\ \hline
\begin{tabular}{l}
$N(2199)$ \\ 
$N(2199)$ \\ 
$\mathbf{N(}\overline{\mathbf{2199}}\mathbf{)}$%
\end{tabular}
& 
\begin{tabular}{l}
$N(2190),450$ \\ 
$N(2220),400$ \\ 
$N(2250),400$ \\ 
$\mathbf{N(}\overline{\mathbf{2220}}\mathbf{),}\overline{\mathbf{417}}$%
\end{tabular}
& $\ \ \ \Delta (2379)$ & $\ \Delta (2420),400$ \\ \hline
\ \textbf{3N(2649)} & \ \textbf{N(2600), 650} & 
\begin{tabular}{l}
$4\Delta (2649)$ \\ 
\textbf{4}$\Delta (2739)$ \\ 
$\mathbf{\Delta (}\overline{\mathbf{2694}}\mathbf{)}$%
\end{tabular}
& $\mathbf{\Delta (2750)}^{\ast },$400 \\ \hline
\ \textbf{4N(2739)} & \ \textbf{N(2700)}$^{\ast }$,\textbf{600} &  &  \\ 
\hline
\ 1N(2919) & \ Prediction & \ 3$\mathbf{\Delta (3099)}$ & 
\begin{tabular}{l}
$\mathbf{\Delta (2950)}^{\ast },$500 \\ 
$\mathbf{\Delta (3000)}^{\ast },$1000 \\ 
$\mathbf{\Delta (}\overline{\mathbf{2975}}\mathbf{)}^{\ast },$ $\overline{750%
}$%
\end{tabular}
\\ \hline
\end{tabular}%
\end{tabular}
\ 

\begin{tabular}{l}
\ \ Table C2. The Strange Baryons $\Lambda $ and $\Sigma $ (S = -1, C = b =
0) \\ 
\begin{tabular}{|l|l||l|l|}
\hline
\ Deduced & Experiment, $\Gamma $ & \ Deduced & Experiment, $\Gamma $ \\ 
\hline
$\ \ \mathbf{\Lambda (1119)}$ & $\ \mathbf{\Lambda (1116)}$ & $\ \ \mathbf{%
\Sigma (1209)}$ & $\ \mathbf{\Sigma (1193)}$ \\ \hline
{\small \ \ }$\Lambda (1399)$ & $\ \Lambda (1406),50\ \ \ $ & $\ \ \Sigma (%
\mathbf{1399})\mathbf{\ \ \ \ \ \ }$ & $\ \ \mathbf{\Sigma (1385),37}$ \\ 
\hline
\begin{tabular}{l}
$\Lambda (1659)$ \\ 
$\Lambda (1659$ \\ 
$\Lambda (1659)$ \\ 
$\mathbf{\Lambda (}\overline{\mathbf{1659}}\mathbf{)}$%
\end{tabular}
& 
\begin{tabular}{l}
$\Lambda (1520),16$ \\ 
$\Lambda (1600),150$ \\ 
$\Lambda (1670),35$ \\ 
$\Lambda (1690),60$ \\ 
$\mathbf{\Lambda (}\overline{\mathbf{1620}}\mathbf{),}\overline{\mathbf{65}}$%
\end{tabular}
& 
\begin{tabular}{l}
$\Sigma (1659)$ \\ 
$\Sigma (1659)$ \\ 
$\Sigma (1659)$ \\ 
2$\Sigma ${\small (1829)} \\ 
$\mathbf{\Sigma }\mathbf{(}\overline{\mathbf{1727}}\mathbf{)}$%
\end{tabular}
& 
\begin{tabular}{l}
$\Sigma (1660),100$ \\ 
$\Sigma (1670),60$ \\ 
$\Sigma (1750),90$ \\ 
$\Sigma (1775),120$ \\ 
$\mathbf{\Sigma (}\overline{\mathbf{1714}}\mathbf{),}\overline{\mathbf{93}}$%
\end{tabular}
\\ \hline
\begin{tabular}{l}
$\Lambda (1829)$ \\ 
$\Lambda (1829)$ \\ 
$\Lambda (1929)$ \\ 
$\Lambda (1929)$ \\ 
$\Lambda (1929)$ \\ 
$\mathbf{\Lambda (}\overline{\mathbf{1889}}\mathbf{)}$%
\end{tabular}
& 
\begin{tabular}{l}
$\Lambda (1800),300$ \\ 
$\Lambda (1810),150$ \\ 
$\Lambda (1820),$ $80$ \\ 
$\Lambda (1830),$ $95$ \\ 
$\Lambda (1890),100$ \\ 
$\mathbf{\Lambda (}\overline{\mathbf{1830}}\mathbf{),}\overline{\mathbf{145}}
$%
\end{tabular}
& 
\begin{tabular}{l}
$\Sigma (1929)$ \\ 
$\Sigma (1929)$ \\ 
$\Sigma (1929)$ \\ 
$\mathbf{\Sigma }\mathbf{(}\overline{\mathbf{1929}}\mathbf{)}$%
\end{tabular}
& 
\begin{tabular}{l}
$\Sigma (1915),120$ \\ 
$\Sigma (1940),220$ \\ 
$\mathbf{\Sigma (}\overline{\mathbf{1928}}\mathbf{),}\overline{\mathbf{170}}$%
\end{tabular}
\\ \hline
\begin{tabular}{l}
2$\Lambda $(2019) \\ 
$\Lambda $(2039) \\ 
$\Lambda $(2129) \\ 
$\Lambda $(2139) \\ 
$\Lambda $(2229) \\ 
$\mathbf{\Lambda (}\overline{\mathbf{2095}}\mathbf{)}$%
\end{tabular}
& 
\begin{tabular}{l}
$\Lambda (2100),200$ \\ 
$\Lambda (2110),200$ \\ 
$\mathbf{\Lambda (}\overline{\mathbf{2105}}\mathbf{),}\overline{\mathbf{200}}
$%
\end{tabular}
& 
\begin{tabular}{l}
$\Sigma $(2019) \\ 
$\Sigma $(2019) \\ 
$\Sigma $(2129) \\ 
$\Sigma $\textbf{(}$\overline{\mathbf{2056}}$\textbf{)}%
\end{tabular}
$\ $ & 
\begin{tabular}{l}
$\Sigma (2000)^{\ast },200$ \\ 
$\Sigma (2030),180$ \\ 
$\Sigma (2070)^{\ast },300$ \\ 
$\Sigma (2080)^{\ast },200$ \\ 
$\mathbf{\Sigma (}\overline{\mathbf{2045}}\mathbf{),}\overline{\mathbf{220}}$%
\end{tabular}
\\ \hline
$\ \Lambda (2379)$ & $\ \ \ \Lambda (2350),150$ & $\ \ \ \Sigma $(2229) & $\
\Sigma (2250),100$ \\ \hline
$%
\begin{tabular}{l}
2$\Lambda (2549)$ \\ 
$\Lambda (2559)$ \\ 
4$\Lambda (2639)$ \\ 
4$\Lambda (2649)$ \\ 
$\mathbf{\Lambda (}\overline{\mathbf{2619}}\mathbf{)}$%
\end{tabular}
\ \ \ \ \ $ & $\ \ \ \mathbf{\Lambda (2585)}^{\ast },225$ & 
\begin{tabular}{l}
$\mathbf{\Sigma (2379)}$ \\ 
2$\Sigma $(2549) \\ 
$\Sigma $(2492)%
\end{tabular}
& $\ \Sigma (2455)^{\ast },140$ \\ \hline
\ {\small 5}$\Lambda \text{({\small 3099})}$ & \ Prediction & 
\begin{tabular}{l}
4$\Sigma $(2639) \\ 
4$\Sigma $(2649) \\ 
$\mathbf{\Sigma (}\overline{\mathbf{2644}}\mathbf{)}$%
\end{tabular}
& $\ \Sigma $(2620)$^{\ast },200$ \\ \hline
$\ \Lambda ${\small (3369)} & Prediction & \ 5$\mathbf{\Sigma (3099)}$ & 
\begin{tabular}{l}
$\Sigma (3000)^{\ast }$, \ \ ? \\ 
$\Sigma (3170)^{\ast }$, \ \ ? \\ 
$\mathbf{\Sigma (}\overline{\mathbf{3085}}\mathbf{)}^{\ast }$, \ ?%
\end{tabular}
\\ \hline
\end{tabular}%
\end{tabular}
\qquad \qquad

\ \ 
\begin{tabular}{l}
$\ \ \ \ \ \text{Table C3. }$The Heavy Unflavored Mesons with\ S=C=b=Q=I=0
\\ 
$%
\begin{tabular}{|l|l|l|l|l|}
\hline
\ \ {\small d}$_{S}${\small (9613)}$\overline{d_{S}(m)}$ & \ {\small E}$%
_{bind}$ & \ Deduced & {\small Exper., }$\Gamma $ (Mev) & $\frac{\Delta M}{M}%
\%$ \\ \hline
\ \ {\small q}$_{b}^{1}${\small (4913)}$\overline{q_{b}^{1}(4913)}$ & \ \ 
{\small - 576} & $\ \ \Upsilon ${\small (9389)} & $\ \ \Upsilon ${\small %
(9460), 53 kev} & 0.75 \\ \hline
\begin{tabular}{l}
{\small d}$_{S}^{\text{-1}}${\small (9613)}$\overline{d_{S}^{-1}(493)_{D}}$
\\ 
{\small d}$_{S}^{\text{-1}}${\small (9613)}$\overline{d_{S}^{0}(493)_{F}}$%
\end{tabular}
& 
\begin{tabular}{l}
{\small -250} \\ 
{\small -150}%
\end{tabular}
& 
\begin{tabular}{l}
$\eta ${\small (9856)} \\ 
$\eta ${\small (9956)} \\ 
$\eta ${\small (}$\overline{9906}${\small )}%
\end{tabular}
& 
\begin{tabular}{l}
$\chi ${\small (9860)} \\ 
$\chi ${\small (9893)} \\ 
$\chi ${\small (9913)} \\ 
$\chi ${\small (}$\overline{9889}${\small )}%
\end{tabular}
& 0.17 \\ \hline
\begin{tabular}{l}
{\small d}$_{S}^{\text{-1}}${\small (9613)}$\overline{d_{S}^{1}(493)_{G}}$
\\ 
{\small d}$_{S}^{\text{-1}}${\small (9613)}$\overline{d_{S}^{1}(493)_{\Sigma
}}$%
\end{tabular}
& 
\begin{tabular}{l}
{\small -50} \\ 
{\small -50}%
\end{tabular}
& 
\begin{tabular}{l}
$\eta ${\small (10056)} \\ 
$\eta ${\small (10056)}%
\end{tabular}
& $\ \ \Upsilon ${\small (10023), 43 kev} & 0.13 \\ \hline
\begin{tabular}{l}
{\small d}$_{S}^{\text{-1}}${\small (9613)}$\overline{d_{S}^{-1}(773)_{%
\Delta }}$ \\ 
{\small d}$_{S}^{\text{-1}}${\small (9613)}$\overline{d_{S}^{0}(773)_{\Delta
}}$ \\ 
{\small d}$_{S}^{\text{-1}}${\small (9613)}$\overline{d_{S}^{1}(773)_{G}}$%
\end{tabular}
& 
\begin{tabular}{l}
{\small - 212} \\ 
-112 \\ 
{\small -12}%
\end{tabular}
& 
\begin{tabular}{l}
$\eta ${\small (10174)} \\ 
$\eta ${\small (10274)} \\ 
$\eta ${\small (10374)} \\ 
$\eta ${\small (}$\overline{10274}${\small )}%
\end{tabular}
& 
\begin{tabular}{l}
$\chi ${\small (10232)} \\ 
$\chi ${\small (10255)} \\ 
$\chi ${\small (10269)} \\ 
$\chi ${\small (}$\overline{10252}${\small )}%
\end{tabular}
& 0.21 \\ \hline
\ \ \ {\small u}$_{C}${\small (6073)}$\overline{\text{{\small u}}_{C}\text{%
{\small (6073)}}}$ & \ {\small -1637} & $\ \ \ \psi ${\small (10509)} & $%
\Upsilon ${\small (10355), 26 kev} & 1.5 \\ \hline
\begin{tabular}{l}
2{\small d}$_{S}^{\text{-1}}${\small (9613)}$\overline{d_{S}^{0}(1203)}_{F}$%
{\small \ } \\ 
{\small d}$_{S}^{\text{-1}}${\small (9613)}$\overline{d_{S}^{0}(1303)_{F}}$
\\ 
{\small \ d}$_{\text{S}}^{-1}${\small (1503)}$\overline{\text{{\small d}}%
_{S}^{-1}\text{{\small (9613)}}}$%
\end{tabular}
& 
\begin{tabular}{l}
-241 \\ 
-271 \\ 
-431%
\end{tabular}
& 
\begin{tabular}{l}
2$\eta ${\small (10575)} \\ 
$\eta ${\small (10645)} \\ 
$\eta ${\small (10685)} \\ 
$\overline{\eta \text{{\small (10620)}}}$%
\end{tabular}
& $\ \Upsilon ${\small (10580), 20} & 0.38 \\ \hline
\begin{tabular}{l}
2{\small d}$_{S}^{\text{-1}}${\small (9613)}$\overline{d_{S}^{0}(1393)_{F}}$
\\ 
{\small d}$_{\text{S}}^{-1}${\small (1603)}$\overline{\text{{\small d}}%
_{S}^{-1}\text{{\small (9613)}}}$ \\ 
{\small d}$_{\text{S}}^{1}${\small (1413)}$\overline{\text{{\small d}}%
_{S}^{-1}\text{{\small (9613)}}}$ \\ 
{\small d}$_{\text{S}}^{1}${\small (1513)}$\overline{\text{{\small d}}%
_{S}^{-1}\text{{\small (9613)}}}$%
\end{tabular}
{\small \ } & 
\begin{tabular}{l}
-296 \\ 
-461 \\ 
-204 \\ 
-234%
\end{tabular}
& 
\begin{tabular}{l}
2$\eta ${\small (10708)} \\ 
$\eta ${\small (10755)} \\ 
$\eta ${\small (10822)} \\ 
$\eta ${\small (10892)} \\ 
$\eta ${\small (}$\overline{10777}${\small )}%
\end{tabular}
& $\ \ \Upsilon ${\small (10865), 110} & 0.73 \\ \hline
\ \ {\small d}$_{\text{S}}^{0}${\small (1923)}$\overline{\text{{\small d}}%
_{S}^{-1}\text{{\small (9613)}}}$ & \ \ -457 & $\ \ \ \eta ${\small (11080)}
& $\ \ \Upsilon ${\small (11020), 79} & 0.54 \\ \hline
\begin{tabular}{l}
4{\small d}$_{\text{S}}^{0}${\small (2013)}$\overline{\text{{\small d}}%
_{S}^{-1}\text{{\small (9613)}}}$ \\ 
{\small d}$_{\text{S}}^{0}${\small (2023)}$\overline{\text{{\small d}}%
_{S}^{-1}\text{{\small (9613)}}}$ \\ 
{\small d}$_{\text{S}}^{1}${\small (1933)}$\overline{\text{{\small d}}%
_{S}^{-1}\text{{\small (9613)}}}$%
\end{tabular}
& 
\begin{tabular}{l}
-483 \\ 
-487 \\ 
-360%
\end{tabular}
& 
\begin{tabular}{l}
4$\eta ${\small (11143)} \\ 
$\eta ${\small (11149)} \\ 
$\eta ${\small (11186)} \\ 
$\eta ${\small (}$\overline{\text{{\small 11151}}}${\small )}%
\end{tabular}
& \ \ \ Prediction &  \\ \hline
\end{tabular}
$%
\end{tabular}

\begin{tabular}{l}
\ \ \ \ \ \ \ \ \ Table C4. The Strange Mesons (S = \ $\pm $1, C = b = 0) \\ 
$%
\begin{tabular}{|l|l|l|l|}
\hline
\ \ \ {\small q}$_{\text{N}}${\small (313)}$\ \overline{\ d_{S}(m)}$ & \ \ E$%
_{Bind}$ & \ \ {\small Deduced} & \ \ {\small Exper.} \\ \hline
\begin{tabular}{l}
{\small q}$_{\text{N}}${\small (313)}$\overline{\text{{\small d}}_{S}^{\text{%
-1}}\text{(773)}}$ \\ 
{\small q}$_{\text{N}}${\small (313)}$\overline{\text{{\small d}}_{S}^{0}%
\text{(773)}}$ \\ 
{\small q}$_{\text{N}}${\small (313)}$\overline{\text{{\small d}}_{S}^{\text{%
1}}\text{(773)}}$%
\end{tabular}
& 
\begin{tabular}{l}
{\small - 170} \\ 
{\small - 270} \\ 
{\small - 170}%
\end{tabular}
& 
\begin{tabular}{l}
{\small K(916)} \\ 
{\small K(816)} \\ 
{\small K(916)} \\ 
{\small K}$\overline{(\text{{\small 883}}}${\small )}%
\end{tabular}
& \ \ {\small K}$^{\pm }${\small (892), 50} \\ \hline
\begin{tabular}{l}
2{\small q}$_{\text{N}}${\small (313)}$\overline{\text{{\small d}}_{S}^{0}%
\text{{\small (1203)}}}$ \\ 
{\small q}$_{\text{N}}${\small (313)}$\overline{\text{{\small d}}_{S}^{0}%
\text{{\small (1303)}}}$%
\end{tabular}
& 
\begin{tabular}{l}
-270 \\ 
-270 \\ 
-170%
\end{tabular}
& 
\begin{tabular}{l}
2K(1246) \\ 
K(1346) \\ 
K$(\overline{\text{1279}})$%
\end{tabular}
& \ \ {\small K(1273), 90 } \\ \hline
\begin{tabular}{l}
{\small q}$_{\text{N}}${\small (313)}$\overline{\text{{\small d}}_{S}^{\text{%
0}}\text{{\small (1393)}}}$ \\ 
{\small q}$_{\text{N}}${\small (313)}$\overline{\text{{\small d}}_{S}^{0}%
\text{(1393{\small )}}}$%
\end{tabular}
& 
\begin{tabular}{l}
-270 \\ 
-270%
\end{tabular}
& 
\begin{tabular}{l}
K(1436) \\ 
K(1436) \\ 
K($\overline{\text{1436}}$)%
\end{tabular}
& 
\begin{tabular}{l}
K$_{1}$(1402),174 \\ 
K$^{\ast }$(1414),232 \\ 
K$_{0}^{\ast }$(1412),294 \\ 
K$_{2}^{\ast }$(1429),100 \\ 
$\overline{\text{K(1414)}},\overline{\text{200}}$%
\end{tabular}
\\ \hline
\begin{tabular}{l}
{\small q}$_{\text{N}}${\small (313)}$\overline{\text{{\small d}}_{S}^{1}%
\text{(1413{\small )}}}$ \\ 
{\small q}$_{\text{N}}${\small (313)}$\overline{\text{{\small d}}_{S}^{1}%
\text{(1513{\small )}}}$ \\ 
2{\small q}$_{N}^{0}${\small (1303)}$\overline{\text{{\small d}}_{S}^{\pm 1}%
\text{{\small (493)}}}$ \\ 
3{\small q}$_{N}^{1}${\small (1303)}$\overline{\text{{\small d}}_{S}^{0}%
\text{{\small (493)}}}$%
\end{tabular}
& 
\begin{tabular}{l}
-170 \\ 
-170 \\ 
-190 \\ 
-190%
\end{tabular}
& 
\begin{tabular}{l}
K(1556) \\ 
K(1656) \\ 
2K(1606) \\ 
3K(1606) \\ 
K($\overline{\text{1606}}$)%
\end{tabular}
& 
\begin{tabular}{l}
K$_{2}$(1580)$^{\#}$, 110 \\ 
K(1616)$^{\#}$, 16 \\ 
K$_{1}$(1650)$^{\#}$, 150 \\ 
K($\overline{\text{1615}}$)$^{\#}$, \ $\overline{\text{92}}$%
\end{tabular}
\\ \hline
\begin{tabular}{l}
{\small q}$_{N}^{1}${\small (1393)}$\overline{\text{{\small d}}_{S}^{0}\text{%
{\small (493)}}}$ \\ 
3{\small q}$_{N}^{1}${\small (1303)}$\overline{\text{{\small d}}_{S}^{-1}%
\text{{\small (493)}}}$ \\ 
\textbf{q}$_{\text{N}}$\textbf{(313)}$\overline{\text{{\small d}}_{S}^{-1}%
\text{(\textbf{1603}{\small )}}}$ \\ 
{\small q}$_{N}^{0}${\small (1573)}$\overline{\text{{\small d}}_{S}^{0}\text{%
{\small (493)}}}$ \\ 
{\small q}$_{N}^{1}${\small (1393)}$\overline{\text{{\small d}}_{S}^{-1}%
\text{{\small (493)}}}$ \\ 
{\small q}$_{N}^{0}${\small (1573)}$\overline{\text{{\small d}}_{S}^{1}\text{%
{\small (493)}}}$%
\end{tabular}
& 
\begin{tabular}{l}
-190 \\ 
-90 \\ 
-170 \\ 
-290 \\ 
-90 \\ 
-190%
\end{tabular}
& 
\begin{tabular}{l}
K{\small (1696)} \\ 
K(1706) \\ 
K(1746) \\ 
K(1776) \\ 
K(1796) \\ 
K(1876) \\ 
K($\overline{\text{1766}}$)%
\end{tabular}
& 
\begin{tabular}{l}
K$^{\ast }$(1717), 322 \\ 
K$_{2}$(1773), 186 \\ 
K$_{3}^{\ast }$(1776), 159 \\ 
K$_{2}$(1816), 276 \\ 
K($\overline{\text{1771}}$), $\overline{\text{236}}$%
\end{tabular}
\\ \hline
\begin{tabular}{l}
{\small q}$_{\text{N}}${\small (313)}$\overline{\text{{\small d}}_{S}^{0}%
\text{(1923{\small )}}}$ \\ 
{\small q}$_{\text{N}}${\small (313)}$\overline{\text{{\small d}}_{S}^{0}%
\text{(1923{\small )}}}$%
\end{tabular}
& \ \ -270 & \ \ K(1966) & 
\begin{tabular}{l}
K$_{0}^{\ast }$(1950)$^{\#}$, 201 \\ 
K$_{2}^{\ast }$(1973)$^{\#}$, 373 \\ 
K($\overline{\text{1962}}$)$^{\#}$, $\overline{\text{287}}$%
\end{tabular}
\\ \hline
\begin{tabular}{l}
4{\small q}$_{\text{N}}${\small (313)}$\overline{d_{S}^{0}(2013)}$ \\ 
{\small q}$_{\text{N}}${\small (313)}$\overline{d_{S}^{0}(2023)}$ \\ 
{\small q}$_{\text{N}}${\small (313)}$\overline{\text{{\small d}}_{S}^{1}%
\text{(1933{\small )}}}$%
\end{tabular}
& 
\begin{tabular}{l}
-270 \\ 
-270 \\ 
-170%
\end{tabular}
& 
\begin{tabular}{l}
\textbf{4K(2056)} \\ 
K(2066) \\ 
K(2076) \\ 
K($\overline{\text{2066}}$)%
\end{tabular}
& \ {\small K}$_{4}^{\ast }${\small (2045), 198} \\ \hline
\end{tabular}
\ \ \ \ \ \ $%
\end{tabular}
\ \ \ \ \ 

\ \ 
\begin{tabular}{l}
\ \ Table C5. The Light Unflavored Mesons (S = C = b = 0) I = 0 \\ 
$%
\begin{tabular}{|l|l|l|l|l|}
\hline
\ \ \ {\small q}$_{N}${\small (313)}$\overline{q_{N}(m)}$ & {\small E}$%
_{bind}$ & Deduced & {\small Experiment, }$\Gamma $ & $\frac{\Delta M}{M}\%$
\\ \hline
$\ \text{q}_{S}\text{(493)}\overline{\text{q}_{S}\text{(493)}}$ & \ {\small %
-437} & $\ \ \eta ${\small (549)} & $\ \ \eta ${\small (548), 1.29} & 
{\small 0.4} \\ \hline
\begin{tabular}{l}
{\small q}$_{N}${\small (313)}$\overline{\text{q}_{N}\text{{\small (583)}}}$
\\ 
{\small q}$_{N}${\small (583)}$\overline{\text{q}_{N}\text{{\small (583)}}}$
\\ 
q$_{\Sigma }$(583)$\overline{\text{{\small q}}_{\Sigma }\text{{\small (583)}}%
}$ \\ 
q$_{\Delta }$(673)$\overline{\text{{\small d}}_{\Delta }\text{{\small (673)}}%
}$ \\ 
q$_{\Xi }$(673)$\overline{\text{{\small d}}_{\Xi }\text{{\small (673)}}}$%
\end{tabular}
& 
\begin{tabular}{l}
{\small -320} \\ 
{\small -528} \\ 
{\small -676} \\ 
{\small -938} \\ 
{\small -538}%
\end{tabular}
& 
\begin{tabular}{l}
$\eta ${\small (576)} \\ 
$\eta ${\small (640)} \\ 
$\eta ${\small (490)} \\ 
$\eta ${\small (408)} \\ 
$\eta \left( \text{{\small 808}}\right) $ \\ 
$\eta \left( \overline{\text{{\small 584}}}\right) $%
\end{tabular}
& 
\begin{tabular}{l}
$\eta ${\small (400-1200)} \\ 
$\Gamma ${\small =600-1000}%
\end{tabular}
& \ / \\ \hline
\begin{tabular}{l}
{\small q}$_{N}${\small (313)}$\overline{q_{N}^{1}(583)}$ \\ 
{\small q}$_{N}${\small (313)}$\overline{q_{N}^{1}(673)}$ \\ 
{\small q}$_{N}${\small (673)}$\overline{q_{N}(673)}$ \\ 
{\small q}$_{N}${\small (313)}$\overline{q_{N}^{0}(853)}$%
\end{tabular}
& 
\begin{tabular}{l}
{\small -220} \\ 
{\small -220} \\ 
{\small -538} \\ 
{\small -320}%
\end{tabular}
& 
\begin{tabular}{l}
$\eta ${\small (676)} \\ 
$\eta ${\small (766)} \\ 
$\eta \left( \text{808}\right) $ \\ 
$\eta ${\small (846)} \\ 
$\eta ${\small (}$\overline{\text{{\small 774}}}${\small )}%
\end{tabular}
& $\ \varpi $(783), 8.49 & {\small 1.15} \\ \hline
\ {\small d}$_{\text{S}}^{1}${\small (493)}$\overline{\text{{\small d}}_{S}^{%
\text{1}}\text{(773)}}$ & \ \ {\small -290} & $\ \ \eta ${\small (976)} & 
\begin{tabular}{l}
$\eta ^{\prime }${\small (958), 0.202} \\ 
{\small f}$_{0}${\small (980), 70.} \\ 
$\eta (\overline{\text{{\small 969}}}),\overline{\text{{\small 35}}}$%
\end{tabular}
& 0.93 \\ \hline
\begin{tabular}{l}
{\small d}$_{S}^{\text{1}}${\small (773)}$\overline{\text{{\small d}}_{S}^{%
\text{1}}\text{{\small (773)}}}$ \\ 
{\small d}$_{S}^{0}${\small (773)}$\overline{\text{{\small d}}_{S}^{0}\text{%
{\small (773)}}}$ \\ 
{\small d}$_{S}^{-\text{1}}${\small (773)}$\overline{\text{{\small d}}_{S}^{-%
\text{1}}\text{{\small (773)}}}$%
\end{tabular}
& \ {\small -505} & $\ \eta ${\small (1041)} & $\ \ \phi (1020),${\small %
4.26.} & 2.1 \\ \hline
\begin{tabular}{l}
{\small d}$_{\text{S}}^{1}${\small (493)}$\overline{\text{{\small d}}_{S}^{%
\text{-1}}\text{(773)}}$ \\ 
{\small d}$_{\text{S}}^{1}${\small (493)}$\overline{\text{{\small d}}_{S}^{%
\text{0}}\text{(773)}}$ \\ 
2{\small q}$_{N}${\small (313)}$\overline{q_{N}^{0}(1213)}$%
\end{tabular}
\  & 
\begin{tabular}{l}
{\small - 90} \\ 
{\small -190} \\ 
{\small -320}%
\end{tabular}
& 
\begin{tabular}{l}
$\eta $(1176) \\ 
$\eta (1077)$ \\ 
2$\eta $(1206) \\ 
$\eta (\overline{1166})$%
\end{tabular}
& \ {\small h}$_{1}${\small (1170), 360} & {\small 0.26} \\ \hline
\begin{tabular}{l}
\textbf{q}$_{N}$\textbf{(313)}$\overline{\mathbf{q}_{N}^{0}\mathbf{(1303)}}$
\\ 
{\small q}$_{\Sigma }${\small (1033)}$\overline{\text{{\small q}}_{\Sigma }%
\text{{\small (1033)}}}$%
\end{tabular}
& 
\begin{tabular}{l}
\textbf{-320} \\ 
{\small -758}%
\end{tabular}
& 
\begin{tabular}{l}
$\eta $\textbf{(1296 )} \\ 
$\eta ${\small (1308)}%
\end{tabular}
& 
\begin{tabular}{l}
{\small f}$_{2}${\small (1275), 185} \\ 
{\small f}$_{1}${\small (1282), 24} \\ 
$\eta $(1294),{\small 55} \\ 
$\eta (\overline{1284}),\overline{88}$%
\end{tabular}
& {\small 0.94} \\ \hline
\begin{tabular}{l}
{\small q}$_{N}${\small (313)}$\overline{q_{N}^{1}(1303)}$ \\ 
{\small q}$_{N}${\small (313)}$\overline{q_{N}^{1}(1303)}$%
\end{tabular}
& 
\begin{tabular}{l}
{\small -220} \\ 
{\small -220}%
\end{tabular}
& 
\begin{tabular}{l}
$\eta ${\small (1396 )} \\ 
$\eta ${\small (1396 )}%
\end{tabular}
& 
\begin{tabular}{l}
{\small f}$_{0}${\small (1350), 350} \\ 
$\eta ${\small (1410), 51} \\ 
$\varpi ${\small (1425), 215} \\ 
{\small f}$_{1}${\small (1426), 55} \\ 
$\eta (\overline{1403})${\small , }$\overline{\text{{\small 168}}}$%
\end{tabular}
& {\small 0.50} \\ \hline
\end{tabular}
\ \ \ \ \ \ $%
\end{tabular}
\ \ 

\begin{tabular}{l}
\ \ Table C5 (Continuation). The Light Unflavored Mesons I = 0 \\ 
$%
\begin{tabular}{|l|l|l|l|}
\hline
\ {\small q}$_{N}${\small (313)}$\overline{q_{N}(m)}$ & \ \ {\small E}$%
_{bind}$ & \ \ {\small Deduced} & \ \ {\small Experiment} \\ \hline
\begin{tabular}{l}
{\small q}$_{N}${\small (313)}$\overline{q_{N}^{1}(1393)}$ \\ 
{\small q}$_{N}${\small (313)}$\overline{q_{N}^{0}(1573)}$%
\end{tabular}
& 
\begin{tabular}{l}
{\small -220} \\ 
{\small -320}%
\end{tabular}
& 
\begin{tabular}{l}
$\eta ${\small (1486)} \\ 
$\eta ${\small (1566)} \\ 
$\eta ${\small (}$\overline{1526}${\small )}%
\end{tabular}
& 
\begin{tabular}{l}
$\eta $(1476), 87 \\ 
f$_{0}$(1507),{\small \ 109} \\ 
f$_{2}^{\prime }$(1525), {\small 73} \\ 
$\eta $($\overline{1503}$), $\overline{90}$%
\end{tabular}
\\ \hline
\begin{tabular}{l}
{\small d}$_{\text{S}}^{1}${\small (493)}$\overline{\text{{\small d}}_{S}^{0}%
\text{{\small (1303)}}}$ \\ 
{\small 2d}$_{\text{S}}^{1}${\small (493)}$\overline{\text{{\small d}}%
_{S}^{0}\text{({\small 1393)}}}$ \\ 
{\small d}$_{\text{S}}^{1}${\small (493)}$\overline{\text{{\small d}}_{S}^{1}%
\text{{\small (1413)}}}$ \\ 
{\small d}$_{\text{S}}^{1}${\small (493)}$\overline{\text{{\small d}}_{S}^{1}%
\text{{\small (1513)}}}$%
\end{tabular}
& 
\begin{tabular}{l}
{\small -190} \\ 
{\small -190} \\ 
{\small -290} \\ 
{\small -290}%
\end{tabular}
& 
\begin{tabular}{l}
$\eta $(1606) \\ 
2$\eta $(1696) \\ 
$\eta $(1616) \\ 
$\eta $(1716) \\ 
$\eta (\overline{1666})$%
\end{tabular}
& 
\begin{tabular}{l}
$\eta _{2}${\small (1617), 181} \\ 
$\varpi ${\small (1670), 315} \\ 
$\varpi _{3}${\small (1667), 168} \\ 
$\phi ${\small (1680), 150} \\ 
{\small f}$_{0}${\small (1714), 140} \\ 
$\eta (\overline{1670}${\small ),}$\overline{\text{{\small 191}}}$%
\end{tabular}
\\ \hline
\begin{tabular}{l}
4d$_{\text{S}}^{0}$(1203)$\overline{\text{{\small d}}_{S}^{0}\text{{\small %
(1203)}}}$ \\ 
d$_{\text{S}}^{1}$(493)$\overline{\text{{\small d}}_{S}^{\text{-1}}\text{%
{\small (1503)}}}$ \\ 
4d$_{\text{N}}^{0}$(1303)$\overline{\text{{\small d}}_{N}^{0}\text{{\small %
(1303)}}}$%
\end{tabular}
& 
\begin{tabular}{l}
-601 \\ 
-90 \\ 
-680%
\end{tabular}
& 
\begin{tabular}{l}
4$\eta (1806)$ \\ 
$\eta (1906)$ \\ 
4$\eta (1926)$ \\ 
$\eta (\overline{\text{{\small 1870}}})$%
\end{tabular}
& 
\begin{tabular}{l}
$\phi _{3}${\small (1854), 87} \\ 
{\small f}$_{2}${\small (1945), 163} \\ 
$\eta ${\small (}$\overline{\text{{\small 1899}}}${\small ), }$\overline{125}
$%
\end{tabular}
\\ \hline
\begin{tabular}{l}
{\small d}$_{\text{S}}^{1}${\small (493)}$\overline{\text{{\small d}}_{S}^{%
\text{-1}}\text{{\small (1603)}}}$ \\ 
\textbf{q}$_{N}$\textbf{(313)}$\overline{q_{N}^{0}(2023)}$%
\end{tabular}
& 
\begin{tabular}{l}
{\small -90} \\ 
{\small -320}%
\end{tabular}
& 
\begin{tabular}{l}
$\eta $(2006) \\ 
$\eta $\textbf{(2016)}%
\end{tabular}
& 
\begin{tabular}{l}
f$_{2}$(2011), 202 \\ 
f$_{4}$(2034), 222 \\ 
$\eta (\overline{\text{2023}})$, $\overline{212}$%
\end{tabular}
\\ \hline
\begin{tabular}{l}
{\small q}$_{N}${\small (1393)}$\overline{q_{N}^{0}(1393)}$ \\ 
\textbf{4q}$_{N}$\textbf{(313)}$\overline{q_{N}^{1}(2023)}$ \\ 
{\small d}$_{S}^{1}${\small (1413)}$\overline{d_{S}^{1}(1413)}$%
\end{tabular}
& 
\begin{tabular}{l}
{\small -707} \\ 
-{\small 220} \\ 
-663%
\end{tabular}
& 
\begin{tabular}{l}
$\eta $(2079) \\ 
\textbf{4}$\eta $\textbf{(2116)} \\ 
$\eta $(2163)%
\end{tabular}
& 
\begin{tabular}{l}
f$_{0}$(2103)$^{\#}$,206 \\ 
f$_{2}$(2156)$^{\#}$,167 \\ 
$\eta (\overline{\text{{\small 2130}}})^{\#}$, $\overline{187}$%
\end{tabular}
\\ \hline
\begin{tabular}{l}
q$_{N}$(313)$\overline{q_{N}(2293)}$ \\ 
\textbf{4d}$_{\text{S}}^{1}$\textbf{(493)}$\overline{\text{d}_{S}^{0}\text{%
(2013)}}$ \\ 
{\small d}$_{\text{S}}^{1}${\small (493)}$\overline{\text{{\small d}}_{S}^{0}%
\text{{\small (2023)}}}$%
\end{tabular}
& 
\begin{tabular}{l}
{\small -320} \\ 
\textbf{-190} \\ 
-190%
\end{tabular}
& 
\begin{tabular}{l}
$\eta $(2286) \\ 
\textbf{4}$\eta $\textbf{(2316)} \\ 
$\eta $(2326)%
\end{tabular}
& 
\begin{tabular}{l}
f$_{2}$(2297), 149 \\ 
f$_{2}$(2339), 319 \\ 
$\eta $($\overline{\text{2318}}$), $\overline{\text{234}}$%
\end{tabular}
\\ \hline
\begin{tabular}{l}
{\small q}$_{\text{N}}^{0}${\small (1573)}$\overline{q_{\text{N}}^{0}(1573)}$
\\ 
{\small d}$_{\text{S}}^{-1}${\small (1603)}$\overline{d_{\text{S}}^{-1}(1603)%
}$%
\end{tabular}
& 
\begin{tabular}{l}
-768 \\ 
--728%
\end{tabular}
& 
\begin{tabular}{l}
$\eta $(2378) \\ 
$\eta $(2478) \\ 
$\eta $($\overline{2428}$)%
\end{tabular}
& \ f$_{6}$(2465)$^{\#}$, 255 . \\ \hline
\end{tabular}
\ \ \ \ \ \ \ $%
\end{tabular}

\ \ \ \ \ \ \ \ \ \ \ \ \ \ \ \ \ \ \ \ \ \ \ \ \ \ \ \ \ \ \ \ \ \ \ \ \ \
\ \ \ \ \ \ \ \ \ \ \ \ \ \ \ \ \ \ \ \ \ \ \ \ \ \ \ \ \ \ \ \ \ \ \ \ \ \
\ \ \ \ \ \ \ \ \ \ \ \ \ \ \ \ \ \ \ \ \ \ \ \ \ \ \ \ \ \ \ \ \ \ \ \ \ \
\ \ \ \ \ \ \ \ \ \ \ \ \ 

\begin{tabular}{l}
$\text{\ Table C6.\ }$The Light Unflavored Mesons (S=C=b=0) with I = 1 \\ 
$%
\begin{tabular}{|l|l|l|l|}
\hline
{\small \ \ \ q}$_{i}${\small (m}$_{i}${\small )}$\overline{q_{j}(m_{j})}$ & 
\ {\small E}$_{bind}$ & {\small Phenomen.} & {\small Experiment, }$\ \Gamma $
\\ \hline
\ \ {\small q}$_{N}${\small (313)}$\overline{q_{N}(673)}$ & {\small \ \ -487}
& $\ \pi ${\small (139)} & {\small \ }$\ \pi ${\small (138)} \\ \hline
\begin{tabular}{l}
{\small d}$_{\text{S}}^{1}${\small (493)}$\overline{\text{{\small d}}%
_{\Sigma }^{0}\text{{\small (583)}}}$ \\ 
{\small d}$_{\text{S}}^{0}${\small (493)}$\overline{\text{{\small d}}%
_{\Sigma }^{0}\text{{\small (583)}}}$ \\ 
{\small d}$_{\text{S}}^{-1}${\small (493)}$\overline{\text{{\small d}}%
_{\Sigma }^{0}\text{{\small (583)}}}$%
\end{tabular}
& 
\begin{tabular}{l}
{\small -267} \\ 
{\small -367} \\ 
{\small -267} \\ 
$\overline{\text{-300}}$%
\end{tabular}
& 
\begin{tabular}{l}
$\pi ${\small (808)} \\ 
$\pi ${\small (708)} \\ 
$\pi ${\small (808)} \\ 
$\pi ${\small (}$\overline{775}${\small )}%
\end{tabular}
& {\small \ }$\ \pi (${\small 776), 150} \\ \hline
{\small \ q}$_{N}^{0}${\small (313)}$\overline{q_{\Delta }^{0}(673)}$ & 
{\small \ \ \ -19} & {\small \ }$\ \pi ${\small (}$\overline{967}${\small )}
& {\small \ \ a}$_{0}${\small (985), 75} \\ \hline
\begin{tabular}{l}
{\small d}$_{\text{S}}^{1}${\small (493)}$\overline{\text{{\small d}}%
_{\Sigma }^{0}\text{{\small (1033)}}}$ \\ 
{\small d}$_{\text{S}}^{0}${\small (493)}$\overline{\text{{\small d}}%
_{\Sigma }^{0}\text{{\small (1033)}}}$ \\ 
{\small d}$_{\text{S}}^{-1}${\small (493)}$\overline{\text{{\small d}}%
_{\Sigma }^{0}\text{{\small (1033)}}}$%
\end{tabular}
& 
\begin{tabular}{l}
{\small -268} \\ 
{\small -368} \\ 
{\small -268}%
\end{tabular}
& 
\begin{tabular}{l}
$\pi ${\small (1258)} \\ 
$\pi ${\small (1158)} \\ 
$\pi ${\small (1258)} \\ 
$\pi ${\small (}$\overline{1225}${\small )}%
\end{tabular}
& {\small \ } 
\begin{tabular}{l}
$b_{1}(${\small 1230), 142} \\ 
$a_{1}(${\small 1230), 425} \\ 
$\pi $($\overline{1230}$),$\overline{284}$%
\end{tabular}
\\ \hline
\begin{tabular}{l}
2q$_{N}^{0}$({\small 313})$\overline{q_{\Delta }^{0}(1033)} $%
\end{tabular}
& \ -19 & $\ \ \pi (\overline{1327})$ & 
\begin{tabular}{l}
$\pi $(1300), 400 \\ 
$a_{2}(${\small 1318), \ }107 \\ 
$\pi _{1}$({\small 1376}), 300 \\ 
$\pi $($\overline{1331}$), $\overline{269}$%
\end{tabular}
\\ \hline
\begin{tabular}{l}
\textbf{d}$_{\text{S}}^{1}$\textbf{(493)}$\overline{\text{{\small d}}%
_{\Sigma }^{0}\text{(1303)}}$ \\ 
{\small d}$_{\text{S}}^{0}${\small (493)}$\overline{\text{{\small d}}%
_{\Sigma }^{0}\text{(1303)}}$ \\ 
{\small d}$_{\text{S}}^{-1}${\small (493)}$\overline{\text{{\small d}}%
_{\Sigma }^{0}\text{(1303)}}$%
\end{tabular}
& 
\begin{tabular}{l}
-268 \\ 
-368 \\ 
-268%
\end{tabular}
& 
\begin{tabular}{l}
$\pi $(1528) \\ 
$\pi $(1428) \\ 
$\pi $(1528) \\ 
$\pi $($\overline{1495}$)%
\end{tabular}
& 
\begin{tabular}{l}
$\rho $(1465), $\Gamma $=400 \\ 
a$_{0}$(1474),$\Gamma $=265 \\ 
$\pi (\overline{1470}${\small ), }$\overline{333}$%
\end{tabular}
\\ \hline
\begin{tabular}{l}
q$_{N}^{0}$({\small 313)}$\overline{q_{\Delta }^{0}(1393)}$%
\end{tabular}
{\small \ } & \ -19 & $\ \ \pi (\overline{1687}${\small )} & 
\begin{tabular}{l}
$\pi _{1}$(1596), 312 \\ 
$\pi _{2}$(1672), 259 \\ 
$\rho _{3}$(1689), 161 \\ 
$\rho $(1720), 250 \\ 
$\pi (\overline{1669}),\overline{246}$%
\end{tabular}
\\ \hline
\begin{tabular}{l}
{\small q}$_{\text{N}}${\small (1213)} $\overline{\text{{\small q}}_{\text{N}%
}\text{{\small (1213)}}}$ \\ 
{\small d}$_{\text{S}}^{0}${\small (493)}$\overline{\text{{\small d}}%
_{\Sigma }^{0}\text{(1753)}}$%
\end{tabular}
& 
\begin{tabular}{l}
-654 \\ 
-368%
\end{tabular}
& 
\begin{tabular}{l}
$\eta $(1772) \\ 
$\pi $(1878) \\ 
$\pi $($\overline{\text{1825}}$)%
\end{tabular}
& 
\begin{tabular}{l}
$\pi $\textbf{(1812), 207} \\ 
$\rho $(1900)$^{\ast }$, 29%
\end{tabular}
\\ \hline
\begin{tabular}{l}
{\small d}$_{\text{S}}^{1}${\small (493)}$\overline{\text{{\small d}}%
_{\Sigma }^{0}\text{(1753)}}$ \\ 
{\small d}$_{\text{S}}^{-1}${\small (493)}$\overline{\text{{\small d}}%
_{\Sigma }^{0}\text{(1753)}}$ \\ 
q$_{N}^{0}$({\small 313)}$\overline{q_{\Delta }^{0}(1753)}$%
\end{tabular}
& 
\begin{tabular}{l}
-268 \\ 
-268 \\ 
-19%
\end{tabular}
& 
\begin{tabular}{l}
$\pi $(1978) \\ 
$\pi $(1978) \\ 
$\pi $(2047) \\ 
$\pi (\overline{2001}${\small )}%
\end{tabular}
& 
\begin{tabular}{l}
$\rho _{3}$(1990)$^{\ast }$,188 \\ 
\textbf{a}$_{4}$\textbf{(2010), 353}%
\end{tabular}
\\ \hline
\begin{tabular}{l}
3{\small d}$_{\text{S}}^{1}${\small (493)}$\overline{\text{{\small d}}%
_{\Sigma }^{0}\text{(2023)}}$%
\end{tabular}
& 
\begin{tabular}{l}
-268%
\end{tabular}
& 
\begin{tabular}{l}
$\pi (\overline{2248}${\small )}%
\end{tabular}
& 
\begin{tabular}{l}
$\rho _{3}(2250${\small )}$^{\ast }${\small , }$200$%
\end{tabular}
\\ \hline
4q$_{N}^{0}$({\small 313)}$\overline{q_{\Delta }^{0}(2113)}$ & \ -19 & $\ \
\pi $(2407) & \ a$_{6}$(2450), 400 \\ \hline
\end{tabular}
\ \ \ \ \ \ $%
\end{tabular}

\begin{tabular}{l}
$\ \ \text{Table C7.\ \ }$The Charmed Strange Mesons (S = \ $\pm $1, C = b =
0)\ \ \  \\ 
$%
\begin{tabular}{|l|l|l|l|l|}
\hline
{\small \ }$\ \ \text{u}_{\text{C}}^{1}${\small (m)}$\overline{\text{d}_{%
\text{S}}\text{{\small (493)}}}${\small \ } & {\small \ \ \ E}$_{bind}$ & 
{\small \ Deduced} & {\small Experiment} & $\frac{\Delta M}{M}\%$ \\ \hline
$\ \ \ \text{u}_{\text{C}}^{1}${\small (1753)}$\overline{\text{d}_{\text{S}%
}^{1}\text{{\small (493)}}}${\small \ } & \ {\small -303} & {\small \ \ D}$%
_{S}${\small (1943)} & {\small \ \ D}$_{S}^{\ast \pm }${\small (1968)} & 
{\small 1.3} \\ \hline
\begin{tabular}{l}
$\text{u}_{\text{C}}^{1}${\small (1753)}$\overline{\text{d}_{\text{S}}^{0}%
\text{{\small (493)}}}$ \\ 
$\text{u}_{\text{C}}^{1}${\small (1753)}$\overline{\text{d}_{\text{S}}^{-1}%
\text{{\small (493)}}}$%
\end{tabular}
& 
\begin{tabular}{l}
{\small -203} \\ 
{\small -103}%
\end{tabular}
& {\small \ } 
\begin{tabular}{l}
{\small D}$_{S}${\small (2043)} \\ 
{\small D}$_{S}${\small (2143)} \\ 
{\small D}$_{S}${\small (}$\overline{2093}${\small )}%
\end{tabular}
& {\small \ \ D}$_{S}${\small (}$2112${\small )} & {\small 0.90} \\ \hline
{\small \ \ u}$_{C}^{1}${\small (1753)}$\overline{\text{{\small d}}_{S}^{1}%
\text{{\small (773)}}}$ & {\small \ \ -215} & {\small \ \ D}$_{S}${\small %
(2311)} & {\small \ \ D}$_{S_{j}}${\small (2317)} & {\small 0.04} \\ \hline
\begin{tabular}{l}
{\small u}$_{C}^{1}${\small (1753)}$\overline{\text{{\small d}}_{S}^{0}\text{%
{\small (773)}}}$ \\ 
$\text{u}_{\text{C}}^{1}\text{(2133)}\overline{\text{d}_{\text{S}}^{1}\text{%
{\small (493)}}}$ \\ 
{\small u}$_{C}^{1}${\small (1753)}$\overline{\text{{\small d}}_{S}^{-1}%
\text{{\small (773)}}}$%
\end{tabular}
& 
\begin{tabular}{l}
{\small -115} \\ 
{\small -184} \\ 
{\small -15}%
\end{tabular}
& 
\begin{tabular}{l}
{\small D}$_{S}${\small (2411)} \\ 
{\small D}$_{S}${\small (2442)} \\ 
{\small D}$_{S}${\small (2511)} \\ 
{\small D}$_{S}${\small (}$\overline{\text{{\small 2455}}}${\small )}%
\end{tabular}
& {\small \ \ D}$_{S_{j}}${\small (2460)} & {\small 0.20} \\ \hline
\begin{tabular}{l}
$\text{u}_{\text{C}}^{1}\text{(2133)}\overline{\text{d}_{\text{S}}^{0}\text{%
{\small (493)}}}$ \\ 
{\small u}$_{C}^{1}${\small (2333)}$\overline{\text{d}_{\text{S}}^{1}\text{%
{\small (493)}}}$%
\end{tabular}
& 
\begin{tabular}{l}
{\small -84} \\ 
{\small -174}%
\end{tabular}
& 
\begin{tabular}{l}
{\small D}$_{S}${\small (2542)} \\ 
{\small D}$_{S}${\small (2652)} \\ 
{\small D}$_{S}${\small (}$\overline{\text{{\small 2597}}}${\small )}%
\end{tabular}
& 
\begin{tabular}{l}
{\small D}$_{S_{1}}${\small (2535)} \\ 
{\small D}$_{S_{j}}${\small (2573)} \\ 
{\small D}$_{S_{j}}${\small (}$\overline{2554}${\small )}%
\end{tabular}
& {\small 1.70} \\ \hline
\ {\small u}$_{C}^{1}${\small (2533)}$\overline{\text{d}_{\text{S}}^{1}\text{%
{\small (493)}}}$ & {\small \ \ - 164} & {\small \ \ D}$_{S}${\small (2862)}
& {\small \ \ Prediction} &  \\ \hline
\end{tabular}
\ \ \ \ \ $%
\end{tabular}

\ 

\end{document}